\documentclass[a4paper, openany, 12pt]{book}
\usepackage{PhD}
\title{Langevin equation and fractional dynamics}
\author{Jakub \'Sl\k{e}zak}
\bg{document}
\thispagestyle{empty}
\bg{center}
\vspace*{1cm}
\Huge \textbf{Langevin equation and fractional dynamics}
\vspace{2cm}

{Jakub \'Sl\k{e}zak}

\Large
 A thesis presented for the degree of\\
 Doctor of Philosophy\nn
 
 \raisebox{-0.5\height}{\includegraphics[height=5.5cm]{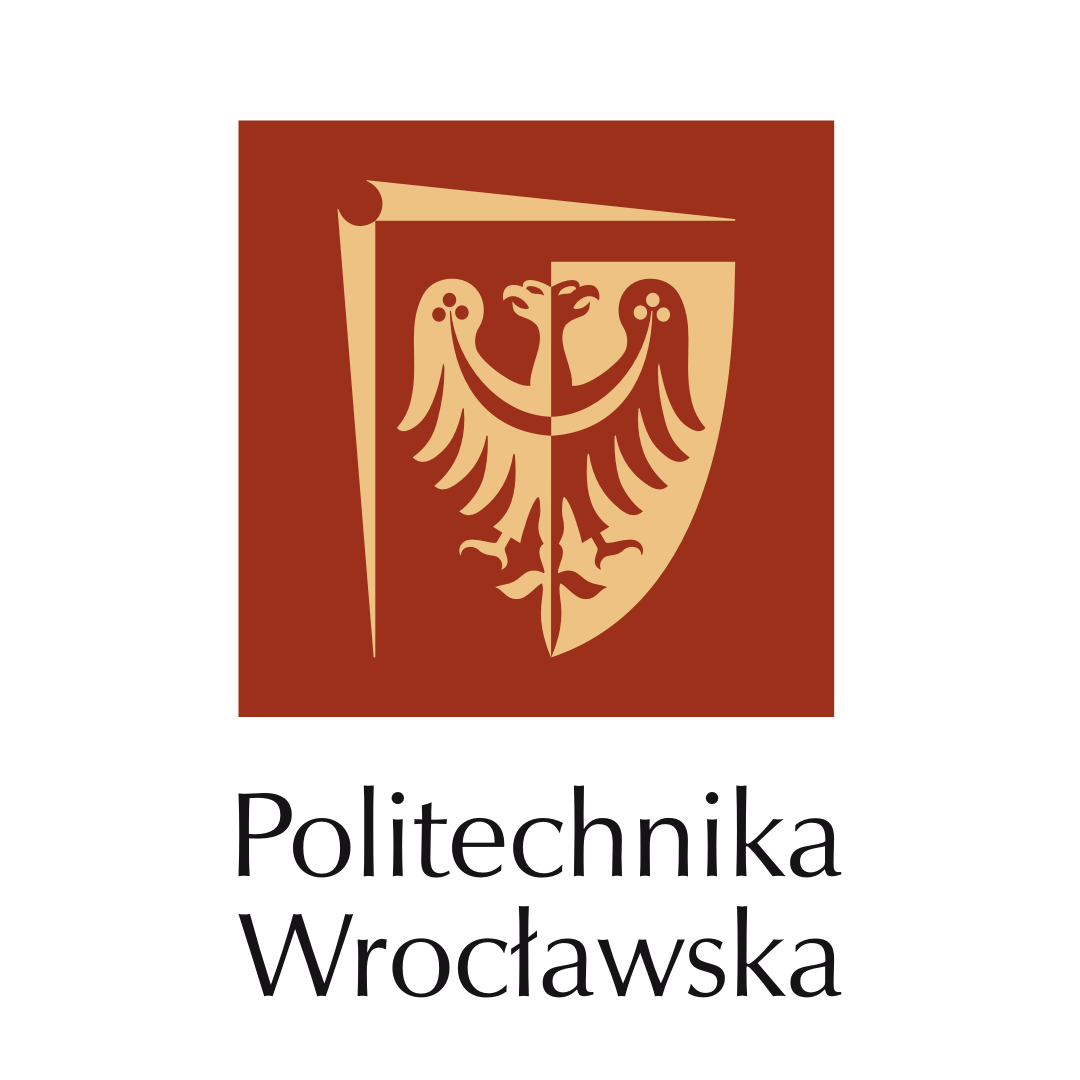}} \hspace{2cm} \raisebox{-0.5\height}{\includegraphics[height=5cm]{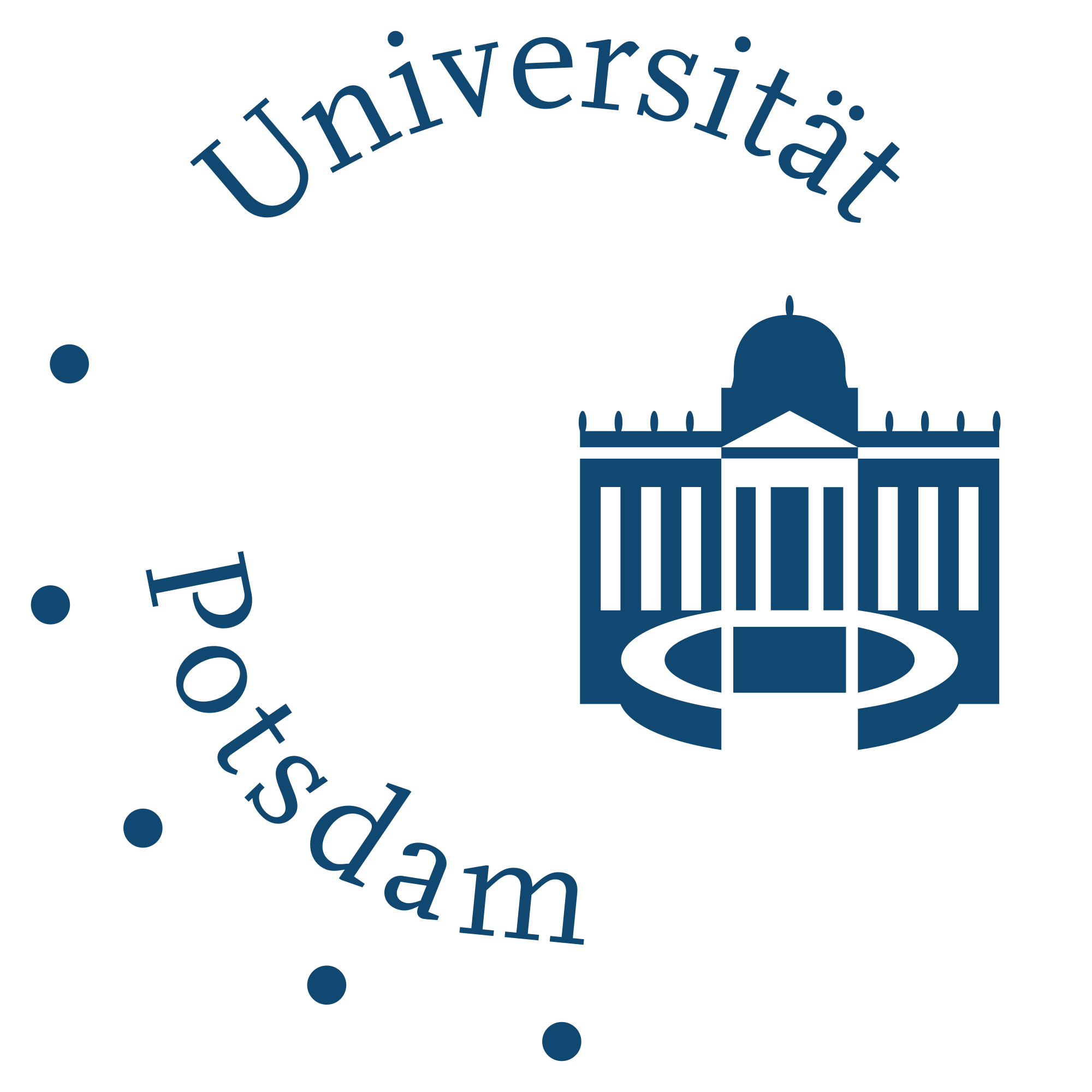}} 
 \vfill
 
\begin{tabular}[t]{ c l}
\hspace{5cm} &Supervisors: \\
 &prof. R. Metzler \\
 &prof. M. Magdziarz\\
\end{tabular}

 \begin{flushright}
 
 \end{flushright}
\vspace{0.8cm}
\end{center}

\addtocontents{toc}{\protect\enlargethispage{2\baselineskip}}
\tableofcontents
\chapter*{Notation}
\addcontentsline{toc}{chapter}{Notation}

\noindent
Physical symbols\\

\begin{tabular}{p{2cm} l}
	$X$ & position\\
	$V$ & velocity\\
	$P$ & momentum\\
	$Q$ & electrical charge\\
	$I$ & electrical current\\
	$\mathcal T$ & temperature\\
	$k_B$ & Boltzmann constant\\
	$\ham$ & Hamiltonian
\end{tabular}\nn	

\noindent
Spaces\\
\begin{tabular}{p{2cm} l}
	$\R^d$ & $d$-dimensional cartesian space\\
	$\ell^2$ & square-summable sequences \\
	$L_B^2(A)$ & square integrable functions $A\to B$,\\
	&  if not given $B$ is $\R$, $A$ suitable subset of $\R^d$\\
	$\mathcal C_0^\infty$ & smooth functions decaying at infinity\\
\end{tabular}\nn

\noindent
Operations\\

\begin{tabular}{p{2cm} l}
	$f*g$ & two-sided convolution of functions $f$ and $g$\\
	$a\cdot b$ & dot/scalar product of vectors $a,b\in\R^d$\\
	$\langle \bd\cdot,\bd\cdot \rangle$ & general scalar product, in the complex case first-argument linear\\
	$a^*$ & complex conjugate of $a\in\mathbb C$\\
	$\Vert\bd\cdot\Vert$ & norm \\
	%$\proj_{v_1,v_2,\ldots}$ & Hilbert space projection onto subspace generated by $v_1,v_2,\ldots$\\
	%$\mathcal{L}$ & Laplace transform as an operator \\
	$f^\#$ & Laplace transform of function $f$\\
	$\widehat f $ & Fourier transform of function $f$,\\
	& the variant without normalisation constant is used \\
	%$\check f $ & inverse Fourier transform of function $f$: $\check f=\mathcal F^{-1}[f]$\\
	$\widehat f^\#$ & Fourier-Laplace transform of function $f$,\\
	&  the Laplace transform is computed  with respect to the first argument\\
	$\overline f$ & time average of function $f$\\
	$\nabla$ & gradient\\
	$\nabla\cdot$ & divergence \\
	$\nabla^2$ & Laplace operator\\
	%$\mathcal I^\alpha$ & fractional integral of order $\alpha$\\
	$\mathcal S_\tau$ & time-shift operator, $\mathcal S_\tau f = t\mapsto f(\tau+t)$\\
\end{tabular}
\newpage 

\noindent
Relations\\

\begin{tabular}{p{2cm} l}
	$f\propto g$ & proportionality, $f=\const\cdot g$\\
	$f\sim g$ & asymptotic equality, $f/g\to 1$\\
	$f\lesssim g$ & asymptotic inequality, $f\sim \tilde f\le g$\\
	$f\asymp g$ & $f$ and $g$ have the same order, $\const_1\cdot g\lesssim f\lesssim \const_2\cdot g$\\
\end{tabular}\nn
\noindent

Probability\\

\begin{tabular}{p{2cm} l}
	$\mathcal N(\mu,\Sigma)$ & Gaussian distribution with mean  vector $\mu$ and covariance operator $\Sigma$\\
	$\mathcal G(\alpha,\beta)$ & gamma distribution with shape parameter $\alpha$ and rate $\beta$\\
	$\mathcal E(\beta) $ & exponential distribution, $\mathcal E(\beta)=\mathcal G(1,\beta)$\\
	$a^\text{est}$ & estimate of variable $a$\\
	$\pr $ & default probability measure \\
	$p_X$ & probability density (pdf) of variable $X$\\
	$p_X(\bd\cdot|C)$ & conditional probability density of variable $X$\\
	$\E $ & expected value, $\E[X]\defeq\int\dd P\ X$ \\
	$\E[\bd\cdot|C]$ & conditional expected value\\
	
	$\var $ & variance\\
	$\corr$ & correlation\\
	$r_X$ & covariance function of process $X$\\
	$r_X(\bd\cdot|C)$ & covariance function of process $X$ conditioned by $C$\\
	$\rho_X$ & partial autocorrelation (pacf) of process $X$\\
	$S_X$ & spectral process corresponding to $X$\\
	$\sigma_X$ & spectral measure of process $X$\\
	$s_X$ & power spectral density  (psd) of process $X$\\
	$\delta^2_X$ & mean square displacement (msd) of process $X$\\
	$\delta^2_X(\bd\cdot|C)$ & mean square displacement (msd) of process $X$ conditioned by $C$\\
	$\deq$ & equality in distribution\\
	$\xrightarrow{d}$ & limit in distribution\\
\end{tabular}

\chapter*{Synopsis}
\addcontentsline{toc}{chapter}{Synopsis}
Recent rapid advances in single particle tracking and supercomputing techniques resulted in an unprecedented abundance of diffusion data exhibiting complex behaviours, such the presence of power law tails of the msd and memory functions, commonly referred to as ``fractional dynamics''. Anomalous diffusion was extensively studied in numerous physical systems \citep{weissAnDiff,anTrans}. In particular ``subdiffusion'' was observed in the cytoplasm of living biological cells \citep{tabei}, various \textit{in vitro} crowded fluids 
\citep{szymanski,lene1} and lipid bilayer membrane systems
\citep{kneller}; and ``superdiffusion'' was reported in systems related to active biological transport \citep{goychukmotorpccp,reverey} or turbulence \citep{budaev,perriSun}. Interesting examples of normal diffusion are also still observed, in which the Brownian-like behaviour can be present together with non-Gaussian distributions \citep{wang} or weak ergodicity breaking \citep{weakErgBreak}.

Motivated by these developments, we study the stationary solutions of the classical and generalized Langevin equation as models of the contemporarily observed phenomena. 

In order to clarify the physical context of this research, in  the first chapter we sketch the historical background of the generalized Langevin equation, stressing the deep relations between the experimental evidence, physical models and mathematical theory. The next chapter is devoted to the brief overview of the theory of the Gaussian variables and processes. Its goal is to introduce sufficient information about the the theory of stationary Gaussian processes, which is extensively used in the further parts of the thesis.

In the third chapter we derive the generalized Langevin equation from Hamilton's equations of motion, using model as general as possible, which provides additional insight into the physical interpretation of this equation. Furthermore, this derivation shows constraints on the possible types of stochastic force which generates the motion and in natural way represents it in Fourier space. The study of the links between Langevin equation and Fourier representation is continued in the next chapters.

The fourth chapter concentrates more on the relations between the models and the data. There, the series of propositions and theorems shows that a large class of Langevin equations has a solutions that, sampled in discrete time, are the moving average autoregressive processes, which can be analysed using large number of available statistical methods \citep{slezakWeron}. We provide formulas for the coefficients of these discrete-time models and illustrate our results in the simple case using the optical tweezers recordings \citep{slezakDrobczynski, drobczynskiSlezak}. 
\newpage
The fifth chapter starts with a short introduction into the basic notions of ergodic theory, which is necessary to understand the behaviour of various time averages calculated from the data modelled by the solution of the Langevin equation. In order to provide a possibly general description, we prove a generalized version of the classical Maruyama's theorem, which describes the time averages of all stationary Gaussian processes with a finite number of spectral atoms \citep{gaussErg}.

In the last, sixth chapter we continue the study of the non-ergodic solutions of the Langevin equation, this time introducing the ergodicity breaking through the random parametrisation of the equation itself, the so-called ``superstatistics'' \citep{superstat}. This leads to a specific form of non-Gaussianity which affects only many-dimensional distributions of the process. The main result of this chapter is the series of propositions which describes the second order structure of solutions of the superstatistical Langevin equation.

\chapter{Historical background}

Diffusion is a fundamental example of a phenomenon linking microscopic and macroscopic properties of physical system. It was observed even in ancient times, it was in fact what at around 50 BCE Roman natural philosopher Lucretius depicted in the 2nd of his 5 books ``De Rerum Natura''  \citep[``On the Nature of Things'',][translation]{lucretius}:\\

\noindent\emph{An image, a type goes on before our eyes\\
	Present each moment; for behold whenever\\
	The sun's light and the rays, let in, pour down\\
	Across dark halls of houses: thou wilt see\\
	The many mites in many a manner mixed\\
	Amid a void in the very light of the rays,\\
	And battling on, as in eternal strife,\\
	(...)\\
	Namely, because such tumblings are a sign\\
	That motions also of the primal stuff\\
	Secret and viewless lurk beneath, behind.\\
	For thou wilt mark here many a speck, impelled\\
	By viewless blows, to change its little course,\\
	And beaten backwards to return again,\\
	Hither and thither in all directions round.\\
	Lo, all their shifting movement is of old,\\
	From the primeval atoms; (...)}\\

It is an impressive account of a scientific intuition, as Lucretius correctly links the erratic motion of dust in the air with microscopic motions of atoms. Unfortunately, for the next nearly two thousand years, no significant progress was made in explaining this phenomena. The modern interest in the area was sparked by Scottish botanist Robert Brown, who in 1827, during the study of fertilization, observed irregular motions of small pollen particles suspended in water. Their movement ``arose neither from currents in the fluid, nor from its gradual evaporation, but belonged to the particle itself''. Brown was initially unsure if the motions are of biological origin or not, but he managed to exclude the former option by repeating the experiment using dead pollen and small mineral particles. Subsequent observations, especially the ones performed by Jean Baptiste Perrin \citep{perrin1}, established the 3 main properties which characterise the Brownian motion:
\bg{itemize}
\item[i)] The mean square displacement (msd) of the particle's position $X$ grows linearly as a function of time
\bgeq
\delta^2_X(t)\defeq \E\big[X(t)^2\big] = 2D t
\eeq
determined by the diffusion constant $D$.
\item[ii)] The probability density (pdf) of the position $p_X$ is Gaussian, and in order to fulfil also the property i), in one-dimension it must have form
\bgeq\label{eq:diffPDF}
p_X(x,t)=\f{1}{\sqrt{4\pi Dt}}\exp\lt(-\f{x^2}{4Dt}\rt).
\eeq
\item[iii)] The motion is chaotic and the observed trajectories seem irregular. At any given moment the particle randomly changes the direction of motion with no regard of what happened up to that moment.
\end{itemize}
In the XIX century there were no mathematical tools that would suffice to describe a motion with such properties. The first one to propose a suitable random model was Louis Bachelier, who analysed the properties of what would be called ``Brownian motion'' in his thesis ``La th{\'e}orie de la sp{\'e}culation'' (``Theory of Speculation'') from 1900 \citep{courtault}. His work was innovative, but lacked a mathematical rigour and was concerned with stock market prices, which was not a recognised mathematical subject at that time. Because of these reasons his results were ignored by contemporary mathematicians and physicists. It was used decades later, when it was rediscovered by Kolmogorov and used as one of the keystones of modern financial mathematics.

Five years after Bachelier, Albert Einstein published a paper which became a foundation of probabilistic statistical mechanics \citep{einstein}. In the first part of his work he used the theory of Adolf Fick, already established in 1855 \citep[reprint]{fick}, to obtain a relation between viscosity of the fluid and diffusion coefficient $D$ by balancing the osmotic pressure with diffusion current. In modern view this approach can be simplified by writing Fick's 1st law for diffusion current $J_\text{diff}$ given concentration $n$
\bgeq
J_\text{diff}=- D\nabla n,
\eeq
and drift current $J_\text{drift}$ caused by the potential $\mathcal V$ acting on the particles characterised by mobility $\mu$
\bgeq
J_\text{drift} = -\mu n\nabla \mathcal V.
\eeq
Using Boltzmann distribution $n= \mathcal{Z}\exp(-\mathcal V/(k_BT))$ we can equate the currents and obtain Einsten's relation
\bgeq\label{eq:einstenRel}
0=J_\text{diff}+J_\text{drift} =-n\nabla\mathcal V\lt(-\f{D}{k_B\mathcal T}+\mu\rt)\implies D=k_B \mathcal T\mu.
\eeq
In the second part of his paper Einstein considers a simple random walk with bounded and independent increments. He then shows that the resulting density of particles solves the diffusion equation
\bgeq\label{eq:heatEq}
\pf{}{t}n = D\nabla^2 n,
\eeq
which was already known as Fick's 2nd law. This result agrees with Eq. \Ref{eq:diffPDF}, and has Gaussian \Ref{eq:diffPDF} acting as the corresponding Green's function. The Einsteins' technique replaced the integral equation for $n$ by a series of derivatives, a method which would later become studied in more formal manner under the name of Kramers-Moyal expansion \citep{risken}.  One of the most important fruits of Einstein's work was a way of determining the Avogadro number using only macroscopic measurement, and 3 years later Jean Baptiste Perrin performed the concluding experiment \citep{perrin1}. For this achievement and his related work about the atomic nature of matter Perrin received a Nobel Prize in 1926.

At the same time as Einstein, Marian Smoluchowski independently developed its own theory of Brownian motion \citep{smol}. His methods were different and more in line with scattering theory: he considered a particle which meets random encounters, each of them changing its velocity by some small and bounded angle, so that it continues its motion inside a cone centred around  pre-collisional velocity. Careful examination of the scales which are important from the experimental point of view allowed Smoluchowski to derive the value of the apparent mean free path and consequently the diffusion constant. In his later works he studied a diffusion under the external force, characterised by the position-dependent pressure $ f$, and obtained partial differential equation for the probability density of the particle
\bgeq
\pf{}{t}p_X= D\nabla^2 p_X-\f{D}{k_B T}\nabla\cdot(f p_X)
\eeq
which is contemporarily called Smoluchowski equation. He was the first to realise that equations of form \Ref{eq:heatEq} may correspond not only to the macroscopic density of particles, but microscopic pdf of each one as well. His work was a major breakthrough in the study of atomic nature of matter and reconciled the unceasing, persistent nature of Brownian motion with the irreversible laws of thermodynamics.

However, Smoluchowski's calculation of the diffusion coefficient differed from Einstein's by a suspicious factor 64/27. The experimental results available at that time were not decisive, so this dispute had to be resolved using theoretical methods. In 1908 Paul Langevin boldly claimed he managed to provide ``une d{\'e}monstration infiniment plus simple'' (``infinitely simpler demonstration'') and settle this dispute \citep[translation]{langevinEng}. Indeed, his approach was straightforward: he just wrote a Newton's equation
\bgeq\label{eq:langOryg}
m\df{}{t}V(t)=-\beta V(t)+F(t)
\eeq
of a system with Stoke's friction force $-\beta V(t)$ and complementary force $F$, which continuously agitates the motion of particle. He argued that assuming symmetry ($F\deq -F$) and high irregularity of $F$, it could be assumed to be independent from current value of particle's velocity, consequently $\E[V(t)F(t)]=0$. Multiplying \Ref{eq:langOryg} by $V(t)$, calculating averages and using the equipartition relation $\E[V(t)^2]=k_B\mathcal T/m$, it is then straightforward to calculate the diffusion coefficient, which agreed with the result of Einstein.

Langevin was the first to directly analyse the dynamical equation of the velocity variable by itself, but his approach was mathematically rough. It was not clear how one should understand the function $F$. The irregularity of Brownian paths made the formal mathematical approach difficult. The first step towards the solution was made by Norbert Wiener, who managed to rigorously describe the Brownian motion process and express it as a random Fourier series \citep{wiener23}. He also proved that the Brownian trajectories $(t,B(t))_t$ are continuous with probability 1, as expected for the model of the particle motion. Nevertheless, Perrin, using geometrical arguments, suggested that the derivative of Brownian motion cannot exist in the classical sense \citep{perrin1,perrin2}. Indeed, Wiener proved that the square-root dependence of the mean amplitude of increments, $\E|B(\tau+t)-B(\tau)|=2\sqrt{D t/\pi}$, implies that the derivative $\dd B/\dd t$ cannot be finite. This property fascinated Paul L{\' e}vy, who significantly expanded the theory of Brownian motion \citep{levy}, and other contemporary mathematicians.

Concurrently, the Langevin equation was thoroughly studied by George Uhlenbeck and Laurence Ornstein \citep{OUoryg} who discussed its relation to the Smoluchowski equation and showed that in this model the second moment of the position $\E|X(\tau+t)-X(\tau)|^2$ was proportional to $t^2$ in short time scales, so the velocity process which appears as the solution of the Langevin equation is well-defined. That was only a partial solution, as the second derivative, the acceleration, still would diverge. One of the widely considered solutions was to modify the Langevin equation in order to get rid of these singularities, but Joseph Doob proposed another approach. He took the idea of Wiener that even if the derivative is not defined, one could still consider the integral, that is write $B(t)=\int_0^t\dd B(\tau)$ understood as a limit of Riemann sums composed of random Gaussian increments \citep{DoobOryg}. He argued that the Langevin equation \Ref{eq:langOryg} should be interpreted as an integral stochastic equation
\bgeq
m\int_0^t\dd V(\tau)=-\beta\int_0^t\dd\tau\ V(\tau)+D\int_0^t\dd B(\tau).
\eeq
At this moment the foundations of stochastic analysis were nearly complete; shortly after Kiyoshi It{\=o} in two very elegant and concise papers defined a notion of stochastic integrals with respect to Brownian motion \citep{ito44} and the corresponding stochastic differential equations \citep{ito46}, providing also the conditions for existence and uniqueness of their solutions.

It does not mean there was no place for further significant progress; on the contrary, many crucial concepts were not yet developed. One of the consequences of the irregular nature of Brownian motion is that its one-dimensional trajectory $(t,B(t))_t$ it is a fractal with Hausdorff dimension 3/2, which was proven by Samuel James Taylor \citep{taylor}. A closely related property is that it is self-similar, that is time-rescaled Brownian motion is statistically equivalent to position-rescaled one, precisely $(B(ct))_t\deq (c^{1/2}B(t))_t$. Beno{\^\i}t Mandelbrot was one of the main creators and a great advocate of this geometric approach to stochastic modelling; he considered a general class of self-similar processes, for which 
\bgeq
(X(ct))_t\deq (c^{H}X(ct))_t.
\eeq
He found out that there are two features, which guarantee this property: the specific form of of the distribution or specific form of the memory, both requiring power-law type behaviour. The first feature led to stable processes, but their diverging second moment rendered them unphysical in many applications; however the second feature helped Mandelbrot to construct self-similar Gaussian process with a power-law tails of covariance: the fractional Brownian motion $B_H$. Subsequently, in 1965 he used increments of this process to explain the so-called ``Hurst phenomenon''  \cite[translation]{mandelbrot}, that is a power-law behaviour of the Rescaled Adjusted Range statistic. Given a window, it is a range (maximum minus minimum, $R$) divided by a scale (standard deviation, $S$), and is denoted $R/S$. Initially, $R/S$ statistic was used to estimate optimal height of the projected dams over Nile River \citep{hurst}, but the Hurst phenomenon was later observed in hundreds of other phenomena related to climate, geology and astronomy.

Mandelbrot's geometrical approach did not seem to be compatible with the philosophy of physics, which stresses the importance of dynamical equations, not geometry (with the notable exception in the form of general relativity). However, Mandelbrot himself proposed a possible bridge, studying properties of his fractional models in Fourier space. Spectral analysis was not considered a natural language in the study of diffusion processes, but it had a good and intuitive interpretation in the modelling of electric phenomena, where the analysed signals often have a form of sinusoids. In 1928, John Bertrand Johnson observed an electromotive force which was linearly dependent on temperature \citep{johnson}. The explanation was immediately provided by Harry Nyquist, who proved that the thermal fluctuations in electric resistors cause the voltage to fluctuate randomly with mean power equal to impedance times $k_B\mathcal T$ \citep{nyquist}. This could be considered an equivalent of Einstein's relation \Ref{eq:einstenRel} for the electric phenomena. These results were further developed by Lars Onsager, who linked the heat conductivity and the correlation function of the energy flow caused by the voltage difference \citep{onsager1,onsager2}. It was called ``Onsager's reciprocity'' and formed a basis of the non-equilibrium thermodynamics of linear systems, the so-called linear response theory. In 1950's rising interest in the area led to discovery of general fluctuation-dissipation relations, the most famous one proven by Ry{\= o}go Kubo \citep{kuboOld}. Eight years later Hajime Mori used projection operator techniques to show that under relatively weak assumptions the microscopic equations of motion can be cast into the macroscopic formula \citep{mori}
\bgeq\label{eq:GLEhist}
m\df{}{t}V = -m\int_{-\infty}^t\!\!\!\dd\tau\ K(t-\tau)V(\tau) + F(t),
\eeq
which is a form of the generalized Langevin equation (GLE). The fluctuation-dissipation relation then assures that the retardation kernel $K$ is, up to a factor, the covariance function of the random force $F$. The GLE was a tool required for the unification of Mandelbrot's fractal models and diffusion processes. Now the stochastic force $F$ could reflect the fractional properties of the environment, and the observed coordinate would be governed by the dynamical equation \Ref{eq:GLEhist}. This relation was further strengthened when Peter Mazur, Mark Kac and George Ford proposed a model of discrete harmonic bath, which led to the GLE \citep{kacGLE}; the same system was also later studied by Robert Zwanzig \citep{zwanzigGLE} and is now known as Kac-Zwanzig model. In this system it was possible to precisely track the relationship between the harmonic bath and the spectrum of the stochastic force $F$, in particular to indicate what type of environment would lead to power-law spectra which appeared in Mandelbrot's works.

These achievements were fully taken use of in recent years, when the unprecedented advancements in microscopic tracking shed a new light onto the rich word of diffusion in complex media.  Diffusion which has the power-law msd, 
\bgeq
\delta_X^2(t)\sim 2D t^\alpha,
\eeq
called anomalous diffusion, has actually been known since 1926, when Lewis Richardson  detected it in the turbulent atmospheric flows \citep{richardson}. Starting in 1960s this phenomenon has been studied in the transport theory, but it took a long time to fully incorporate ideas of Kubo and Mandelbrot. The interest in modelling anomalous diffusion by the GLE became noticeable in 1990s \citep{muralidhar}, when the significant works of Ke-Gang Wang and his collaborators were published \citep{KGWang92,porra96,KGWang99}. This area of research became increasingly attractive, because in the GLE the power-law memory is naturally related through fluctuation-dissipation relation to a convolution with a power-law kernel, and this operation is a basis of the fractional integrals and fractional derivatives, a form of generalisation of classical analysis for fractional models \citep{samko, hilfer}. 

In 2000s significant experimental progress was made \citep{caspi,seisenberger,banks}. One of the most famous new examples of anomalous diffusion was the motion of mRNA molecules in \textit{E. coli} cells, measured by Ido Golding and Edward Cox, where the subdiffusion $\delta_X^2(t)\sim D t^\alpha, \alpha\approx 0.7$ was found \citep{goldingCox}. Modelling of this and similar observations was performed using a variety of models such as continuous time random walks, L{\' e}vy flights, L{\' e}vy walks, diffusion on fractals, and, of course, the GLE \citep{rwGuide,sokolovAnDiff}.

In recent years the study of the GLE with power-law kernel was continued e.g. in influential works of Eric Lutz \citep{lutz} and Samuel Kou \citep{kou}. The former used Laplace methods to provide exact formulas for covariance and msd; the latter analysed thoroughly spectrum of the process and showed convincing experimental confirmation analysing conformational dynamics of proteins.

\chapter{Gaussian variables and processes}\label{ch:gaussVar}
In this chapter we introduce the notions of various types of Gaussian variables, vectors and the corresponding spaces \citep{gaussHS,gaussHilb}. These are objects whose distribution after some linear manipulations can be reduced to the classical Gaussian function $\e^{-x^2/2}$.  Time-indexed collections of such objects, called Gaussian processes \citep{doob,dym}, can describe the dynamics. They can be fully characterised by their mean and correlation structure \citep{yaglom}, which greatly simplifies both modelling and statistical analysis. Not only is  the Gaussian distribution the most observed one in diffusion phenomena, there are also strong theoretical explanations of its prevalence, especially in physics \citep{fox}. The most significant ones are three observations:
\bg{itemize}
\item[i)] It is a distribution which maximises entropy under the constraint of fixed mean and variance. In physical systems the constant value of variance and mean is often forced by the conservation of energy, at least in the linear approximation. At the same time the global increase of entropy is guaranteed by the Second Law of Thermodynamics. \citep{mandelbrot}
\item[ii)] It is a Gibbs distribution for quadratic Hamiltonians, that is the invariant measure of the corresponding dynamical systems which should be observed in  thermal equilibrium.
\item[iii)] It is a limit of a rescaled sum of many sufficiently weakly dependent (inf fact, only finite-range dependence suffices) microscopic variables, as given by Central Limit Theorem. Therefore the macroscopic variables, after proper rescaling, are very likely to have this distribution.
\end{itemize}
All three of these observations are deeply related: the increase of entropy can be used to explain the form of Gibbs measure and entropy is a Lyapunov function of the transformation $(X+Y)/\sqrt{2}$ for i.i.d. $X,Y$ which can be used to determine the fixed point of the rescaled sum in the Central Limit Theorem.

\section{Gaussian vectors}
Let $H$ be a Hilbert space with scalar product $\langle\bd\cdot,\bd\cdot \rangle$. We will assume that $H$ is separable. For non-separable spaces, the theorem of H. Sato \citep{satoGauss} guarantees that any Gaussian measure would be concentrated on separable subspace. Therefore, this assumption is technical and for convenience only.

\bg{dfn}[Gaussian vector, Gaussian measure] We call a vector $X\in H$ Gaussian, if any projection $\langle X, y\rangle$ is a random variable with Gaussian distribution $\mathcal N(m_y,\sigma^2_y)$. The corresponding distribution on $H$ is called a Gaussian measure.
\end{dfn}
As a side note, we allow for zero variance Gaussian variables, i.e. treat deterministic constants as a subclass of Gaussian variables. This assumption simplifies many results.

It would be very convenient if Gaussian measures on Hilbert spaces could be characterised in a manner similar to finite-dimensional Gaussian distributions. The simplest approach makes use of the characteristic function. It is always defined as a function on the continuous dual of the state space, which in the case of Hilbert space is isomorphic to the space itself. Therefore it is a functional $\phi\colon H\to \mathbb C$  given by
\bgeq
\phi_X(\theta)\defeq\E\lt[\e^{\I\langle X,\theta\rangle}\rt]=\exp\lt(\I m_\theta +\f{1}{2}\sigma^2_\theta \rt).
\eeq
In the finite-dimensional case it is sufficient for a function to be positive-definite to be a characteristic function of some random variable, as stated in the famous Bochner's theorem. For infinite-dimensional spaces the necessary condition is more complicated and is stated in the Milnos-Sazanov theorem.
\bg{thm}[Milnos-Sazanov] A functional on Hilbert space $\phi\colon H\mapsto \mathbb C$ is a characteristic function of a probabilistic measure on $H$ if, and only if
\bg{itemize}
\item[-] it is positive definite,
\item[-] $\phi(0)=1$,
\item[-] there exists a positive trace-class operator $Q$ which generates the seminorm,
\bgeq
\Vert x\Vert_Q\defeq\sqrt{\langle Q x, x\rangle}
\eeq
and $\phi$ is a continuous function with respect to this seminorm.
\end{itemize}
\end{thm}
For positive symmetric operator $Q$ the trace is
\bgeq
\tr\, Q = \sum_j \langle Q e_j,e_j\rangle,
\eeq
where $(e_j)_j$ is any Hilbert base; if the operator $Q$ is trace class this quantity is finite, base-independent and equal to the sum of the operator's eigenvalues.

\bg{thm}[Characterisation of Gaussian measures]
Any Gaussian measure on a Hilbert space is uniquely determined by the mean vector $\mu\in H$ and covariance operator $\Sigma\colon H\mapsto H$ which is positive, symmetric and trace-class. Conversely, for any such $\mu$ and $\Sigma$ there exists a corresponding Gaussian measure witch the characteristic function
\bgeq
\phi(\theta)=\exp\lt(\I \langle \mu, \theta\rangle -\f{1}{2}\langle \Sigma \theta,\theta\rangle\rt)
\eeq
Moreover,
\bgeq
\E[X] = \mu, \quad \E\Vert X-\mu\Vert^2 = \tr \Sigma.
\eeq
\end{thm}

\bg{proof}For given vector $\mu$ and positive, symmetric, trace class operator $Q$ the function
\bgeq
\phi(\theta)=\exp\lt(\I \langle \mu, \theta\rangle -\f{1}{2}\langle \Sigma \theta,\theta\rangle\rt)
\eeq
fulfils the conditions of the Milnos-Sazanov theorem with operator $Q=\Sigma$. Additionally, any scalar product $\langle X, y\rangle$ has a characteristic function of Gaussian variable, so $X$ is a Gaussian vector. Let us choose a Hilbert base $(e_j)_j$. We know that $\langle X,e_j\rangle\deq\mathcal N(\langle \mu,e_j\rangle, \langle \Sigma e_j,e_j\rangle)$. So, the mean of $X$ is
\bgeq
\E[X] = \E \sum_j\langle X,e_j\rangle e_j =  \sum_j\E \langle X,e_j\rangle e_j = \sum_j \langle \mu, e_j\rangle e_j= \mu.
\eeq
The mean of the squared norm of $X-\mu$ is
\bgeq
\E\Vert X-\mu\Vert^2 = \E\sum_j|\langle  X-\mu,e_j\rangle|^2=\sum_j\E|\langle X-\mu,e_j\rangle|^2 = \sum_j\langle\Sigma e_j,e_j\rangle = \tr \Sigma
\eeq
The vector $X$ can be also constructed explicitly. Let $(\lambda_j)_j$ be a sequence of eigenvalues of $\Sigma$, take some Hilbert base $(e_j)_j$ and a sequence of independent variables $\xi_j\deq\mathcal N(0,\lambda_j)$. The vector
\bgeq
X=\mu+\sum_j \xi_j e_j
\eeq
has all the required properties.

Now we prove that any Gaussian vector $X$ can be characterised in that way. Let us fix one such $X$. The operator $Q$ from the Milnos-Sazanov theorem is trace-class, therefore it is also compact and continuous. The real and imaginary parts of the function $\phi$ are continuous, which forces the functions $m_\theta$ and $\sigma^2_\theta$ to be continuous, as well.

The variable $\langle \alpha X,y\rangle, \alpha\in \mathbb C$ has distribution $\alpha \mathcal N(m_y,\sigma^2_y)=\mathcal N(\alpha m_y, |\alpha|^2\sigma^2_y)$. Therefore $m_y$ is a linear continuous function of $y$ and $\sigma^2_y$ is a continuous quadratic form of $y$. The Riesz representation theorem states that in such a case there exist a vector $\mu$ and a continuous, symmetric positive linear operator $\Sigma$ such that
\bgeq
m_y = \langle \mu, y\rangle,\quad \sigma^2_y = \langle \Sigma y, y\rangle. 
\eeq
Further on we can consider only zero mean vectors because the mean can be simply subtracted, i.e. we can analyse $X-\mu$. We want to prove that $\Sigma$ for such $X$ is trace class. From the Milnos-Sazanov theorem, by rescaling we can choose $Q$ such that
\bgeq
\langle Q \theta, \theta \rangle \le 1 \implies |1-\phi(\theta)| = \lt|1 -\exp\lt(-\f{1}{2}\langle \Sigma \theta, \theta\rangle\rt)\rt|\le \epsilon
\eeq
for  arbitrarily small $\epsilon > 0$. Now, for any $\theta\in H$ outside of kernel of $Q$ (kernels of $Q$ and $\Sigma$ must agree) consider 
\bgeq
\theta'\defeq\f{\theta}{\sqrt{\langle Q \theta, \theta\rangle}}, \quad \langle Q \theta', \theta'\rangle = 1.
\eeq
We have the inequality
\bgeq
|1-\phi(\theta')| =  \lt|1 -\exp\lt(-\f{1}{2}\f{\langle \Sigma \theta, \theta\rangle}{\langle Q \theta, \theta\rangle}\rt)\rt|\le \epsilon
\eeq
and for $\epsilon$ sufficiently small
\bgeq
\langle \Sigma \theta, \theta\rangle \le 2|\ln(1-\epsilon)|\langle Q \theta, \theta\rangle.
\eeq
In particular for any base $\langle \Sigma e_j, e_j\rangle \le 2|\ln(1-\epsilon)|\langle Q e_j, e_j\rangle$ and
\bgeq
\tr \Sigma \le 2|\ln(1-\epsilon)|\tr Q
\eeq
\end{proof}
We also want to characterise the space of Gaussian vectors itself. First, we present a simple result about the linear transformation of a Gaussian vector.
\bg{prp}[Transformations of Gaussian vectors]\leavevmode

\bg{itemize}
\item[i)] For two vectors $X\deq \mathcal N(\mu_X,\Sigma_X)$ and independent $Y\deq \mathcal N(\mu_Y,\Sigma_Y)$
\bgeq
X+Y\deq\mathcal N (\mu_X+\mu_Y,\Sigma_X+\Sigma_Y).
\eeq
\item[ii)] For $X\deq \mathcal N(\mu_X,\Sigma_X)$, vector $b\in H$ and bounded operator $A$
\bgeq
AX+b \deq\mathcal N(A\mu+b,A\Sigma A^\dagger).
\eeq
\end{itemize}
\end{prp}
\bg{proof} The variables $\langle X,\theta\rangle$ and $\langle Y,\theta\rangle$ are also independent, therefore the characteristic function of $X+Y$ is
\begin{align}
\phi_{X+Y}(\theta) &= \exp\lt(\langle \mu_Y,\theta\rangle+\langle \mu_X,\theta\rangle-\f{1}{2}\langle \Sigma_X\theta,\theta\rangle-\f{1}{2}\langle \Sigma_Y\theta,\theta\rangle\rt)\nonumber\\
&=\exp\lt(\langle \mu_X+\mu_Y,\theta\rangle-\f{1}{2}\langle (\Sigma_X+\Sigma_Y)\theta,\theta\rangle\rt).
\end{align}
The characteristic function of the variable $AX+b$ is
\begin{align}
\phi_{AX+b}&=\E\lt[\e^{\I\langle AX+b,\theta\rangle}\rt]=\e^{\I\langle b,\theta\rangle}\E\lt[\e^{\I\langle X,A^\dagger\theta\rangle}\rt]=\exp\lt(\langle \mu, A^\dagger\theta\rangle+\langle b, \theta\rangle-\f{1}{2}\langle \Sigma A^\dagger\theta,A^\dagger\theta\rangle\rt)\nonumber\\
&=\exp\lt(\langle A\mu+b,\theta\rangle-\f{1}{2}\langle A\Sigma A^\dagger\theta,\theta\rangle\rt).
\end{align}
Because $A$ is bounded, the dual operator $A^\dagger$ is also bounded, $\Vert A^\dagger\Vert =\Vert A\Vert$ and $\tr (A \Sigma A^\dagger)=\tr (A^2\Sigma) \le \Vert A\Vert^2\tr \Sigma$. So $A\Sigma A^\dagger$ trace-class operator, it is also clearly positive.
\end{proof}

The space of Gaussian vectors is a vector space by itself and a subspace of $L^2_H(\Omega,\mathcal F,\pr)$ - the Hilbert space of all $H$-valued random variables with the scalar product $\langle \bd\cdot,\bd\cdot\rangle_{L^2_H}=\E\langle \bd\cdot,\bd\cdot\rangle_H$. This subspace is a Hilbert space of its own.
\bg{thm}The vector space of all $H$-valued Gaussian vectors is a Hilbert space with respect to scalar product $\E\langle \bd\cdot,\bd\cdot\rangle_H$.
\end{thm}
\bg{proof} We fix a Cauchy sequence of Gaussian variables $(X_n)_n$. It is also a Cauchy sequence in $L^2_H$ and this space is complete, so $X_n\to X\in L^2_H$. Because $\E\Vert X_n-X\Vert^2\to 0$ also, for every $y\in H$,
\bgeq
\E \langle X_n,y\rangle\to \E \langle X,y\rangle ,\quad \var \langle X_n,y\rangle \to \var \langle X,y\rangle.
\eeq
Therefore $\langle X_n,y\rangle $ is a sequence of Gaussian variables, whose means and variances converge. It means that the limit $\langle X,y\rangle$ is also Gaussian for every $y$ and $X$ is a Gaussian vector.
\end{proof}

We end this section with a simple proposition, which links the geometry of the Gaussian Hilbert space to the conditional expectancy.
\bg{prp}\label{prp:gaussProj}
In a Hilbert space of zero-mean Gaussian variables, any projection onto a given subspace is the same as the conditional expectancy under $\sigma$-algebra generated by this subspace.
\end{prp}
\bg{proof}
The reasoning is simple if we start from considering the projection. Given vector $X$ and subspace $V$ we can uniquely decompose $X$ into orthogonal and parallel components
\bgeq
X = X_\bot +X_\parallel,\quad X_\parallel \in V, \quad X_\bot \bot V.
\eeq
Because $X_\parallel\in V$, this variable is $\sigma(V)$-measureable. On the other hand, the variable $X_\bot$ is uncorrelated with any linear combination of variables from $V$. Because all these variables are Gaussian, it implies $X_\bot$ is independent from $\sigma$-algebra $\sigma(V)$. Therefore the conditional expectancy is
\bgeq
\E[X|V] = \E[X_\bot|V]+\E[X_\parallel|V] = \E[X_\bot]+X_\parallel = X_\parallel.
\eeq
\end{proof}

\section{Gaussian processes}\label{s:gaussProc}
Let $(I,+)$ be an ordered semi-group of indices and $X=(X(t))_{t\in I}$ a collection of random variables. We call this collection a stochastic process. If set $I$ is continuous (for our purposes later on some interval on $\R$, $\R_+$ or $\R$ itself) we use the notation $X(t)$ for a random variable corresponding to the index $t$. Whenever we want to stress that the indices are discrete, we will rather use the notation $X_k$.

\bg{dfn}\label{dfn:gaussProc}
We call a stochastic process $(X(t))_{t\in I}$ Gaussian if any vector
\bgeq
[X(t_1),X(t_2),\ldots, X(t_N)]
\eeq
for any choice of $t_k\in I$ and $N\in \mathbb N$ has Gaussian distribution. In other words the so-called finite-dimensional distributions of the process $X$ must be Gaussian.
\end{dfn}
Directly from the definition it is clear that any linear transformation of a Gaussian process is still a Gaussian process. It is also true for many Gaussian processes as long as they are only linearly dependent.

Gaussian processes have a very useful characterisation.
\bg{prp}[Mean and covariance functions] The distribution of a Gaussian process $X$ is fully determined by the mean function $m_X$
\bgeq
m_X(t)\defeq \E[X(t)]
\eeq
and the covariance function $r_X$
\bgeq
r_X(s,t)\defeq\E\big[(X(s)-m_X(s))(X(t)-m_X(t))\big]
\eeq
Conversely, for any positive semi-definite function $(s,t)\mapsto r(s,t)$ and mean function $t\mapsto m(t)$ there exists a Gaussian process $X$ with mean function $m$ and covariance function $r$.
\end{prp}
\bg{proof} It is only required to determine the distribution of any vector
$$[X(t_1),X(t_2),\ldots,X(t_N)].$$
As it is a Gaussian vector, it is fully determined by the mean vector and covariance matrix. The mean vector is $[m_X(t_j)]_{j=1}^N$ and the covariance matrix is $[r_X(t_i,t_j)]_{i,j=1}^N$.

Conversely, for a given positive semi-definite function $r$ for any $t_1,t_2,\ldots, N$ and any $N$ the matrix $[r_X(t_i,t_j)]_{i,j=1}^N$ is a covariance matrix so there exists a corresponding Gaussian measure. The obtained measures fulfil the consistency conditions of the Kolmogorov extension theorem  and there exists a stochastic process $(X(t))_t$ with the exact finite-dimensional distributions. To obtain process with the right covariance and mean functions it is sufficient to consider $X(t)+m(t)$.
\end{proof}

\noindent Later on we will very intensively use the notion of a stationary stochastic process
\bg{dfn}[Stationary process]\label{dfn:statProc}
A process $X$ is stationary if it has the same finite-dimensional distributions as the shifted process $\mathcal S_\tau X$ for any $\tau\in I$.
\bgeq
\mathcal S_\tau X\defeq t\mapsto X(t+\tau)\deq X.
\eeq
\end{dfn}

\bg{prp}[Stationary Gaussian process] A Gaussian process is stationary if, and only if
\bgeq\label{eq:rmCond}
m_X(t)=\const,\quad r_X(s,t)=r_X(s+\tau,t+\tau).
\eeq
\end{prp}
For index sets $I=\R$ or $I=\mathbb Z$ the covariance of a stationary process is a function of the difference of the arguments only and for simplicity we will use the notation $r_X(s,t) = r_X(t-s)$.
\bg{proof}
If the conditions \Ref{eq:rmCond} hold, the vectors
\bgeq
[X(t_1),X(t_2),\ldots, X(t_N)]\text{  and  } [X(t_1+\tau),X(t_2+\tau),\ldots, X(t_N+\tau)]
\eeq
have the same mean vector and covariance matrix. Therefore they have the same Gaussian distribution and the process $X$ is stationary.

On the other hand, for a stationary Gaussian process $X$ the mean function is
\bgeq
m_X(t_1)=\E[X(t_1)]=\E[X(t_2)]=m_X(t_2)
\eeq
and the covariance function is
\begin{align}
r_X(s,t)&=\E\big[X(s)-m_X(s))(X(t)-m_X(t))\big] \nonumber\\
&= \E\big[X(s+\tau)-m_X(s+\tau))(X(t+\tau)-m_X(t+\tau))\big]\nonumber\\
&= r_X(s+\tau,t+\tau)
\end{align}
\end{proof}

The stationary zero mean Gaussian processes have another very elegant description, which uses the Fourier space representation. Let us start with recalling the classical result.
\bg{thm}[Bochner's theorem] For any positive definite function $r\colon \R\to \R$ there exists a measure $\sigma$ on $\R$ such that
\bgeq
r(t)=\int_\R\sigma(\dd\omega)\ \e^{\I\omega t}.
\eeq
\end{thm}
This result suggests that the process $X$ could also be expressed in Fourier space. However, for such a representation to hold, some regularity conditions are required.

\bg{dfn}[Continuity in the mean-sense] We call a process $(X(t))_t$ mean-square continuous if
\bgeq
\lim_{t\to s} \E |X(t)-X(s)|^2=0
\eeq
for every $t,s$. 
\end{dfn}
For a zero mean process $\E[X(t)]=0$ this is equivalent to the condition that the covariance function $r_X(t,s)$ is continuous. Indeed,
\bgeq
\E |X(t)-X(s)|^2 = r_X(t,t)+r_X(s,s)-2r_X(t,s)
\eeq
and 
\begin{align}
&|r_X(t',s')-r_X(t,s)|=|\E[X(t')X(s')]-\E[X(t)X(s)]|\nonumber\\
&=|\E[(X(t')-X(t))X(s')]-\E[X(t)(X(s)-X(s'))]|\nonumber\\
&\le \big(\E|X(s')|^2\E|X(t')-X(t)|^2\big)^{1/2}+\big(\E|X(t)|^2\E|X(s')-X(s)|^2\big)^{1/2}
\end{align}

Now we give the aforementioned characterisation of the Gaussian stationary processes.

\bg{thm}[Harmonizable representation]\label{thm:harmRep}
Every stationary zero mean Gaussian process  can be expressed, up to a distribution, as an integral, understood in the mean square sense, which is
\bg{itemize}
\item[i)] In the continuous-time case $(X(t))_t$
\bgeq\label{eq:harmRep}
X(t) =\mathrm{re} \int_\R \dd S(\omega)\ \e^{\I\omega t},
\eeq
\item[ii)] In the discrete-time case $(X_k)_k$
\bgeq\label{eq:harmRepD}
X_k =\mathrm{re} \int_{-\pi}^\pi \dd S(\omega)\ \e^{\I\omega k}.
\eeq
\end{itemize}

In the above the process $S$, called the spectral process, is a Gaussian complex, mean square left continuous process which has independent real and imaginary parts and independent increments with variance
\bgeq
\E|\dd S(\omega)|^2 = \sigma_X(\dd\omega),
\eeq
where $\sigma_X$ is called the spectral measure. Also, for any measure $\sigma_X(\R)<\infty$ the integral \Ref{eq:harmRep} defines a stationary Gaussian process.
\end{thm}
We note that instead of taking the real part in $\Ref{eq:harmRep}$ one could consider the spectral process for which $S(-\omega)=S(\omega)^*$. Without taking the real part, the resulting process $X$ has i.i.d. real and imaginary parts. For the proof we will consider the latter case.

\bg{proof} For the continuous-time case i) the integral in $\Ref{eq:harmRep}$ is well-defined, because
\bgeq
\E\lt|\int_\R \dd S(\omega)\ \e^{\I\omega t}\rt|^2\le \E\lt[\int_\R |\dd S(\omega)|^2\rt]=\sigma_X(\R)<\infty,
\eeq
also
\begin{align}
\E[X(s)X(t)^*] &= \E\lt[\int_\R\dd S(\omega)\int_\R\dd S(\omega')^*\ \e^{\I(\omega s-\omega' t)}\rt]\nonumber\\
&=\E\lt[\int_\R |\dd S(\omega)|^2\ \e^{\I\omega(s-t)}\rt] = \int_\R\sigma_X(\dd\omega)\ \e^{\I\omega(s-t)}
\end{align}
which is a covariance function of some stationary process.

Starting from a stationary process $X$, given its covariance function $r_X$ the Bochner's theorem uniquely determines the spectral measure $\sigma_X$ and, consequently, the spectral process $S$.

The proof for the discrete-time case ii) is essentially the same. The Bochner's theorem is valid for a positive definite function on locally compact Abelian groups, so we could consider any process indexed by elements of such a group. The discrete time setting corresponds to a group $(\mathbb Z,+)$ and in this case the Bochner's theorem relates uniquely the series $(r_X(k))_k$ and the measure on the Pontryagin dual, the interval $[-\pi,\pi]$.
\end{proof}
It is worth to add that for a given measure $\sigma$ the integral
\bgeq
\int_\R \dd S(\omega)\ f(\omega)
\eeq
may be viewed as an isomorphism between $L^2_\mathbb{C}(\R,\mathcal B(\R),\sigma)$ and the Gaussian Hilbert space within $L^2_\mathbb{C}(\Omega,\mathcal F, \pr)$ which maps $\e^{\I t(\bd\cdot)}\mapsto X(t)$.

Unfortunately, the spectral process $S$ cannot be expressed by the Fourier inversion formula, because the stationary process $X$ cannot decay at infinity and the integral $\int_\R \dd t\ X(t)\e^{-\I\omega t}$ diverges. However, there is an explicit formula for $S$ if $X$ is mean square continuous.

For such Gaussian process $X$ one can define the Fourier transform in the weak sense \citep{yaglom}
\bgeq\label{eq:specSForm}
S(\omega)\defeq\lim_{T\to\infty}\f{1}{2\pi}\int_{-T}^T\dd t\  \f{\e^{-\I\omega t}-1}{-\I t} X(t).
\eeq
In a discrete-time setting the integral above should be replaced by the sum. This formula converges at the points of continuity of $\sigma_X$. For the other points $S$ can be taken so that it has left limits in mean square sense. The process $S$ defined above fulfils the equality
\bgeq
\int_\R \dd S(\omega)\ \bd{1}_{[\omega_1,\omega_2]}(t)=S(\omega_2)-S(\omega_1)=\lim_{T\to\infty}\int_{-T}^T\dd t\f{1}{2\pi}\f{\e^{-\I\omega_2t}-\e^{-\I\omega_1 t}}{-\I t} X(t).
\eeq
The integral kernel on the right is the inverse Fourier transform of $\bd{1}_{[\omega_1,\omega_2]}$, which shows that $S$ is indeed the dual of $X$ in the Fourier sense. Therefore, any mean square continuous $X$ has a harmonic representation not only in the sense of equality of distribution, but also in the stronger, mean-square sense. This stronger representation is unique \citep{procFourier}.

Formula \Ref{eq:specSForm} is well-defined even for non-stationary processes under broader conditions \citep{yaglom}. But, as one may suspect, the obtained process $S$ has independent increments if and only if $X$ is stationary. For other cases the memory structure of $S$ may be complex, e.g. by taking the process $X$ to have independent increments we can obtain arbitrary stationary $\dd S(\omega)$.
%%%%%%%%%%%%%%%

The harmonic representation is very useful in studying the linear transformations of the Gaussian processes. The spectral measure of the transformed process can be often obtained explicitly as a measure multiplied by the response function.

\bg{dfn}[Measure multiplied by function] For a given measure $\sigma$ and $f\in L^1(\sigma)$ we define measure $\sigma f$ as
\bgeq
(\sigma f)(A) = \int_A\sigma(\dd \omega) \ f(\omega)
\eeq
\end{dfn}

Let us start from the simplest type of transformation.
\bg{prp}\label{prp:YdiscSum}
If process $X$ has spectral process $S$ (i.e. is given by Eq. \Ref{eq:harmRep}), then the combination of time-shifted values 
\bgeq\label{eq:YdiscSum}
Y(t) = \sum_k a_k X(t-t_k)
\eeq
with deterministic $t_k,a_k$ and $\sum_k |a_k|^2<\infty$, has spectral representation
\bgeq
Y(t) = \int_\R\dd S(\omega)\ \sum_k a_k\e^{-\I\omega t_k}\e^{\I\omega t} ,
\eeq
and spectral measure $\sigma(\dd\omega)\lt|\sum_k a_k\e^{-\I\omega t_k}\rt|^2$.
\end{prp}
\bg{proof}
The time-shifted process $t\mapsto X(t-t_k)$ has harmonizable representation
\bgeq
X(t-t_k) = \int_\R\dd S(\omega)\ \e^{-\I\omega t_k}\e^{\I\omega t},
\eeq
in other words its spectral process has increments  $\dd S(\omega)\e^{-\I\omega t_k}$. Consequently, it has the same spectral measure and the same distribution as $X$. The process $Y$ is a sum of such shifted process, obtained in the mean-square sense, so the result fallows.
\end{proof}
One significant consequence of the above is that the ensemble averaged mean-square displacement can be expressed as
\bgeq
\delta^2_X(t)=\E|X(\tau+t)-X(\tau)|^2 = \int_\R \sigma(\dd\omega)\ \lt|\e^{\I \omega t}-1\rt|^2 = 4 \int_\R \sigma(\dd\omega)\ \sin\lt(\f{\omega t}{2}\rt)^2.
\eeq 

Moreover, taking the mean-square limit $\lim_{h\to0} (X(t+h)-X(t))/h$ one obtains the harmonizable representation of the mean-square derivative
\bgeq
\df{}{t}X(t) = \I\int_\R \dd S(\omega)\  \omega\e^{\I\omega t},
\eeq
which exists if and only if $\int \sigma(\dd\omega)\ \omega^2<\infty$. Analogically to the last proposition, the fallowing fact is true
\begin{prp}\label{prp:YDerSum}
Any process given by
\bgeq
Y(t) = \sum_k a_k\df{^k}{t^k} X(t),
\eeq
which exists if $\sigma(\dd \omega) \sum_k |a_k|^2\omega^{2k}$ is a finite measure, has harmonic representation
\bgeq
Y(t) = \int_\R\dd S_X(\omega)\ \sum_k a_k (\I\omega)^k\e^{\I\omega t} ,
\eeq
and spectral measure $\sigma(\dd\omega) \lt| \sum_k a_k (\I\omega)^k\rt|^2$.  
\end{prp}

A similar reasoning generalises formula \Ref{eq:YdiscSum} for a convolution
\begin{prp}\label{prp:Yconv}
Let  $g$ be a function in the Schwartz class and 
\bgeq
Y(t) = g*X(t) = \int_\R\dd s\ g(t-s)X(s),
\eeq
then the process $Y$ has harmonic representation
\bgeq\label{eq:Yconv}
Y(t) = \int_\R\dd S_X(\omega)\  \widehat{g}(\omega)\e^{\I\omega t}.
\eeq
\end{prp}
\begin{proof}
The process $Y$ is well-defined in the mean-square sense because
\begin{align}
\E|Y(t)|^2&=\int_\R\dd s_1\int_\R\dd s_2 \ g(t-s_1)g(t-s_2)\E[X(s_1)X(s_2)]\nonumber\\
& = \int_\R\dd s_1\int_\R\dd s_2 \ g(t-s_1)g(t-s_2)r_X(s_2-s_1)\nonumber\\
& = \langle g,g*r_X\rangle =\int_\R \sigma_X(\dd\omega)\  |\widehat g|^2.
\end{align}
In the last line we applied Parseval's theorem using the fact that $g$ and $\widehat g$ decay rapidly and $r_X$ is bounded.

This mean square convergence allows for commutating the order of integrals, and
\begin{align}
Y(t)&=\int_\R\dd s\ g(t-s)\int_\R\dd S(\omega)\ \e^{\I\omega s} = \int_\R\dd S(\omega)\int_\R\dd s\ g(t-s)\e^{\I\omega s} \nonumber\\
&= \int_\R\dd S_X(\omega)\ \widehat g(\omega)\e^{\I \omega t} .
\end{align}
\end{proof}
The above representation also holds under weaker conditions if the integrals in derivation are absolutely convergent.

\section{Generalized Gaussian processes}\label{s:genGauss}
The theory of stationary Gaussian processes does not describe many notions, which appear in the applied theory of the generalised Langevin equation. Objects like white noise or fractional noise are natural choices for the force term in the Langevin equation, but cannot be described as a real-valued stochastic processes. There are ways around this problem such as the It\={o} interpretation or pathwise interpretation in the sense of integrals. We will briefly describe the approach based on distribution theory. We denote by $\mathcal D_{\mathbb C}(\R)$ a class of $\mathbb C$-valued smooth functions with compact support on $\R$ and call a continuous linear functional $X\colon \mathcal D_{\mathbb C}(\R)\to L^2(\pr)$ a generalized process. If $X[\phi]$ is Gaussian for any test function $\phi$ we call the process Gaussian. We will consider only the zero mean case $\E[X[\phi]]=0$.

For a mean-square continuous Gaussian process $X$ formula
\bgeq
X[\phi]=\int_\R\dd t \ X(t)\phi(t),
\eeq
defines a linear functional on $\mathcal D_{\mathbb C}(\R)$, it is continuous if the covariance function $r_X$ is continuous and does not grow rapidly. Therefore a Gaussian process under some weak assumptions can be interpreted as generalized process.
\bg{dfn}We call a generalized process $X$ stationary if
\bgeq
X[\phi] = \int_\R\dd S(\omega) \ \widehat \phi(\omega)
\eeq
for some spectral process $\dd S$ with a spectral measure $\sigma$ which is tempered, i.e.
\bgeq
\int_\R\sigma(\dd\omega) \ \f{1}{(1+\omega^2)^N}<\infty
\eeq
for some $N\in\mathbb N$.
\end{dfn}
The above formula for $X[\phi]$ defines a linear functional well-defined for $\phi\in \mathcal D_{\mathbb C}(\R)$ because Fourier transform $\widehat \phi$ is in Schwartz class, in particular $\widehat\phi(\omega)\le C (1+\omega^2)^{-n}$ for any $n\in\mathbb N$. Moreover
\bgeq
\E|X[\phi]|^2=\int_\R\sigma(\dd\omega)\ |\widehat\phi(\omega)|^2\le C^2\int_\R\sigma(\dd\omega)\ \f{1}{(1+\omega^2)^{2N}}<\infty
\eeq
Now, take a test function shifted to the right, $\mathcal S_\tau\phi(t)=\phi(t+\tau)$, then $\widehat{\mathcal S_\tau\phi}(\omega)=\widehat \phi(\omega)\e^{-\I\omega \tau}$ and calculating the variance we obtain
\bgeq
\E|X[\mathcal S_\tau\phi]|^2 = \int_\R\sigma(\dd\omega)\ \lt|\widehat\phi(\omega)\e^{-\I\omega \tau}\rt|^2=\int_\R\sigma(\dd\omega)\ \lt|\widehat\phi(\omega)\rt|^2 = \E|X[\phi]|^2.
\eeq
Therefore, because of Gaussian distribution of $X[\phi]$, also $ X[\mathcal S_T\phi]\deq X[\phi]$, which explains the adjective ``stationary''.

Also note that any generalized stationary process with a finite spectral measure $\sigma(\R)<\infty$ corresponds to a classical one. In fact, if we take
\bgeq
\phi(x) = \f{1}{\sqrt{2\pi}s}\e^{-(x-t)^2/(2s^2)}\eta(x)
\eeq
with $\eta\in \mathcal D_{\mathbb C}(\R), \eta(t)=1$, in the limit $s\to 0$ we can `pinpoint' the value $X(t)$
\bgeq
X[\phi] = \int_\R\dd S(\omega) \ \e^{-\f{s^2\omega^2}{2}}\e^{\I\omega t}*\widehat \eta(\omega)\xrightarrow{s\to0}\int_\R\dd S(\omega) \ \e^{\I\omega t}*\widehat \eta(\omega) = \int_\R\dd S(\omega) \ \e^{\I\omega t}.
\eeq
On the other hand, if $X$ is a stationary process with continuous covariance function $r_X$, consider
\bgeq\label{eq:genClass}
X[\phi]=\int_\R \dd t \ X(t)\phi(t).
\eeq
It is a mean-square continuous process because $\E(X(t+h)-X(t))^2= 2r_X(0)-2r_X(h)$, the covariance function is bounded by $r_X(0)$, so indeed
\bgeq
X[\phi]=\int_\R \dd t \int_\R\dd S(\omega)\ \phi(t)\e^{\I\omega t} = \int_\R\dd S(\omega)\ \widehat \phi(\omega)
\eeq
because $\sigma(\R)<\infty$. Any stationary Gaussian process can be interpreted as a generalized process in the sense of Eq. \Ref{eq:genClass}.

New generalized processes can be defined as a transformation of the known ones. The following definitions are natural if we take into account results from Section~\ref{s:gaussProc}.
\bg{dfn}\label{dfn:genDer} Let $X$ be a generalized process. The derivative $\df{}{t} X$ is defined as a functional
\bgeq\label{eq:genProcInt}
\df{}{t}X[\phi]\defeq-X\lt[\df{}{t}\phi\rt].
\eeq
\end{dfn}
Because $\df{}{t}\phi\in\mathcal D_\mathbb{C}(\R)$ such generalized derivative always exists. Moreover, for the stationary case, we define
\bgeq
\df{}{t}X[\phi]\defeq\I\int_\R\dd S(\omega)\ \omega\widehat\phi(\omega),
\eeq
which is equivalent to the definition above. Because $\sigma(\dd\omega)\ |\omega|^N$ is a truncated measure for any $N\in\mathbb N$, the derivative of a stationary generalized process is stationary.
\bg{dfn}  Let $X$ be generalized process. A convolution with kernel $K$ is defined as functional
\bgeq
K*X[\phi]\defeq X[K*\phi]
\eeq
for $K\in\mathcal D_\mathbb{C}(\R)$. For stationary $X$ we may define
\bgeq
K*X[\phi]\defeq \int_\R\dd S(\omega)\ \widehat K(\omega)\widehat \phi(\omega),
\eeq
whenever $\sigma |\widehat K|^2$ is a truncated measure.
\end{dfn}
The requirement in the stationary case is much weaker than $K\in\mathcal D_\mathbb{C}(\R)$. In particular, we can convolve $K$ from the Schwartz class with any stationary $X$ and for a specific $X$ with $\sigma_X$ decaying sufficiently fast, $K$ can be outside $L^2$ and be interpreted as generalized (distributional) Fourier pair with $\widehat K$.

These two notions allow us to consider the solutions of stochastic differential equations and convolution-type integro-differential equations in the generalized sense, especially in the case when the force can be interpreted as a stationary generalized process. Examples of such forces, which are not real processes, include
\bg{itemize}
\item Derivative of Ornstein-Uhlenbeck process. It has $\sigma(\dd\omega) = \dd\omega\ \omega^2 (\lambda^2+\omega^2)^{-1} $.
\item White noise $\df{}{t}B$. It has $\sigma(\dd\omega)=\dd\omega$.
\item Fractional noise $\df{}{t}B_H$. It has $\sigma(\dd\omega)=\dd\omega\ |\omega|^{1-2H}, 0<H<1$.
\end{itemize}
The first example is a straightforward application of the definition of a generalized derivative. The second and third one can be viewed as a definitions, or derived, if we define generalized, in this case non-stationary, process $B_H[\phi]$ by Eq. \Ref{eq:genProcInt}. It is well-defined in the mean-square sense because the covariance function of $B_H$ is continuous and majorized by a polynomial, to obtain the formula for the spectral we may use the well-known representation of $B_H$
\bgeq
B_H(t)=\int_\R\dd S(\omega)\f{\e^{\I\omega t}-1}{\I\omega}|\omega|^{1/2-H},\quad \E|\dd S(\omega)|^2=\dd\omega.
\eeq

\chapter[Derivation and solutions of the GLE]{Derivation and solutions of the generalized Langevin equation}\label{ch:der}

Generalised Langevin equation is special among diffusion models, as it can be strictly derived from statistical mechanics. Various proofs derive it analysing the Hamiltonian systems with the use of general operator theory \citep{mori,chandler}, or considering more specific models, such as  interaction with a finite number of discrete coordinates \citep{kacGLE,zwanzigGLE} or interaction with a smooth field \citep{rey-bellet,pavliotis}. It also appears as the solution of the Rouse model describing the conformational dynamics of a monomer in a polymeric bead spring model of mass points connected by harmonic springs \citep{rouse,panja}.

Here we use a heat-bath Hamiltonian model which we try to make as general as possible: the system has an infinite number of discrete coordinates and an ensemble of smooth, field coordinates. The derivation links the obtained form of the GLE and its parameters to the properties of the heat bath, which has implications in later chapters.

After that we provide some elementary theory of the solutions of the GLE. As we are interested mostly in specific examples of solutions with established physical interpretation, we will concentrate on the form and properties of the solutions more than the requirements necessary for their existence in a wider range of cases. The general theory of stochastic Volterra equations, which the GLE is and example for, is by itself an interesting and rich topic \citep{karczewska}.

\section{Hamiltonian model}\label{s:hamMod}
Our subject of interest is a particle described by the macroscopic coordinate $X\in \R$ and its conjugated momentum $P \in \R$. In reality the motion is often three-dimensional, but without any specific form of coupling the coordinates along different axes would evolve independently, therefore we can consider them separately. Subsystem $X,P$ is coupled to a larger space  with infinitely many degrees of freedom, some of which are microscopic (unobservable individually), some mesoscopic (small, but having observable individual influence).

We denote the mesoscopic generalised coordinates by $\bd{q}=(q_k)_{k=1}^\infty$ and the conjugated  momenta $\bd{p}=(p_k)_{k=1}^\infty$. The model with a finite number of degrees of freedom can be treated as a specific case of this more general case, in which we take $q_l=0=p_l$ for $l>N$. The coordinates $\bd{q}$ can be real positions (e.g. of the particles within the liquid) or more abstract quantities, e.g. phonon degrees of freedom.

The microscopic coordinates are a pair of conjugated fields $\varphi,\pi$ which are functions $\R^d\mapsto \R$. The function $\pi$ is momentum-like, the function $\varphi$ is position-like. In particular, $\nabla \varphi$ can be interpreted as a field of deviations and a field equivalent of $\bd q$

We assume that the energy associated with the observed macroscopic coordinates $X,P$ has the typical form
\bgeq
\ham_{S} = \f{P^2}{2M}+\mathcal{V}(X),
\eeq
where $M$ is a mass or similar quantity and the potential function $\mathcal{V}$ is some differentiable function $\R\to\R$. The particle described by $X,P$ is interacting with the surrounding media, which is a harmonic bath, containing both discrete and field degrees of freedom

\begin{align}\label{eq:HB}
\ham_{B} &= \ham_{B_d}+\ham_{B_f}, \nonumber\\
\ham_{B_d} & = \f{1}{2}\sum_{k=1}^\infty\lt(\f{p_k^2}{m_k}+m_k\omega_k^2q_k^2\rt), \nonumber\\
\ham_{B_f} &= \f{1}{2}\int_{\R^d}\dd x\ \big(|\pi(x)|^2 +  |\nabla\varphi(x)|^2\big).
\end{align}

We require that $\ham_B<\infty$ for at least some dense subspace of $\bd q,\bd p, \varphi, \pi$. That is, we assume $\bd q,\bd p \in \ell^2, \pi\in L^2(\R^d)$ and $\varphi \in H^1(\R^d)$ where $H^1(\R^d)$ is the completion of $C^\infty_0(\R^d)$ with respect to $\Vert\bd\cdot\Vert_{L^2}$. These requirements will become more clear later, when we will impose initial conditions.

The bath Hamiltonian $\ham_B$ can be expressed using the scalar products of the used spaces.
In order to write it in a simple manner let us introduce a notation, in which formulas which contain bold vectors are understood component wise, e.g. $\bd y \sin(\bd x^2)$ is a vector with components $y_k\sin(x_k^2)$. The discrete bath Hamiltonian is finite on the domain $\dom(\ham_{B_d})=\{\bd{p}\oplus\bd{q}\in\ell^2\oplus\ell^2\colon \bd m^{-1}\bd{p}\in\ell^2, \bd m\bd\omega^2\bd{q}\in\ell^2\}$. 

Note that $\bd m^{-1}$ and $\bd m\bd \omega$ can be understood as positive and self-adjoint multiplication operators acting on $\bd q$ and $\bd q$, respectively. Now, the bath Hamiltonian is

\bgeq
\ham_B = \f{1}{2}\big(\Braket{\bd m^{-1}\bd{p},\bd{p}}_{\ell^2} + \Braket{\bd m\bd\omega^2\bd{q},\bd{q}}_{\ell^2}  + \Braket{\pi,\pi}_{L^2} + \Braket{\nabla \varphi,\nabla \varphi}_{L^2} \big).
\eeq 
There is a visible asymmetry in $\ham_B$ in the lack of physical constants associated with fields $\varphi,\pi$. More natural choices would be terms like $|\pi(x)^2|/m(x)$ and $m(x)\omega(x)|\nabla \varphi(x)|^2$ for some $\R^d\mapsto \R$ functions $m$ and $\omega$. However, because of the vastly different behaviour of the multiplication operators on $\ell^2$ and $L^2$ spaces that would greatly complicate the analysis. Instead, we treat the fields $\pi$ and $\varphi$ as if they were already normalised by substitutions $\pi(x)/\sqrt{m(x)}\to \pi(x)$ and  $\sqrt{m(x)\omega(x)}\nabla \varphi(x)\to \nabla \varphi(x)$. This procedure is also possible for $\bd q$ and $\bd q$, but there would be a price for that, which will become apparent at the end of Sec. \ref{s:fieldBath}.

The last term in the total Hamiltonian is the coupling between $X$ and the bath, which we choose to be linear
\bgeq
\ham_I = \ham_{I_d}+\ham_{I_f} = -X\sum_{k=1}^\infty m_k\gamma_k q_k -X\int_{\R^d}\dd x\ \nabla\varphi(x)\cdot\rho(x).
\eeq
Here $\rho$ is a function $\R^d \to\R^d$ and the dot ``$\cdot$'' is a scalar product of $\R^d$. We assume $\bd \gamma \in\ell^2$ and $\rho\in L^2_{\R^d}(\R^d)$. In the equations of motion the gradient $\nabla\cdot\rho$ will appear, so it seems $\rho$ should be differentiable, but one can argue that the result will be affected only by the Fourier transform $\widehat\rho$, so $\nabla\cdot\rho$ can be interpreted in the weak sense. It becomes more visible if we express $\ham_I$ using scalar products
\bgeq
\ham_I = -X\Braket{\bd m\bd{\gamma},\bd{q}}_{\ell^2} - X\Braket{\rho, \nabla \varphi}_{L^2_{\R^d}}
\eeq

All of the variables $X,P,\bd q,\bd p,\varphi,\pi$ are time-dependent, which for $\bd q,\bd p,\varphi,\pi$ we can denote by a subscript when necessary, i.e. $\bd q_t,\bd p_t,\varphi_t,\pi_t$, in order to avoid confusion with other parameters. The time evolution of these variables is governed by the Hamilton's equations of the system $\ham_S+\ham_B+\ham_I$ are
\begin{subequations}
	\begin{align}\label{eq:firstPair}
	\df{}{t}X&=\pf{}{P}\ham,\quad \df{}{t}P=-\pf{}{X}\ham,\\
	\label{eq:sndPair}
	\df{}{t}\bd{q}&=\f{\delta}{\delta \bd{p}}\ham,\quad\df{}{t}\bd{p}=-\f{\delta}{\delta \bd{q}}\ham\\
	\label{eq:thrdPair}
	\df{}{t}{\varphi}&=\f{\delta}{\delta \pi}\ham,\quad\df{}{t}\pi=-\f{\delta}{\delta \varphi}\ham
	\end{align}
\end{subequations}

The first pair of equations \Ref{eq:firstPair} can be easily combined obtaining
\bgeq\label{eq:Xmot}
M\df{^2}{t^2} X=-\df{}{x}\mathcal{V}(X)+\Braket{\bd m\bd{\gamma},\bd{q}}_{\ell^2}+\Braket{\rho, \nabla \varphi}_{L^2_{\R^d}}.
\eeq

The two other pairs of equations \Ref{eq:sndPair} and \Ref{eq:thrdPair} should be understood in terms of a functional derivative.
\bg{dfn}[Functional derivative]\label{dfn:funDer}
Let $H$ be a Hilbert space and $F$ a functional on this space, $F\colon H\to \R$. The functional derivative $\f{\delta}{\delta g}F$ is also a functional, which equals
\bgeq
\f{\delta}{\delta g}F[h]=\lim_{\epsilon\to 0} \f{F[g+\epsilon h]-F[g]}{\epsilon},
\eeq
for all $g,h\in H$ for which the above limit exists.
\end{dfn}
In our case for all Hamilton equations this functional will be a scalar product with some vector. Additionally, this type of derivative, like a classical one, is a linear operator
\bgeq
\f{\delta}{\delta g}(aF_1+bF_2) = a \f{\delta}{\delta g}F_1+b\f{\delta}{\delta g}F_2,
\eeq
so we can reduce the remaining Hamilton equations to
\begin{subequations}
\begin{align}\label{eq:discFunDer}
\df{}{t}\bd{q}&=\f{\delta}{\delta \bd{p}}\ham_{B_d},\quad\df{}{t}\bd{p}=-\f{\delta}{\delta \bd{q}}\ham_{B_d}-\f{\delta}{\delta \bd{q}}\ham_{I_d}\\\label{eq:fieldFunDer}
\df{}{t}\bd{\varphi}&=\f{\delta}{\delta \pi}\ham_{B_f},\quad\df{}{t}\pi=-\f{\delta}{\delta \varphi}\ham_{B_f}-\f{\delta}{\delta \varphi}\ham_{I_f}.
\end{align}
\end{subequations}
Without surprise, the equations for the field and discrete parts of the bath are coupled only through the variable $X$. They can be solved separately, which we perform in two short sections below. They require some assumptions on $X$ as a function of time. Equation \Ref{eq:Xmot} requires the existence of the second derivative, but in practice it can be interpreted in the weak sense and the requirements lowered. For the calculations below it is sufficient for $X$ to be integrable on finite intervals, e.g. in the Riemann-Stieltjes sense. In this case the Laplace transform and convolution with bounded functions are well-defined.

\section{Discrete heat bath}\label{s:dhb}
We start from calculating the functional derivatives in Eq. \Ref{eq:discFunDer}. We take an arbitrary vector $\bd\eta$ in the domain of the Hamiltonian and use Definition \ref{dfn:funDer},
\begin{align}
\Braket{\f{\delta}{\delta \bd{p}}\ham_{B_d},\bd \eta}_{\ell^2} &=  \f{1}{2}\lim_{\epsilon\to 0} \f{\Braket{\bd m^{-1}(\bd p+\epsilon\bd{\eta}),(\bd p+\epsilon\bd{\eta})}_{\ell^2}-\Braket{\bd m^{-1}\bd{p},\bd{p}}_{\ell^2}}{\epsilon}\nonumber\\
&= \f{1}{2}\lim_{\epsilon\to 0} \f{\epsilon\Braket{\bd m^{-1}\bd{p},\bd{\eta}}_{\ell^2}+\epsilon\Braket{\bd m^{-1}\bd{\eta},\bd{p}}_{\ell^2}+\epsilon^2\Braket{\bd m^{-1}\bd{\eta},\bd{\eta}}_{\ell^2}}{\epsilon}\nonumber\\
&= \Braket{\bd m^{-1}\bd{p},\bd{\eta}}_{\ell^2}
\end{align}
so
\bgeq
\df{}{t}\bd{q}=\bd m^{-1}\bd{p}.
\eeq
In a similar manner we obtain
\bgeq
\Braket{\f{\delta}{\delta \bd{q}}\ham_{B_d},\bd \eta}_{\ell^2} =\Braket{ \bd m\bd\omega^2\bd{q},\bd \eta}_{\ell^2}
\eeq
and
\bgeq
\Braket{\f{\delta}{\delta \bd{q}}\ham_{I_d},\bd \eta}_{\ell^2}=-X\lim_{\epsilon\to 0}\f{\Braket{\bd m\bd\gamma,\bd q+\epsilon\bd\eta}_{\ell^2}-\Braket{\bd m\bd\gamma\bd q,\bd q}_{\ell^2}}{\epsilon} = -X\Braket{\bd m\bd\gamma,\bd\eta}_{\ell^2},
\eeq
therefore
\bgeq
\df{}{t}\bd{p}=-\bd m\bd\omega^2\bd{q}+ X\bd m\bd{\gamma}.
\eeq

We solve these equations using the Laplace transform, which we denote by the symbol ``$^\#$''. In the Laplace image these equations state that
\bgeq
s\bd{q}^\#_s-\bd{q}_0=\bd m^{-1}\bd{p}^\#_s,\quad s\bd{p}^\#_s-\bd{p}_0=-\bd m\bd\omega^2\bd{q}^\#_s+X^\#\bd m\bd{\gamma},
\eeq
where $\bd p_0\in\ell^2$ and $\bd q_0\in\ell^2$ are initial conditions of $\bd p$ and $\bd q$.
After merging the above equations we get
\bgeq
(s^2+\bd\omega^2)\bd{q}^\#_s=s\bd{q}_0+\bd m^{-1}\bd{p}_0+X^\#\bd{\gamma}.
\eeq
The factor $(s^2+\bd\omega^2)$ is a multiplication operator. To proceed further, we make assumptions on $\bd m$ and $\bd m\bd\omega^2$

\bgeq\label{eq:regCond}
\sum_{k=1}^\infty m_k <\infty,\quad \sum_{k=1}^\infty \f{1}{m_k\omega_k^2}<\infty.
\eeq
The first condition of the finiteness of total mass is very natural. The second is more abstract, it limits the possible dispersion relations stating that $\omega_k\to\infty$ sufficiently fast comparing to the decay of the masses $m_k\to 0$. It prohibits e.g. the situation when $\omega_k$ are  scattered in some bounded interval.

Assuming the above, $\omega_k\to\infty$ and the inverse $(s^2+\bd\omega^2)^{-1}$ is bounded regarded as $\ell^2\to\ell^2$ operator. We may inverse the Laplace transform and get

\bgeq\label{eq:qSol}
\bd q_t =\bd q_0\cos(\bd\omega t) +\bd p_0\f{\sin(\bd\omega t)}{\bd m\bd\omega}+\f{\bd\gamma}{\bd\omega}\int_{0}^t\dd \tau\ X(\tau)\sin(\bd\omega(t-\tau)).
\eeq
This formula shows that $\omega_k$, as we suspect from the form of the Hamiltonian, can be interpreted as the oscillation frequency of the $k$-th degree of freedom. Because $\cos(\omega_k)$ are bounded by $1$ and $\f{1}{m_k\omega_k}\to 0$ first to terms on the right side are in $\ell^2$. The last term contains $\bd\gamma/\bd\omega$ which also is in $\ell^2$ (because $\bd{\gamma}\in\ell^2$ and $\omega_k\to\infty$).

Integrating by parts the integral on the right of Eq. \Ref{eq:qSol} we obtain the second formula for $\bd{q}_t$

\begin{align}\label{eq:qSol2}
\bd q_t &=\bd q_0\cos(\bd\omega t) +\bd p_0\f{\sin(\bd \omega t)}{\bd m\bd\omega}+\f{\bd\gamma}{\bd \omega^2}X-\f{\bd\gamma}{\bd\omega^2}X_0\cos(\bd\omega t)\nonumber\\
&-\f{\bd\gamma}{\bd\omega^2}\int_{0}^t\dd X(\tau)\ \cos(\bd\omega(t-\tau)),
\end{align}
where the integral over $\dd X(\tau)$ is understood in the Riemann sense. To obtain the equation of motion for $X$ \Ref{eq:Xmot} we need to know the scalar product $\Braket{\bd m\bd \gamma, \bd q_t}_{\ell^2}$. Substituting the obtained formulas for $\bd{q}_t$ 
we get two possible forms. Using \Ref{eq:qSol} we find
\bgeq
\Braket{\bd m\bd \gamma, \bd q_t}_{\ell^2}=\int_0^t\dd \tau\ X(\tau) \tilde{K}(t-\tau)+\tilde{F},
\eeq
where
\begin{subequations}
\begin{align}
\tilde{K}(t)&=\Braket{\bd m\bd\gamma,\f{\bd\gamma}{\bd\omega}\sin(\bd\omega t)}_{\ell^2},\\
{F}(t)&=\Braket{\bd m\bd\gamma,\lt(\bd q_0\cos(\bd\omega t)+\bd p_0\f{\sin(\bd\omega t)}{\bd m\bd\omega}\rt)}_{\ell^2}.
\end{align}
\end{subequations}
Both these scalar products are finite, in fact
\begin{align}
|\tilde{K}(t)|&\le  \max_k \f{m_k}{\omega_k} \Vert \bd{\gamma}\Vert^2_{\ell^2},\nonumber\\
|{F}(t)|&\le \max_k m_k \Vert\bd{\gamma}\Vert_{\ell^2}\Vert \bd{q}_0\Vert_{\ell^2}+\max_k\f{1}{\omega_k}\Vert\bd{\gamma}\Vert_{\ell^2}\Vert\bd{p}_0\Vert_{\ell^2} .
\end{align}
Alternatively, using \Ref{eq:qSol2}, we get
\begin{align}\label{eq:GLEtermD}
\Braket{\bd m\bd \gamma, \bd q_t}_{\ell^2}&=X\Braket{\bd m\bd\gamma,\frac{\bd \gamma}{\bd \omega^2}}_{\ell^2} -X_0\Braket{\bd m\bd\gamma,\f{\bd\gamma}{\bd\omega^2}\cos(\bd\omega t)}_{\ell^2}\nonumber\\
&-\int_0^t\dd X(\tau)\ K(t-\tau)+F,
\end{align}
where
\bgeq
K(t)=\Braket{\bd m\bd\gamma,\f{\bd\gamma}{\bd\omega^2}\cos(\bd\omega t)}_{\ell^2},\quad \df{}{t}K=-\tilde{K}.
\eeq

Equation \Ref{eq:GLEtermD} leads the standard GLE model with the force $F$ and memory kernel $K$, see e.g. \cite{zwanzigGLE,kubo}. The first term on the right side of this equation can be incorporated into the effective deterministic potential
\bgeq
\mathcal{V_\text{eff}}(X) \defeq \mathcal{V}(X)-\f{1}{2}X^2\Braket{\bd m\bd\gamma,\frac{\bd \gamma}{\bd \omega^2}}_{\ell^2}.
\eeq

The second term equals to $-X_0 K(t)$ and describes the dependence on the initial condition. We will mostly consider stationary solutions which will not be affected by this term. The third term is the expected convolution with the memory kernel. Finally, the last term $F$ is the driving force, which is a linear, time-dependent functional of the initial conditions $\bd q, \bd p$. If these are random, $F$ is also random. Let us suppose that at time 0 the heat bath was at thermal equilibrium. In this case the initial conditions should have a Gibbs measure given by the formula
\bgeq
\dd \mu_{\bd q_0,\bd p_0}\propto \exp\lt(-\f{1}{k_B\mathcal T}\ham_{B_d}(\bd q_0,\bd p_0)\rt).
\eeq
Because $\bd q_0$ and $\bd p_0$ are vectors in Hilbert space the above formula should be interpreted carefully, there is no Lebesgue measure, so probability densities can be defined only as a Radom-Nikodym derivatives with respect to other suitable measures \cite{SDEinfDim}. In our case we say that the Gibbs distribution is a Gaussian distribution with covariance operator $k_B\mathcal T (m\oplus (\bd m\bd\omega^2)^{-1})$. Eq. \Ref{eq:regCond} guarantees that it is positive and trace-class. Through simple calculation it is easy to see that this distribution is invariant under evolution of the bath itself, as long as it is decoupled. As we have seen, after the coupling the coordinate $X$ becomes disturbed by the force $F$ which is a Gaussian process with covariance function
\begin{align}
r_F(s,t)&=k_B\mathcal T\lt\langle \bd m^2\bd\gamma^2,\f{1}{\bd m\bd\omega^2}\cos(\bd\omega s)\cos(\bd\omega t)+\f{\bd m}{\bd m^2 \bd \omega^2}\sin(\bd\omega s)\sin(\bd\omega t)\rt\rangle_{\ell^2}\nonumber\\
&=k_B\mathcal T\lt\langle \bd m^2\bd\gamma^2,\f{1}{\bd m\bd\omega^2}\cos(\bd\omega(s-t))\rt\rangle_{\ell^2} = k_B\mathcal TK(s-t).
\end{align}
This is the most typical form of the famous fluctuation-dissipation relation \citep{kubo}.

\section{Field heat bath}\label{s:fieldBath}
In order to solve the Hamilton's equations of the field coordinates of freedom \Ref{eq:fieldFunDer} we must calculate the distributional derivatives
\begin{align}
\Braket{\f{\delta \ham_{B_f}}{\delta \pi},\eta}_{L^2} &=\f{1}{2} \lim_{\epsilon\to0}\f{\Braket{\pi+\epsilon\eta,\pi+\epsilon\eta}_{L^2}-\Braket{\pi,\pi}_{L^2}}{\epsilon} = \Braket{\pi,\eta};\nonumber\allowdisplaybreaks\\
\Braket{\f{\delta \ham_{B_f}}{\delta \varphi},\eta}_{L^2} &=\f{1}{2} \lim_{\epsilon\to0}\f{\Braket{\nabla(\varphi+\epsilon\eta),\nabla(\varphi+\epsilon\eta)}_{L_{\R^d}^2}-\Braket{\nabla\varphi,\nabla\varphi}_{L_{\R^d}^2}}{\epsilon}\nonumber\\
& = \Braket{\nabla\varphi,\nabla\eta}_{L^2_{\R^d}} = -\Braket{\nabla^2\varphi,\eta}_{L^2};\nonumber\allowdisplaybreaks\\
\Braket{\f{\delta \ham_{I_f}}{\delta \varphi},\eta}_{L^2} &=-X\lim_{\epsilon\to 0}\f{\Braket{\rho,\nabla(\varphi+\epsilon\eta)}_{L^2_{\R^d}}-\Braket{\rho,\nabla\varphi}_{L^2_{\R^d}}}{\epsilon} =-X\Braket{\rho,\nabla\eta}_{L^2_{\R^d}}\nonumber\\
& = X\Braket{\nabla\cdot \rho,\eta}_{L^2}.
\end{align}
Therefore, the Hamilton's equations are 

\bgeq
\df{}{t}\varphi_t = \pi_t,\quad \df{}{t}\pi_t=\nabla^2\varphi_t-X\nabla\cdot\rho,
\eeq
and we assume the initial conditions $\varphi_0\in H^1(\R^d),\pi_0\in L^2(\R^d)$. To solve this equation we will translate the problem into Fourier-Laplace space. Let us denote by $\widehat f^\#$ the Laplace transform of $f$ with respect to the first (time) argument followed by the Fourier transform with respect to the rest (space) arguments. We perform a Laplace transform for the equation for $X$ and $P$, then Fourier and Laplace transform for $\varphi,\pi$. The Hamilton's equations become
\begin{align}
s\widehat{\varphi}_s^\# &= \widehat{\pi}_s^\#-\widehat{\varphi}_0,\nonumber\\
s\widehat{\pi}_s^\# &= -|\xi|^2 \widehat{\varphi}_s^\# -\I X^\# \xi\cdot \widehat\rho- \widehat{\pi}_0,
\end{align}
which we solve for $\widehat\varphi^\#_s$
\bgeq
\widehat\varphi_s^\# = -\I X^\#\f{\xi\cdot\widehat\rho}{s^2+|\xi|^2}-\widehat\varphi_0\f{s}{s^2+|\xi|^2}-\widehat\pi_0\f{1}{s^2+|\xi|^2}.
\eeq
Inverting the Laplace transform we obtain
\bgeq
\widehat\varphi_t = -\I\f{\xi\cdot\widehat \rho}{|\xi|}\int_0^t\dd\tau\ X(\tau)\sin(|\xi|(t-\tau))-\widehat\varphi_0\cos(|\xi|t)-\widehat\pi_0\f{1}{|\xi|}\sin(|\xi|t).
\eeq
In order to determine the form of the GLE we need to calculate the scalar product from  Eq. \Ref{eq:Xmot}. We use the Plancherel theorem and obtain the familiar formula
\bgeq\label{eq:fieldSP}
\Braket{\rho,\nabla\varphi_t}_{L^2_{\R^d}} =\I\Braket{(\bd\cdot)\cdot\widehat\rho,\widehat\varphi_t}_{L^2_\mathbb{C}} = \int_0^t\dd\tau\ X(\tau)\tilde K(t-\tau) + F(t),
\eeq
where we use the notation $f(\bd\cdot)\defeq \xi\mapsto f(\xi)$, e.g. $(\bd\cdot)\cdot x = \xi\mapsto \xi\cdot x$ for fixed $x$. In the above equation
\bg{align}
\tilde K(t) = \lt\langle\f{|(\bd\cdot)\cdot\widehat\rho|^2}{|\bd\cdot|},\sin(|\bd\cdot|t)\rt\rangle_{L^2}
\end{align}
and 
\bgeq
F(t)=-\lt\langle\I(\bd\cdot)\cdot \widehat\rho,\widehat\varphi_0\cos(|\bd\cdot|t)+ \widehat \pi_0\f{1}{|\bd\cdot|}\sin(|\bd\cdot|t)\rt\rangle_{L^2_\mathbb{C}}.
\eeq
In \Ref{eq:fieldSP} we can integrate by parts, similarly as in the discrete case, and obtain
\bgeq
\Braket{\rho,\nabla\varphi_t}_{L^2_{\R^d}} =X\Vert\widehat \rho\Vert^2-X_0K(t)-\int_0^t\dd X(\tau)\ K(t-\tau)+F,
\eeq
where
\bgeq
K(t)=\lt\langle\f{|(\bd\cdot)\cdot\widehat\rho |^2}{|\bd\cdot|^2},\cos(|\bd\cdot|t)\rt\rangle_{L^2},
\eeq
note that because $|\xi\cdot\widehat \rho|^2\le |\xi|^2|\widehat\rho|^2$ this scalar product is finite. Similarly as in the last section, the randomness of the force $F$ is caused by the random fields $\widehat\pi_0, \widehat\varphi_0 $, which are most often assumed to have the Gibbs distribution
\begin{align}
\dd \mu_{\pi_0,\varphi_0}&\propto \exp\lt(-\f{1}{k_B\mathcal T}\ham_{B_f}(\pi_0,\varphi_0)\rt)=\exp\lt(-\f{1}{k_B\mathcal T}\langle\pi_0,\pi_0\rangle_{L^2}-\f{1}{k_B\mathcal T}\langle\nabla^2\varphi_0,\varphi_0\rangle_{L^2}\rt)\nonumber\\
&=\exp\lt(-\f{1}{k_B\mathcal T}\langle\widehat\pi_0,\widehat\pi_0\rangle_{L^2_\mathbb{C}}-\f{1}{k_B\mathcal T} \langle|\bd\cdot|\widehat\varphi_0,|\xi|\cdot\varphi_0\rangle_{L^2_\mathbb{C}}\rt)\propto \dd \tilde\mu_{\widehat\pi_0,\widehat\varphi_0}.
\end{align}

This formula looks like a Gaussian distribution on the space $L^2_\mathbb{C}(\R^d)$ of $\widehat\pi_0,\widehat\varphi_0$ with covariance operator $k_B\mathcal T \cdot 1\oplus |\xi|^{-1}$. It is not exactly the case. The first, technical point is that $\widehat\pi_0,\widehat\varphi_0$ are Fourier transforms of real functions, so they must be Hermitian, i.e. $\widehat\pi_0(-\xi)=\widehat\pi_0(\xi)^*,\widehat\varphi_0(-\xi)=\widehat\varphi_0(\xi)^*$ (this is an iff condition). Therefore, the distribution is actually supported on $\R^d_+$ and values on $\R^d_-$ are determined by the Hermitian condition.

The second, more profound problem is that the multiplication operator $1\oplus |\xi|^{-1}$ is positive, but not trace-class. Inclusion of mass and stiffness density in the Hamiltonian would not alleviate this problem, because, in contrast to $\ell^2$, for $L^2$ spaces in general the multiplication operators are not trace-class. This problem reflects the fact that the Hilbert space of $\pi$ and $\varphi$ is, in a sense, too small. We are interested in how $\pi,\varphi$ (equivalently $\widehat\pi,\widehat\varphi$) acts on observed, discrete coordinates of freedom, and it does this only through projections. Therefore the distribution space is a more natural space for the Gibbs distribution. 
\bg{dfn}
The vector $Y$ in Hilbert space $H$ is said to have canonical Gaussian cylinder set measure if for any $x,y\in H$
\bgeq
\E[\langle x,Y\rangle\langle Y,y\rangle]=\langle x,y\rangle.
\eeq
\end{dfn}
Note that this definition is similar to that of a generalized process, in fact for $\rho$, which is a test function, the definitions agree and we could interpret $\widehat\pi_0,\widehat\varphi_0$ as generalized processes. For more general $\rho\in L^2_{\R^d}$ we must use a canonical Gaussian set measure. In any case we can derive the covariance function of $F$
\begin{align}
r_F(s,t)&=\E\lt[\lt\langle\f{(\bd\cdot)\cdot\widehat\rho}{|\bd\cdot|}\cos(|\bd\cdot|s),|\bd\cdot|\widehat\varphi_0\rt\rangle_{L^2_\mathbb{C}}\lt\langle |\bd\cdot|\widehat\varphi_0,\f{(\bd\cdot)\cdot\widehat\rho}{|\bd\cdot|}\cos(|\bd\cdot|t)\rt\rangle_{L^2_\mathcal{C}}\rt]\nonumber\\
&+ \E\lt[\lt\langle\f{(\bd\cdot)\cdot\widehat\rho}{|\bd\cdot|}\sin(|\bd\cdot|s),\widehat\pi_0\rt\rangle_{L^2_\mathbb{C}}\lt\langle \widehat\pi_0,\f{(\bd\cdot)\cdot\widehat\rho}{|\bd\cdot|}\sin(|\bd\cdot|t)\rt\rangle_{L^2_\mathbb{C}}\rt]\nonumber\\
&=k_B\mathcal T\lt\langle\f{|(\bd\cdot)\cdot\widehat\rho|^2}{|\bd\cdot|^2},\cos(|\bd\cdot|(s-t))\rt\rangle_{L^2}=k_B\mathcal TK(s-t).
\end{align}
We again obtain the fluctuation-dissipation relation.
\section{Elementary solutions of the GLE}\label{s:elSol}
After we combine the results from the previous sections, we obtain the GLE
\bgeq\label{eq:Xmot2}
M\df{^2}{t^2} X=-\df{}{x}\mathcal{V}_{\text{eff}}(X)-X_0K(t)-\int_0^t\dd X(\tau)\ K(t-\tau) + F
\eeq
Further on we will consider only the simplified version of this equation. We impose $\mathcal{V}_{\text{eff}} = 0$, i.e. assume that the motion is effectively free. We will not be interested in studying the relaxation, so we set $X_0=0$, which is a neutral position (the interaction integral is zero). Then the GLE depends solely on $V=\df{}{t}X$. The mass $M$ only influences the scale of the solution, the rescaled equation $MV\to V$ corresponds to the equation with force rescaled as $F/\sqrt{M}$. The equation becomes
\bgeq\label{eq:GLEorg}
\df{}{t}V=-\int_0^t\dd s\ V(s)K(t-s)+F.
\eeq
If we move the time axes, such that the evolution starts at time $t_0$, not 0, the limits in the convolution integral change to $\int_{t_0}^t$. If we consider the limit $t_0\to -\infty$, the equation becomes
\bgeq\label{eq:GLEmain}
\df{}{t}V=-\int_{-\infty}^t\!\!\!\dd s\ V(s)K(t-s)+F.
\eeq
This equation describes a coordinate which evolved for sufficient time to reach equilibrium with the environment. Not for all systems this is possible, e.g. if $K$ is a periodic function, the convolution integral diverges for $\lim_{t\to-\infty}\E[V(t)^2]>0$, so a stationary solution is impossible. We note that it would be the case if in Sections \ref{s:hamMod} and \ref{s:dhb} the number of oscillators would be finite.

We will use a definition which may seem trivial, but is quite useful.
\bg{dfn}[Casual function] We call a function $f\colon \R\to\R$ casual if
\bgeq
f(t)=0\quad\text{for } t<0.
\eeq
\end{dfn}
The usefulness of the casual functions stems from the fact that the convolution of the process with the casual function depends only on the past states of the system. Here and further on we will always assume that the memory kernel $K$ is causal.

\bg{prp}\label{prp:fSolGLE}
The stationary solution of \Ref{eq:GLEmain} (if exists) has the spectral representation
\bgeq
V(t)=\int_\R\dd S_F(\omega)\f{1}{\I\omega+\widehat K(\omega)}\e^{\I\omega t},
\eeq
for real-valued $F$, or for generalized $F$ 
\bgeq
V[\phi]=\int_\R\dd S_F(\omega)\f{1}{\I\omega+\widehat K(\omega)}\widehat\phi(\omega).
\eeq
In both cases it has the spectral measure
\bgeq\label{eq:GLEspec}
\sigma_V(\dd\omega) = \f{\sigma_F(\dd\omega)}{\lt|\I\omega+\widehat K(\omega)\rt|^2}
\eeq
which must be finite or tempered, respectively.
\end{prp}
\bg{proof}
The force $F$ is stationary, so it has the spectral process $\dd S_F$. If the solution $V$ is stationary, it has the spectral process $\dd S_V$. Plugging these into Eq. \Ref{eq:GLEmain} we obtain
\bgeq
\df{}{t}\int_\R\dd S_V(\omega)\ \e^{\I\omega t} = -\int_\R \dd s \int_\R \dd S_V(\omega)\ \e^{\I\omega s}K(t-s) + \int_\R\dd S_F(\omega)\ \e^{\I\omega t}.
\eeq
Switching the order of differentiation and convolution with integration, justified by the existence and stationarity of all terms in the equation, we obtain equivalent formula
\bgeq
\int_\R\dd S_V(\omega) \lt(\I\omega+\widehat K(\omega)\rt)\e^{\I\omega t} = \int_\R\dd S_F(\omega) \ \e^{\I\omega t}.
\eeq
The uniqueness of the spectral representation yields the result. In the light of the definitions of derivatives and convolution \Ref{dfn:genDer} from Section \ref{s:genGauss} the reasoning is the same for a generalized process.
\end{proof}

The representation of the solution in Fourier space is useful, but we would like to study the stationary solution also in time space. Let us assume that the force is a derivative of some stochastic impulse $J, F=\df{}{t}J$, $J$ will have stationary increments. The solution should be a linear stationary functional of $\dd J$. We make an ansatz of the form
\bgeq\label{eq:convDefGLE}
V(t)=\int_{-\infty}^t\!\!\!\dd J(\tau)\ G(t-\tau),
\eeq
in other words it is an impulse-response process governed by the causal Green's function $G$. The integral can be interpreted in the mean-square sense, as limit of sums $\sum_k\big(J(\tau_{k+1})-J(\tau_k)\big)G(t-\tau_k)$ for $\tau_k$ becoming dense in the interval $[-T,t]$ and subsequently $T\to-\infty$. Substituting this ansatz into GLE we formally obtain
\bgeq
\int_\R\dd J(\tau)\ \df{}{t}G(t-\tau)=-\int_\R\dd J(\tau)\int_\R\dd \tau'\ G(\tau')K(\tau-\tau') + \df{}{t} J
\eeq
so $G$ fulfils
\bgeq\label{eq:defG}
\df{}{t}G(t) = -\int_\R\dd \tau\ G(\tau)K(t-\tau)+\delta(t),
\eeq
where $\delta$ is Dirac delta distribution and the whole equation should be understood in the weak sense. We call a process $V$ given by Eq. \Ref{eq:convDefGLE} with $G$ solving Eq. \Ref{eq:defG} a mild solution of the GLE equation \Ref{eq:GLEmain}. For non-stationary GLE \Ref{eq:GLEorg} we understand Eq. \Ref{eq:convDefGLE} as a mild solution if the integral is taken from $0$ to $t$. A non-zero initial condition can be accounted for adding term $V_0G(t)$.

If $G$ is an $L^2$ function, Eq. \Ref{eq:defG} leads to a Fourier space solution
\bgeq
\widehat G(\omega) = \f{1}{\I\omega+\widehat K},
\eeq
which is not surprising given the form of the stationary solution \Ref{eq:GLEspec}. Conversely, because $K$ and $G$ are causal, for $t> 0$ we may write
\bgeq
\df{}{t}G(t) = -\int_0^t\dd \tau\ G(\tau)K(t-\tau),\quad G(0)=1,
\eeq
so the Laplace transform of $G$ is given by
\bgeq
G^\#(s)=\f{1}{s+K^\#(s)}.
\eeq
The form of the Green's function is an analytical property: it has no relation to the random dynamics of the  GLE. The memory structure of the solution is however a stochastic property. The fluctuation-dissipation relation provides a link between the analytical and stochastic aspects of the GLE and allows for proving a strong relation between them.
\bg{prp}\label{prp:covGreen}
The covariance function of the stationary solution of the GLE equals
\bgeq
r_V(t)=k_B\mathcal T G(t), \quad t\ge 0
\eeq
and the mean-square displacement of the position process $X(t)=\int_0^t\dd\tau\ V(t)$ is
\bgeq \label{eq:msdGen}
\E[X(t)^2] = 2k_B\mathcal T\int_0^t\dd\tau_1\int_0^{\tau_1}\!\dd\tau_2\ G(\tau_2).
\eeq
\end{prp}
\bg{proof} The solution is stationary and has mean 0, so we can take $t>0$ and	 calculate the covariance function as
\bg{align}
r_F(t)&=\E[V(0)V(t)]=\E\lt[\int_\R \dd J(\tau_1)\ G(-\tau_1)\int_\R\dd J(\tau_2)\ G(t-\tau_2)\rt]\nonumber\\
&= \int_\R\dd\tau_1\int_\R\dd\tau_2\ r_F(\tau_2-\tau_1) G(-\tau_1)G(t-\tau_2),
\end{align}
where as usual we assumed that $G$ is casual. The covariance function $r_F$ is not casual, but can be represented as $r_F(\tau)=k_B\mathcal TK(\tau)+k_B\mathcal TK(-\tau)$. This formula fails only at $\tau=0$, but it does not affect the integration. The integral separates into two parts, the first is
\bg{align}
I_1&= k_B\mathcal T\int_\R\dd\tau_1\int_\R\dd\tau_2\ K(\tau_2-\tau_1) G(-\tau_1)G(t-\tau_2) \nonumber\\
&= k_B\mathcal T\int_\R\dd\tau_1\ G(-\tau_1)\int_\R\dd\tau_2'\ K(t-\tau_1-\tau_2')G(\tau_2')\nonumber\\
&= - k_B\mathcal T\int_\R\dd\tau_1\ G(-\tau_1)\df{}{\tau}G(t-\tau_1),
\end{align}
where we used the defining equation  \Ref{eq:defG} in the interior of the support of $G$ and denoted by $\df{}{\tau} G(t-\tau_1)$ the derivative of $G$ at the point $t-\tau_1$. Similarly for the second integral,
\bg{align}
I_2&= k_B\mathcal T\int_\R\dd\tau_1\int_\R\dd\tau_2\ K(\tau_1-\tau_2) G(-\tau_1)G(t-\tau_2) \nonumber\\
&= k_B\mathcal T\int_\R\dd\tau_2\ G(t-\tau_2)\int_\R\dd\tau_1'\ K(-\tau_2-\tau_1') G(\tau_1')\nonumber\\
&=  - k_B\mathcal T\int_\R\dd\tau_2\ G(t-\tau_2)\df{}{\tau}G(-\tau_2).
\end{align}
In $I_1$ and $I_2$ we can substitute $-\tau_1=\tau$ and $-\tau_2=\tau$. In their sum we recognise the formula for integration by parts, which yields
\bgeq
r_V(t) = -k_B\mathcal TG(t+\tau) G(\tau)\Big|_{\tau=0^+}^{\tau=\infty}= k_B\mathcal T G(t) G(0^+)=k_B\mathcal TG(t).
\eeq
Now, for the position process
\bgeq
\E[X(t)^2] =\int_0^t\dd\tau_1\int_0^t\dd\tau_2 \ r_V(\tau_2-\tau_1) = k_B\mathcal T \int_0^t\dd\tau_1\int_0^t\dd\tau_2 \ \big(G(\tau_2-\tau_1)+G(\tau_1-\tau_2)\big).
\eeq
Because of the symmetry between $\tau_1$ and $\tau_2$ the integral is twice the term with $G(\tau_1-\tau_2)$, after substitution $\tau_1-\tau_2=\tau_2'>0$ we get
\bgeq
\E[X(t)^2] =2k_B\mathcal T \int_0^t\dd\tau_1\int_0^t\dd\tau_2 \ G(\tau_1-\tau_2) = 2k_B\mathcal T \int_0^t\dd\tau_1\int_0^{\tau_1}\!\dd\tau_2' \ G(\tau_2').
\eeq
\end{proof}
The above proposition works not only when $F$ is a proper stationary process, but also in the generalized sense when the impulse $J$ has stationary increments and
\bgeq\label{eq:rFcond}
\E\lt|\int\dd J(\tau)\ \phi(\tau)\rt|^2 =\int\dd\tau_1\int\dd\tau_2\ r_F(\tau_2-\tau_1)\phi(\tau_1)\phi(\tau_2)
\eeq
for bounded $\phi$ and some function $r_F$, which may not be a proper covariance function. 

To end this section, let us show some elementary examples, which are applications of the theory presented above.

\textbf{Infinitely short memory: $K(t)=\lambda \delta(t)$}. In this case, the GLE becomes the classical Langevin equation
\bgeq\label{eq:classLang}
\df{}{t} V(t) = -\lambda V(t) + \sqrt{k_B\mathcal T \lambda}\df{}{t}B.
\eeq
This choice of the kernel can be viewed as a limit of the GLE in which the covariance function of $F$ approaches a Dirac delta times $\lambda$, e.g. $r_F(t)=\lambda c 2^{-1}\exp(-c|t|),c\to 0^+$. There is a nuance here in how to view this limit. In the GLE the kernel $K$ is considered to be casual, so when approximating unity only half of the mass of $K$ will be included in the integration. For this reason it may be sensible to the divide term $-\lambda V(t)$ by 2 in the classical Langevin equation. However, we will leave it as it is, because the result depends on the symmetry of the approximating kernels, which depends on the details of the considered physical system. In fact this form of equation can be also derived independently using methods from hydrodynamics (we will use this interpretation in Section \ref{s:OTAppl}) and then $-\lambda V$, without factor $1/2$, can be interpreted as a Stokes force. In this case $\lambda$ is a function of viscosity of the surrounding medium and the shape of the particle.

The Green's function of the classical Langevin equation is
\bgeq
G^\#(s)=\f{1}{s+\lambda},
\eeq
so $G(t)=\exp(-\lambda t)$. The stationary solution of this equation is given by the convolution
\bgeq
V(t) =\sqrt{k_B\mathcal T\lambda}\int_\R \dd B(\tau)\ G(t-\tau)=\sqrt{k_B\mathcal T\lambda}\int_{-\infty}^t\!\!\! \dd B(\tau)\ \e^{-\lambda(t-\tau)},
\eeq
called the Ornstein-Uhlenbeck process. It has the covariance function
\bgeq
r_V(t)=\f{k_B\mathcal T}{2}\e^{-\lambda t}
\eeq
Note that this result differs by the factor $1/2$ from what would be suggested by Proposition \ref{prp:covGreen}, which works only for real-valued kernels.

The msd of $X$ can be calculated easily by direct integration,
\bgeq\label{eq:msdclassLang}
\delta^2_{X}(t)=\f{k_B \mathcal T}{2\lambda} t +\f{k_B\mathcal T}{2\lambda^2}\lt(\e^{-\lambda t}-1\rt).
\eeq
This is the model of normal diffusion.

The Fourier form of the Ornstein-Uhlenbeck process is also simple, the Fourier transform of the Green function  is
\bgeq
\widehat G(\omega)=\f{1}{\I\omega+\lambda} 
\eeq
so the stationary solution can be represented as
\bgeq
V(t)=\int_\R \dd S(\omega)\  \f{1}{\lambda+\I\omega}\e^{\I\omega t},\quad \E|\dd S(\omega)|^2=\dd\omega\ k_B\mathcal T\lambda .
\eeq
Therefore $V$  has power spectral density $s_V(\omega)=k_B\mathcal T\lambda(\lambda^2+\omega^2)^{-1}$.

\textbf{Exponentially decaying memory: $K(t)=b^2\exp(-2at)$.} This choice of  parametrisation by $a,b>0$ is for convenience only, it simplifies most of the formulas. In this model the stochastic force itself is an Ornstein-Uhlenbeck process.

The Laplace transform of the Green's function can be easily obtained
\bgeq\label{eq:expGLE}
G^\#(s)=\f{1}{s+\f{b^2}{s+2A}}=\f{s+2a}{(s+a)^2-(a^2-b^2)}.
\eeq
Its Laplace inverse is a covariance function, which is a mixture of two exponential functions
\begin{align}
r_{V}(t)=\f{1}{2}\lt(1-\f{a}{\sqrt{a^2-b^2}}\rt)\e^{-(a+\sqrt{a^2-b^2})t}
+\f{1}{2}\lt(1+\f{a}{\sqrt{a^2-b^2}}\rt)\e^{-(a-\sqrt{a^2-b^2})t}.
\end{align}
There are 3 cases here: $a>B, a\to b, a<b$. In the first case the covariance is decaying exponentially. In the second it reduces to
\bgeq
r_V(t)=(1+at)\e^{-at},
\eeq
so it exhibits a similar, although slower decay. In the third case $a<b$ the argument of the exponential becomes complex and the memory is oscillatory

\bgeq
r_{V}(t)= \lt(\cos\big(\sqrt{b^2-a^2}t\big)+\f{a}{\sqrt{b^2-b^2}}\sin\big(\sqrt{b^2-a^2}t\big)\rt)\e^{-at}.
\eeq
It is exactly a trigonometric oscillation truncated by factor $\exp(-at)$.

In any case the msd is given by the formula

\begin{align}
&\delta^2_{X}(t) = 4\f{a}{b^2} t -\f{8a^2}{b^4}+2b^2 \nonumber\\
&+\f{1}{\sqrt{a^2-b^2}}\e^{-at}\lt(\f{\sqrt{a^2-b^2}-a}{(a+\sqrt{a^2-b^2})^2}\e^{\sqrt{a^2-b^2}t}+\f{\sqrt{a^2-b^2}+a}{(b-\sqrt{a^2-b^2})^2}\e^{-\sqrt{a^2-b^2}t}\rt).
\end{align}
As we can see this is again a model of normal diffusion. The spectral representation of this model will be considered in Chapter \ref{ch:discTDyn} below Eq. \Ref{eq:expGLeF}.

\textbf{Power-law memory: $K(t)=z\Gamma(2h-1)^{-1} |t|^{2h-2}$}. This is a case when the force is, up to a scale, a distributional derivative of fractional Brownian motion $B_h, 0<h<1$ \citep{intFBM}. The factor $1/\Gamma(2h-1)$ is added for convenience.  The Green's function of this GLE is given by
\bgeq
G^\#(s)=\f{1}{s+z s^{1-2h}}=\f{s^{2h-1}}{z+s^{2h}}.
\eeq
In the last form we recognise the Laplace transform of a function $E_{2h}(-zt^{2h})$, where $E_{2h}$ is a function from the Mittag-Leffler class defined by its Taylor series \citep{haubold}
\bgeq
E_{\alpha,\beta}(t)\defeq\sum_{k=0}^\infty \f{t^k}{\Gamma(\alpha k+\beta)},\quad E_\alpha\defeq E_{\alpha,1}.
\eeq

The asymptotics of the conditional covariance can be derived from Tauberian theorem or analyzing the Mittag-Leffler function directly \citep{haubold, gorenflo}
\bgeq\label{eq:MLasympt}
r_{V}(t)=k_B\mathcal TG(t)\sim \f{k_B\mathcal T}{z\Gamma(1-2h)}t^{-2h},\quad t\to\infty.
\eeq

Similar reasoning yields the properties of the position process. Eq.\Ref{eq:msdGen} describes the msd as a double integral of the Green's function, so it has the simple form in Laplace space
\bgeq
\delta^{2\#}_{X}(s)= 2s^{-2}r_{V}^\#(s)=k_B\mathcal T\f{s^{2h-3}}{z+s^{2h}}.
\eeq
The inverse transform can be found using tables of two-parameter Mittag-Leffler function, which also determines its asymptotics
\bgeq
\delta^{2}_{X}(t)=k_B\mathcal Tt^2E_{2h,3}(-zt^{2h})\sim \f{k_B\mathcal T}{z\Gamma(3-2h)}t^{2-2h},\quad t\to\infty.
\eeq
This is a model of anomalous diffusion. It is noticeable that when the stochastic force becomes more persistent and superdiffusive, the solution becomes more subdiffusive and antipersistent. This is caused by the minus sign of the convolution $-\int_{-\infty}^t \dd\tau \ K(\tau)V(t-\tau)$. When $r_F=k_B\mathcal T K$ has thicker tails, the friction becomes larger, which causes the changes of velocity to be more negatively correlated to the past values of $V$. This effect outweighs the correlations introduced in the GLE by the stochastic force itself.

\chapter{Langevin dynamics in discrete time}\label{ch:discTDyn}
\newif\ifmain

Physical processes are naturally described as dynamical systems parametrised by a continuous time parameter. But experimental data is always discretised and can only mirror the continuous dynamics in the so-called infill asymptotics, that is when the frequency of measurements increases to infinity. This distinction between statistical and physical models is by itself an interesting and practically important subject.

For the continuous-time case the most common models are those based on differential equations. In the discrete-case case the recursive ARMA equations can be viewed as their counterpart. The most popular discrete-time linear models of dynamics are ARMA, ARFIMA and their various variants \citep{BD,BJ}, which are common in modelling of financial and econometric data \citep{mills,enders}, which resulted in the 2003 Nobel Prize in Economic Sciences for C. W. J. Granger and R. Engel. In recent years it also started being used in analysis of the diffusion phenomena \citep{ARFIMunif,ARFIMAalg}.  The study of relations between these continuous- and discrete-time  classes of models has a long history, the first results were obtained by A. Philips in 60s \citep{phillips} and the area is still being developed \citep{BrockwellCAR}.

In this chapter we study the behaviour of sampled solutions of the wide range of Langevin equations and show how they can be interpreted as ARMA time series. Using this correspondence we provide exact formulas which relate the physical constants and the coefficients of the observed ARMA processes, study their covariance function, spectral density and propose methods of statistical verification \citep{slezakWeron}.

In the last section we use our methods in practice,  analysing optical tweezers data, which are an example of the system modelled by the classical Langevin equation. We map the continuous time physical dynamics to a discrete time ARMA model and modify it, taking into account distortions caused by the CCD camera using during measurements \citep{slezakDrobczynski,drobczynskiSlezak}. We achieve very good agreement between our discrete-time model and real data, which allows for separation of physical dynamics and the external distortions.

\section{Basic intuitions}\label{s:basicInt}
The future evolution of the stationary GLE
\bgeq\label{eq:biGLE}
\df{}{t}X = -\int_{-\infty}^t\!\!\!\dd\tau\ X(\tau)K(t-\tau) + F,
\eeq
depends on both the derivative at time $t$, the force at time $t$ and all past values $X(s),s\le t$  so it seems that the severe loss of information after sampling procedure is inevitable. Later in this section we show rather counterintuitively, that it is not always the case. Some straightforward observations can be made immediately.

We will consider only the time series obtained as the values of the process sampled with constant rate $X_k\defeq X(k\Delta t)$. If we integrate Eq. \Ref{eq:biGLE} from $(k-1)\Delta t$ to $k\Delta t$ we obtain
\begin{align}
X_k-X_{k-1} &= -\int_{-\infty}^{(k-1)\Delta t}\!\!\dd \tau\ (\tilde K(k\Delta t+\tau)-\tilde K((k-1)\Delta t+\tau))X(\tau)\nonumber\\
&+\int_{(k-1)\Delta t}^{k\Delta t}\!\!\dd s\ \tilde K(k\Delta t+\tau) X(\tau),
\end{align}
where
\bgeq\label{eq:GLEStrange}
F_k = \int_{(k-1)\Delta t}^{k\Delta t}\!\!\! \dd \tau\ F(\tau),\quad  \tilde K(t) = \int_0^t\dd \tau\ K(\tau).
\eeq
The sample $(X_k)_k$ does not depend on all values of $F$, but only on discretised $F_k$ from above. Moreover, the expression
\bgeq\label{eq:GLEDiscNaive}
X_k-X_{k-1} \approx \sum_{j=1}^{\infty} \kappa_j X_{k-j} + F_k
\eeq
is an approximation of the convolution integral in \Ref{eq:biGLE} by a discrete sum, or, more strictly, the integral in Eq. \Ref{eq:GLEStrange}, and the dependence on $F_k$ is exact. This is nothing but the well known Euler scheme, the base of most of numerical simulations. We will not spend time considering these approximations, but rather analyse cases in which we can determine the exact dynamics of the sampled process. However, Eq. \Ref{eq:GLEDiscNaive} naturally introduces the most important class of linear discrete models, i.e. ARMA.

In Eq. \Ref{eq:GLEDiscNaive} the present value $X_k$ depends linearly on the past values $X_{k-j}$, such a behaviour is called autoregressive.
\bg{dfn}
We call the stationary time series $(X_k)_k$ an autoregressive process of order $p$, in short AR(p), if it fulfils the recursive relation
\bgeq\label{eq:ARdef}
X_k-\sum_{j=1}^p\phi_j X_{k-j}= \xi_k,
\eeq
where $(\xi_k)_k$ is discrete white noise.
\end{dfn}
As a side note, there also exist autoregressive non-stationary processes defined only for $k\ge 0$, but the cases most often seen in applications can be reduced to the stationary ARMA through differencing, i.e considering $X_{k+1}-X_k$. Examples will be shown in Section \ref{s:physAppl}.

AR processes are of great statistical importance. The first reason is that their covariance function is a sum of exponential decays, which is a case often met in practice. To see that consider $k>p$, multiply Eq. \Ref{eq:ARdef} by $X_0$ and calculate the expected value to obtain the homogeneous equation
\bgeq
r_X(k)-\sum_{j=1}^p\phi_j r_X(k-j)=0.
\eeq
Using the backward shift operator $\mathcal S_{-1} f(n) = f(n-1)$ the above formula can be written as a polynomial of $\mathcal S_{-1}$ acting on the function $r_X$
\bgeq
\lt(1-\sum_{j=1}^p\phi_j\mathcal S_{-1}\rt)r_X=\prod_l\big(a_l-\mathcal S_{-1}\big)^{n_l}r_X=0,
\eeq
where $a_l$ are roots of the polynomial $x\mapsto 1-\sum_{j=1}^p x^j$. The general solution is a linear combination of solutions of functions $(a_l-\mathcal S_{-1})^{n_l}$. For distinct roots, that is $n_l=1$ these are geometrical decays $k\mapsto a_k^l$. For repeated roots the differentiation of the function $1/(1-x)$  yields solutions of the form $k\mapsto l a_k^{l-1}, k\mapsto l(l-1)a_k^{l-2}$ and so on. The coefficients of the specific solution that we need are determined by $p$ linear inhomogeneous equations obtained for $0\le k <p$.

The second reason for using AR(p) model is that, when the first $p+1$ values of the covariance function are fixed, the AR(p) time series with this covariance maximizes the entropy \citep{frankeEntr}. Therefore it may be considered a model with the minimal statistical assumptions, given the short-range values of covariance.

The above remarks are important mostly when AR is an effective model which approximates the behaviour of the system. In our case there will be an even more direct reason: after the sampling procedure the obtained time series can be exactly the AR or ARMA process. Let us show a specific example: the classical Langevin equation \Ref{eq:classLang}
\bgeq
\dd X = -\dd t \lambda X +D\dd B
\eeq
with its stationary solution, the Ornstein-Uhlenbeck process
\bgeq
X(t) = D\int_{-\infty}^t\!\!\!\dd B(s)\ \e^{-\lambda(t-s)}.
\eeq
We propose the following:

\bg{prp}\label{prp:sampledOU}
The sampled Ornstein-Uhlenbeck process is an AR(1) time series.
\end{prp}
This is a very simple statement if we think in the language of the covariance. The covariance of the Ornstein-Uhlenbeck process is an exponential decay. The covariance of the AR(1) process is a geometrical decay. The sampled exponential decay is a geometrical decay. Nonetheless, we present a different proof, which will allow for a simple generalisation of the above proposition.
\bg{proof} We multiply the Langevin equation by the integrating factor $\e^{\lambda t}$ and obtain
\bgeq
\dd \lt(X(t)\e^{\lambda t}\rt) = D\dd B(t)\e^{\lambda t}.
\eeq
Now we integrate only from $(k-1)\Delta t$ to $k\Delta t$
\bgeq
X_k \e^{\lambda k\Delta t}-X_{k-1}\e^{\lambda (k-1)\Delta t} = D\int_{(k-1)\Delta t}^{k\Delta t}\!\!\!\dd B(s)\ \e^{\lambda t}.
\eeq
After dividing by $\e^{\lambda k \Delta t}$, this yields 
\bgeq
X_k-\phi_1X_{k-1} = \xi_k,
\eeq
where
\bgeq
\phi_1=\e^{-\lambda\Delta t},\quad \xi_k=D\int_{(k-1)\Delta t}^{k\Delta t}\!\!\dd B(s)\ e^{-\lambda(k\Delta t- t)}.
\eeq
As different $\xi_k$ use values $\dd B(s)$ from different intervals $((k-1)\Delta t, k\Delta t)$, they are independent, moreover, by It\={o} isometry,
\bgeq
\E\big[\xi_k^2\big] = \f{D^2}{2\lambda}\lt(1-\e^{-2\lambda\Delta t}\rt)
\eeq
and so, $(\xi_k)_k$ is a white noise.
\end{proof}
An exemplary trajectory of the Ornstein-Uhlenbeck and the sampled AR(1) time series are shown in Fig \ref{f:sampl}. Note that if we approximated the solution of the Langevin equation using the naive Euler scheme, we would also obtain an AR(1) process, but with different coefficients $\phi_1=1-\lambda\Delta t, \E\big[\xi_k^2\big]=D^2\Delta t$ which are linear approximations of the true values.

\begin{figure}[h]\centering
\includegraphics[width=12cm]{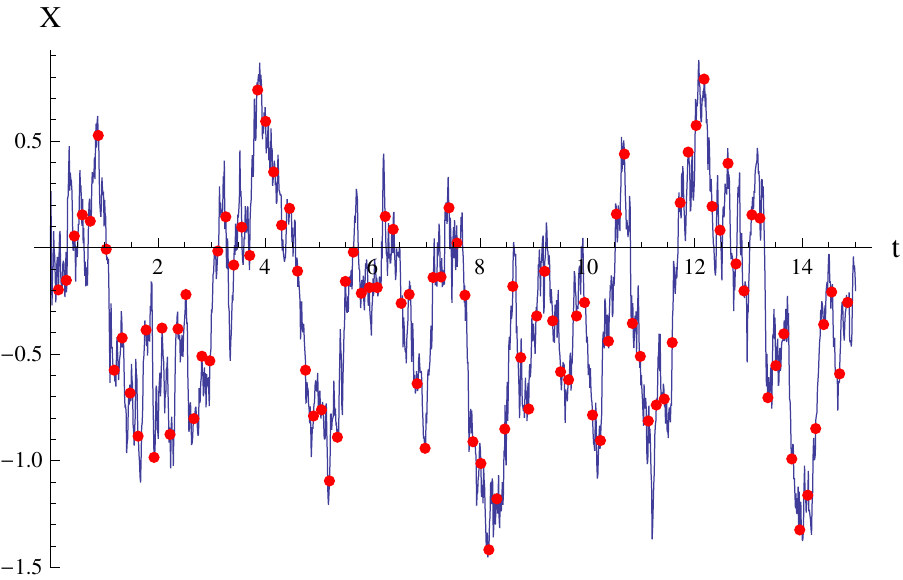}
\caption{Continuous-time orignal process (blue line) together with the sampled process (red dots).}\label{f:sampl}
\end{figure}

Our above considerations were visibly limited by the assumption that the stochastic force was required to be white. In practice, when we increase the sampling frequency during the experiment, often some previously unseen correlation comes into view. This situation corresponds to a model in which $r_X(t)=0$ for $t>\Delta t$. For time series this behaviour is characteristic for the moving average processes (MA).
\bg{dfn} We call the stationary time series $(X_k)_k$ a moving average process of order $q$, in short MA(q), if it fulfils the relation
\bgeq
X_k = \xi_k+\sum_{j=1}^q \theta_j\xi_{k-j},
\eeq
where $(\xi_k)_k$ is discrete white noise.
\end{dfn}

Directly from the above definition we get that the covariance function of the MA(q) time series is
\bgeq\label{eq:MACov}
r_X(k) = \bg{cases}
\sum\limits_{i=0}^{q-k}\theta_i\theta_{i+k}, & k\le q;\\
0, & k>q,
\end{cases}
\eeq
where we additionally denoted $\theta_0\defeq1$. Indeed, this covariance function has finite support.

Immediately the question appears if any finite-range time-series can be expressed as some MA(q). In other words we ask if, given finite $K$ and values $r_X(k), k\le K$ there exist a set of coefficients $\theta_j$ such that \Ref{eq:MACov} is fulfilled. This is a non-linear set of equations with no obvious way of solving them, so the question is non-trivial, but the answer is yes.

\bg{prp}\label{prp:finRange}
Any sampled finite-range memory stationary zero mean Gaussian process is an MA(q) process for $q\le \lceil R\rceil$, where $R$ is the memory range.
\end{prp}

\bg{proof} We use a concept of innovations, very important in the signal theory. Innovation $\xi_k$ is defined as a new information brought into system by the value $X_k$ calculated using the Hilbert space projection onto a linear subspace generated by the variables $\{X_j\}_{j<k}$. Proposition \ref{prp:gaussProj} shows that any such projection is the same as the conditional expected value, so we can define $\xi_k$ as
\bgeq\label{eq:innSeries}
\xi_k = X_k-\E[X_k|\{X_j\}_{j<k}].
\eeq
In other words it is a part of $X_k$ orthogonal to the history of the process $\{X_j\}_{j<k}$. Directly from the definition, $(\xi_k)_k$ is a stationary time series, as it is a linear transformation of a stationary time series. Moreover the values $\xi_i,\xi_j$ are independent because the one with the larger index, let it be $j$, is orthogonal to all past values $\{X_k\}_{k<j}$, in particular $X_i$ and $\E[X_i|\{X_l\}_{l<i}]$. So $(\xi_k)_k$ is a white noise series.

Now we will show that it is exactly the white noise from the definition of the MA(q) process. Because $X_k = \xi_k+\E[X_k|\{X_j\}_{j<k}]$ we can write
\bg{align}
\spn\{X_{k-1},X_{k-2},\ldots\} &= \spn\{\xi_{k-1},X_{k-2},X_{k-3}\ldots\}=\spn\{\xi_{k-1},\xi_{k-2},X_{k-3},\ldots\}\nonumber\\
& = \spn\{\xi_{k-1},\ldots,\xi_{k-q},X_{k-q-1},\ldots\}.
\end{align}
The families $\{\xi_{k-j}\}_{j=1}^p$ and $\{X_{k-j}\}_{j>q}$ are independent from the definition of $\xi_k$. Moreover, $X_k$ is independent from $\{X_{k-j}\}_{j>q}$. The orthogonality of the corresponding subspaces show that
\bgeq
\E[X_k|\{X_j\}_{j<k}] = \E[X_k|\xi_{k-1},\ldots,\xi_{k-q},X_{k-q-1},\ldots] = \E[X_k|\xi_{k-1},\ldots,\xi_{k-q}].
\eeq
The last projection is $\sum_{j=1}^q \theta_j \xi_{k-j}$ and
\bgeq
X_k = \xi_k +\sum_{j=1}^q \theta_j \xi_{k-j}.
\eeq

\end{proof}

The proof unfortunately is not constructive in a sense that it relates the process and its MA representation through $\E[\bd\cdot|\{X_j\}_{j<k}]$, which is a rather abstract quantity. What is even worse, they are not uniquely determined, which becomes apparent when we consider their spectral density (See Section \ref{s:physAppl}, Proposition \ref{prp:ARMApsd} and comments below). However given one choice of coefficients, all others can be determined, so we can treat them as if defined up to a gauge. In practice the coefficients which correspond to the innovation series \Ref{s:physAppl} are estimated using the recursive Durbin-Levinson algorithm. For simple cases some solutions can be calculated directly, which will be shown later in this section.

With the above propositions in mind, we are ready to formulate the main point of this subsection: the correspondence between an ARMA time series and the classical Langevin equation governed by a force with finite-range memory. First, we define
\bg{dfn} We call a stationary time series $(X_k)_k$ autoregressive moving average process of order (p,q), in short ARMA(p,q) if it is an AR(p) process governed by an MA(q). In other words if it fulfils the recursive equation
\bgeq
X_k-\sum_{j=1}^p\phi_j X_{k-j} = \xi_k +\sum_{j=1}^q\theta_j \xi_{k-j}
\eeq
for white noise $(\xi_k)_k$ and set of coefficients $\phi_j,\theta_j$.
\end{dfn}
Under this definition of course AR(p)$\equiv$ ARMA(p,0) and MA(q)$\equiv$ARMA(0,q).
\bg{thm} If the classical Langevin equation
\bgeq
\dd X = -\dd t \lambda X +\dd F,
\eeq 
understood as stochastic equation in the mean-square sense, is governed by stationary increments of a Gaussian process $\dd F$, which have a finite-range $R$ of the memory, then the sampled process is ARMA(1,q) with $q\le R/\Delta t + 1$.
\end{thm}

\bg{proof} It is a consequence of Prop. \ref{prp:sampledOU} and Prop.\ref{prp:finRange}. From Prop. \ref{prp:sampledOU} the sampled process has form
\bgeq
X_k = \phi_1X_{k-1}+\xi_k
\eeq
and the noise $\xi_k$ is given by the integral
\bgeq
\xi_k=\int_{(k-1)\Delta t}^{k\Delta t}\!\!\dd F(s)\ e^{-\lambda(k\Delta t- t)},
\eeq
which cannot have a memory range longer than $k = \lceil R/\Delta t\rceil$ as $\xi_k$ uses only values of $\dd F$ from one interval of length $\Delta t$. 
\end{proof}
\section{Differential systems in discrete time}
In general we would like to study more complex equations than the classical equation, e.g. the full Langevin equation
\bgeq\label{eq:langFull}
m\df{^2}{t^2}X = -\kappa X-\beta\df{}{t}X + F,
\eeq
which is a good model of normal diffusion in short, but not extremely short, time scales \citep{BmotShortT}. It is a second order stochastic differential equation, which can be interpreted as a first order vector equation for the variable of position $X$ and momentum $m V$. The dynamics itself is deterministic and linear, but the system is disturbed by the stochastic noise $F$, as shown in Fig. \ref{f:phase}
\begin{figure}[h]\centering
\includegraphics[width=12cm]{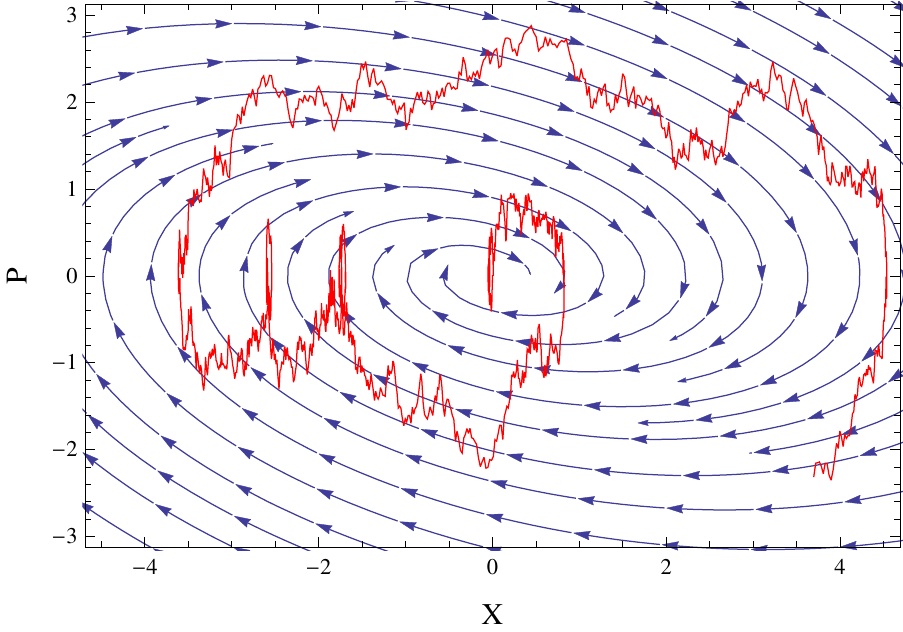}
\caption{Phase plot (blue lines with arrows) for coordinates $X, P$ and stochastic solution of the Langevin equation \Ref{eq:langFull} (red line), $m=1,\kappa=1/4,\beta=1/4$.}\label{f:phase}
\end{figure}
Motivated by this example, we will consider a more general class of a linear stochastic system with an $N$-dimensional state vector $\bd X=[X^1,X^2,\ldots,X^N]^\mathrm{T}$, parametrised by discrete or continuous time parameter, as before.

\bg{dfn}[VAR,VMA,VARMA] We call a stationary vector time series $(\bd X_k)_k$ a vector ARMA, in short VARMA(p,q), process if it fulfils the recursive vector equation
\bgeq
\bd X_k-\sum_{j=1}^p\Phi_j \bd X_{k-j} = \bd \xi_k +\sum_{j=1}^q\Theta_j \bd\xi_{k-j},
\eeq
where $\Phi_j$ and $\Theta_j$ are $N\times N$ matrices and $(\bd \xi_k)_k$ is vector white noise, i.e. series of i.i.d. Gaussian vectors. As before VAR(p)=VARMA(p,0) and VMA(q)=VARMA(0,q).
\end{dfn}
Proposition \ref{prp:sampledOU} describing the relation between Ornstein-Uhlenbeck process and AR(1) time series has a straightforward analogue in the vector case.

\bg{lem} The sampled stationary solution of the linear stochastic vector equation
\bgeq\label{eq:Ssol}
\dd\bd X(t)= \dd t\Lambda\bd X(t)+ \dd \bd F(t)
\eeq
fulfils the matrix recursive equation
\bgeq\label{eq:Sn}
\bd X_{k}=E\bd X_{k-1}+\bd\xi_{k},\quad E=\e^{\Lambda\Delta t},\bd\xi_k=\int_{(k-1)\Delta t}^{k\Delta t}\!\!\!\! \dd \bd F(s)\ \e^{\Lambda(k\Delta t-s)}
\eeq
and, for white noise $\dd\bd F$, is a VAR(1) process.
\end{lem}
The proof is practically identical as the proof of Proposition \ref{prp:sampledOU}, it only uses matrices instead of scalars, so we skip it. From the point of view of dynamical system theory the above lemma states that for the continuous-time system with the generator $A$, the embedded discrete-time subsystem has the generator $E=\exp(\Lambda\Delta t)$. Because of the above correspondence, the statistical methods available for VAR(1) model \citep{pfaff} can be used in physical applications. Using estimators for matrix $E$ and taking the matrix logarithm we obtain estimates for the underlying the matrix $\Lambda=\ln(E)/\Delta t$. The properties of the force $\dd \bd F(t)$ can be studied through analysis of $\bd\xi_k$, which can be estimated as $\bd\xi_k=\bd X_{k}-E\bd X_{k-1}$.

But, in practical applications, having access to the whole vector $\bd X$ is rarely the case. For example, in diffusion phenomena the position $X$ is often easily measured, but the velocity $V$ can be only approximated, often as $V_k = (X_k-X_{k-1})/\Delta t$ or similar quantity. In electrical circuits on the contrary, the current $I$ is easier to observe than the electrical charge $Q$ and the latter requires using less precise methods or some kind of numerical integration of the current. For this reason it would be very practical to determine the dynamical model of any given degree of freedom. The evolution of the separate coordinates is linearly coupled through matrix $E$; the proper decoupling procedure requires the use of a particular theorem from algebra.

\bg{thm}[Cayley-Hamilton] If an $N\times N$ matrix $E$ has the characteristic equation
\bgeq
\det(E-\lambda I) = \lambda^N-\sum_{j=1}^{N}\phi_j\lambda^{N-j} = 0,
\eeq
then the matrix itself fulfils the polynomial equation with the same coefficients
\bgeq\label{eq:charCH}
E^N-\sum_{j=1}^{N-1}\phi_jE^{N} = 0.
\eeq
\end{thm}
This interesting result provides means to remove the action of $E$ on the vector $\bd X$ and is a crucial ingredient of the following theorem.

\bg{thm}\label{thm:disc}
Let $\bd X$ be a stationary solution of Eq. \Ref{eq:Ssol}, where $\dd\bd F(t)$ is a finite-range memory stationary process with memory range $R\Delta t, R\in\mathbb N$. Then any one dimensional projection projection $ \bd X\cdot \bd x, \bd x\in {\R^N}$ sampled with $\Delta t$ is an ARMA(N,N+R-1) process with AR coefficients given by the characteristic equation of matrix $E$, formula \Ref{eq:charCH}.
\end{thm}
\bg{proof}

We fix $k$ and recursively use the relation $\bd X_r = E\bd X_{r-1}+ \xi_r$ to  express the variables $\bd X_{k-j}$ as functions of $\bd X_{k-N}$ starting  from $\bd X_{k-N}$
\begin{align}
\bd X_{k-N\phantom{+1}}&=\phantom{E^2}\bd X_{k-N}\nonumber\\
\bd X_{k-N+1}&=E\phantom{^2}\bd X_{k-N}+\phantom{E^2}\bd \xi_{k-N+1}\nonumber\\
\bd X_{k-N+2}&=E^2\bd X_{k-N}+E\phantom{^2}\bd \xi_{k-N+1}+\phantom{E}\bd \xi_{k-N+2}\nonumber\\
\bd X_{k-N+3}&=E^3\bd X_{k-N}+E^2\bd \xi_{k-N+1}+E\bd \xi_{k-N+2}+\bd\xi_{k-N+3}\nonumber\\
\phantom{\bd X_{k-N+2}}&\ \ \vdots\nonumber\\
\bd X_{k\phantom{-N+1}} &= E^{N}\bd X_{k-N}+\sum_{j=0}^{N-1}E^j\bd\xi_{k-j},
\end{align}
Now we multiply the above equations by the coefficients from Eq. \Ref{eq:charCH} such that each $\bd X_{k-i}$ on the left is multiplied by $\phi_i$. Using this procedure we obtain the terms  $\phi_iE^{N-i}\bd X_{k-N}$ on the right. After subtracting all $\phi_i\bd X_{k-i}$ from $\bd X_k$ we get a formula which allows us to use the Cayley-Hamilton theorem

\bgeq\label{eq:Snvorg}
\bd X_k - \sum_{i=1}^N \phi_j\bd X_{k-i} = \lt(E^N-\sum_{i=1}^N\phi_i E^{N-i}\rt)\bd X_{k-N} + \bd\eta_k = \bd\eta_k.
\eeq
The transformed noise $(\bd\eta_k)_k$ is given by the complicated, but explicit formula
\bgeq\label{eq:discNoiseExpl}
\bd{\eta}_k=\bd\xi_{k}+\sum_{l=1}^{N-1}R_{l}\bd\xi_{k-l},\quad R_l = E^l-\sum_{j=1}^l\phi_jE^{l-j}.
\eeq
After taking the projection, the left hand side of \Ref{eq:Snvorg} is clearly an AR(N) equation

\bgeq
\bd X_k\cdot\bd x - \sum_{i=1}^N \phi_j\bd X_{k-i}\cdot\bd x  =\bd\eta_k\cdot\bd x
\eeq

For the right hand side, any value of the noise $(\bd\eta_k)_k$ is a linear combination of the last $N-1$ values of $(\bd\xi_k)_k$. Therefore, $(\bd\eta_k)_k$ is a finite-range memory process with zero covariance function after the first $N+R-1$ values. Using Proposition \ref{prp:finRange}, $\bd\eta_k\cdot\bd x$ is an MA($N+R-1$) process. This concludes the proof.
\end{proof}
Using Eq. \Ref{eq:charCH} the AR($N$) coefficients can be given explicitly
\bgeq
\phi_k=(-1)^{k+1}\sum_{D_k}\e^{\Delta t\sum_{i\in D_k}\nu_i},
\eeq
where $\nu_i$ are eigenvalues of the evolution matrix $A$ and $D_k$ denotes the family of all $k$-element subsets of the set $\{1,2,\ldots, N\}$.
These numbers do not depend on the direction of the projection $\bd x$ or the exact form of $(\bd\eta_k)_k$'s memory. The global, deterministic dynamics of the whole state $\bd X$ is reflected in the AR coefficients. The random dynamics and specific choice of the observed quantity affects only the MA coefficients. These can be calculated in two steps, first calculating the covariance of the noise from Eq. \Ref{eq:discNoiseExpl} and then solving the formula for the covariance \Ref{eq:MACov}. The solutions are not unique, see the remark after Proposition \ref{prp:ARMApsd}. For systems with a small number of degrees of freedom these solutions are explicit, but often complicated.

The above theorem can be directly used to study the solutions of higher-order stochastic equations
\bg{prp}
Let $X$ be a stationary solution of the linear stochastic equation
\bgeq
a_q\df{^q}{t^q}X+a_{q-1}\df{^{q-1}}{t^{q-1}}X +\ldots a_0X=f,\quad a_q\neq 0
\eeq
where $f$ is a Gaussian stochastic stationary force. Then:
\bg{itemize}
\item[i)] If $f$ has a covariance with range $R\Delta t$ then the sample $(X_k)_k$ is ARMA($q,q+R-1$).
\item[ii)] If $f$ has a covariance which is a sum of $K$ exponential functions then the sample $(X_k)_k$ is ARMA($q+K,q+K-1$).
\end{itemize}
\end{prp}
\bg{proof}
The above differential equation can be expressed as a $q$ dimensional system of equations with $X_1=X$, $X_2=\dd/\dd t X,\ldots$ 
\begin{align}
\dd X_{q-1}&=\dd t\lt(\f{a_{q-1}}{q_q}X_{q-2}+\ldots \f{a_0}{a_q}X_1\rt) + \f{1}{a_q}\dd t f\nonumber\\
\dd X_{q-2}&=\dd t X_{q-1}\nonumber\\
\vdots\nonumber\\
\dd X_2&=\dd t X_1,
\end{align}
Point i) follows immediately. As for ii), a process $f$ with a covariance function given by $K$ exponents can be decomposed as a sum of $K$ Ornstein-Uhlenbeck processes $Y_1,\ldots, Y_K$. Therefore the whole systems can be described by $q+K$ equations
\bg{align}
\dd X_{q-1}&=\dd t\lt(\f{a_{q-1}}{q_q}X_{q-2}+\ldots \f{a_0}{a_q}X_1+\f{1}{a_q}Y_1+\ldots + \f{1}{a_q} Y_K\rt)\nonumber\\
\dd X_{q-2}&=\dd t X_{q-1}\nonumber\\
\vdots\nonumber\\
\dd X_2&=\dd t X_1\nonumber\\
\dd Y_1 &= -\dd t\lambda_1 Y_1+D_1\dd B_1\nonumber\\
\vdots \nonumber\\
\dd Y_K&=-\dd t \lambda_K Y_K + D_K\dd B_K.
\end{align}
Because all noises $\dd B_k$ are white, the result follows.
\end{proof}
\section{Behaviour of sampled Langevin equations}\label{s:physAppl}
The results from the last section have immediate applications. Many physical systems are commonly modelled using stochastic linear equations, the most important ones related to the Langevin and generalized Langevin equation. Few important examples are given below.\\

\textbf{The classical Langevin equation.} The case when the driven noise is white was already considered in Proposition \ref{prp:sampledOU}, but different memory models can also be studied. One of the most important ones is when the driving force is the Ornstein-Uhlenbeck process, i.e. the coordinate fulfils a set of equations
\begin{align}
\dd X &= -\dd t\lambda_1 X +\dd t F,\nonumber\\
\dd F &= -\dd t \lambda_2 F + \dd B.
\end{align}
The eigenvalues of the evolution matrix are very simple in this case
\bgeq
\nu_1 = -\lambda_1,\quad \nu_2 = -\lambda_2.
\eeq
The sampled system is ARMA(2,1) with  AR coefficients equal to
\begin{align}
\phi_1&=\e^{-\Delta t\lambda_1}+\e^{-\Delta t\lambda_2}\nonumber,\\
\phi_2&=-\e^{-\Delta t(\lambda_2+\lambda_1)}.
\end{align}
The inverse relations, useful for estimation, are
\bgeq
\lambda_{1,2}=-\f{1}{\Delta t}\ln\lt(\f{1}{2}\lt(\phi_1\pm\sqrt{\phi_1^2+4\phi_2}\rt)\rt).
\eeq
An interesting property, visible from the above formulas, is that $\lambda_1$ and $\lambda_2$ are not physically distinguishable from AR coefficients. It is a consequence of the fact that they depend only on the content of the spectrum, not the order of eigenvalues. To break this symmetry the additional information contained in MA(1) coefficient must be used.\\

\textbf{The full form of classical Langevin equation.} In the short time scales the motion of particle diffusing in water can be modelled by equation
\bgeq
m\df{^2}{t^2}X = -\kappa X-\beta\df{}{t}X + F,
\eeq
governed by the finite-range memory force $F$. It was mentioned as a motivation at the beginning of the last section. The equation for the charge $Q$ in a linear RLC circuit with stochastic electromotive force $\mathcal E$ has the same form
\bgeq
L\df{^2}{t^2}Q = -\f{1}{C} Q-R\df{}{t}Q + \mathcal E.
\eeq
The eigenvalues of the corresponding evolution matrix for both equations, up to a change of letters, are
\bgeq
\nu_{1,2}=-\frac{\beta}{2m}\pm\sqrt{\left(\frac{\beta}{2m}\right)^2-\frac{\kappa}{m}}.
\eeq
The sampled system is ARMA(2,1) with  AR coefficients equal to
\begin{align}\label{eq:langARcoef}
\phi_1&=2\exp\left(-\Delta t\frac{\beta}{2m}\right)\cosh\lt(\Delta t\sqrt{\left(\frac{\beta}{2m}\right)^2-\frac{\kappa}{m}}\rt)\nonumber,\\
\phi_2&=-\exp\left(-\Delta t\frac{\beta}{m}\right).
\end{align}
The inverse relations are
\begin{align}\label{eq:langARcoefInv}
\f{\beta}{m}&=-\f{1}{\Delta t}\ln(-\phi_2)\nonumber,\\
\f{\kappa}{m} &= -\f{1}{2\Delta t}\ln(-\phi_2)-\f{1}{\Delta t^2}\lt(\cosh^{-1}\lt(\f{\phi_1}{2\phi_2^2}\rt)\rt)^2.
\end{align}
The mass $m$ or inductance $L$ can be determined exclusively using the variance of the process, as the MA(1) coefficient also depends only on $\beta/m$ and $\kappa/m$. It is often hard to determine from the experimental data, because it depends on the units and scale of the system.\\

\textbf{RLC circuit with leakage. } The loss of charge is determined by the constant $G$. In more realistic model the charge may escape from the system. In this situation the equation for the current $I$ changes and the whole system is described by
\begin{align}
\dd Q &= \dd t I - \dd t \f{G}{C}Q\nonumber\\
\dd I &= -\dd t \f{1}{LC}Q-\dd t \f{R}{L} I +\dd \mathcal E \f{1}{L}
\end{align} 

The eigenvalues have a similar form as for the simple RLC circuit
\bgeq
\nu_{1,2}=-\lt(\f{G}{C}+\frac{R}{L}\rt)\pm\sqrt{\lt(\f{1}{2}\lt(\f{G}{C}+\f{R}{L}\rt)\rt)^2-\frac{1}{CL}},
\eeq
therefore the formulas \Ref{eq:langARcoef} and \Ref{eq:langARcoefInv} for the AR(2) coefficients can be used after a simple substitution $\frac{R}{L}+\f{G}{C}\to\f{\beta}{m}$ and $\f{1}{LC}\to\f{\kappa}{m}$. These two numbers are not sufficient to determine $R,C$ and $G$, the additional information contained in the MA(1) process must be used during the estimation.\\

\textbf{Generalised Langevin equation with exponential-type kernel}. Precisely, the kernel and the covariance function of the force can be a finite sum of exponents multiplied by polynomials and sine or cosine functions. This example is a realisation of our motivation stated at the beginning of Section \Ref{s:basicInt}. However, the proper procedure that translates the sampled solution of the GLE into an ARMA time series is much more complicated than the simple Euler discretisation would suggest.

To see this we need to show that the solution of the GLE as stated above can be described in terms of a system of linear equations with constants coefficients, which will allow us to use Theorem \ref{thm:disc}.
\bg{thm}
Let $X$ be a stationary solution of the GLE
\bgeq
\df{}{t}X = -\int_{-\infty}^t\!\!\!\dd\tau\ X(\tau)K(t-\tau) + F,
\eeq
where the kernel $K$ and the covariance function $r_F$ are finite sums of exponentials, which can be multiplied by polynomials and sine or cosine functions. In such case $X$ is equivalent to $\bd Y\cdot \bd q$, where $X$ is a solution of a system of linear stochastic equations with constant coefficients, and $\bd q$ come fixed vector.
\end{thm}
\textit{Proof.}
The Fourier transform of $\sin(v t)\e^{-\lambda t}$ on $\R_+$ is $v\big(v^2+(\lambda-i\omega)^2\big)^{-1}$. The Fourier transforms of $K$ and $r_F$ are sums of derivatives and polynomial times this quantity, so they are rational functions. From our considerations in Section \ref{s:gaussProc} and \ref{s:elSol} we know that the power spectral density of the stationary solution is
\bgeq\label{eq:psdRat}
s_X(\omega) =\f{s_F}{|\I\omega +\widehat K|^2}=\f{P(\omega)}{Q(\omega)}
\eeq
i.e., as a ratio of two rational functions, it is also a rational function and it may be assumed that it is a ratio of the polynomials $P$ and $Q$ with no common roots. Moreover, this function must be even, like any power spectral density. It is even if and only if  $P$ and $Q$ are even, because $P(\omega)Q(-\omega)=P(-\omega)Q(\omega)$ implies that $P(\omega)$ and $P(-\omega)$ have the same roots. 

It may not be immediately clear, but both $P$ and $Q$ can be taken to be square modulus of some other complex valued polynomials. This property holds because they are positive and even, so they have only complex roots that come in conjugate pairs. They are products of terms like $|(\omega-z_k)(\omega-z_k^*)|$, and as $\omega$ is real, any such term is equal to  $|\omega-z_k|^2 = |\omega - z_k^*|^2$. Moreover, in this decomposition we can only choose roots $z_k$ with positive imaginary part.

Now, let us consider $P$. We can write $P(\omega)=\big|\sum_{k=0}^n(a_k+\I b_k)\omega^k\big|^2,a_k,b_k\in\R$. Therefore
\bgeq
P(\omega) =\big(a_n\omega^n+a_{n-1}\omega^{n-1}+\ldots+a_0\big)^2+\big(b_n\omega^n+b_{n-1}\omega^{n-1}+\ldots+b_0\big)^2
\eeq
and because $P$ is even, the above formula is a function of $\omega^2$. The left and right parentheses also must be a polynomials of $\omega^2$ which is only possible when any of them contains only odd or even powers of $\omega$ inside. In other words $a_{2k}=0$ and $b_{2k+1}=0$ or $a_{2k+1}=0$ and $b_{2k}=0$. So $P$ has the form
\bgeq
P(\omega) = \lt|\sum_{k=0}^n  p_k(\I\omega)^k\rt|^2,\quad \tilde p_{2k}=(-1)^ka_{2k},p_{2k+1}=(-1)^kb_{2k+1}, p_k\in\R.
\eeq

The same goes for $Q(\omega)=\big|\sum_{j=0}^m q_j(\I\omega)^j\big|^2,q_j\in\R $.  Consider a process $X$ defined by two equations
\bg{gather}\label{eq:YdiffEq}
p_n\df{^n}{t^n}Y+p_{n-1}\df{^{n-1}}{t^{n-1}}Y+\ldots+ p_0 Y = \df{}{t}B,\\\label{eq:YdiffEq2}
X = q_m\df{^m}{t^m}Y+q_{m-1}\df{^{n-1}}{t^{n-1}}Y+\ldots+ q_0 Y. 
\end{gather}
We have chosen the polynomial $\omega\mapsto\sum_{k=0}^n p_k(\I\omega)^k$ such that it has only roots with positive imaginary parts. This implies that the polynomial $s\mapsto\sum_{k=0}^n p_ks^k$ has roots with negative real parts. It guarantees that Eq. \Ref{eq:YdiffEq} has a stationary solution. The process $X$ is well-defined as $m<n$ and also stationary. It has exactly the psd \Ref{eq:psdRat} and has the same distribution as the solution of given GLE.

Equations \Ref{eq:YdiffEq} and \Ref{eq:YdiffEq2} are equivalent to a vector equation for $\bd Y$ made of $Y$ and its $n-2$ derivatives, additionally $X=\bd Y\cdot\bd q$ where $\bd q = [0,0,\ldots, 0, q_m,q_{m-1},\ldots,q_0]$. 
\begin{flushright}
$\blacksquare$
\end{flushright}
The hard part of the proposition above is proving that the reduction of the GLE to a system of linear equations is always possible, given the assumptions. The procedure itself is not so complex, it requires mainly calculating polynomials $P, Q$ and their roots. 

As an example, let us consider the GLE with exponential kernel: $K(t)=b^2\e^{-2a t}$ for which the fluctuation-dissipation theorem holds. We have already provided some basic information about this model following Eq. \Ref{eq:expGLE}. Now we consider its Fourier space representation,
\bgeq \label{eq:expGLeF}
\widehat K(\omega) =\f{b^2}{2a-\I\omega},\quad s_F(\omega)= \widehat r_F(\omega) =\f{4ab^2}{4a^2+\omega^2},
\eeq
consequently
\bgeq
s_X(\omega) = \f{4ab^2}{4a^2+\omega^2}\f{|2a-\I\omega|^2}{|\I\omega(2a-\I\omega)+b^2|^2} = \f{4ab^2}{|\omega^2+\I 2a\omega+b^2|^2}.
\eeq
The roots of polynomial $\omega^2+2\I a\omega+b^2$  do not both have positive imaginary part
\bgeq
\omega^2+2\I a\omega+b^2=\lt(\omega -\I\big(\lambda+\sqrt{\lambda^2+b^2}\big)\rt)\lt(\omega -\I\big(\lambda-\sqrt{\lambda^2+b^2}\big)\rt).
\eeq
The first root has positive imaginary part, the second one negative, so it should be conjugated,
\bgeq
\lt(\omega -\I\big(\lambda+\sqrt{\lambda^2+b^2}\big)\rt)\lt(\omega -\I\big(-\lambda+\sqrt{\lambda^2+b^2}\big)\rt)=\omega^2-2\I\sqrt{\lambda^2+b^2}\omega -b^2.
\eeq
The solution of  the GLE is equivalent to a solution of the stochastic differential equation
\bgeq
\df{^2}{t^2}X +2\sqrt{\lambda^2+b^2}\df{}{t}X+b^2X=2b\sqrt{a}\df{}{t}B
\eeq
which is a full form of the classical Langevin equation with stiffness $\kappa=b^2$ and friction coefficient $\beta= 2\sqrt{a^2+b^2}$. The sampled solution of this GLE is an ARMA(2,1) process with AR coefficients
\begin{align}\label{eq:langARcoefEx}
\phi_1&=2\exp\left(-\Delta t\sqrt{a^2+b^2}\right)\cosh\lt(\Delta t a\rt)\nonumber,\\
\phi_2&=-\exp\lt(-2\Delta t\sqrt{a^2+b^2}\rt).
\end{align}
As we see during a statistical analysis $a$ can be estimated using $\phi_1^2/\phi_2$.\nn

Studying models, such as the examples  given above, can be performed in both continuous- and discrete-time. To some degree both approaches are equivalent. For example, the covariance function of the sampled series $(X_k)_k$ is simply the covariance of the continuous-time $(X(t))_t$ at points $\Delta t k$, i.e. $(r_X(\Delta t k))_k$. But there are also differences.

One method, which has no variant in the continuous-time case, is studying a partial autocorrelation function (pacf) $\rho_X$. It is the memory function measuring a dependence between $X_i$ and $X_{i+k}$ with influences from in-between $X_{j}, i<j<i+k$ removed. This removal is performed by subtraction of the projection onto the subspace of the variables ${X_{i+1},\ldots,X_{i+k}}$. Explicitly: 

\bg{dfn} The partial autocorrelation function of the stationary time series $X$, $\rho_X(k)$ is given by formula
\bgeq
\rho_X(k)\defeq\corr(X_i-\E[X_i|\{X_j\}_{j=i+1}^{i+k-1}],X_{i+k}-\E[X_{i+k}|\{X_j\}_{j=i+1}^{i+k-1}]),
\eeq
where we additionally  use the convention that $\E[\bd\cdot|\varnothing]=0$.
\end{dfn}
The partial autocorrelation is a useful quantity, because by removal of these projections we get rid of the influence of the time evolution between instants $i$ and $j$. This procedure helps to clarify the analysis of the memory. The pacf is a measure of the direct dependence between the values of the process and is well-suited to study AR time series.
\bg{prp}
The partial autocorrelation of an AR($p$) time series is zero for lag equal or greater than $p$.
\end{prp}
\bg{proof}
The representation
\bgeq
X_k = \phi_1X_{k-1}+\ldots + \phi_pX_{k-p}+\xi_k
\eeq
decomposes $X_k$ as a sum of $\xi_k$, which is independent from the $\mathrm{span}\{X_{k-1},X_{k-2},\ldots\}$ and $\phi_1X_{k-1}+\ldots + \phi_pX_{k-p}$, which is clearly a variable in $\mathrm{span}\{X_{k-1},X_{k-2},\ldots,X_{k-p}\}$. So, for $r\ge p$
\bgeq
\E[X_k|X_{k-1},\ldots,X_{k-r}] = \phi_1X_{k-1}+\ldots + \phi_pX_{k-p}
\eeq
and
\bgeq
\rho_X(r)=\corr(\xi_k,X_{k-r}-\E[X_{k-r}|X_{k-1},\ldots,X_{k-r+1}])=0
\eeq
due to the aforementioned independence.
\end{proof}
For the AR(1) process the one significant non-zero value of partial autocorrelation is easy to calculate, $\rho_X(1)=\corr(\phi_1X_{k-1}+\xi_k,X_{k-1})=\phi_1$, so
\bgeq\label{eq:pacfAR1}
\rho_X(k) =  \bg{cases}
1, & k = 0; \\
\phi_1, & k=1; \\
0, & k> 1.
\end{cases}\eeq
In the general case the partial autocorrelation can be estimated using the Yule-Walker equations, which relate it to the covariance function \citep{BD}. This set of equations can be easily solved, at least numerically.

Apart from the partial autocorrelation function, there is a significant difference in the spectral form of the continuous and discrete-time processes. The latter, instead of the whole $\R$ has a spectral measure defined on a torus (a natural dual space), or, equivalently, on the interval $[-\pi,\pi]$, see Theorem \ref{thm:harmRep}. Indeed, a short calculation yields
\begin{align}\label{eq:sigmaCsigmaD}
X_k&=X(\Delta t k)=\int_\R\sigma(\dd\omega)\ \e^{\I\omega\Delta t} = \int_{-\f{\pi}{\Delta t}}^{\f{\pi}{\Delta t}} \lt(\sum_{j=-\infty}^{\infty} \sigma\lt(\dd\omega+\f{2\pi}{\Delta t}k\rt)\rt)\ \e^{\I\omega \Delta t k}\nonumber\\
&= \f{1}{\Delta t}\int_{-\pi}^\pi  \lt(\sum_{j=-\infty}^{\infty} \sigma\lt(\f{1}{\Delta t}(\dd\omega+2\pi k)\rt)\rt)\ \e^{\I\omega k},
\end{align}
which relates the spectral measure of a process and the corresponding sampled time series. Only the discrete variant can be estimated from the data, but, as one can see, the full information about the process can be regained in the infill limit $\Delta t\to 0$. Equation \Ref{eq:sigmaCsigmaD} can be used to calculate the spectral measure of the sampled process, but in most of the cases the infinite sum within is hard to study analytically. For an ARMA processes there exists a much stronger result.

\bg{prp}\label{prp:ARMApsd}
The power spectral density $s$ of an ARMA($p,q$) process is
\bgeq
s(\omega) =c\f{\lt|1+\sum_{k=1}^q\theta_k\e^{-\I\omega k t}\rt|^2}{\lt|1-\sum_{k=1}^p\phi_k\e^{-\I\omega k t}\rt|^2},
\eeq
where $c$ is a constant spectral density of $(\xi_k)_k$.
\end{prp}
\bg{proof}
This proposition is similar to results for linear filters in Section \ref{s:gaussProc}. Namely, we use the harmonic representation for the series $(X_k)_k$, as stated in Theorem \ref{thm:harmRep}
\bgeq
X_k = \int_{-\pi}^\pi\dd S_X(\omega)\ \e^{\I\omega k}.
\eeq
From this formula it is clear that the process $(X_{k-1})_k$ has the spectral process $\dd S_X(\omega)\e^{-\I\omega}$. A similar shift occurs for $(\xi_k)_k$, so the following relation holds
\bgeq
\int_{-\pi}^\pi\dd S_X(\omega) \lt(1-\sum_{k=1}^p\phi_k\e^{-\I\omega k t}\rt)\e^{\I\omega t} = \int_{-\pi}^\pi\dd S_\xi(\omega) \lt(1+\sum_{k=1}^q\theta_k\e^{\I\omega k t}\rt)\e^{\I\omega t}.
\eeq
The uniqueness of the harmonic representation guarantees that
\bgeq
\dd S_X(\omega)=\dd S_\xi(\omega)\f{1+\sum_{k=1}^q\theta_k\e^{-\I\omega k t}}{1-\sum_{k=1}^p\phi_k\e^{-\I\omega k t}}.
\eeq
The power spectral density is the variance of $\dd S_X$, so the result holds.
\end{proof}
A consequence of this proposition is that the coefficients of the MA processes are not uniquely determined, which complicates their physical interpretation. A distribution of the process is uniquely determined by its spectral measure, which for MA process is $\lt|1+\sum_{k=1}^q\theta_k\e^{-\I\omega k t}\rt|^2$. It is a polynomial of the argument $\e^{-\I\omega}$, which has $q$, not necessarily distinct, zeros $z_k$. It can be factorised into a product of $q$ terms $\lt|z_k-\e^{-\I\omega}\rt|^2$. However, because of the modulus, if the zero $z_k$ is real, the corresponding factor can be transformed as
\bgeq 
\lt|z_k-\e^{-\I\omega}\rt| =\lt|1-z_k\e^{\I\omega}\rt|= |z_k|\lt|z_k^{-1}-\e^{-\I\omega}\rt|.
\eeq
Note that factor $|z_k|$ only changes the variance of the resulting process. For complex zeros this procedure would result in a complex polynomial and a complex MA process if $z_k$ and its complex conjugate $z_k^*$ are not inversed at the same time. Because we consider only real-valued MA processes, the number of possibilities is $2^{q'}$, where $q'$ is a number of $z_k$ which are real plus half of the number of complex ones.

For a real-valued MA(1) process there are always exactly two options: $\theta_1$ and $\theta_1^{-1}$. The corresponding processes differ only by the variance, which changes by the ratio $\theta_1^2(1+\theta_1^2)$.
\section{Application to modelling of optical tweezers}\label{s:OTAppl}
An example of a system in which our approach is especially viable are optical tweezers. These are a versatile tool allowing to manipulate micrometer-sized particles in liquids \citep{OTintr} non-invasively and to measure forces even on the picoNewton scale. For this reason they have a wide range of applications in many fields of biology and soft condensed matter physics including, e.g. stretching of DNA \citep{wang} and other polymers \citep{poly2}, molecular motors \citep{molMot} research or analysis of colloidal suspensions \citep{OTcoloid}.

The most classical form of the measurements related to optical tweezers are series of of positions of a bead trapped in the tweezers. Observations performed using CCD camera are especially useful, because the camera provides a lot of diverse information and allows for the tracking of many objects simultaneously. The data that we use was measured using this setup by S\l awomir Drobczy\' nski at Wroc\l aw University of Science and Technology. As we will show in our analysis below,  in the case of high-frequency measurements the camera has a significant influence on the measurements, which is possible to analyse using discrete-time methods.

According to the classical Einstein theory of diffusion, a trajectory $(X(t))_{t\in\R}$ of a colloidal particle trapped in a
viscous fluid by optical tweezers is a solution of the equation of the force balance
\bgeq 0 = {F}_S+{F}_O + {F}_T, 
\eeq
where

\begin{itemize}
\itemsep1em
\item[-] $F_S = -\beta \f{\dd X}{\dd t}$ is the Stokes force (friction of liquid) acting on the spherical bead with radius $r$ within the liquid with viscosity $\eta$, $\beta = 6\pi\eta r$;
\item[-] $F_O = -\kappa X$ is the force caused  by the optical tweezers, we use an harmonic approximation, i.e., assume that the potential is harmonic with stiffness $\kappa $;
\item[-] $F_T=\sqrt{k_B\mathcal T\beta}{\dd B}/{\dd t}$ is the {thermal force}: it models the exchange of momenta with particles of the liquid.
\end{itemize}

Substitution of the explicit  formulas  assures that  the above  force balance condition is equivalent to the Langevin equation
\bgeq \dd X = - \dd t \lambda X + D\dd B, \eeq
where $\lambda = \kappa/\beta$ and $D = \sqrt{k_B\mathcal T/\beta}.$
So, the stationary solution is the Ornstein-Uhlenbeck process, and the sampled time series is an AR(1) process. The AR(1) coefficient $\phi_1$ can be estimated using the autoregressive nature of the process. One can use the classical regression with the sequence of $(X_k)_k$ taken as a variable y and shifted sequence of $(X_{k-1})_k$ taken as a variable x. The well-known least-squares estimator of the slope in this case reads %\citep{pol}
\bgeq
{\phi}_1^\text{est} = \f{\sum_k X_kX_{k-1}}{\sum_kX_k^2}.
\eeq

However, using a partial autocorrelation is a more robust method, because the estimated $\rho_X(1)$ is an estimate of $\phi_1$ (Eq. \Ref{eq:pacfAR1}) and checking if $\rho_X(k)\approx 0$ for $k>1$ is a form of the test of validity of the model.

However, the pure physical model of the optical tweezers is not sufficient to analyse typical recordings because it does not take into account the distortions which are caused by the video camera. The most basic and well-researched type of distortion is an addition of the noise generated by the camera's CMOS matrix. In this case we do not observe $X_k$, but the series $X'_k=X_k+w_k$, where $w_k$ is a noise independent from the values of $X_k$. The independence of $X_k$ and $w_k$ implies that the resulting covariance is the sum $r_{X'}=r_X+r_w$, the same for the spectral measure $\sigma_{X'}=\sigma_X+\sigma_w$. In order to understand the structure of $(X'_k)_l$ let us look what linear filter governs its evolution

\begin{align}
X'_k&=X_k+w_k=\phi_1 X_{k-1}+\xi_k+w_k \nonumber\\
&=\phi_1 X'_{k-1}+\xi_k+w_k-\phi_1 w_{k-1}=\phi_1 X'_{k-1}+\xi_k'.
\end{align}
Now, we take the common assumption that the additive noise $(w_k))k$ is Gaussian and white, with a variance $s^2$. Then the new effective noise $\xi'_k=\xi_k+w_k-\phi_1 w_{k-1}$ has the covariance function
\bgeq\label{eq:acovxi'}
r_{\xi'}(h)=
\bg{cases}
\f{D^2}{2\lambda}\lt(\phi_1^{-2}-1\rt)+\lt(1+\phi_1^2\rt)s^2,& h=0;\\
-\phi_1 s^2,& h =  1;\\
0, & h> 1
\end{cases}
\eeq
As we see it is a time series with a short-term negative memory. As stated before, any such process can be expressed as a linear filter acting on a Gaussian white noise, in this case it is an MA(1) time series
\bgeq\label{eq:eta}
\xi'_k=\eta_k-a\eta_{k-1},
\eeq
where $(\eta_k)_k$ is a white noise series with variance $v^2$.
To present the constant $a$ in a simple form it is convenient to introduce a ratio
\bgeq\label{eq:r}
\mathfrak r=-\frac{r_{\xi'}(0)}{r_{\xi'}(1)},
\eeq
which is always greater or equal than two, because $\phi_1>0$ implies $\mathfrak r\ge\phi_1+\phi_1^{-1}\ge 2$. Solving the quadratic equation obtained comparing the covariances, we get only one solution with $|a|<1$, 
\bgeq\label{eq:av}
a=\frac{\mathfrak r-\sqrt{\mathfrak r^2-4}}{2}, \ \ \ v^2=\frac{2 r_{\xi'}(0)}{\mathfrak r(\mathfrak r-\sqrt{\mathfrak r^2-4})}.
\eeq
We may consider the series $\eta_k$ as the noises $\xi_k$ and $w_k$ mixed and orthogonalised. Through this calculation  we have shown that the observed $X'_k$ fulfils
\bgeq
X'_k-\phi_1 X'_{k-1}= \eta_k-a\eta_{k-1}.
\eeq
This is a specific ARMA(1,1) time series.

Let us once again turn back to the influence of the camera. Additive white noise is not the only common type of distortion. The cameras have a tendency to blur the images. Especially when the frequency of sampling is high and the exposure time is long, the CMOS matrix may not have sufficient time to fully refresh. As a result, the remainder of a previous frame is left in the image of the next one. Instead of $X_k$ we observe $\tilde X_k=X_k+bX_{k-1}$, $0<b<1$. It is also possible that $\tilde X_k$ depends on even older images, but the intensity of this effect should decay exponentially. Our measurements show that the observed $b$ is not very large, so we will neglect this effect.

So, let us denote a short-time linear filter $M=1+b \mathcal S_{-1}$. The operators $M$ and $L=1-\phi_1 \mathcal S_{-1}$ commute and
\bgeq
L \tilde X_k=LMX_k=MLX_k=M\xi_k
\eeq
or, equivalently,
\bgeq
\tilde X_k-\phi_1 \tilde X_{k-1}=\xi_k+b\xi_{k-1}.
\eeq
Once again the observed process is ARMA(1,1), but this time with the positive MA(1) coefficient equal to $b$. We have come to the conclusion that both possible distortions caused by the digital camera cause the observed series to fulfil the ARMA(1,1) equation, the difference being the sign of the MA(1) part.

Let us denote the observed series by $\mathcal X_k$, which may be $X'_k$ or $\tilde X_k$.  The direct dependence between $\mathcal X_k$ and the noise $\eta_k$ can be studied by applying the inverse operator $(1-\phi_k \mathcal S_{-1})^{-1}=\sum_{h=0}^\infty \phi_1^{h}\mathcal S_{-h}$ to both sides of the above equation, yielding
\bgeq
\mathcal X_k=\eta_k+\lt(\phi_1+\theta_1\rt)\sum_{j=1}^\infty \phi_1^{j-1}\eta_{k-j}.
\eeq 

The covariance of this series is straightforward to calculate and equal to
\bgeq
r_{\mathcal X}(h) =v^2\cdot
\bg{cases}
1+\f{(\phi_1+\theta_1)^2}{1-\phi_1^2},& h=0;\\
\phi_1^{|h-1|}(\phi_1+\theta_1)\f{1+\phi_1\theta_1}{1-\phi_1^2},& h\neq 0;
\end{cases}
\eeq
Hence, the ARMA(1,1) process also exhibits an exponential decay of the covariance function with one special value at $h=0$. It is a natural insight for the case $\mathcal X_k=X'_k$ when we can simply add the covariance function of the original process and white noise, but is also true in a more general situation. It can be interpreted as an effect of distortion which affects only the shortest one-time-step memory.

Similarly to the previously considered problems, the power spectral density of ARMA(1,1) process can be obtained using Proposition \ref{prp:ARMApsd}, and equals 
\bgeq
s_{\mathcal X}(\omega) =  v^2\f{1+\theta_1^2+2\theta_1\cos(\omega\Delta t)}{1+\phi_1^2-2\phi_1\cos(\omega\Delta t)}.
\eeq
A comparison of the spectra for different processes relevant to the optical tweezers measurements is shown in Fig. \ref{f:psdComp}. One must be careful, because the psd of the Ornstein-Uhlenbeck process $\propto 1/(\omega^2+\lambda^2)$ is for low frequencies nearly identical to the truly observed psd disturbed by a blur described by a small MA coefficient. This is an unfortunate coincidence which can easily misguide the analysis, creating a false resemblance of the lack of distortions.

\begin{figure}[h!] \centering
\includegraphics[width=12cm]{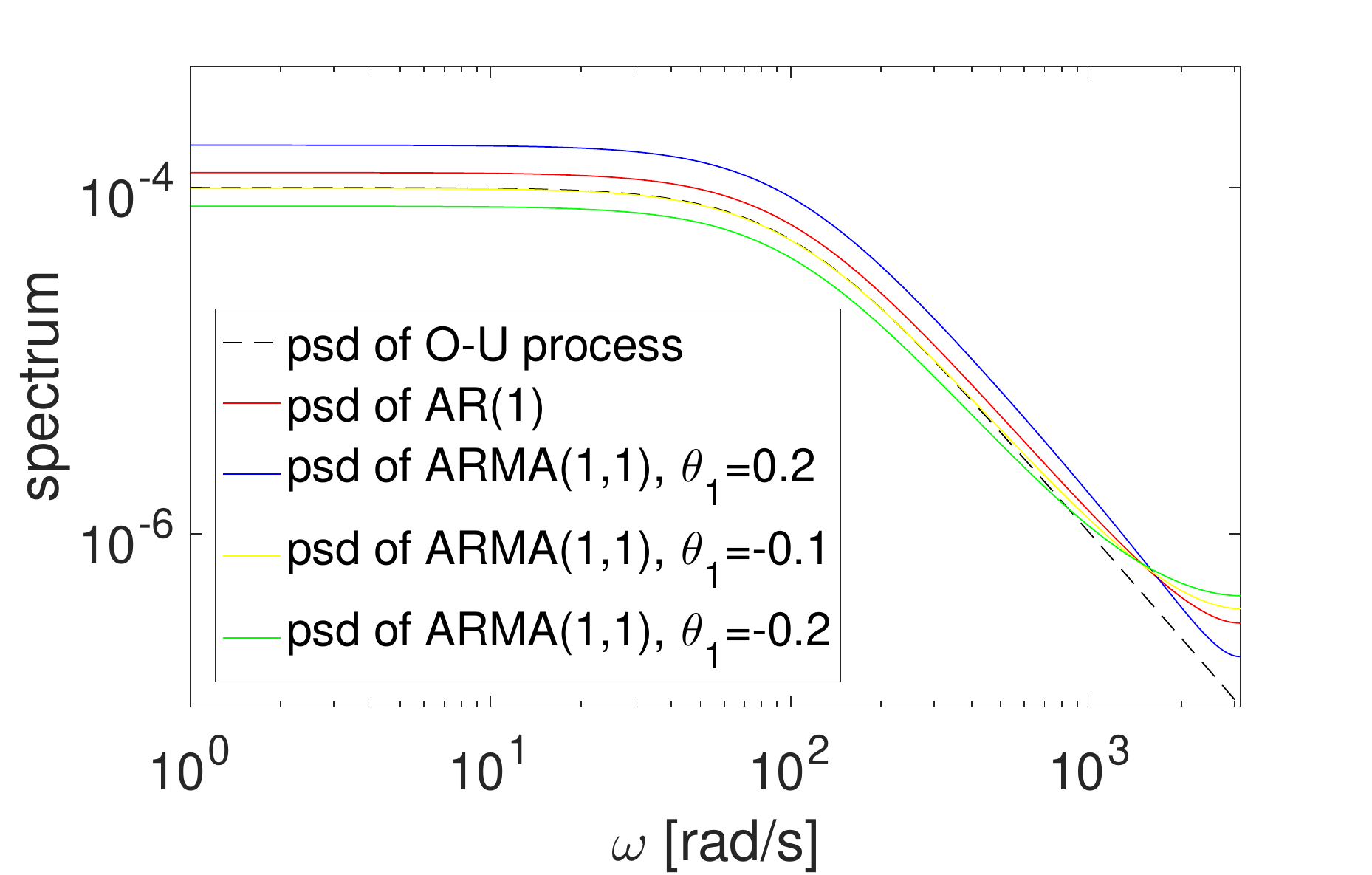}
\caption{Comparison of psd functions for continuous-time Ornstein-Uhlenbeck process, discrete-time  undisturbed  and disturbed ARMA(1,1) time series. }
\label{f:psdComp}
\end{figure}

To study the approximate behaviour of the partial autocorrelation function we inverse the operator $1+\theta_1 \mathcal S_{-1}$, expressing $\mathcal X_k$ directly as a function of past $\mathcal X_{k-j}$,
\bgeq
\mathcal X_k = (\phi_1+\theta_1)\sum_{j=1}^\infty (-\phi_1)^{j-1}\mathcal X_{k-j}+\eta_k.
\eeq
It shows that the direct dependence between $\mathcal X_k$ and $\mathcal X_{k-j}$ is proportional to the constant $(\phi_1+\theta_1)(-\phi_1)^{j-1}$; the partial autocorrelation of $\mathcal X$ is a geometrical decay but its rate depends on the MA coefficient $\theta_1$ instead of the AR coefficient $\phi_1$ as for the covariance function. As  $\theta_1$ is close to one for common sampling frequencies and $\phi_1$ is not large even for highly disturbed data, we have $\phi_1+\theta_1>0$ and the sign of $\rho_{\mathcal X}(2)$ is the same as the sign of $-\phi_1$. Thus, the estimation of the partial autocorrelation function immediately determines the presence and type of distortions in the data. The values of the pacf for larger $h$ have little practical importance in our case; because of the fast decay these values quickly become unobservable. The first values of the pacf can be calculated analytically
\bgeq
\rho_{\mathcal X}(h)=\begin{cases}
\frac{(\phi_1+\theta_1)(1+\phi_1\theta_1)}{1+2\phi_1\theta_1+\phi_1^2},& h=1;\\
-\phi_1 \frac{(\phi_1+\theta_1)(1+\phi_1\theta_1)}{(1+2\phi_1\theta_1+\phi_1^2)^2}, & h=2.
\end{cases}
\eeq
The partial autocorrelation function estimated from the data very closely follows the described behaviour, an example is shown in Fig. \ref{f:pacf1}. From the rate of the exponential decay we can assess that $\theta_1\approx 0.17$. This is a very good preliminary estimate, even compared to the more precise methods that we present later.
\begin{figure}[h!] \centering
\includegraphics[width=12cm]{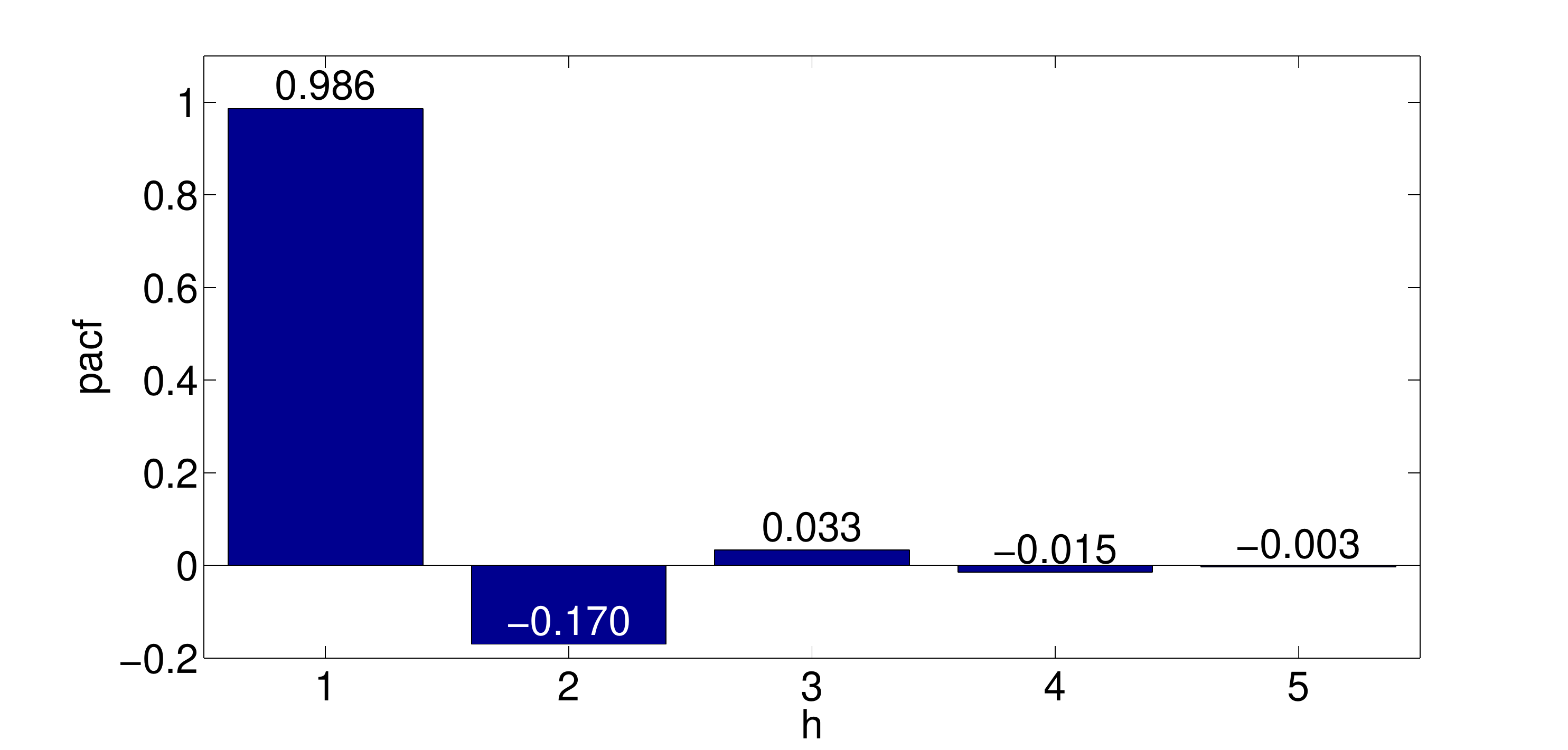}
\caption{Estimated pacf of the optical tweezers time series; sampling frequency is 5 [kHz], camera exposure time is 100 [$\mu$s]. }
\label{f:pacf1}
\end{figure}

Both distortions considered above are very common, so the natural question appears: what happens if they are present simultaneously? Their joint effect depends on the sequence in which video camera imperfections distort the recording. One possibility is that the camera records images disturbed by the white noise, which situation corresponds to $X'_k$, and the film is blurred later, so that we observe $ \tilde X'_k=MX'_k$. In this case
\begin{align}\label{eq:ARMA(1,2)}
L \tilde X'_k&=M\xi_k'=(1+b\mathcal S_{-1})(1-a\mathcal S_{-1})\eta_k\nonumber\\
&=\eta_k+(b-a)\eta_{k-1}-ba\eta_{k-2}.
\end{align}
This is an ARMA(1,2) process. Note that the first MA coefficient $\theta_1=b-a$ has a sign determined by the relative strengths of both distortions; the second coefficient $\theta_2=-ba$ is relatively small and its presence may be hard to detect.

The situation is different if the camera adds the additive noise to the already blurred recording, e.g. during processing of the electrical signal from the CMOS matrix. Then we observe ${\tilde X_k}'=\tilde X_k+w_k$. This process fulfils the relation
\begin{align}
{\tilde X_k}'&=\phi_1 \tilde X_k+\xi_k+b\xi_{k-1}+w_k\nonumber\\
&=\phi_1 {\tilde X}'_{k-1}+\xi_k+b\xi_{k-1}+w_k-\phi_1 w_{k-1}.
\end{align}
The new effective noise $\tilde\xi_k=\xi_k+b\xi_{k-1}+w_k-\phi_1 w_{k-1}$ has the covariance function
\bgeq\label{eq:acovxi*}
r_{\tilde\xi}(h)=
\bg{cases}
\lt(1+b^2\rt)\f{D^2}{2\lambda}\lt(\phi_1^{-2}-1\rt)+\lt(1+\phi_1^2\rt)s^2,& h=0;\\
-\phi_1 s^2+b\frac{D^2}{2\lambda}\lt(\phi_1^{-2}-1\rt),& h =  1;\\
0, & h> 1.
\end{cases}
\eeq
This is again an MA(1) process; its $\theta_1$ parameter is given by Eqs. \Ref{eq:r} and \Ref{eq:av} with $r_{\tilde\xi}$ inserted instead of $r_{\xi'}$. The time series $({\tilde X_k}')_k$ is ARMA(1,1) but with MA parameter different from that of $X_k'$ or $\tilde X_k$.

The additive noise and blur have opposite influences, which can be seen looking at the case $h=1$ in \Ref{eq:acovxi*}. It is even possible for them to cancel each other in which case the recording behaves like an undisturbed trajectory of the stochastic oscillator, i.e. an AR(1) time series. This effect can be created involuntarily during calibration of the experimental setup. For the case ``blur+noise'' (${\tilde X}'$) there is a possibility for exact cancelling of the blur and additive noise effects, for the ``noise+blur'' case $\tilde X'$ there will be a small second MA(2) coefficient left equal to $\theta_2=-a^2=-b^2$. For both cases the only large statistical change would be an increase of the generating noise variance ($h=0$ in \Ref{eq:acovxi*}) and a subsequent increase of the variance for the process $X$ itself. As measuring the variance of $X$ is one of the popular methods of estimating the stiffness of the trap, special care must be taken to avoid the discussed possibility.

We have come to the conclusion that the presence and type of the experimental distortions observed in the data can be recognised by the order of the ARMA time series and the value of the MA coefficient. At the same time, the AR coefficient $\phi_1$ does not depend on the distortions. Hence, the stiffness coefficient of the tweezers' harmonic trap can always be accurately determined through fitting of the ARMA model.

The simplest method to start the analysis is to fit parameters to an estimated memory function: covariance, partial autocorrelation or power spectral density. The covariance function was estimated as the sample covariance (e.g. method \texttt{acf} in R package). The sample pacf was obtained by substituting into the Yule-Walker equations the values of the sample covariance. This method is already implemented in the Matlab environment as a \texttt{parcorr} function, and in the R package as \texttt{pacf}. To estimate the discrete-time psd we used the periodogram defined as
\bgeq\label{eq:period}
I(\omega) = \frac{\Delta t}{N}\lt|\sum_{k=1}^N X_k\e^{\I \Delta t k\omega}\rt|^2.
\eeq
The periodogram itself is an ubiased estimator, but is not consistent. For this reason  we take a smoothed periodogram as an estimate, that is the convolution of the function \Ref{eq:period} with a smoothing kernel. For a sufficient choice of smoothing the result has better statistical properties than the pure periodogram. In Matlab this method is implemented as command \texttt{periodogram} (uses a rectangular smoothing function) and in the R package as the function \texttt{spec.pgram} with more options available allowing to regulate the calculation method and smoothing.

The values of $v^2, \phi_1$ and $\theta_1$ obtained from fitting one of the memory functions are suboptimal estimates. Better results are obtained using the least-squares method, i.e. looking for the values ${\phi}_1^\text{est} ,{\theta}_1^\text{est} $ for which the sum
\bgeq
\sum_{k=1}^N({\eta}_k^{\text{est}})^2 = \sum_{k=1}^N \lt(X_k-{\phi}_1^\text{est}  X_{k-1}-{\theta}_1^{\text{est}}  {\eta}_{k-1}^{\text{est}} \rt)^2
\eeq
has a minimal value. Here ${\eta}_k^\text{est} $ are calculated recursively; as an initial value ${\eta}_0^\text{est} $ we take 0. If our estimates ${\phi}_1^\text{est} $ and ${\theta}_1^\text{est} $ are close to the real values, the estimated series $({\eta}_k^\text{est} )$ is close to $(\eta_k)_k$ and is similar to white noise. If they are not close, the variables ${\eta}_k^\text{est} $ are correlated, which causes the sum of squares to increase. Minimising the above sum, given initial estimates for $\phi_1$ and $\theta_1$ is an accurate and computationally fast method. In Matlab it is provided by the command \texttt{estimated} and in the R package by \texttt{arima} or, using maximum likelihood method, in the library FitARMA. The estimate of the psd obtained from the data and the psd of the fitted ARMA(1,1) process is shown in Fig. \Ref{f:psdEst}. This figure proves that the least-squares fit of an ARMA model provides estimates which fully explain the spectral properties of the measurements.
\begin{figure}[h!] \centering
\includegraphics[width=12cm]{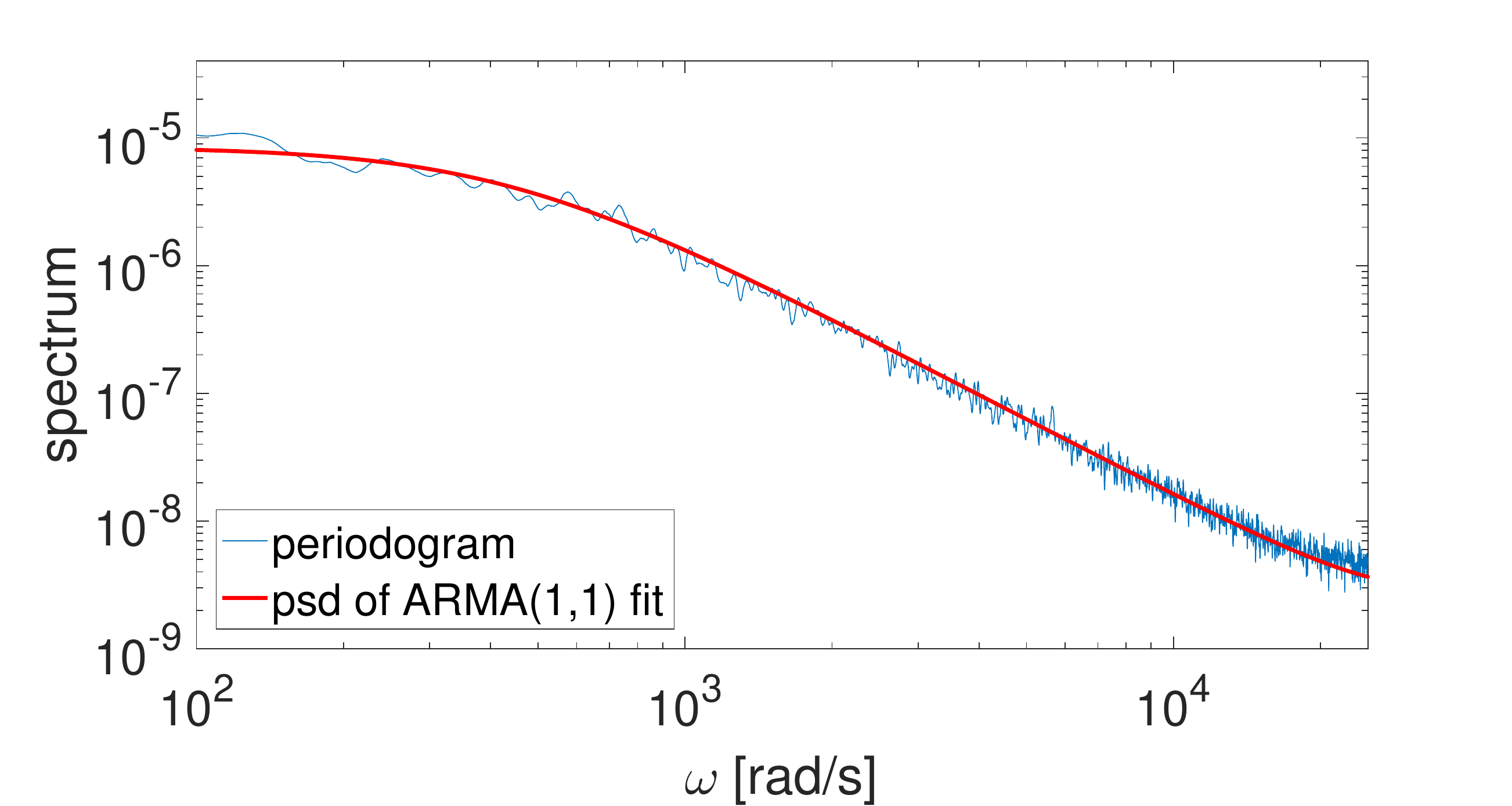}
\caption{Smoothed periodogram (blue thin line) of the data and the dpsd of the ARMA(1,1) process (thick red line), ${\phi}^\text{est} _1=0.9580$, ${\theta}_1^\text{est} =0.0472$.}
\label{f:psdEst}
\end{figure}

We have experimentally studied the influence of the digital camera for eight different sampling frequencies and seven different exposure times. The results, obtained using least-squares fitting, are shown in Fig \ref{f:main}. The shape of this plot is very natural to explain. For short exposure times not much light goes into camera and additive noise is an important factor compared to the intensity of the undisturbed image. For longer exposure times there is more light, but also less time to refresh the CMOS matrix between subsequent frames, which causes blur to dominate. Increasing sampling frequencies further shortens time when the CMOS matrix can refresh and strengthens this effect. The level for which the MA coefficient $\theta_1\approx 0$ corresponds to the situation when these two effects cancel each other, leading to the false illusion of the distortion absence.

The MA coefficient depends on the sampling frequency $f$ (in kHz) and the exposure time $t_e$ in a regular manner, approximately linearly in the studied range, so that even a simple linear approximation of the obtained dependence: $\theta_1= 0.0075 f+0.001 t_e-0.0798$ provides a good interpolation. However, our measurements for larger frequencies suggest that the studied effect becomes non-linear in the larger range.

During these measurements we also checked the values of the estimated AR(1) coefficient. During few subsequent measurements we obtained estimates ${\phi}_1^\text{est} =0.981\pm 0.003$, which correspond to the normalised stiffness ${\lambda}^\text{est} =95\pm 15$ [$s^{-1}$]. There was no dependence between the obtained values, the sampling frequency and exposure time, which proves that the ARMA fitting allows for the estimation of the trap stiffness free of the studied effects.
\begin{figure}[h!] \centering
\includegraphics[width=12cm]{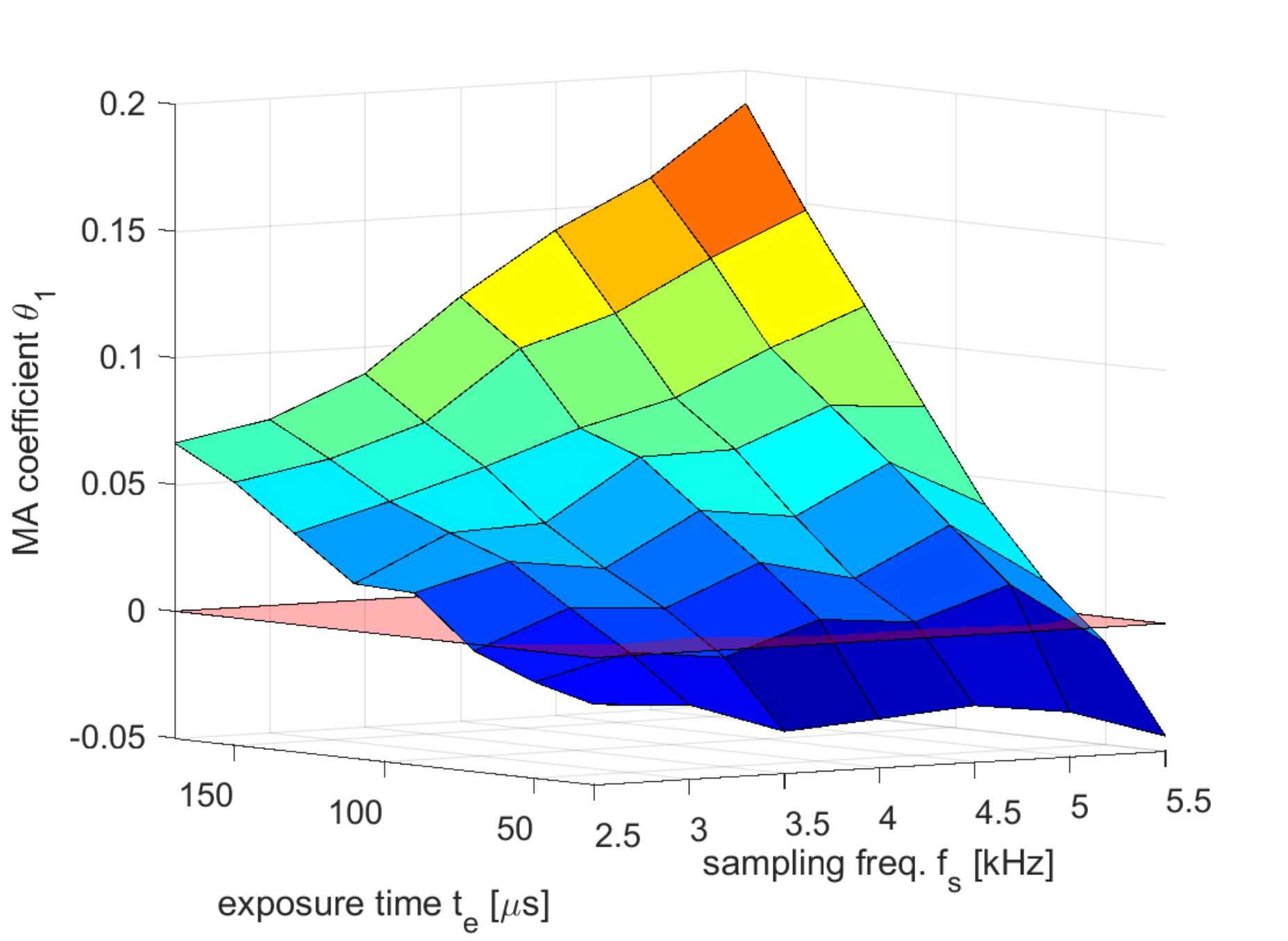}
\caption{MA(1) coefficient for different sampling frequencies and video camera exposure times. The level $\theta_1=0$ is also drawn.}
\label{f:main}
\end{figure}

Given estimates $\phi^\text{est}_1$ and $\theta^\text{est}_1$ we can inverse the AR and MA filters, effectively removing their influence from the data.

The parameter $\phi_1$ is determined by the stiffness of the harmonic trap. For a given trap with known stiffness, we can apply the filter $L=1-\phi^\text{est}_1\mathcal S_{-1}$ to the data, i.e. analyse $X_k-\phi^\text{est}_1X_{k-1}$. This series is an MA part of the ARMA model which provides insight into distortions caused by the experimental setup. If the covariance function of $X_k-\phi^\text{est}X_{k-1}$ has two non-zero values (for $h=0$ and $h=1$), it is confirmed that some short-time disturbance is present in the system. It can be mainly additive noise for the positive estimate of $r_X(1)$ or mainly blur for negative value of $r_X(1)$. We have not observed non-zero values of the covariance for larger values of $h$, see Fig. \ref{f:acf1}. There we present the estimated covariance function  normalised. i.e. divided by the assessed value of the variance; such memory function is often called correlation function. 

The presence of $h>1$ non-zero values of the covariance would suggest the presence of a blur between few adjacent frames or a composition of a blur and an additive noise as in Eq. \Ref{eq:ARMA(1,2)}. The lack of this type of memory confirms that ARMA(1,1) is a model sufficient to fully describe the physical dynamics and distortions.

\begin{figure}[h!] \centering
\includegraphics[width=12cm]{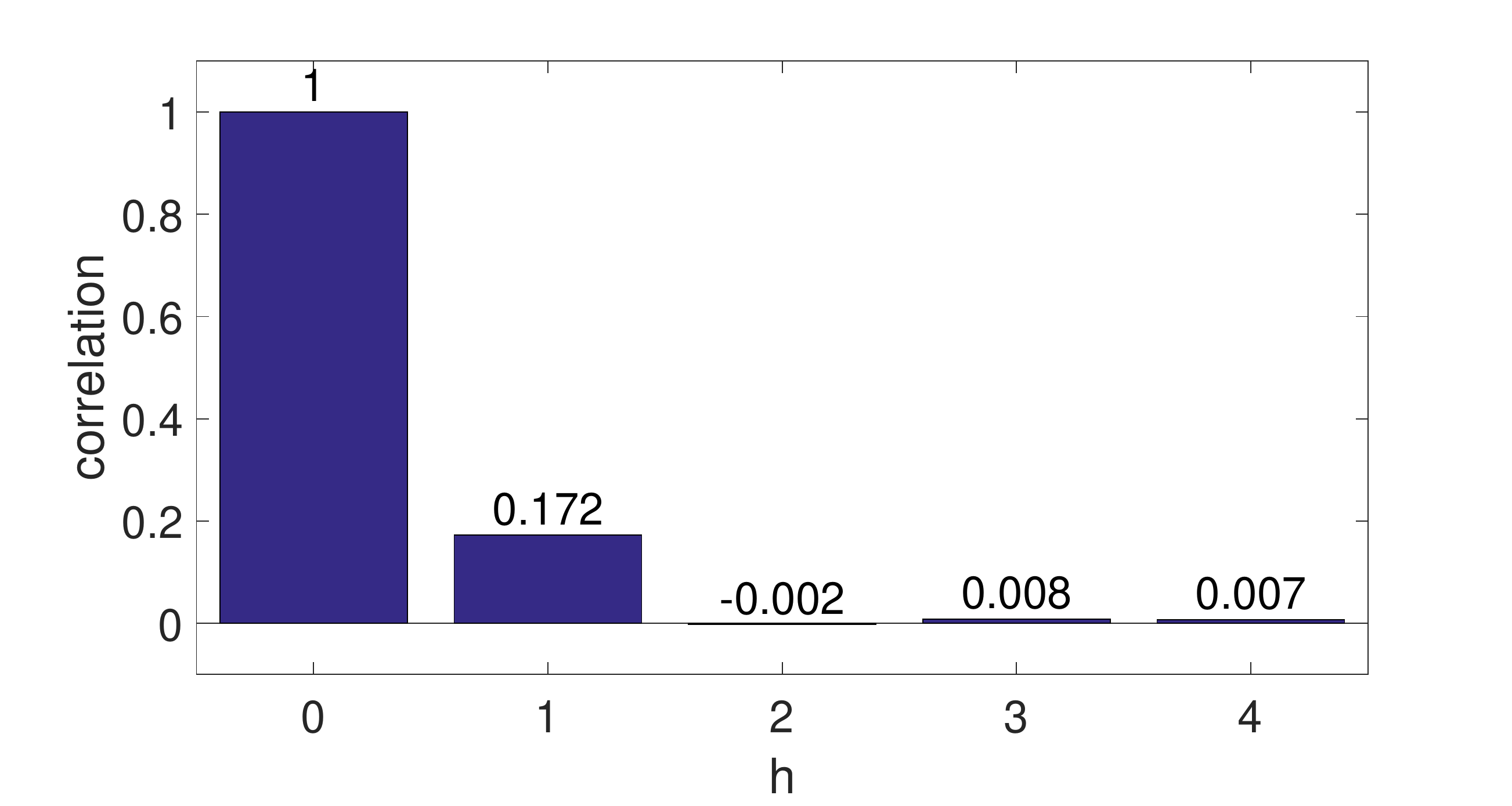}
\caption{ Estimated correlation function of the series $X_k-\phi^\text{est}_1X_{k-1}$, the used estimate for $\phi_1$ was $\phi^\text{est}_1=0.9816 $. }
\label{f:acf1}
\end{figure}

If we have calibrated  the setup, i.e. we have determined the MA part of the model, we can retrieve the original undisturbed series by applying the filter $(1+\theta^\text{est}_1 \mathcal S_{-1})^{-1}$. This operator has the form of the infinite series $\sum_{k=0}^\infty (-\theta^\text{est}_1)^k \mathcal S_{-k}$, so for practical purposes only the first few terms need to be used. The error that we make using this approximation decays exponentially with the number of terms taken along, so it is easy to obtain sufficient accuracy. If $\theta^\text{est}_1\approx \theta_1$, the retrieved series ${X}^\text{org}_k\defeq(1+\theta^\text{est}_1 \mathcal S_{-1})^{-1}X_k$ satisfies the original AR equation
\begin{align}\label{eq:detract}
&{X}^\text{org}_k-\phi_1 {X}^\text{org}_{k-1}=(1+\theta^\text{est}_1 \mathcal S_{-1})^{-1}(X_k-\phi_1 X_{k-1})\nonumber\\
&= (1+\theta^\text{est}_1\mathcal S_{-1})^{-1} (1+\theta_1\mathcal S_{-1})\eta_k\approx \eta_k.
\end{align}
In the case of blur, the calculated $\eta_k$ is exactly the original physical white noise; for the case of the additive white noise it has larger variance than the original process (equal to $\f{D^2}{2\lambda}\lt(\phi_1^{-2}-1\rt)$) which can be accounted for by a proper rescaling.

The memory function of the exemplary calculated ${X}^\text{org}$ is shown in Fig. \ref{f:pacf2}. We have taken the same time series as in Fig. \ref{f:acf1} and using the estimated value $\theta^\text{est}_1=0.1795$ we applied the first 20 terms of the inverse operator $(1+\theta^\text{est}_1 \mathcal S_{-1})^{-1}$. The result is an AR(1) process, which is confirmed by Fig. \ref{f:pacf2}. The first estimated value of pacf is $0.918$, so is exactly equal to the previously obtained estimate $\theta^\text{est}_1$. 
\begin{figure}\centering
\includegraphics[width=12cm]{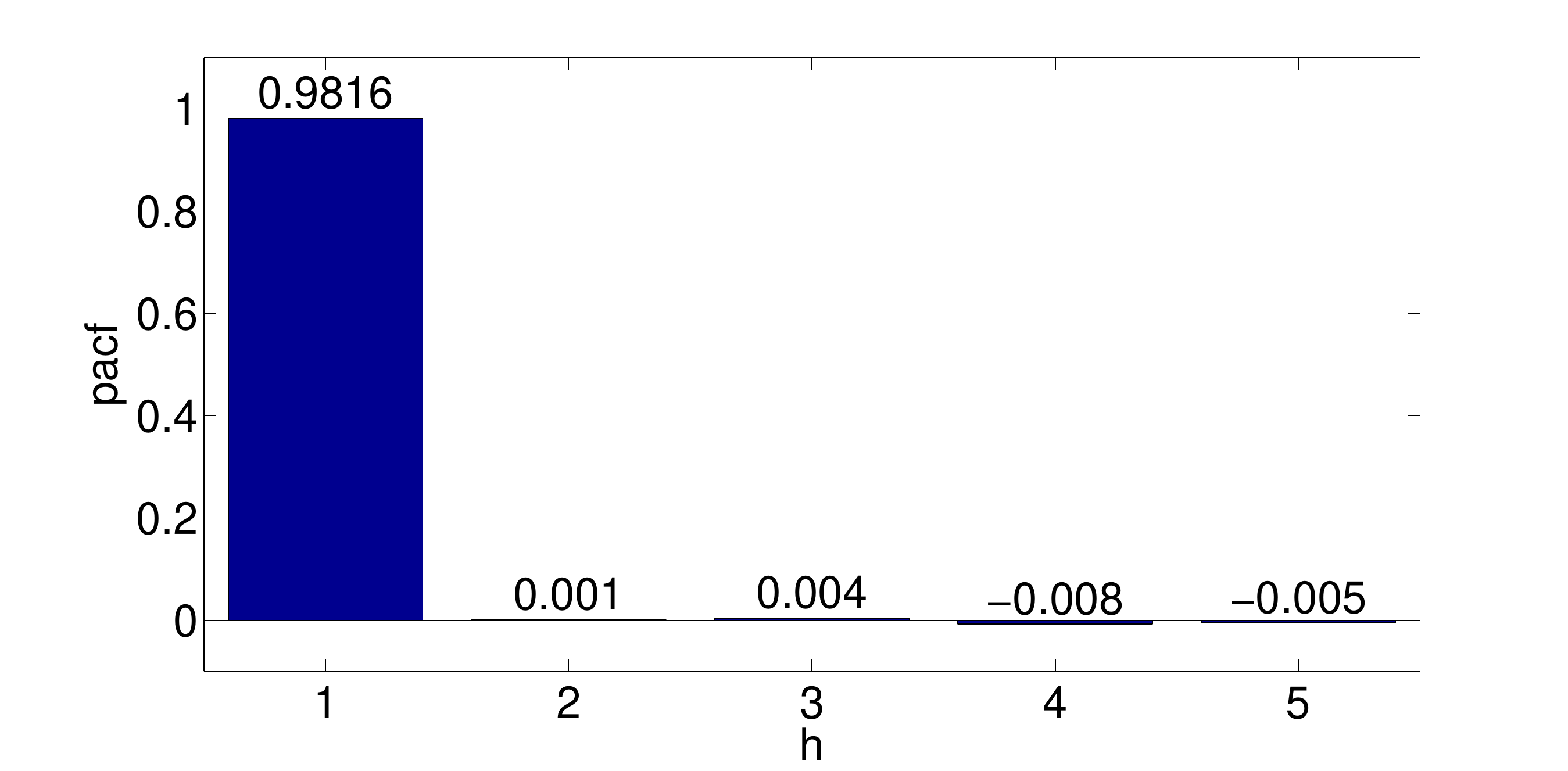}
\caption{Estimated pacf of the approximated series $(1+\theta^\text{est}_1\mathcal S_{-1})^{-1}X_k$, $\theta^\text{est}_1= 0.1795$.}
\label{f:pacf2}
\end{figure}

We can apply this procedure to the measured time-series if only we know the value of MA parameter $\theta_1$ for a given calibration of the used equipment, effectively removing its influence. This can be a powerful technique if we calibrate the tweezers analysing their behaviour in the controlled conditions, e.g. for spherical beads embedded in water, and later remove the disturbances for  data acquired in more complex systems such as biological media.

For our data, the pacf for low frequencies fits to the model, see Fig \ref{fig:pacf1}. By two dashed lines near zero we denote the level of an expected statistical error. Values between these lines are statistically insignificant. 
\bg{figure}[H] \centering
\includegraphics[width=10cm]{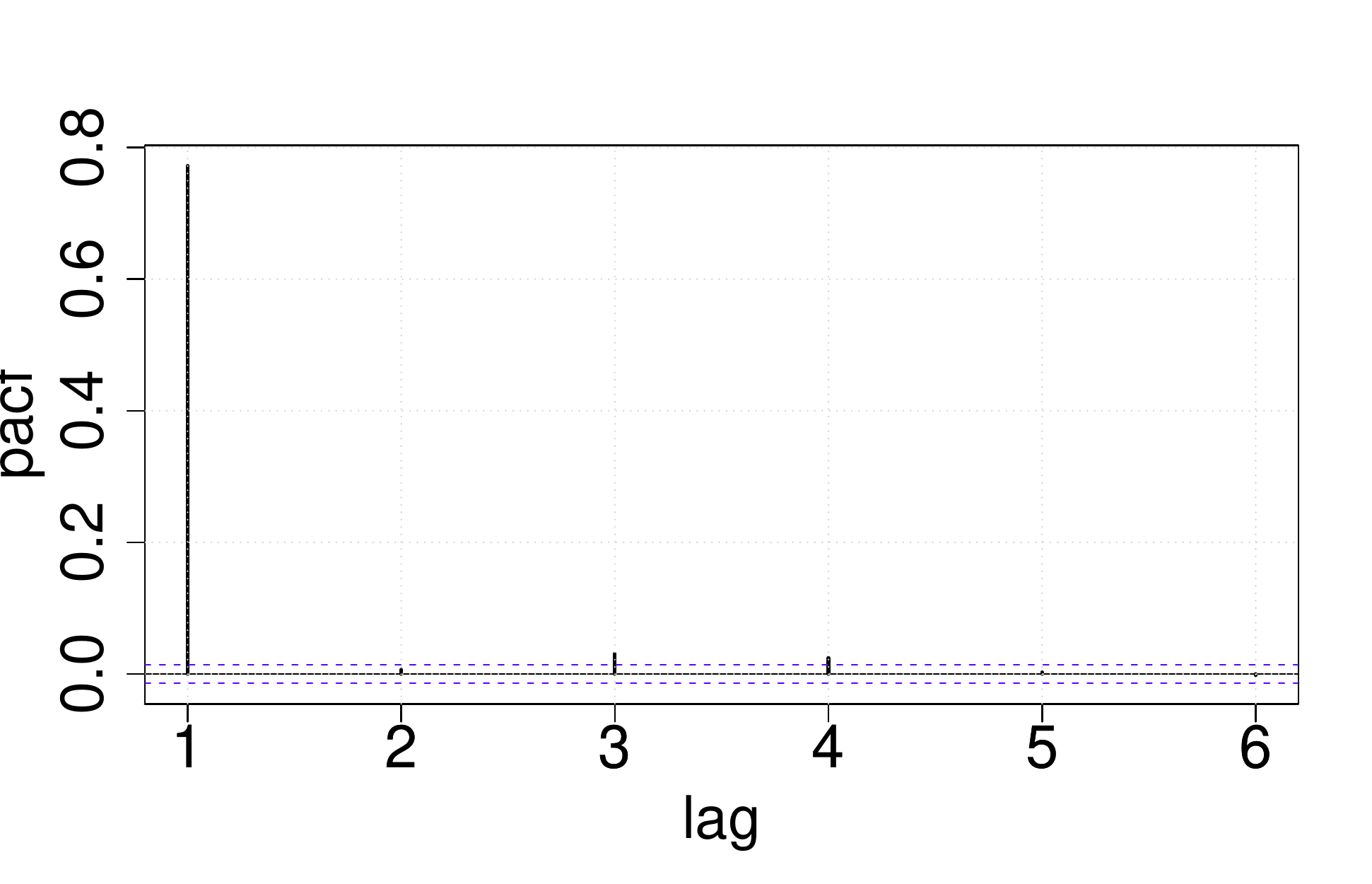}%pacf1.png
\caption{Partial autocorrelation estimated from the bead's trajectory for low frequency of sampling $10^3$ [fps].}\label{fig:pacf1}
\end{figure}
However, the high frequency data have the pacf of different type, see Fig \ref{fig:pacfD}. It looks like a geometric series with a negative rate which indicates the proper adjustment of the  model.
\bg{figure}[H] \centering
\includegraphics[width=10cm]{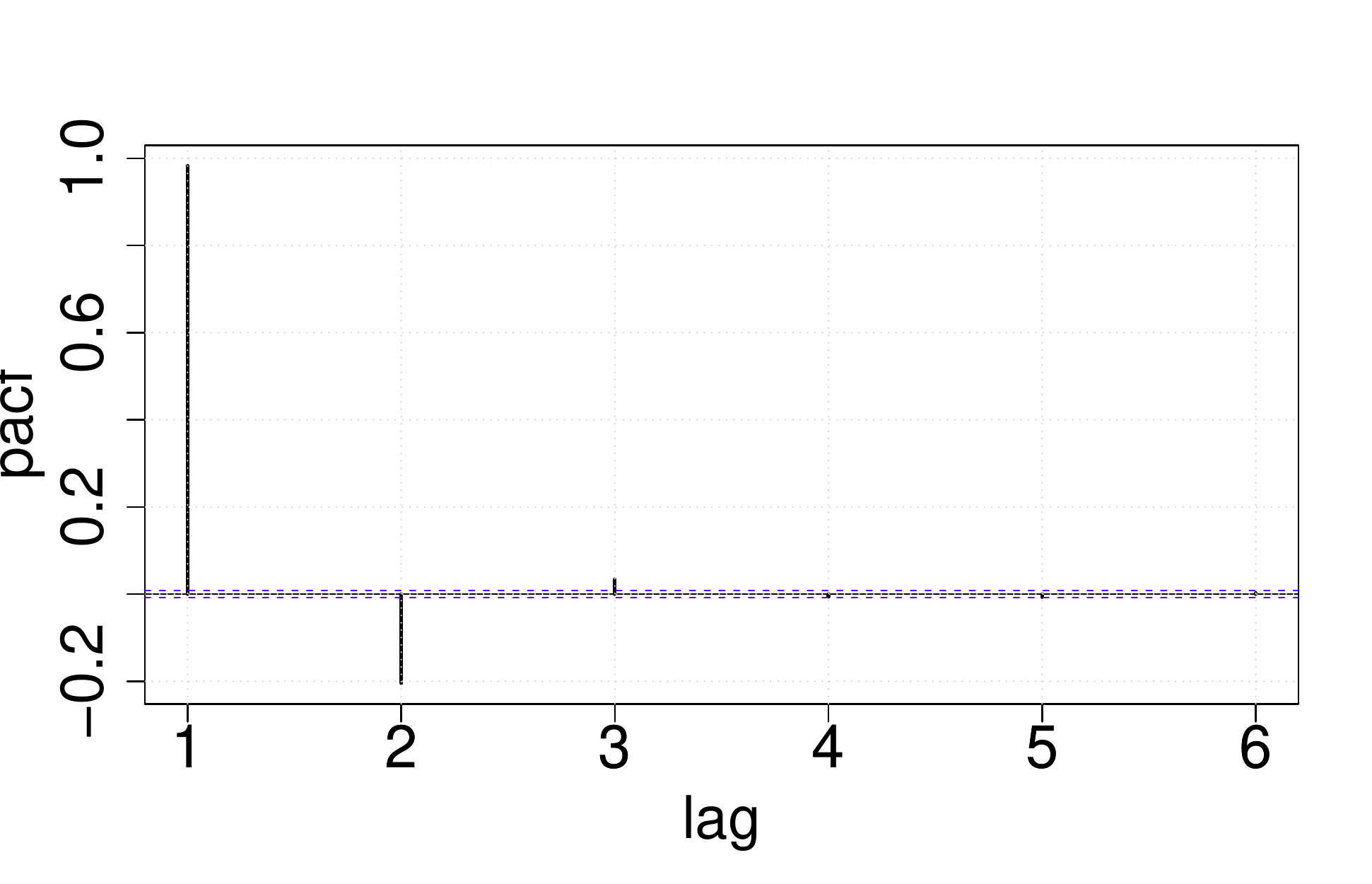}%pacfD.png
\caption{Partial autocorrelation estimated from the bead's trajectory for high frequency of sampling $10^4$ [fps].}\label{fig:pacfD}
\end{figure}

It appears that a proper modification explaining the data is addition of the first-order Moving Average ( MA(1) )  part to the  AR(1) model. The MA(1)  part represents  a process for which the present value depends linearly on one past value of the  external noise (not the process itself), i.e., $\xi_k + \theta_1\xi_{k-1}$. The complete model, with both AR(1) and MA(1) parts, is expressed explicitly as
\bgeq\label{eq:ARMA}
X_k = \phi_1X_{k-1} + \xi_k+\theta_1\xi_{k-1}.
\eeq
Unfortunately, from this form it is not clear what is the direct dependence between $X_k$ and the past values of the process, because the value $\xi_{k-1}$ is not orthogonal to these values. But, using the identity $X_{k-1}=\phi_1X_{k-2}+\xi_{k-1}+\theta_1\xi_{k-2}$ we may write  $\xi_{k-1} = X_{k-1} - \phi_1X_{k-2} - \theta_1\xi_{k-2}$. Substituting this equality into Eq. \Ref{eq:ARMA} we obtain
\bgeq
X_k = (\phi_1+\theta_1)X_{k-1}- \phi_1\theta_1X_{k-2} - \theta_1^2\xi_{k-2}.
\eeq
Now, we may repeat this procedure using the identity for $\xi_{k-2}$ and the above formula. Continuing recursively we obtain

\bgeq
X_k = (\phi_1+\theta_1)\sum_{j=1}^\infty (-\theta_1)^{j-1}X_{k-j}.
\eeq
This formula shows the explicit form of a dependence between $X_k$ and the past values $X_{k-j}$, therefore justifying why the partial autocorrelation has the form of a geometric series fitting the behaviour of the analysed data. The parameters $\phi_1$ and $\theta_1$ can be estimated using least-squares or maximum-likelihood techniques. The parameter $\theta_1$ is not unique, there are two possible choices, see end of the last section. In our context the one with $0<\theta_1<1$ has a proper interpretation and this one was chosen. Performed fit returned $\theta^\text{est}_1 = 0.212\pm 0.005$ and $\phi^\text{est}_1=0.976\pm 0.001$, which perfectly corresponds to  the ARMA(1,1) process with  the pacf drawn in Fig. \ref{fig:pacfD}. Given uncertainties of $\phi^\text{est}_1$ and $\theta^\text{est}_1$ are standard deviations estimated from a sample of ten trajectories with 60 000 observations in each of them. The obtained sample of ten estimates of $\phi_1$ can be considered Gaussian on standard level of significance 0.05 by Shapiro-Wilk, Anderson-Darling and Pearson $\chi^2$ tests, the same is true for $\theta_1$. This agrees with the well-known theory of long-trajectory asymptotics of the used estimators \citep{BD}. These standard deviations are only slightly greater than the deviations for these estimators for such an ARMA(1,1) process (which was checked by Monte Carlo simulation). It means that measurement imperfections do not distort the precision of estimation.

Of course, there is no possibility to rule out that performed estimation is precise, but biased. However, note that the obtained value $\phi^\text{est}_1$ is consistent with the measurements of the stiffness for lower frequencies.

The ARMA(1,1) can be naturally explained as an influence of the  high-frequency CCD camera. When the frequency of the sampling is high, the CCD matrix has no time to fully refresh between subsequent photos. The remainder of the last frame is still visible on the current one (the recording is smudgy and blurred), which causes the centre of the mass position $X_k$ to include part of the value $X_{k-1}$ from the last frame.   Instead of $X_k$ {we observe} $X'_k=X_k+\theta_1X_{k-1}$. Parameter $\theta_1$ is exactly the amount of intensity left from the last frame on the actual one, $0<\theta_1<1$. In this situation
\begin{align} X'_k &= X_k+\theta_1X_{k-1}  \nonumber\\
&= \phi_1(X_{k-1}+\theta_1X_{k-2}) + \xi_k+\theta_1\xi_{k-1} \nonumber\\
&= \phi_1X'_{k-1}+\xi_k+\theta_1\xi_{k-1},
\end{align}
so $X'_k$ is the ARMA(1,1) process.
It is straightforward to obtain the covariance function of this new process
\begin{align}
r_{X'}(j) &= \cov(X_k+\theta_1X_{k-1},X_{k+j}+\theta_1X_{k-1+j}) \nonumber\\
&= (1+\theta_1^2)r_X(j)+\theta_1\,r_X(j-1) +\theta_1\,r_X(j+1).
\end{align}
As we see it is a sum of three geometric sequences.
The form of the revised power spectral density $s_X$ follows from Proposition \ref{prp:ARMApsd}
\bgeq
s_X(\omega) = c\f{1+\theta_1^2+2\theta_1\cos(\omega)}{1+\phi_1^2-2\phi_1\cos(\omega)}.
\eeq
This revised psd fits perfectly the data, see Fig. \ref{fig:psdARMA}. Hence, the  ARMA(1,1) in the case of high-frequency recordings describes better the experimental data than the commonly used AR(1) model.

\bg{figure}[H] \centering
\includegraphics[width=10cm]{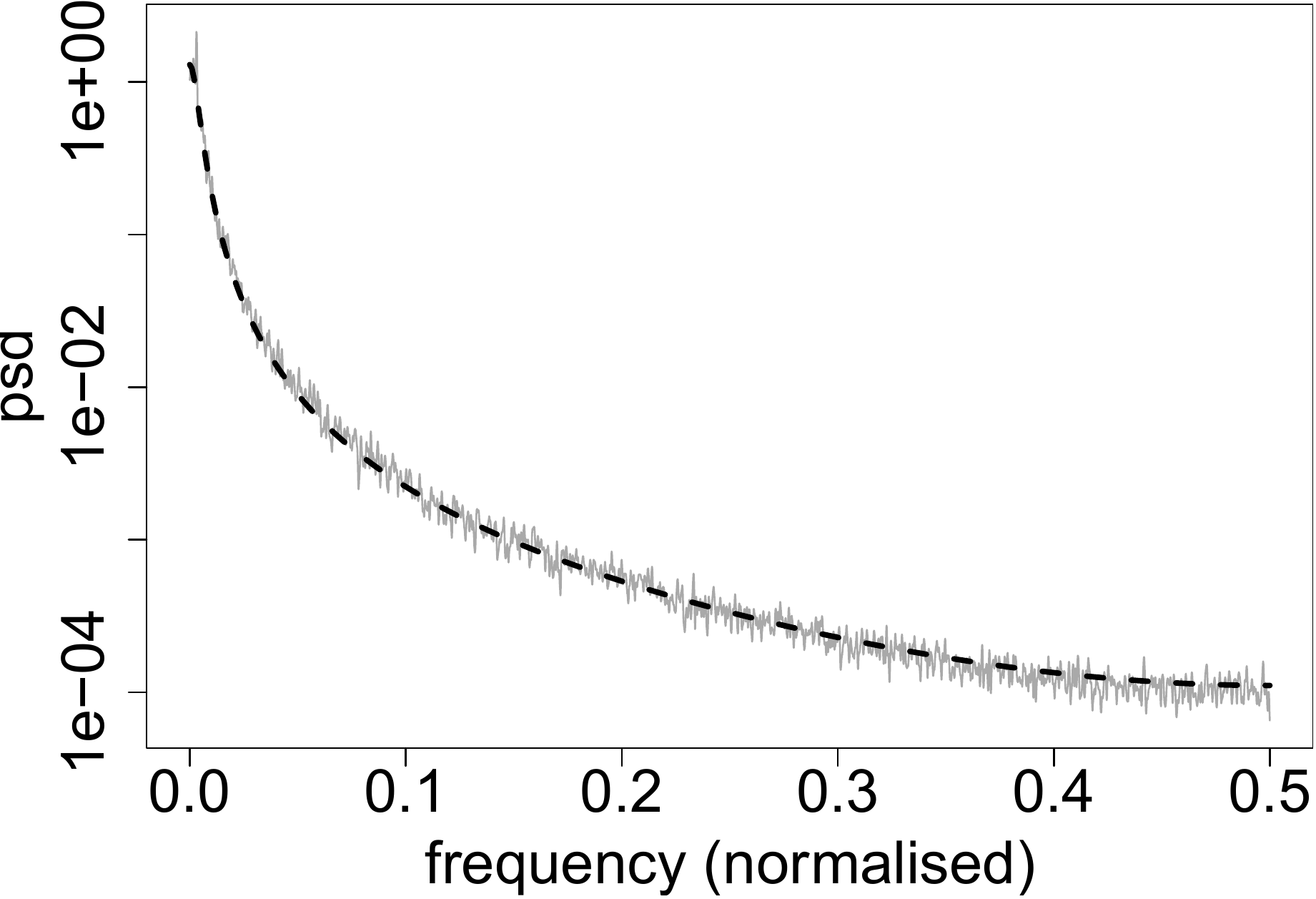}%psdARMA.png
\caption{Estimate of the psd (grey line) and the fitted ARMA(1,1) psd (dashed black line).}\label{fig:psdARMA}
\end{figure}
The previous considerations prove usefulness of the ARMA(1,1) model, but they do not imply that the observed effect is due to the influence of the CCD camera. To justify our claim we provide further experimental evidence. During all our previous measurement the exposure time for taking a frame was 45 [$\mu s$]. The longer exposition is in most of the situations used only when necessary (e.g. in cases of low illumination), since the recording is becoming more blurred. So, if our conclusion is correct, an increase of the exposition time would incorporate an MA(1) part even for low frequencies. This indeed is true, as seen in Fig. \ref{fig:expCh}; compare this result to Figs. \ref{fig:pacf1} and \ref{fig:pacfD}.

\bg{figure}[H] \centering
\includegraphics[width=10cm]{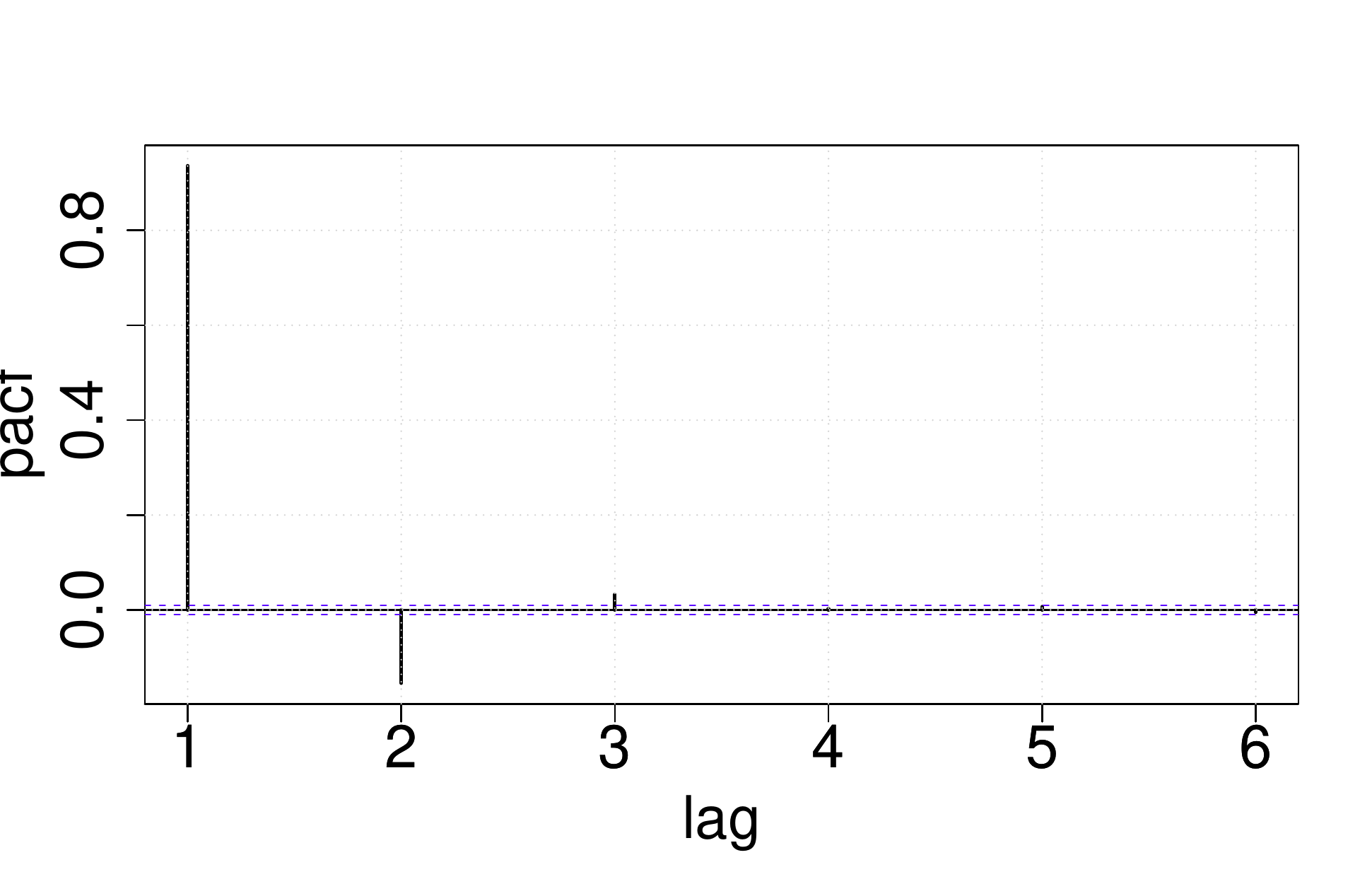}%expCh.jpeg
\caption{Estimated pacf of trajectory taken with frequency of sampling $10^3$ [fps] but exposure time increased to 895 [$\mu s$].}\label{fig:expCh}
\end{figure}
The obtained effect is identical to the MA(1) part (with coefficient $\theta_1^\text{est}=0.172\pm0.005$) present for high frequency and low exposure time data. Analogical operation performed for larger sampling frequencies increases the value of the MA(1) coefficients. Unfortunately, a decrease of the exposition, which would likely diminish the MA(1) part for higher frequencies, is hard to obtain from technical reasons, which leaves described methodology as the only practical way to deal with this influence.  

\chapter{Ergodicity and Fourier space}\label{ch:erg}

The relation between time averages and ensemble averages is still one of the most important topics of statistical physics and this area of research is under intense development. Its mathematical description, abstract ergodic theory is a very wide subject \citep{ergWalters,ergTh}. However, in recent years, a new trend has emerged which concentrates on very practical questions, concerning the behaviour of time-averaged statistics useful in analysis of the real data. The most prominent example of a quantity used in this approach is the time-averaged mean square displacement \citep{weakErgJeon, weakErgBreak}.

In this chapter we use the Maruyma's ergodic theorem \citep{maruyama} to study the long-time behaviour of linear response systems, in particular modelled by the classical and generalized Langevin equations. This theorem allow us to formulate simple criteria which determine the ergodicity of the studied processes and provides interesting insight into the physical origin of ergodicity and non-ergodicity, which relates them to the structure of a heat bath and the Hamiltonian of the system. It also allows to analyse Gaussian models which are ergodic, but non-mixing which is an often overlooked possibility.

Our most important result is a generalization of the Maruyma's theorem \citep{gaussErg}, which determines the behaviour of the time-averages for processes which are non-ergodic but stationary, and have profound physical interpretation based on theory of harmonic oscillations, phonons and Langevin equation. We demonstrate that, in the case of these processes, the disparity between ensemble- and time-averages can be studied using non-linear statistics, such as time-averaged characteristic function.

The obtained result is also interesting from the theoretical standpoint, as its proof links mathematical concepts from the number theory, group theory and second order stochastic processes, which is an unusual, yet insightful, occurrence in the field of applied mathematics.
\newpage
\section{Basics of ergodic theory}
Ergodic theory studies the behaviour of various averages of $f(X)$, where $f$ is a function of the whole trajectory $X$.

\bg{dfn}For a given stochastic process $X$, if for a function $f$ 
\bgeq
\E|f(X)|<\infty,
\eeq
we call this function an observable. If
\bgeq
\E|f(X)|^2<\infty,
\eeq
we call $f$ a second-order observable.
\end{dfn}

Examples include observable of mean position $f(X)=X(t)$, mean square displacement $f(X)= \big(X(t+\Delta)-X(t)\big)^2$, covariance $f(X)=X(t+\Delta)X(t)$, and others. Take note, that for the above examples to make sense, the measured averages must be time independent which will indeed be the case. That means we consider processes which are stationary, or at least have stationary increments (see Definition \ref{dfn:statProc}). Under such assumptions one may take $t=0$ without any lose of generality.  

If the process has a time-varying mean $m(t):=\E[X(t)]\neq \const$ we can always decompose it as a random, zero-mean part and deterministic non-zero part. Because we will study systems in which complex behaviour will be contained in the random part, we will assume $m(t)=0$; the case $m(t)\neq 0$ would be a straightforward generalisation.

For every process there exists an associated family of time-shift operators $\mathcal{S}_\tau$ which describe the temporal evolution of the system, i.e. $\mathcal{S}_\tau X = (X(t+\tau))_{t\in\R}$. The time-shift operator was already used as a tool to analyse discrete-time ARMA processes, now we consider its continuous-time counterpart.

With that in mind we can formulate the ergodic theorem. We will use general form of this result, which gives deeper insight into behaviour of non-ergodic processes.
\bg{thm}[Birkhoff's ergodic theorem] For a stationary stochastic process $X$, that is, for a process with the measure-preserving shift operators $\mathcal{S}_\tau$, and any observable $f$
\bgeq
\lim_{T\to\infty}\f{1}{T}\int_0^T\dd\tau\ f\big(\mathcal{S}_\tau X\big) = \E[f(X)|\mathcal{C}],
\eeq
where by $\E[\bd{\cdot}|\mathcal{C}]$ we understand conditional expected value under condition $\mathcal{C}$, with $\mathcal{C}$ being the $\sigma$-algebra of sets invariant under family of transformations $\{\mathcal{S}_\tau\}_{\tau\in\R}$.
\end{thm}
Essentially we calculate the expected value assuming that all time invariant properties of $X$ are fixed. The physical interpretation of $\mathcal{C}$ is that this is a set of constants of motion associated with $X$.

By the measure preserving transformation we mean that
\bgeq
\pr(f(X)\in A) = \pr( f(\mathcal{S}_\tau X)\in A)
\eeq
for all measurable $A$. This essentially means that $X$ and time shifted $\mathcal{S}_\tau X$ are statistically undistinguishable, which coincides with the previously introduced notion of stationarity.

Therefore the Birkhoff's ergodic theorem states that if only the process $X$ is stationary, the time-average of any observable converges to a random variable $\E[f(X)|\mathcal{C}]$ which we can precisely determine if we can identify all time invariant properties of $X$.

It is now clear that for the classical ergodic theorem to hold, the process should be stationary and the condition $\mathcal{C}$ must be sufficiently weak, so that $\E[f(X)|\mathcal{C}] = \E[f(X)]$; there should be no significant time-invariant properties of $X$.

One immediate and important consequence of the Birkhoff's theorem is that, because of the so-called tower property $\E\big[E[\bd{\cdot}|\mathcal{C}]\big] = \E[\bd{\cdot}]$, for a stationary process the ensemble average of the time average of any observable is equal to the ensemble average
\bgeq
\E\lt[\lim_{T\to\infty}\f{1}{T}\int_0^T\dd\tau\ f\big(\mathcal{S}_\tau X\big)\rt] = \E[f(X)],
\eeq
in particular there is no possibility for a phenomena such as the weak ergodicity breaking 
\bgeq
\E\lt[\overline\delta{}^2_X(t)\rt]\neq \E\lt[\big(X(\tau+t)-X(\tau)\big)^2\rt]
\eeq
studied e.g. in \citep{weakErgBreak}.

Another concept closely related to the ergodicity is mixing. We say that the system is mixing if for any observables $f,g$ such that $\E|f(X)g(X)|<\infty$,
\bgeq
\lim_{T\to\infty}\E[f(X)g(\mathcal{S}_T X)] = \E[f(X)]\E[g(X)].
\eeq
This is basically a statement that the process $X$ and the shifted process $\mathcal{S}_T X$ are asymptotically independent; the dynamics of $X$ leaves no persisting memory. It implies ergodicity and is often easier to study than the ergodicity itself; however, there are examples of non-mixing ergodic processes, even in the Gaussian case, which we will show in Section \ref{s:specMaruyama}.
\section{Classical Maruyama's theorem}
The main part of our considerations is true only for the class of Gaussian processes. The precise definition \ref{dfn:gaussProc} was stated in Chapter \ref{ch:gaussVar}, but intuitively speaking the sufficient and necessary condition for the process to be Gaussian is that all $X(t)$ are Gaussian (we also admit the case when the process is non-random as a degenerate Gaussian with variance 0) and they are only linearly dependent \citep{gaussHS}. The presence of non-linear dynamics excludes Gaussianity. This topic will be discussed in more detail in Chapter \ref{ch:superstat}.

The limitation on the possible memory type has large consequences: the Gaussian variables are fully described by their linear dependence structure which is reflected in second moments. For stationary Gaussian process the description is further simplified, because then $m_X(t)=\E[X(t)]=\const$ and $r_X(s,t)=r_X(t-s)$ (Proposition \ref{eq:rmCond}). Non-zero constant mean does not change any significant properties of the process, so further on in this chapter we will consider only zero-mean case. The above facts, and the linear structure of Gaussian processes allow the mixing condition to be greatly reduced. 
\bg{prp}
A stationary Gaussian process is mixing if and only if
\bgeq
\lim_{T\to\infty}r_X(T) = 0.
\eeq
\end{prp}
This condition is often straightforward to check.
\begin{proof}
If the process is mixing, taking $f(X)=g(X)=X(t)$ we obtain
\bgeq
\lim_{T\to\infty}r_X(T)=\lim_{T\to\infty}\E[f(X)g(\mathcal S_TX)] = \E[f(X)]\E[g(X)]=0.
\eeq
To prove another implication we first show that 
\bgeq\label{eq:ggErg}
\lim_{T\to\infty}\E[f(X)f(\mathcal S_T X)]=\big(\E[f(X)]\big)^2
\eeq
for any $f$ such that $\E|f(X)|^2<\infty$. It is sufficient to show this property for the subclass of functions which are dense in the space of observables. A natural choice is the space of finite-dimensional characteristic functions, which contains the full information about the distribution of $X$
\bgeq
f(X)=\exp\lt(\I\sum_{j=1}^L \theta_j X(t+t_j)\rt).
\eeq
The considered expected value is
\begin{align}
&\E[f(X)f(\mathcal S_T X)]=\E\lt[\exp\lt(\I \sum_{j=1}^L\theta_j\big(X(t+t_j)+X(t+t_j+T)\big)\rt)\rt]\nonumber\\
&=\exp\lt(\sum_{i,j=1}^L\theta_i\theta_j\big(2r_X(t_j-t_i)+r_X(t_j-t_i+T)+r_X(t_j-t_i+T)\big)\rt).
\end{align}
Taking the limit $T\to\infty$ we can use the fact that $r_X(t+T)\to 0$ and get
\bgeq
\lim_{T\to\infty}\E[f(X)f(\mathcal S_T X)]=\exp\lt(2\sum_{i,j=1}^L\theta_i\theta_jr_X(t_j-t_i)\rt)=\big(\E[f(X)]\big)^2.
\eeq

Now, consider the observable $f+g$. Using the above result and mixing,
\begin{align}
&\lim_{T\to\infty}\E[(f(X)+g(X))(f(\mathcal S_T X)+g(\mathcal S_T X))]=\big(\E[f(X)]+\E[g(X)]\big)^2\nonumber\\
&=\big(\E[f(X)]\big)^2+2\E[f(X)]\E[g(X)]+\big(\E[g(X)]\big)^2\nonumber\\
&= \lim_{T\to\infty}\big(\E[f(X)f(\mathcal S_T X)]+\E[f(X)g(\mathcal S_T X)]+\E[g(X)f(\mathcal S_T X)]+\E[g(X)g(\mathcal S_T X)]\big)
\nonumber\\
&=\big(\E[f(X)]\big)^2+\lim_{T\to\infty}\big(\E[f(X)g(\mathcal S_T X)]+\E[g(X)f(\mathcal S_T X)]\big)+\big(\E[g(X)]\big)^2.
\end{align}
If we compare the second and fourth line we get the equality
\bgeq
\lim_{T\to\infty}\big(\E[f(X)g(\mathcal S_T X)]+\E[g(X)f(\mathcal S_T X)]\big)=2\E[f(X)]\E[g(X)].
\eeq
The two terms on the left are in fact equal. These expected values depend only on the distribution of $X$. But the distribution of the process reflected in time and process shifted in time is identical to the original one, so
\bgeq
\E[g(X)f(\mathcal S_T X)]=\E[g(X)f(\mathcal S_{-T} X)]=\E[g(\mathcal S_T X)f(X)]=\E[f(X)g(\mathcal S_T X)].
\eeq
Finally
\bgeq
\lim_{T\to\infty}\E[f(X)g(\mathcal S_T X)]=\E[f(X)]\E[g(X)].
\eeq

\end{proof}

The ergodicity itself can also be expressed in the language of the covariance function. Initially, the notion of metric transitivity was used in this context \citep{grenander, dym}. The most important theorem in this field was proven by Gishiro Maryuama \citep{maruyama}.
\bg{thm}[Maruyama]A Gaussian stationary process $X$ is ergodic if and only if
\bgeq
\lim_{T\to\infty}\f{1}{T}\int_0^T\dd\tau\ |r_X(\tau)| = 0.
\eeq
\end{thm}
The presence of the modulus $|r_X(\tau)|$ is crucially important, because it excludes periodic oscillations of $r_X$. Generally this condition may also be easy to check, as it is enough to know the asymptotic tail behaviour of the covariance function. But, at the same time, this theorem does not give much insight into the memory structure of non-ergodic Gaussian processes which are studied in Section \ref{s:statProp}. In fact it can be viewed as a consequence of the more general theorem shown there.

\section{Harmonic processes}\label{s:harmonic}
Consider elementary example of a motion in the harmonic potential, governed by the equation
\bgeq
\df{}{t}{X}=-\omega_0^2 X,\quad X(0)=X_0,\df{}{t}{X}(0)=V_0,
\eeq 
with has the solution
\bgeq
X(t)= X_0\cos(\omega_0 t) +\f{V_0}{\omega_0} \sin(\omega_0 t).
\eeq
Under the assumption that at the beginning of the evolution, the system interacted with the heat bath, $X_0$ and $V_0$ have Gibbs distribution given by the density
\bgeq
p(x_0,v_0)\propto \exp\lt(- \omega_0^2\f{x_0^2}{2k_B\mathcal T}\rt)\exp\lt( - \f{v_0^2}{2 k_B\mathcal T}\rt).
\eeq
The resulting stochastic process has the covariance function
\bgeq
r_X(t) = \f{k_B\mathcal T}{\omega_0^2}\cos(\omega_0 t),
\eeq
so it is a stationary Gaussian process. Its spectral representation is therefore given by the measure
\bgeq
\sigma_X(\dd\omega) = \f{k_B\mathcal T}{2\omega_0^2}\big( \delta(\dd\omega-\omega_0) +\delta(\dd\omega+\omega_0)\big)
\eeq
which is concentrated in 2 points, which we denote by two Dirac deltas. In a natural way a question about ergodicity arises. Whereas the time average of the observable of the position $f(X)=X(t)$ 
\bgeq
\lim_{T\to\infty}\f{1}{T}\int_0^T\dd\tau\ X(\tau) = \lim_{T\to\infty}\f{1}{T}\lt(\f{X_0}{\omega_0}\sin(\omega_0 T) = \f{V_0}{\omega_0^2}(\cos(\omega_0 T)-1)\rt) = 0
\eeq
converges to the ensemble mean $0=\E[X(t)]$, the observable of the mean square displacement does not, as
\bgeq\label{eq:msdta}
\lim_{T\to\infty}\f{1}{T}\int_0^T\dd\tau\ \big(X(\tau+\Delta)-X(\tau)\big)^2 = 2\lt(X_0^2+\f{V_0^2}{\omega_0^2}\rt)\lt(\sin\lt(\f{\omega_0\Delta}{2}\rt)\rt)^2
\eeq
differs from the ensemble average
\bgeq\label{eq:msdaa}
\E\lt[\big(X(t+\Delta)-X(t)\big)^2\rt]=4 \f{k_B\mathcal T}{\omega_0^{2}}\lt(\sin\lt(\f{\omega_0\Delta}{2}\rt)\rt)^2.
\eeq

From the point of view of Hamiltonian mechanics this lack of ergodicity is expected; after the initial contact with a heat bath the system evolves as a microcanonical ensemble and the trajectories are trapped on the surface of constant energy, which prohibits ergodicity. Indeed, the term $X_0^2+V_0^2/\omega_0^2$ on the left side of Eq. \Ref{eq:msdta} is the total energy of the system, generally random, but constant on the trajectories of $X$; whereas $2k_B\mathcal T/\omega_0^2$ in \Ref{eq:msdaa} is the mean total energy.

A similar reasoning applies to the more general process of the form
\bgeq\label{eq:harm}
X(t) = \sum_k A_k \e^{\I\omega_k t},
\eeq
where the sum can even be infinite if $A_k$ are independent, complex Gaussian variables and $\sum_k\E|A_k|^2<\infty$. Random functions of this class, called harmonic processes, are appearing e.g. in phonon theory. Their covariance function and spectral measure are
\bgeq
r_X(t)=\sum_k\E|A_k|^2\cos(\omega_k t),\quad\sigma_X(\dd\omega)=\sum_k \f{\E|A_k|^2}{2}\big( \delta(\dd\omega-\omega_k) +\delta(\dd\omega+\omega_k)\big).
\eeq
If one calculates the ensemble- and time-average of the mean-square displacement for such a process, the different nodes of oscillation prove to be uncoupled in both time- and ensemble- average sense; the corresponding formulas are sums of terms as in Eq. \Ref{eq:msdaa} or \Ref{eq:msdta}. Therefore, any process within this class is stationary, but non-ergodic. The next section will show that it is the only case of non-ergodic Gaussian stationary process.

\section{Spectral form of Maruyama's theorem}\label{s:specMaruyama}
All the properties of a Gaussian process can be described interchangeably by its covariance function or its spectral measure; this very specific property of the Gaussian class is caused by its linear structure. For sure Maruyama's theorem can be expressed in the language of the spectral measure, and this reformulation leads to a surprisingly elegant statement \citep{dym}
\bg{thm} A stationary Gaussian process is ergodic if and only if its spectral measure has no atoms.
\end{thm}
Any measure can be decomposed as a sum of three distinct components: the absolutely continuous, singular and discrete measures. For a stochastic process the corresponding decomposition of the spectral measure $\sigma=\sigma_{\text{ac}}+\sigma_\text{s}+\sigma_\text{d}$ causes also the process itself to decompose into three independent components
\bgeq\label{eq:ergDec}
X(t) = X_\text{ac}(t)+X_\text{s}(t)+X_\text{d}(t),
\eeq
which is guaranteed by the harmonic representation \Ref{eq:harmRep}.
\bg{itemize}
\item[i)] The component $X_\text{ac}$ is mixing. It has a power spectral density. The Riemann-Lebesgue lemma shows that in this situation the covariance of $X_\text{ac}$, and all memory functions  decay at infinity, i.e. the values of the process become asymptotically independent at long time scales.
\item[ii)] The component $X_\text{s}$ is ergodic, but its memory structure may be complex. Its covariance function does not necessarily decay to 0. It may oscillate, but must be aperiodic and the high correlation events must become more scarce as $t\to\infty$.
\item[iii)] The last component $X_\text{d}$ is non-ergodic and is a Gaussian harmonic process.
\end{itemize}

The case ii) is most mathematically challenging, because the set of spectral measures, for which the corresponding covariance function decays, called Rajchman measures, does not have a convenient description \citep{70y}. As a demonstration let us consider an example using the best-known singular measure: the Cantor measure.

The Cantor set is obtained by removing from the middle one-third the interval $[0,1]$, then repeating this procedure for the two remaining intervals $[0,1/3]$, $[1/3,1]$ and recursively applying this procedure infinitely many times. The points which will remain are Cantor points. They can be more easily characterised as points in $[0,1]$ which have no 1s in their ternary representation, i.e. have the representation $\sum_{k=1}^\infty d_k 3^{-k}, d_k\in\{0,2\}$.

The Cantor measure $\sigma_C$ is the uniform measure on the Cantor set. We move it by -1/2 to the interval $[-1/2,1/2]$ so that the corresponding process will be real-valued. Elementary calculation proves that the length of the intervals removed during construction of the Cantor set is 1, therefore the Cantor measure cannot have a density and must be singular. It is probably the simplest to understand it as a discrete uniform distribution on the i.i.d. series $D_k$, $\pr(D_k=-1)=\pr(D_k=1)=1/2$ mapped onto interval $[-1/2,1/2]$ using the formula $Y=\sum_{k=1}^\infty D_k 3^{-k}$. The process $C$ which has the Cantor measure as a spectral measure has the covariance function
\bgeq\label{eq:rC}
r_{C}(t)=\int_\R\sigma_C(\dd \omega)\ \e^{\I\omega t}=\E\lt[\e^{\I t\sum_{k=1}^\infty D_k3^{-k}}\rt]=\prod_{k=1}^\infty \E\lt[\e^{\I t D_k 3^{-k}}\rt] = \prod_{k=1}^\infty \cos\lt(t3^{-k}\rt).
\eeq
In contrast to the better-known classes of the covariance functions, $r_{C}$ has a specific property close to a self-similarity
\bgeq
r_{C}(3t) = \cos(t) r_{C}(t),
\eeq
which also guarantees that $r_{C}$ does not decay to zero. The extremal points of $r_C$ are located at $t_k= 3^k\pi$, $r_C(t_k)=(-1)^{k+1}r_C(\pi)$, where it attains the values $\approx \pm 0.47$, see Fig. \ref{f:rC}. It may not be clear that the function $r_C$ can be easily calculated numerically, but the convergence is actually quite fast.
\bg{prp}
For the numerically approximated covariance function of the Cantor process 
\bgeq
\tilde r(t) \defeq \prod_{k=1}^N \cos\big(3^{-k}t\big),
\eeq
the above function converges to the Cantor covariance function $r_C$ and for $N\ge \log_3 t$ it is bounded by
\bgeq
(1+ct^29^{-N}) r_C(t)\le \tilde r(t) \le \e^{dt^29^{-N}} r_C(t),\quad c,d>0.
\eeq
This inequality holds for those values of $t$ for which $r_C(t)$ is positive, and the reverse inequality holds for  those $t$ when it is negative.
\end{prp}
\bg{proof}
For $N\ge \log_3 t$ and any $j\in\mathbb{N}$ we have $x=3^{-(N+j)}t\le \pi/2$, $\cos\big(3^{-(N+j)}t\big)>0$ and
\bgeq
1 - x^2/2\le \cos(x) \le 1- 4/\pi^2 x^2.
\eeq
Expressing $\tilde r$ in terms of $r_C$
\bgeq
\tilde r(t)=r_C(t)\prod_{j=1}^\infty \f{1}{\cos\big(3^{-(N+j)}t\big)}
\eeq
and assuming that  $r_C$ is positive for a given $t$,
\bgeq
r_C(t)\prod_{j=1}^\infty \f{1}{1-3^{-2(N+j)}t^24/\pi^2} \le \tilde r(t)\le r_C(t)\prod_{j=1}^\infty \f{1}{1-3^{-2(N+j)}t^2/2}.
\eeq
We need to approximate the product of terms
\bgeq
\f{1}{1-a 9^{-N}9^{-j}} = 1 + \f{1}{a^{-1}9^N9^j-1}= 1+9^{-N}\f{1}{a^{-1}9^j-9^{-N}}\mathop{=\colon} 1+p_j.
\eeq
To use the inequality valid for positive $p_j$'s
\bgeq
1+\sum_{j=1}^\infty p_j\le \prod_{j=1}^\infty (1+p_j)\le \exp\lt(\sum_{j=1}^\infty p_j\rt)
\eeq
we estimate the sum of $p_j$'s by
\bgeq
\sum_{j=1}^\infty p_j\ge a9^{-N}\sum_{j=1}^\infty \f{1}{9^j}\ge a9^{-N}\f{1}{8}
\eeq
from the bottom, and writing
\bgeq
\sum_{j=1}^\infty p_j= 9^{-N}\f{1}{a^{-1}-9^{-N}}+9^{-N}\sum_{j=1}^\infty\f{1}{9a^{-1}9^j-9^{-N}}\le \f{1}{9^Na^{-1}-1}+\f{1}{9}\sum_{j=1}^\infty p_j
\eeq
we estimate
\bgeq
\sum_{j=1}^\infty p_j\le \f{9}{8}\f{1}{9^Na^{-1}-1}
\eeq
from the top. Substituting the proper $a$ we obtain the final result
\bgeq
\lt(1+\f{t^2}{2\pi^2}9^{-N}\rt)r_C(t)\le \tilde r(t)\le \exp\lt(\f{9}{8}\f{1}{2t^{-2}-9^{-N}}9^{-N}\rt)r_C(t).
\eeq
For the negative $r_C(t)$ the above inequality is reversed.
\end{proof}

\begin{figure}[h]\centering
\includegraphics[width=\columnwidth]{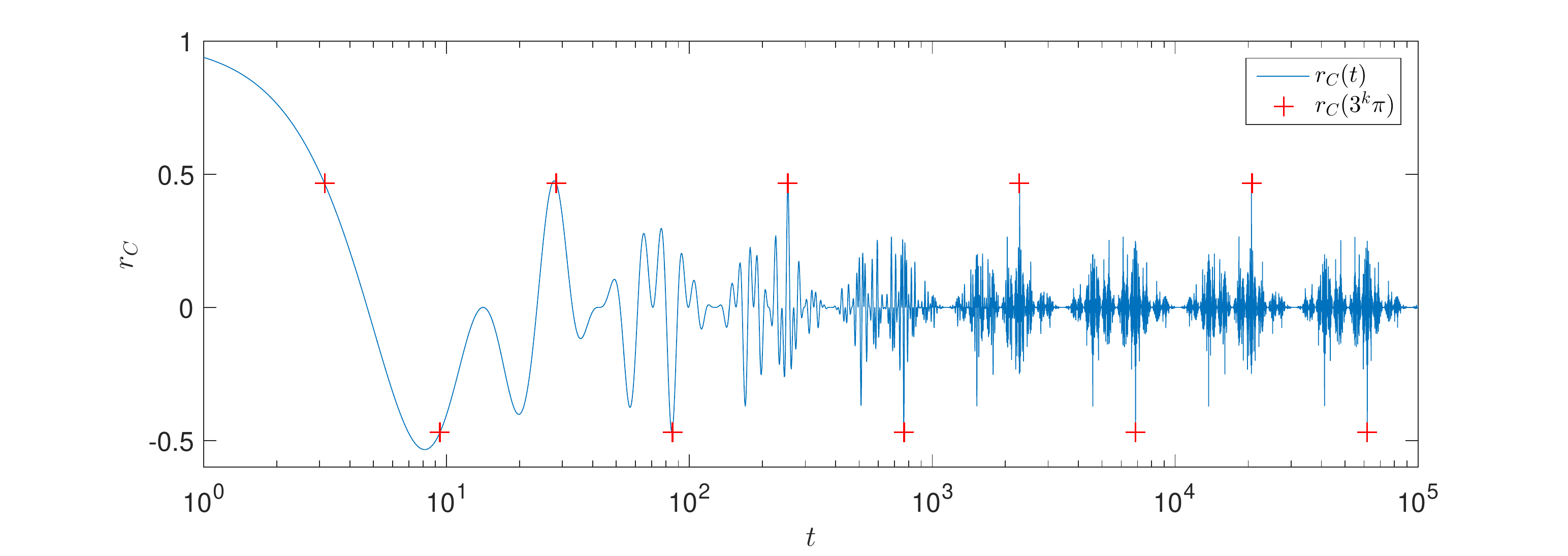}
\caption{Plot of the covariance function $r_C$ of the process with the Cantor spectral measure.}\label{f:rC}
\end{figure}

The process $(C(t))_t$ exhibits recurring correlations that prohibit mixing; however, events of high-dependence are becoming exponentially  rare as the time delay increases, which allows for the emergence of ergodicity.

The situation becomes more complicated with even slight generalisations of the model. If instead we perform the recursive removal procedure such that at any step the remaining intervals on the left and right have the length one-$\eta$th of the previous one ($\eta$ being real number bigger than $2$), the obtained singular measure and the process  is non-mixing for natural $n$, but it is mixing for any $\eta$ which is not a Pisot-Vijayaraghavan number \citep{salem}. The Pisot-Vijayaroghavan numbers are a closed countable set which causes even infinitesimally small changes of $\eta$ to change the mixing behaviour.

The complex ergodic behaviour complicates the analysis of the models with singular spectral measures, but it is worth stressing that the erratic behaviour of the covariance functions may be useful for describing observations which could be otherwise accounted for as experimental errors. It is also worth noting that the singular measures are gaining attention for their relation to fractal dynamics and self-similarity \citep{taqqu,mandelbrot, embrechts}.

\section{Generalised Maruyama's theorem}\label{s:statProp}

Any real stationary Gaussian process can be written as
\bgeq
X(t)=X_\text{erg}(t)+\sum_{k=1}^N R_k\cos(\Theta_k+\omega_k t) +X_0,\quad \omega_k\neq0,
\eeq
which follows from the ergodic decomposition \Ref{eq:ergDec}, after taking the real part of the harmonic process \Ref{eq:harm}. The variables $R_k\defeq|A_k|$ are amplitudes of the spectral points at frequencies $\omega_k$ and have Rayleigh distribution with densities
\bgeq
p_{R_k}(x) = \f{x^2}{\sigma_k^2}\exp\lt(-\f{x^2}{2\sigma_k^2}\rt)
\eeq
and scale parameters $\sigma_k^2=\E|A_k|^2$. Variables $\Theta_k$ are phases of $A_k$ and have uniform distribution on $[0,2\pi)$, a consequence of rotational invariance of i.i.d. Gaussian vectors. In full generality the number of spectral points may be infinite $N=\infty$, however in this case the process $X$ may exhibit a complicated aperiodic behaviour; as it is not very important for the applicational purposes, most of our subsequent result will be true only in the case $N<\infty$.

The decomposition into non-ergodic and ergodic components yields a useful and straightforward description of the statistical properties of Gaussian processes. It is made possible by the full characterisation of the invariant sets of this dynamical system.

\bg{thm} For any stationary Gaussian process $X$ with $N<\infty$ spectral points at rationally incommensurable frequencies $\{\omega_k\}_{k=1}^N$, the family of invariant sets $\mathcal{C}$ is the $\sigma$-algebra $\sigma(X_0,\{R_k\}_k)$
\end{thm}
\bg{proof}
We will combine methods for trigonometric series presented in \citep{kimme} and Gaussian ergodic theorem from Section 3.9 of \citep{dym}. 

We will use the representation
\bgeq
X(t)=X_\text{erg}(t)+\sum_{k=1}^N R_k\cos(\Theta_k+ \omega_k t) +X_0,\quad \omega_k\neq0,
\eeq
where the $\omega_k$ are distinct. The process $X_\text{erg}(t)$ has no spectral points. The full distribution of $X$ is generated by the values $X_\text{erg}(t_l),R_j\cos(\Theta_j+ \omega_j t_j), X_0$. Therefore it is sufficient to study the time-average distribution of the sum 

\bgeq
\sum_{l=1}^L\theta_lX_\text{erg}(t+t_l)+  \sum_{j=1}^N\lambda_j R_j\cos(\Theta_j+\omega_jt_j+\omega_jt)+\lambda_0X_0.
\eeq
For brevity we denote $\tilde\Theta_j\defeq\Theta_j+\omega_jt_j$. As the distribution is uniquely determined by the corresponding characteristic function, we will compare the time-averaged characteristic function and the ensemble-averaged one given the condition $\sigma(X_0,\{R_k\})$. Their equality will prove the theorem.

First we calculate the conditional ensemble-averaged characteristic function. The variables $\tilde\Theta_j$ have the same distribution as  $\Theta_j$ modulo $2\pi$, because they are independent from each other, $R_j$'s, $X_0$, $X_\text{erg}$, and have marginal uniform distribution. We get
\begin{align}
&\ \E\lt[\e^{\I\lt(\sum_{l=1}^L\theta_lX_\text{erg}(t+t_l)+  \sum_{j=1}^N\lambda_j R_j\cos(\tilde\Theta_j+\omega_jt)+\lambda_0X_0\rt)}|X_0,\{R_k\}\rt]\nonumber\\
& = \E\lt[\e^{\I\lt(\sum_{l=1}^L\theta_lX_\text{erg}(t_l)\rt)}\rt]\prod_{j=1}^N\E\lt[\e^{\I R_j\cos(\tilde\Theta_j)}|R_j\rt]\e^{\I\lambda_0 X_0}\nonumber\\
& = \phi_{\theta_1,\ldots,\theta_L}\prod_{j=1}^NJ_0(R_j\lambda_j)\e^{\I\lambda_0 X_0},\quad \phi_{\theta_1,\ldots,\theta_L}\defeq\E\lt[\e^{\I\lt(\sum_{l=1}^L\theta_lX_\text{erg}(t_l)\rt)}\rt],
\end{align}
where $J_0$ are Bessel functions of the first kind and order $0$; they stem from the formula
\bgeq
\f{1}{2\pi}\int_{-\pi}^\pi\dd x\ \e^{\I\lambda R\cos(x)}= J_0(\lambda R).
\eeq
The rest of the proof will be the calculation of the time-average.

We denote
\bgeq\label{eq:Phi}
\Phi(t)\defeq\e^{\I\lt(\sum_{l=1}^L\theta_lX_\text{erg}(t+t_l)\rt)}
\eeq
which appears in the integral used during calculations of the time-average
\bgeq
I_T\defeq\int_0^T\dd\tau\ \Phi(t) \e^{\I\lt(\sum_{j=1}^L\lambda_j R_j\cos(\tilde\Theta_j+\omega_jt)\rt)}\e^{\I\lambda_0 X_0}.
\eeq
The factor $\e^{\I\lambda_0 X_0}$ already agrees with the conditional average, so we will assume $X_0=0$ later on for brevity.

Next we expand each exponent of cosine using the Jacobi-Anger identity
\bgeq\label{eq:JA}
\e^{\I z\cos(w)}=\sum_{m=-\infty}^\infty \I^m\e^{\I mw}J_m(z),
\eeq
obtaining
\begin{align}\label{eq:IT}
I_T &=\int_0^T\dd\tau\ \Phi(\tau)\prod_{j=1}^N\sum_{m=-\infty}^\infty \I^m\e^{\I m(\tilde\Theta_j+\omega_j\tau)}J_m(\lambda_jR_j)\nonumber\\
& = \int_0^T\dd\tau\ \Phi(\tau)\sum_{S\in M_N}\prod_{m_j\in S}\e^{\I m_j(\tilde\Theta_j+\omega_j\tau)}\I^{m_j}J_{m_j}(\lambda_jR_j)\nonumber\\
&=\int_0^T\dd\tau\ \Phi(\tau)\sum_{S\in M_N} \exp \lt(\I\sum_{m_j\in S} m_j(\tilde\Theta_j+\omega_j\tau)\rt)  \prod_{m_j\in S}\I^{m_j}J_{m_j}(\lambda_jR_j)\nonumber\\
&=\sum_{S\in M_N}\exp\lt(\I\sum_{m_j\in S} m_j\tilde\Theta_j\rt)\prod_{m_j\in S}\I^{m_j}J_{m_j}(\lambda_jR_j) \int_0^T\dd\tau\ \Phi(\tau)\exp\lt(\I \tau\sum_{m_j\in S} m_j\omega_j\rt)
\end{align}
where $M_N$ is the family of all $N$-element subsets of integers $S$. Exchanging the order of infinite sum and integral is possible because the integrated function is bounded by $1$, which also justifies commuting the sum and the limit in the next step. We shall denote $\Omega_S\defeq\sum_{m_j\in S} m_j\omega_j$ and check that
\bgeq\label{eq:decay}
\lim_{T\to\infty}\f{1}{T}\int_0^T\dd\tau\ \Phi(\tau)\exp\lt(\I \tau\Omega_S\rt) = 0, \text{ if } \Omega_S\neq 0
\eeq
The proof of this statement is given in the lemma below. Let us conclude the whole proof. Because $\{\omega_k\}$ are rationally incommensurable, the equality $\Omega_S=\sum_{m_j\in S} m_j\omega_j=0$ holds for integer $m_j$ only when $m_1=m_2=\ldots=m_N=0$. In the sum $\sum_{S\in M_N}$ only one element $S=\{0,0,\ldots,0\}$ remains and
\bgeq
\lim_{T\to\infty}\f{1}{T}I_T=\prod_{j=1}^NJ_0(\lambda_jR_j)\lim_{T\to\infty}\f{1}{T}\int_0^T\dd\tau\ \Phi(\tau)= \prod_{j=1}^NJ_0(\lambda_jR_j)\phi_{\theta_1,\ldots,\theta_L},
\eeq 
where the last equality holds due to the ergodicity of $X_\text{erg}$. This is the desired conditional mean.
\end{proof}

\bg{lem}
\bgeq \label{eq:TPhiLim}
\lim_{T\to\infty}\f{1}{T}\int_0^T\dd\tau\ \Phi(\tau)\exp\lt(\I \tau\Omega \rt) = 0, 
\eeq
for $\Omega\neq 0$ and $\Phi$ given by \Ref{eq:Phi}.
\end{lem}
\bg{proof}
First note that $\Phi$ is a strictly stationary random process. Take $U\deq\mathcal{U}([0,2\pi))$ independent from $\Phi$. The process $t\mapsto \Phi(t)\exp(\I U+\I\Omega t)$ is also stationary and has a finite first moment equal to 1. Therefore, the Birkhoff's ergodic theorem guarantees the time-average exists almost surely and equals
\bgeq
\lim_{T\to\infty}\f{1}{T}\int_0^T\dd\tau\ \Phi(\tau)\exp\lt(\I U +\I \tau\Omega \rt) =\exp(\I U) \lim_{T\to\infty}\f{1}{T}\int_0^T\dd\tau\ \Phi(\tau)\exp\lt(\I\tau\Omega \rt)=\widehat X(\Omega).
\eeq
Thus, the limit \Ref{eq:TPhiLim} also exists almost surely and equals to a random variable $\exp(-\I U) \widehat X(\Omega)$ (for more details on $\widehat X(\Omega)$ see \citep[Chapter XI.2, page 516]{doob}). We will prove that it is 0.

Let $r$ be the covariance function of $X_\text{erg}$ and $\sigma$ its continuous spectral measure.  We will study $\E|\bd{\cdot}|^2$ of the above time-average and show its mean-square convergence to 0, which suffices to prove also the almost sure convergence to the same limit:
\begin{align}
& \ \f{1}{T^2}\int_0^T\dd\tau_1 \int_0^T\dd\tau_2 \E\lt[\exp\lt(\I\sum_{j=1}^L\theta_j \big(X_\text{erg}(t_j+\tau_1)-X_\text{erg}(t_j+\tau_2)\big)\rt)\rt]\e^{\I\Omega(\tau_1-\tau_2)} \nonumber\\
& = \f{C}{T^2}\int_0^T\dd\tau_1 \int_0^T\dd\tau_2 \exp\lt( \sum_{j,k=1}^L\theta_j\theta_k r(t_k-t_j+\tau_1-\tau_2)\rt)\e^{\I\Omega(\tau_1-\tau_2)}\nonumber\\
& = \f{C}{T^2} \int_0^T\dd\tau_1 \int_0^T\dd\tau_2  \exp\lt(\int_\R\dd\tilde\sigma(\omega)\e^{\I\omega(\tau_1-\tau_2)}\rt)\e^{\I\Omega(\tau_1-\tau_2)},
\end{align}
where we denoted by $C$ the factor before the integral and by $\tilde \sigma$ the modified spectral measure; it is just multiplied by a continuous function.
\bgeq
C\defeq\exp\lt(2\sum_{j,k=1}^L\theta_i\theta_j r(t_k-t_j)\rt),\quad \dd\tilde\sigma(\omega)\defeq \dd\sigma(\omega) \lt|\sum_{j=1}^L\theta_j\e^{\I\omega t_j}\rt|^2
\eeq
Next, we expand external $\exp(\bd\cdot)$ into the Taylor series, obtaining
\begin{align}
&\ \f{C}{T^2}\int_0^T\dd\tau_1 \int_0^T\dd\tau_2 \lt(1+\sum_{n=1}^\infty\f{1}{n!}\lt(\int_\R\dd\tilde\sigma(\omega)\e^{\I\omega(\tau_1-\tau_2)}\rt)^n\rt)\e^{\I\Omega(\tau_1-\tau_2)}\nonumber\\
&= \f{C}{T^2}\int_0^T\dd\tau_1 \int_0^T\dd\tau_2 \lt(1+\sum_{n=1}^\infty\f{1}{n!}\int_\R\dd\tilde\sigma^{*n}(\omega)\e^{\I\omega(\tau_1-\tau_2)}\rt)\e^{\I\Omega(\tau_1-\tau_2)},
\end{align}
where $\tilde\sigma^{*n}$ is the $n$-fold convolution power of $\tilde\sigma$. This Taylor series is uniformly convergent. We commute the limit $T\to\infty$ with the sum and calculate the integrals; for the term $n=1$ we have
\bgeq
\f{1}{T^2}\int_0^T\dd\tau_1 \int_0^T\dd\tau_2 \e^{\I\Omega(\tau_1-\tau_2)}=\f{1}{T^2}\f{1}{\Omega^2}\lt|\e^{\I\Omega T}-1\rt|^2\xrightarrow{T\to\infty}0,
\eeq
where the assumption $\Omega\neq0$ is crucial. For any other term
\begin{align}
&\ \lim_{T\to\infty}\f{1}{T^2}\int_0^T\dd\tau_1 \int_0^T\dd\tau_2 \int_\R\dd\tilde\sigma^{*n}(\omega)\e^{\I(\omega+\Omega)(\tau_1-\tau_2)}\nonumber\\
&=\lim_{T\to\infty}\f{1}{T^2} \int_\R\dd\tilde\sigma^{*n}(\omega-\Omega)\int_0^T\dd\tau_1 \int_0^T\dd\tau_2\e^{\I\omega(\tau_1-\tau_2)}\nonumber\\
& = 2\lim_{T\to\infty}\int_\R\dd\tilde\sigma^{*n}(\omega-\Omega)\f{1-\cos(\omega T)}{(\omega T)^2}.
\end{align}
In the last line one recognises the functional which returns the jump of the measure $\tilde\sigma^{*n}$ at the point $\Omega$. But, the measure $\sigma$ is continuous and $\tilde\sigma^{*n}$ is also continuous; the result is $\tilde\sigma^{*n}(\{\Omega\})=0$.
\end{proof}

The assumption that spectral points at $\{\omega_k\}_{k=1}^N$ are rationally incommensurable means that they cannot be represented as $\omega_k= q_k \alpha$ for any rational $q_k$'s. It is trivially fulfilled in most real physical systems, in which $\omega_k$ are self-frequencies of the harmonic oscillators and depend on the complex set of the system's parameters.

In such a case, the aforementioned theorem guarantees that for any observable $f$, the time average converges to
\bgeq
\lim_{T\to\infty}\f{1}{T}\int_0^T\dd\tau\ f(\mathcal{S}_\tau X)=\E\big[f(X)|,X_0,\{R_k\}_k\big],
\eeq
i.e. to the ensemble average calculated under the condition that the amplitudes of the spectral points and the constant term $X_0$ are fixed. For an observable which depends on the one time moment of $X$ only, $f(X)=f(X(t))$, the above formula simplifies to the explicit integral
\begin{align}
& \lim_{T\to\infty}\f{1}{T}\int_0^T\dd\tau\ f(X(\tau))\nonumber\\
&=\int_0^{2\pi}\dd\theta_1\int_0^{2\pi}\dd\theta_2\ldots \int_\R \dd x\f{1}{\sqrt{2\pi}c}\e^{-\f{x^2}{2c^2}} f\big( x +\sum_kR_k\cos(\theta_k)+X_0\big),
\end{align}
which depends only on the variance of the ergodic component $c^2=\E\big[X_\text{erg}(t)^2\big]$,  $R_k$'s and $X_0$, which are random, but fixed for each trajectory.

In particular, using the cumulative distribution function method, one can calculate the non-ergodic time-averaged probability density $\overline p_{X_k}$ of any given discrete spectral component with amplitude $R_k=R$
\bgeq\label{eq:pdfnerg}
\overline p_{X_k}(x) = p_{X_k}(x|R)=\f{\big(\pi R\big)^{-1}}{\sqrt{1-\big(x/R\big)^2}},\quad -R\le x\le R.
\eeq 
This quantity should be observed, if one uses the time-average estimation of the probability density, e.g. the kernel density estimators or the histogram \citep{denEst}.

This is also an interesting example of a singularity caused by the non-ergodicity. The flat extrema of the cosine function are responsible for the probability density  divergence of type $x^{-1/2}$ at points $-R$ and $+R$. However, this unusual behaviour is not easy to directly observe as typical data contain an ergodic component, e.g. some kind of noise from the experimental setup. In such a case, the observed empirical probability distribution $\overline p$ is a convolution of the ergodic component's Gaussian density with a given stationary variance $\sigma^2$ and some number of densities of type \Ref{eq:pdfnerg}, see Figure \ref{f:pdfComp}. The singular concentration of the probability mass around $-R$ and $+R$ distorts the tails of the distribution in some specific way: it thickens them by moving the original distribution, but thins them through dividing by a square root factor. More precisely, we recognise that the convolution of the densities is an example of a Laplace transform and use Abelian theorem \citep{taub} to obtain the asymptotic behaviour of the left tail

\begin{align}\label{eq:pdfAsympt}
\overline p(-x)&=\f{1}{\sqrt{2\pi^3}\sigma R}\int_{-R}^R\dd y \f{1}{\sqrt{1-\big(x/R\big)^2}}\e^{-\f{(y+x)^2}{2\sigma^2}}\nonumber \\
&=\e^{-\f{(x-R)^2}{2\sigma^2}}\f{1}{\sqrt{2\pi^3 R}\sigma}\int_0^{2R}\dd y \f{1}{\sqrt{y}\sqrt{2-y/R}}\e^{-\f{y^2}{2\sigma^2}}\e^{-\f{y(x-R)}{\sigma^2}}\nonumber\\
&\sim \f{1}{4\pi \sqrt{1-x/R}}\e^{-\f{(x-R)^2}{2\sigma^2}},\quad x\to \infty;
\end{align}
symmetrically for the right tail. This result does not change significantly for a finite number of spectral points, as it depends only on the presence of singularities of the convolved densities.

If there is only one discrete spectral component, that is $X(t)=X_\text{erg}(t)+R\cos(\Theta+\omega t),\E[X_\text{erg}^2]=\sigma^2$, the time-averaged pdf can be expressed as a series of Bessel functions,
\bg{align}\label{eq:pdfSerExact}
\overline p_X(x)&=\f{1}{\sqrt{2\pi\sigma^2}}\f{1}{2\pi}\int_0^{2\pi}\dd\theta\ \e^{-\f{(x-R\cos(\theta))^2}{2\sigma^2}}=\f{1}{\sqrt{2\pi\sigma^2}}\e^{-\f{x^2}{2\sigma^2}}\f{1}{2\pi}\int_0^{2\pi}\dd\theta\ \e^{\f{xR\cos(\theta)}{\sigma^2}}\e^{-\f{R^2\cos(\theta)^2}{2\sigma^2}}\nonumber\\ 
&=p_{X_\text{erg}}(x) \f{1}{2\pi}\int_0^{2\pi}\dd\theta\ \e^{\f{xR\cos(\theta)}{\sigma^2}}\sum_{k=0}^\infty\f{(-1)^k}{k!2^k}\f{R^{2k}}{\sigma^{2k}}\cos(\theta)^{2k}\nonumber\\
&=p_{X_\text{erg}}(x)\sum_{k=0}^\infty\f{(-1)^k}{k!2^k}\df{^{2k}}{x^{2k}}\f{1}{2\pi}\int_0^{2\pi}\dd\theta\ \e^{\f{xR\cos(\theta}{\sigma^2}}=p_{X_\text{erg}}(x)\sum_{k=0}^\infty\f{(-1)^k}{k!2^k}\df{^{2k}}{x^{2k}}I_0\lt(\f{xR}{\sigma^2}\rt).
\end{align}
By $I_0$ we denoted the modified Bessel function of the first kind and order $0$; the derivatives of this function can be expressed using linear combinations of $I_n$ if needed.

Both asymptotic formula \Ref{eq:pdfAsympt}  and series \Ref{eq:pdfSerExact} differ considerably from the normal distribution; for most realisations the amplitudes $R_k$ are large enough to strongly affect the time-averaged probability density. This behaviour is presented in Fig. \ref{f:pdfComp}. For nearly all realisations statistical tests also show significant non-Gaussianity (e.g. Shapiro and Kolmogorov-Smirnov tests). However, for small amplitudes of the $R_k$'s this effect could be less noticeable.

\begin{figure}[h]\centering
\includegraphics[width=\columnwidth]{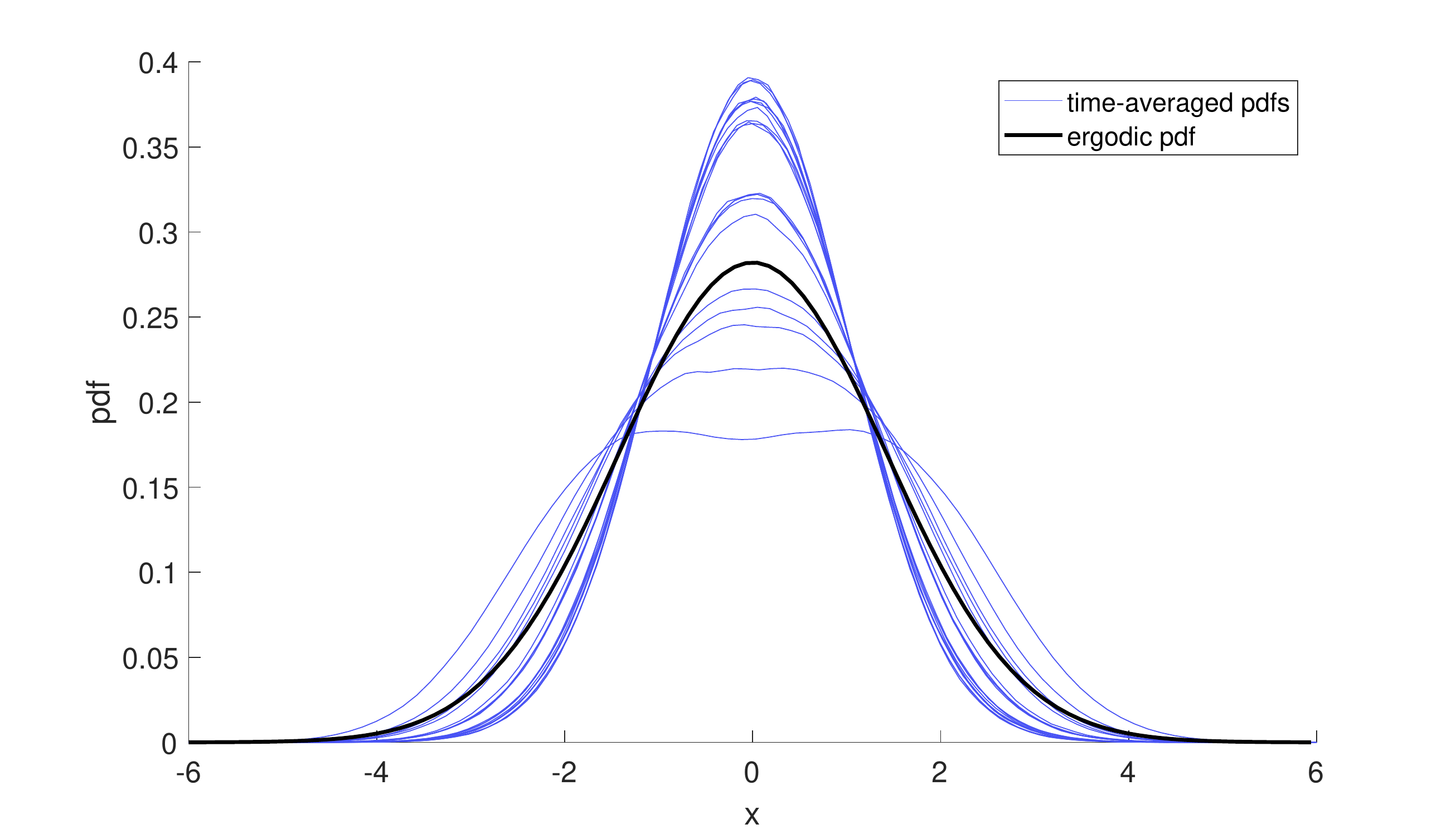}
\caption{Time-average kernel density estimation and ergodic density for a Gaussian process with one discrete spectral component, $\E\big[R_k^2\big]=\E\big[X_\text{erg}(t)^2\big]=1$.}\label{f:pdfComp}
\end{figure}

The fast decay of the function $\exp(-x^2)$ may complicate the analysis of non-ergodicity through a density estimation. A more convenient method is to use the time-averaged characteristic function. It is a time-average of the observable $f(X)=\exp(\I\theta X(t))$, which for a non-ergodic component equals
\bgeq
\lim_{T\to\infty}\f{1}{T}\int_0^T\dd\tau \ \e^{\I\theta X(\tau)} = \f{1}{2\pi}\int_{-\pi}^\pi\dd x\ \e^{\I\theta R\cos(x)}= J_0(R \theta),
\eeq
where $J_0$ is a Bessel function of the first kind and order $0$. Therefore the time-averaged characteristic function $\overline\phi$ of any stationary Gaussian process with incommensurable frequencies has form
\bgeq\label{eq:phiErg}
\overline\phi(\theta)= \e^{-c^2\theta^2/2}\prod_{k}J_0\big(R_k\theta\big).
\eeq
Unlike for the ergodic case, this function has zeros determined by the zeros of the function $J_0$, which may be approximated numerically, the first being $x_1\approx 2.405$, the second $x_2\approx 5.520$. Therefore the location of zeros for the time-averaged characteristic function may serve to preliminary estimate the number and values of the amplitudes $R_k$. A more precise estimation requires least-squares fitting with the function in Eq. \Ref{eq:phiErg}. Exemplary results are shown in Fig. \ref{f:estBoxPlot} which prove that it is a viable method of estimation for a small number of spectral points.

The estimators of $R_k$ have the tendency to return some undershoot values which cause negative bias, especially for lower lengths of the trajectories, but are generally reliable. The results depend on the sample length $N$, but do not depend on the sampling time $\Delta t$ as long as $\pi\Delta t$ is incommensurable with $\{\omega_k\}_k$. If this is not true, the measured time series has a periodic component and the proper values of the time-averaged observables are obtained in the infill asymptotics, i.e. $\Delta t\to0$, and $\Delta t N\to\infty$. Such requirements guarantee that the calculated mean converges to the time integral $1/T\int_0^T\dd \tau$ and $T\to\infty$.

\begin{figure}[h]\centering
\includegraphics[width=\columnwidth]{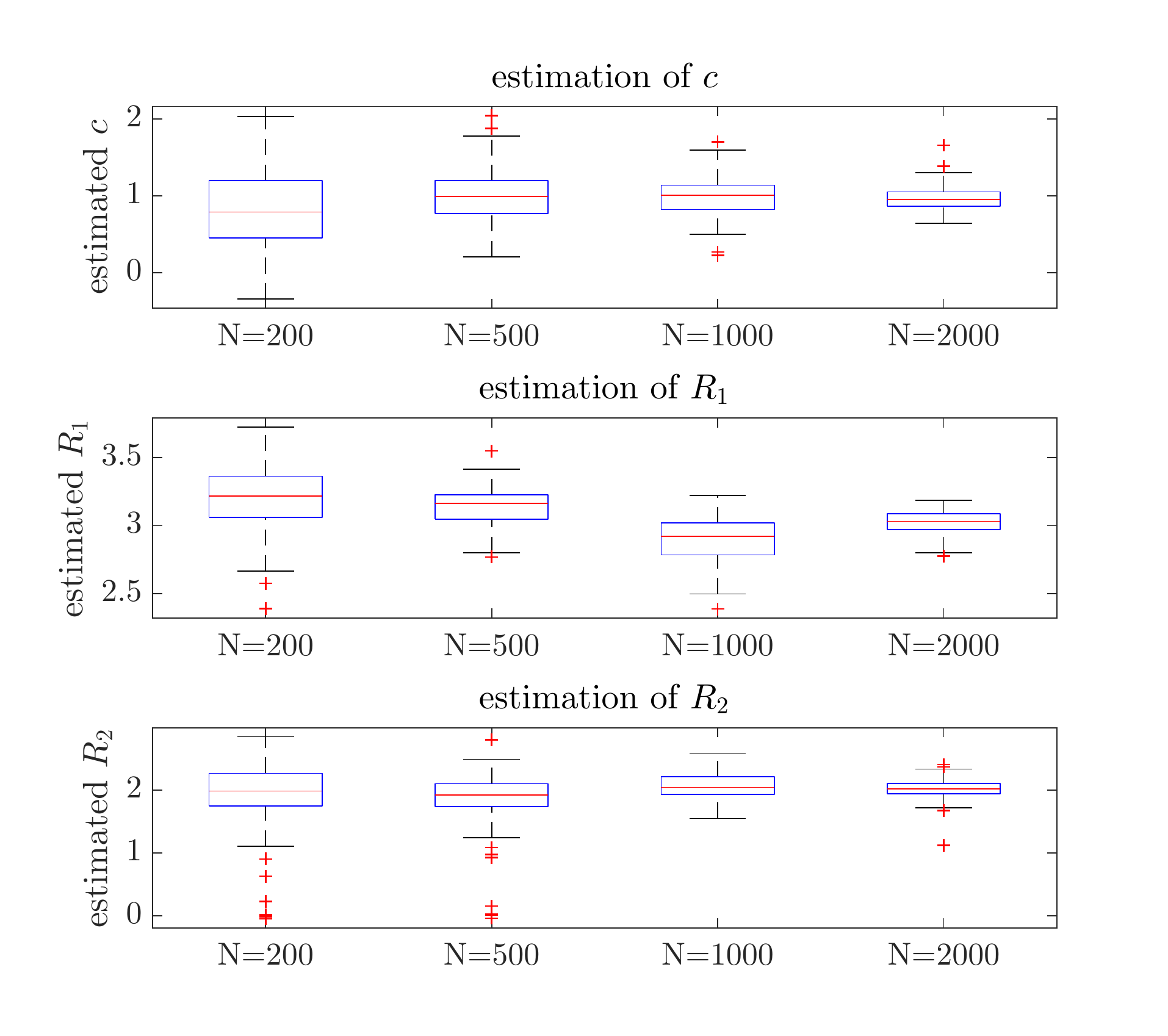}
\caption{Estimation of $c=\E[X_\text{erg}(t)^2],R_1$ and $R_2$ for different sample sizes $N$, using a least-square fit of the time-averaged characteristic function for the process $X_\text{erg.}(t)+3\cos(t)+2\cos(\sqrt{2}t)$.}\label{f:estBoxPlot}
\end{figure}

Let us consider such models in which it may happen that $\{\omega\}_k$ are commensurable. This situation may appear practically when $\{\omega_k\}$ by coincidence or due to some symmetry are close to being commensurable, that is they can be expressed as $\omega_k=\alpha p_k/q_k+\epsilon_k$ where $p_k,q_k$ are small natural numbers and $\epsilon_k$ is small compared to $1/T$. 

For commensurable $\{\omega_k\}$ the harmonic process is in fact periodic and the length of its period is proportional to the lowest common denominator of $\omega_k$'s. We show an example in Fig. \ref{f:fChar}, where the empirical characteristic function of the process $R_1\cos(\Theta_1\omega_1 t){+R_2\cos(\Theta_2+\omega_2 t)}$ is presented. For simplicity we fixed $\omega_1=1$ and $R_1=1, R_2=2$, as the  $R_k$'s are constants of motion in this case. Because of the periodicity, the calculated time-average depends on the random initial phases $\Theta_1,\Theta_2$.

\begin{figure}[h]\centering
\includegraphics[width=\columnwidth]{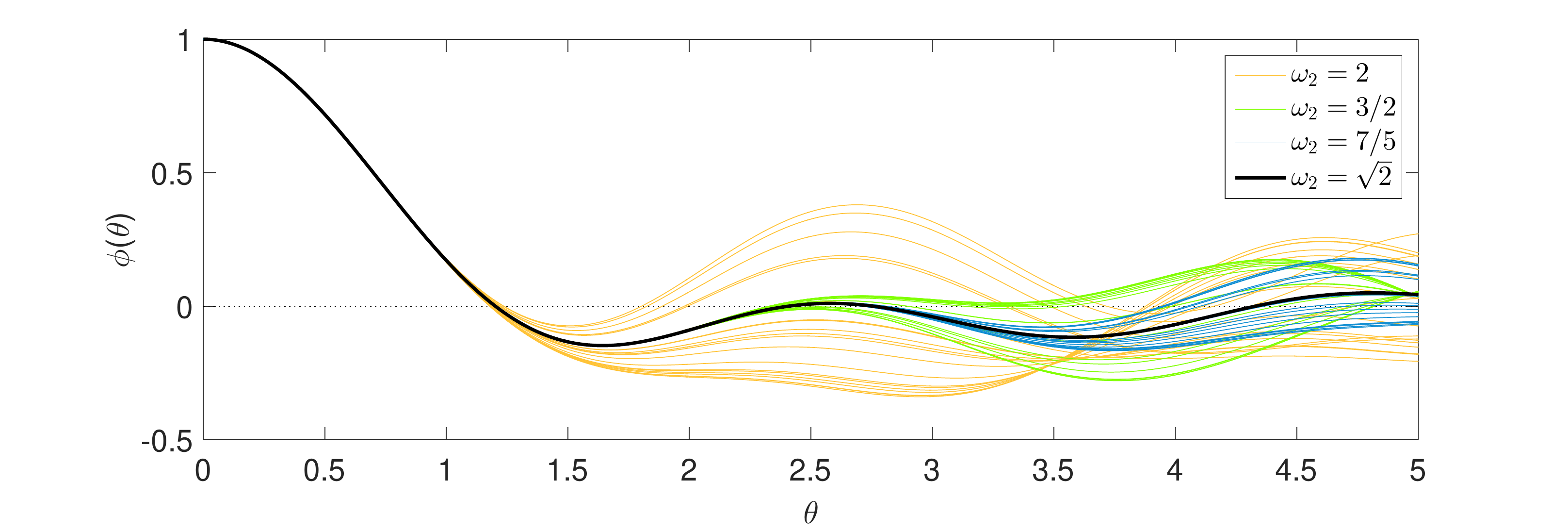}
\caption{Time-averaged characteristic function of the process $2\cos(\Theta_1+t)+2\cos(\Theta_2+\omega_2 t)$ for different $\omega_2$ and 20 realisations of $\Theta_1,\Theta_2$. The black line, corresponding to the simulation for $\omega_2=\sqrt{2}$ up to numerical accuracy $\epsilon\approx 10^{-16}$, agrees perfectly with ergodic average $J_0(t)J_0(2 t)$, see Eq. \Ref{eq:phiErg}.}\label{f:fChar}
\end{figure}

The exact dependence on both random amplitudes and angles is actually known; the time-averaged characteristic function of any stationary Gaussian process is given by the formula (see proof of the theorem below)
\bgeq\label{eq:TACharF}
\overline{\phi}(\theta)
=\e^{-c^2\theta^2/2}\sum_{S\in G}\exp\lt(\I\sum_{m_j\in S} m_j \Theta_j\rt)\prod_{m_j\in S}\I^{m_j}J_{m_j}(\theta R_j) ,
\eeq
where $G$ is the family of sets of $m_j$'s for which $\sum_jm_j\omega_j=0$. The result depends on linear combinations of the random phases $\sum_{j}m_j\Theta_j\Mod{2\pi}$. This is true not only for the time-averaged characteristic function, but any observable, which is stated in the following theorem.

\bg{thm} For any stationary Gaussian process $X$ with $N<\infty$ spectral points  $\{\omega_k\}_{k=1}^N$ (some of which may be commensurable), the family of invariant sets $\mathcal{C}$ is the $\sigma$-algebra $\sigma(X_0,\{R_k\}_k,\mathcal{M})$, where  $X_0$ is the constant term, $R_k$ are the amplitudes of the atoms of the spectral measure and $\mathcal{M}$ is a family
\bgeq\label{eq:famM}
\mathcal M=\lt\{\sum_{m_j}m_j\Theta_j \Mod{2\pi} \colon \sum_{j}m_j\omega_j=0\rt\}.
\eeq
Moreover, $\mathcal M$ can be reduced to contain at most $N-1$ integer linear combinations.
\end{thm}
\bg{proof}
Combining \Ref{eq:IT} and \Ref{eq:decay} we obtain the formula of the time-averaged characteristic function  of finite-dimensional distribution in the general case, which is
\bgeq\label{eq:finDimExp}
\phi_{\theta_1,\ldots,\theta_L}\e^{\I\lambda_0 X_0}\sum_{S\in G_N}\exp\bigg(\I\sum_{m_j\in S} m_j\tilde\Theta_j\bigg)\prod_{m_j\in S}\I^{m_j}J_{m_j}(\lambda_jR_j);
\eeq
here $G_N$ are all $N$-element subsets of integers $m_j$ for which $\sum_{j=1}^Nm_j\omega_j=0$. What is left is to show that the above quantity equals the conditional ensemble-average characteristic function
\begin{align}
&\E\lt[\e^{\I\lt(\sum_{l=1}^L\theta_lX_\text{erg}(t+t_l)+  \sum_{j=1}^N\lambda_j R_j\cos(\tilde\Theta_j+\omega_jt)+\lambda_0X_0\rt)}|X_0,\{R_k\},\mathcal M\rt]\nonumber\\
& = \phi_{\theta_1,\ldots,\theta_L}\e^{\I\lambda_0 X_0}\E\lt[\e^{\I\lt(  \sum_{j=1}^N\lambda_j R_j\cos(\tilde\Theta_j+\omega_jt)\rt)}|\{R_k\},\mathcal M\rt]
\end{align}
The factor $\phi_{\theta_1,\ldots,\theta_L}\e^{\I\lambda_0 X_0}$ already agrees, so we will omit it later on. Next, we expand the remaining expected value using the Jacobi-Anger identity \Ref{eq:JA}, obtaining

\bgeq\label{eq:expTrans}
\sum_{S\in M_N}\prod_{m_j\in S}\I^{m_j}J_{m_j}(\lambda_jR_j)\E\lt[\e^{\I \sum_{m_k\in S}m_k(\tilde\Theta_k+\omega_k t)}|\mathcal M\rt].
\eeq
Because $\theta\mapsto\e^{\I \theta}$ is an injection on $[0,2\pi)$ and $\tilde\Theta_k$ differ from $\Theta_k$ only by deterministic constants, the $\sigma$-algebra $\sigma(\mathcal M)$ is equivalent to $\sigma(\tilde{\mathcal M})$ generated by
\bgeq
\tilde{\mathcal M}=\lt\{\exp\bigg(\sum_{m_j}m_j\tilde\Theta_j\bigg) \colon \sum_{j}m_j\omega_j=0\rt\}.
\eeq
For terms with $S\in G_N\subset M_N$ the random phases in \Ref{eq:expTrans} are $\tilde{\mathcal M}$-measurable, therefore also $\mathcal M$-measurable, moreover they agree with the corresponding terms in \Ref{eq:finDimExp}. What is left is to show that the expected value of the remaining terms for $S\notin G_N$ is zero.

Now, for any incommensurable $\omega_j$ the corresponding $m_j=0$. Commensurable $\omega_j$'s can be divided into subsets of jointly commensurable numbers, i.e. into blocks $\{\omega_{k_i}\}$ for which $\omega_{k_i}=\alpha q_{k_i}/p_{k_i},q_{k_i}\in \mathbb{Z},p_{k_i}$ and  $q_{k_i}$ are coprime; different blocks have different incommensurable factors $\alpha$. Each such block corresponds to a different subset of independent $\tilde\Theta_j$, therefore they can be considered separately.

Let us choose one such block and for simplicity of notation, change indices such that these are $\{\omega_1,\omega_2\ldots,\omega_r\}$. The condition $\sum_{j=1}^rm_j\omega_j=0$ is equivalent to the condition $\sum_{j=1}^rm_j\eta_j=0$, where $\eta_j$ are relatively prime integers obtained by multiplying $\omega_j$ by the least common multiple of $p_j$'s. The equation
\bgeq
\sum_{j=1}^rm_j\eta_j=0
\eeq
has exactly $r-1$ linearly independent solutions in integers \citep{kimme}. For our one chosen block let us name these solutions $\{m_j^1\},\{m_j^2\},\ldots, \{m_j^{r-1}\}$. Any other solution is a linear combination of the elementary solutions
\bgeq
m_j=\sum_{\rho=1}^{r-1} \nu_\rho m_j^\rho,\quad \nu_j\in.\mathbb{Z}
\eeq
Therefore
\bgeq
\sum_{j=1}^r m_j\tilde\Theta_j = \sum_{j=1}^r \sum_{\rho=1}^{r-1} \nu_\rho m_j^\rho\tilde\Theta_j =\sum_{\rho=1} ^{r-1}\nu_\rho\sum_{j=1}^r  m_j^\rho\tilde\Theta_j,
\eeq
and for each block $\mathcal M$ depends actually only on $r-1$ variables
\bgeq
\Xi_\rho=\sum_{j=1}^r  m_j^\rho\tilde\Theta_j \Mod{2\pi},
\eeq
the rest of the variables are linear combinations of the elementary ones. For all blocks together $\mathcal M$ depends on at most $N-1$ such variables.

The factor in the studied conditional expectation corresponding to the chosen block is
\bgeq
\E\lt[\exp\bigg(\I\sum_{j=1}^r m_j\tilde\Theta_j\bigg)|\big\{\Xi_\rho\big\}_{\rho=1}^{r-1}\rt],\quad m_j\in S.
\eeq
Because $m_j\in S\notin G_N$ the sum $\sum_{j=1}^r m_j\tilde\Theta_j \Mod{2\pi}$ is  linearly independent from the set $\{\Xi_\rho\}_{\rho=1}^{r-1}$. We prove in the lemma below that it implies that this sum is also probabilistically independent from $\{\Xi_\rho\}_{\rho=1}^{r-1}$ and has uniform distribution on $[0,2\pi)$. Therefore
\begin{align}
&\E\lt[\exp\bigg(\I\sum_{j=1}^r m_j\tilde\Theta_j\bigg)|\big\{\Xi_\rho\big\}_{\rho=1}^{r-1}\rt]= \E\lt[\exp\bigg(\I\sum_{j=1}^r m_j\tilde\Theta_j\bigg)\rt]\nonumber\\
&=\E\lt[\e^{\I\Theta'}\rt] = \f{1}{2\pi}\int_0^{2\pi}\dd\theta\ e^{\I\theta} = 0, \quad \Theta'\deq \mathcal{U}(0,2\pi).
\end{align}

We have proven that all elements in the sum \Ref{eq:expTrans} which contain combinations of $\Theta_j$ linearly independent from elements of $\mathcal{M}$ are zero. Only $\mathcal{M}$-dependent elements remain, which exactly agrees with the time-average characteristic function \Ref{eq:finDimExp}. This concludes the proof.
\end{proof}

\bg{lem}
For i.i.d. $\{\Theta_j\}_{j=1}^N, \Theta_j\deq\mathcal U (0,2\pi)$ any $N$ linearly independent integer combinations
\bgeq\label{eq:lemmComb}
\Xi_i = \sum_{j=1}^N m_{ij} \Theta_j \Mod{2\pi}, \quad m_{ij}\in\mathbb Z
\eeq
are a set of jointly independent random variables with distribution $\Xi_i\deq\mathcal U(0,2\pi)$.
\end{lem}
\bg{proof}
Because we work in $\Mod{2\pi}$ arithmetic, all variables can be considered to have values in the torus $\mathbb T = \R/2\pi\mathbb Z$, $\mathcal U(0,2\pi)\equiv \mathcal U(\mathbb T)$. The continuous dual of $\mathbb T^N$ is $\mathbb Z^N$ and this a natural space of parameters of the characteristic function of the vector $(\Xi_i)_{i=1}^N$.

For a uniform distribution on torus the characteristic function has a very simple form: it is the Kronecker delta
\bgeq
\E\lt[\e^{\I k \Theta_j}\rt] = \delta_k,\quad k\in\mathbb Z.
\eeq
This is clear if we think about characteristic function as a Fourier series of density $1/(2\pi)$ on $\mathbb T$. We will show that the multidimensional characteristic function of $(\Xi_i)_{i=1}^N$ is the product $\prod_{j=1}^N\delta_{k_j}$ which corresponds to the  distribution $\mathcal U(\mathbb T^N)$ of $(\Xi_i)_{i=1}^N$.

Let us choose any $k_1,\ldots,k_N\in\mathbb Z$ and consider $\sum_i k_i\Xi_i$. We calculate the characteristic function
\begin{align}
&\E\lt[\exp\bigg(\I\sum_{i=1}^Nk_i\Xi_i\bigg)\rt]=\E\lt[\exp\bigg(\I\sum_{i=1}^Nk_i\sum_{j=1}^Nm_{ij}\Theta_j\bigg)\rt]\nonumber\\
&=\E\lt[\exp\bigg(\I\sum_{j=1}^N\Theta_j\sum_{i=1}^Nk_im_{ij}\bigg)\rt]= \prod_{j=1}^N\E\lt[\exp\Big(\I\Theta_j\sum_{i=1}^Nk_im_{ij}\Big)\rt] \nonumber\\
&= \prod_{j=1}^N\delta_{\sum_{i=1}^Nk_im_{ij}}.
\end{align}
The above product equals 1 and not 0 if, and only if for all $j$ we have $\sum_{i=1}^Nk_im_{ij}=0$. But, the linear integer combinations \Ref{eq:lemmComb} are linearly independent which is equivalent to saying that this is true if, and only if $k_1=k_2=\ldots=k_N=0$. The last product is $\prod_{j=1}^N\delta_{k_j}$ which was to be proven.

\end{proof}

This general theorem completely determines the behaviour of time-averaged observables for stationary Gaussian processes and has important practical applications, allowing for the statistical analysis of non-ergodic stationary Gaussian models.

As an example, let us come back to the time-averaged characteristic function \Ref{eq:TACharF}. For the case of a process with two spectral points with frequencies $\omega_1=1,\omega_2=2$, the numerically calculated time-averages are shown shown as yellow lines in Fig. \ref{f:fChar}. The integer combinations in the family $\mathcal M$ from Eq. \Ref{eq:famM} are exactly $m_1=2, m_2=-1$ and multiples of these. Therefore the time-averaged characteristic function depends only on $2\Theta_1-\Theta_2 \Mod{2\pi}$. Indeed, any yellow line in Fig. \ref{f:fChar} corresponds to many different random choices of $\Theta_1, \Theta_2$, and these lines can be parametrised only by the number $c\in[0,2\pi)$ defined as $2\Theta_1-\Theta_2 = c \Mod{2\pi}$. Similarly, green lines on the same figure depend only on $3\Theta_1-2\Theta_2 \Mod{2\pi}$, because we have $m_1=3, m_2=-2$, and so on.

It may seem counter-intuitive that the rationality or irrationality of the number $\omega_2$, which cannot be experimentally studied, affects the numerical simulations and the behaviour of the real systems. This apparent paradox disappears if we carefully analyse the behaviour of the time-average in the two essential cases
\bg{itemize}
\item For $\omega_2=p/q$ with coprime $p,q$ the trajectory will have period $2\pi q$. For $q$ sufficiently larger than the experimental time ($q\gg T$), the periodicity will be unobservable during the measurement and the time-averaged observables will seem independent from initial phases.
\item For irrational $\omega_2 =p/q+\epsilon$ with coprime $p,q$ and $\epsilon q \ll 1$ the time-averaged observables will not depend on initial phases in the long-time limit $T\to\infty$, however the process will be very close to a  periodic one, therefore the convergence will be slow.
\end{itemize}
So, we realise that in the real case with finite time of the experiment $T<\infty$ the practically significant property is how close $\omega_2$ is to a irreducible fraction with a small denominator. Analogously, for multiple $\omega_k$ we only need to determine how close they are to a set of commensurable numbers with simple rational ratios.

In order to illustrate this fact numerically we simulated the process $R_1\cos(\Theta_1+t)+R_2\cos(\Theta_2+\omega_2t)$ with fixed $R_1=2,R_2=3$ and calculated the variance of the time-averaged characteristic function at one point $\overline{\phi}(2.5)$. For irrational $\omega_2$ and in the limit $T\to\infty$ this quantity should be equal to zero, however in the numerical experiment it is a positive function of $\omega_2$ which indeed measures how close $\omega_2$ is to a simple irreducible fraction. See Fig. \ref{fig:freqAnal} where the peaks of the variance indicate the positions of the simplest irreducible fractions like $1/2, 1/3,2/3,1/4,3/4$ \emph{etc}.

\begin{figure}[h]\centering
\includegraphics[width=\columnwidth]{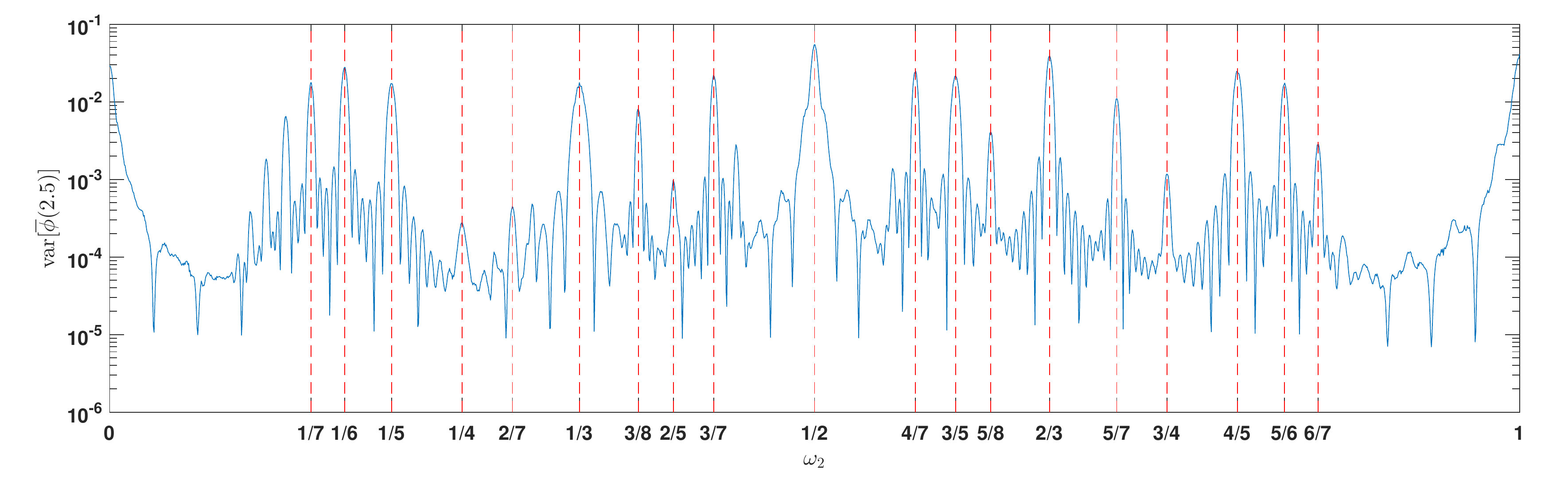}
\caption{Numerical estimation of the variance $\var[\overline\phi(2.5)]$ obtained using $10^3$ samples of trajectories with lengtht $T=200$ and values of $\omega_2$ taken as one thousand uniformly scattered numbers in the interval $[0,1]$ stored in format double.}\label{fig:freqAnal}
\end{figure}

Another result, which may be somehow unexpected, is that according to the calculation in Section \ref{s:harmonic} even for commensurable $\{\omega_k\}_k$ the time-average second-order properties depend only on $\{R_k\}_k$ and the dependence of initial phases $\{\Theta_k\}_k$ is lost.

\bg{prp} For any stationary Gaussian process $X$, the time-average covariance structure is the ensemble-average structure conditioned by the $\sigma$-algebra $\sigma(X_0,\{R_k\}_k)$, where $R_k$ are the amplitudes of the atoms of the spectral measure and $X_0$ is the constant term. The result is true even for rationally commensurable frequencies $\{\omega_k\}$ of spectral points and infinite number of them, $N=\infty$.
\end{prp}
\bg{proof} We begin with calculating the conditional ensemble-average covariance. The conditional mean equals
\bgeq
\E[X(t)|X_0,\{R_k\}]= X_0,\quad\text{as }\E[\cos(\Theta_k)]=0\text{, and }\E[X_\text{erg}(t)]=0
\eeq
Next we fix $t_1,t_2$ and use the independence of the $X_\text{erg}$ and $X_k$'s.
\begin{align}
&\ \E\big[(X(t_1)-X_0)(X(t_2)-X_0)|X_0,\{R_k\}\big]\nonumber\\
&=\E\big[X_\text{erg}(t_1)X_\text{erg}(t_2)|X_0,\{R_k\}\big]+\sum_{k=1}^\infty\E\big[ X_k(t_1)X_k(t_2)|X_0,\{R_k\}\big]\nonumber\\
&=\E\big[X_\text{erg}(t_1)X_\text{erg}(t_2)\big]+\sum_{k=1}^\infty R_k^2\E\big[\cos(\Theta_k+\omega_kt_1)\cos(\Theta_k+\omega_kt_2)\big]   \nonumber\\
&= r(t_2-t_1)+\sum_{k=1}^\infty R_k^2\f{1}{2\pi}\int_0^{2\pi}\dd x\ \cos(x+\omega_k t_1)\cos(x+\omega_k t_2) \nonumber\\
&=r(t_2-t_1)+\f{1}{2}\sum_{k=1}^\infty R_k^2 \cos(\omega_k(t_2-t_1)).
\end{align}
The time-average consist of an integral from the parts $X_\text{erg}(t_i+\tau)X_\text{erg}(t_j+\tau)$,$X_k(t_i+\tau)X_k(t_j+\tau)$, $X_{k_1}(t_i+\tau)X_{k_2}(t_j+\tau), k_1\neq k_2$ and $X_\text{erg}(t_i+\tau)X_k(t_j+\tau)$, $i,j\in\{1,2\}$; we call them $I_1,I_2,I_3, I_4$, respectively. All sums are absolutely convergent: we can commute summation, integration and taking limit $T\to\infty$.

Time-average $T^{-1}I_1$ converges to $r(t_2-t_1)$ because $X_\text{erg}$ is ergodic. For $T^{-1}I_2$ we have
\begin{align}
&\f{1}{T}I_4= \f{1}{T}R_k^2\int_0^T\dd\tau\ \cos(\tau+\omega_kt_i)\cos(\tau+\Theta_k+\omega_kt_j)\nonumber\\
&= \f{1}{2}R_k^2\cos(\omega_k(t_2-t_1))\nonumber\\
&+\f{1}{\omega_kT}\cos(2\Theta_k+\omega_k(t_1+t_2+T))\sin(\omega_kT)\xrightarrow{T\to\infty} \f{1}{2}R_k^2\cos(\omega_k(t_2-t_1)).
\end{align}
Therefore we need to prove that $T^{-1}I_3$ and $T^{-1}I_4$ decay to 0. For $T^{-1}I_3$ it is straightforward, denoting $\omega_\pm \defeq\omega_{k_2}\pm\omega_{k_1},\Theta_\pm\defeq\Theta_{k_2}\pm\Theta_{k_1}$ we get
\begin{align}
&\ \f{1}{T}I_3=\f{1}{2T\omega_-}\big(\sin(\omega_{k_1}t_i-\omega_{k_2}t_j-\Theta_-)-\sin(\omega_{k_1}t_i-\omega_{k_2}t_j-\Theta_--T\omega_-)\big)\nonumber\\
&+\f{1}{2T\omega_+}\big(-\sin(\omega_{k_1}t_i+\omega_{k_2}t_j+\Theta_+)-\sin(\omega_{k_1}t_i+\omega_{k_2}t_j+\Theta_++T\omega_+)\big)\xrightarrow{T\to\infty}0.
\end{align}
As for $T^{-1}I_4$, we will show that the time-average of $X_\text{erg}(t_i+\tau)R_k\exp(\I\Theta_k+\I\omega_k\tau)$ converges, which is an equivalent condition. The factor $R_k\exp(\I\Theta_k)$ can be brought outside the integral, therefore only showing the convergence of time-average of $X_\text{erg}(t_i+\tau)\exp(\I\omega_k\tau)$ is required. The latter is
\bgeq\label{eq:intX}
\f{1}{T}\int_0^T\dd\tau X_\text{erg}(\tau)\e^{\I\omega_k\tau}.
\eeq
The limit $T\to\infty$ of the above formula exists almost surely, the argument is the same as at the beginning of the lemma. We will prove it is 0. Let us calculate $\E|\bd\cdot|^2$ of \Ref{eq:intX}
\begin{align}
&\ \f{1}{T^2}\int_0^T\dd\tau_1\int_0^T\dd\tau_2\ r(\tau_2-\tau_1)\e^{\I\omega_k(\tau_2-\tau_1)}\nonumber\\
& = \f{1}{T^2}\int_0^T\dd\tau_1\int_0^T\dd\tau_2\int_\R\dd\sigma(\omega)\e^{\I\omega(\tau_2-\tau_1)}\e^{\I\omega_k(\tau_2-\tau_1)}\nonumber\\
&= \f{1}{T^2}\int_\R\dd\sigma(\omega)\int_0^T\dd\tau_1\int_0^T\dd\tau_2\e^{\I(\omega+\omega_k)(\tau_2-\tau_1)}\nonumber\\
&= 2\int_\R\dd\sigma(\omega-\omega_k)\f{1-\cos(\omega T)}{(\omega T)^2}\nonumber\xrightarrow{\T\to\infty}\sigma(\{\omega_k\})=0.
\end{align}
That shows the mean-square and almost sure convergence to 0.
\end{proof}

It means that the additional memory structure induced by the commensurability of $\{\omega_k\}_k$ is purely non-linear and it is possible to detect it only using higher-order statistics, e.g. the time-averaged characteristic function. This fact do not contradict the purely linear dependence structure of the Gaussian processes, because it applies only to the time-averages, which in the non-ergodic case make use only of a part of the full information contained in the process.

The exemplary covariance estimation is shown in Fig. \ref{f:autocov}, where we sampled the stationary Ornstein-Uhlenbeck process with mean-returning parameter $\lambda=3$, and the addition of two spectral components $2\cos(t)$  and $\cos(2 t)$. The covariance conditioned by $R_1=2,R_2=1$ is 
\bgeq
r(t)=\e^{-\lambda t}+2\cos(t)+\f{1}{2}\cos(2 t)
\eeq
and it agrees with the estimated time-averaged covariance  shown in Fig.  \ref{f:autocov}.

\begin{figure}[h]\centering
\includegraphics[width=\columnwidth]{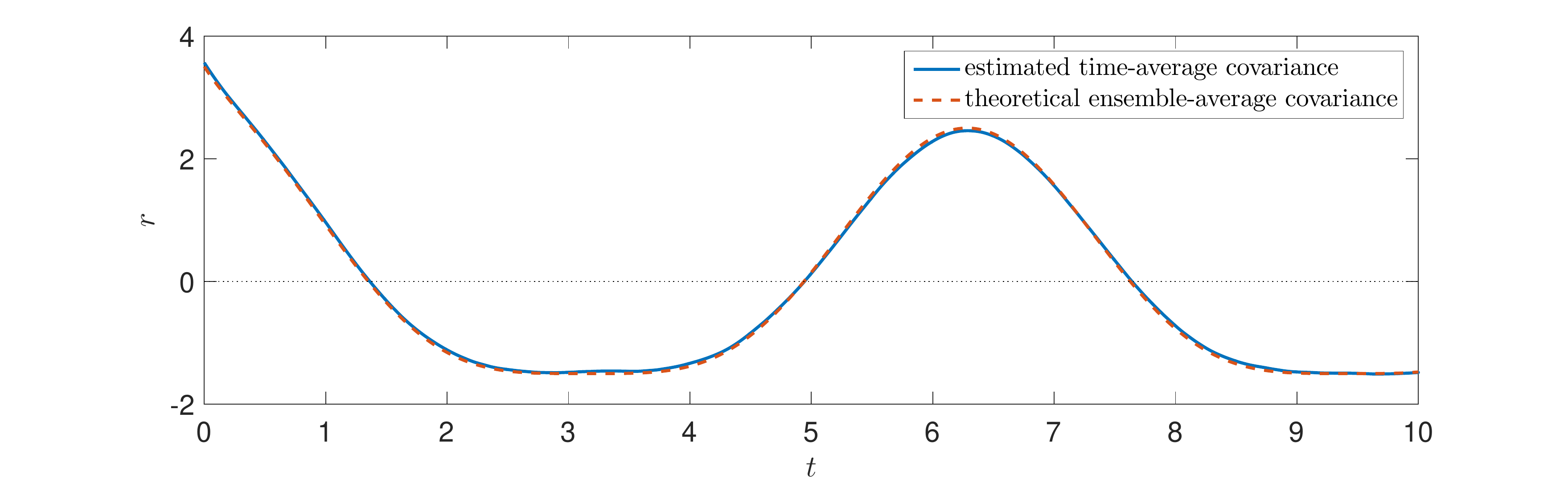}
\caption{Comparison between estimated time-average covariance function and a theoretical ensemble one for a stationary non-ergodic process with commensurable frequencies of the spectral points.}\label{f:autocov}
\end{figure}

\section{Ergodicity of Langevin equations}\label{s:LangEq}
%\section{Linear filters}\label{s:linFilt}
In this section we will make use of three basic facts.
\bg{prp}
Let us consider any finite measure $\sigma$ and a measurable function $f$, defined $\sigma$-almost everywhere. In this case:
\bg{itemize}
\item[i)] If $\sigma$ is absolutely continuous, then $\sigma f$ is absolutely continuous.
\item[ii)] If $\sigma$ is Rajchmann, then $\sigma f$ is Rajchmann; equivalently: if $\widehat{\sigma}$ decays, then $\widehat{\sigma f}$ also decays.
\item[iii)] If $\sigma$ is continuous, then $\sigma f$ is continuous.
\end{itemize}
The function $f$ is often called the spectral gain.
\end{prp}
\bg{proof}
Fact i) fallows directly from the definition of $\sigma f$. The proof of Fact iii) is also simple: $\sigma(\{x_0\})f(x_0)\neq 0$ only if $\sigma(\{x_0\})\neq 0$.

For Fact ii) the argument is a little more complicated. Because the measure $\sigma$ is finite, trigonometric functions are dense in $L^1(\R,\sigma)$. We take a sequence
\bgeq
f_k(\omega)=\sum_{j=1}^{N_k}A_j \e^{\I t_j\omega},
\eeq
such that $\Vert f-f_k\Vert_{L^1(\R,\sigma)}\to 0$. Note that
\bgeq
\widehat{\sigma f_k}(t) =\sum_{j=1}^{N_k}A_j\widehat{\sigma}(\omega+\omega_j)\xrightarrow{t\to\infty}0,
\eeq
in other words $\sigma f_k$ are Rajchamann. Additionally
\bgeq
\big|\widehat{\sigma f}(t)-\widehat{\sigma f_k}(t)\big|=\lt|\int_\R\sigma(\dd\omega)\ \e^{\I\omega t}\big(f(\omega)-f_k(\omega)\big)\rt|\le \Vert f-f_k\Vert_{L^1(\R,\sigma)},
\eeq
so sequence of functions $\widehat{\omega f_k}$ converge uniformly to $\widehat{\omega f}$. Therefore we can commute the limits $k\to\infty,t\to\infty$ and obtain
\bgeq
\lim_{t\to\infty}\widehat{\sigma f}(t) = 0.
\eeq
\end{proof}
Considered together, these three facts guarantee that if a Gaussian process $X$ has spectral measure $\sigma$, then the Gaussian process $Y$ with spectral measure $\sigma f$ inherits all ergodic properties (ergodicity, mixing) from $X$. In fact, in some cases the time-averages of $Y$ can behave more regularly than those of $X$, which will be shown later in this section.

The above proposition can be used to determine the ergodic behaviour of various transformations of a given process. The process of type
\bgeq
Y(t) = \sum_k a_k X(t-T_k),
\eeq
discussed in Chapter \ref{ch:gaussVar}, appears directly e.g. in the biological systems in which delayed responses are to be expected \citep{barbuti}.

Proposition \ref{prp:YdiscSum} shows that $Y$ has a spectral measure of $X$ multiplied by a continuous function, so it inherits the ergodic properties of the process $X$. The case
\bgeq
Y(t) = \sum_k a_k\df{^k}{t^k} X(t),
\eeq
considered in Proposition \ref{prp:YDerSum} is similar. Such a $Y$ inherits the ergodic properties of $X$, but can also be ergodic whereas $X$ is not. The function $\omega\mapsto \lt| \sum_k a_k (\I\omega)^k\rt|^2$ may have zeros and if such zero agrees with the position of the spectral point of $X$, the process $Y$ does not contain this spectral point. The simplest such case is when $X$ has exactly one spectral point at $\omega=0$, that is it contains a time-independent Gaussian constant $X_0$. In this situation any time-averaged observable which depends on the mean of $X$ does not converge to the ensemble-average. However, the derivative $\df{}{t}X(t)$, corresponding to the spectral gain function $\omega\mapsto\omega^2$, does not contain $X_0$ and is ergodic.

For $Y(t)=g*X(t)$ from Proposition \ref{prp:Yconv} the result is also analogous, $Y$ inherits the ergodic properties of $X$. If the zeros of $\widehat g$ agree with the spectral points of $X$, the process $Y$ may be ergodic whereas $X$ is not. Moreover, if the singular non-mixing measure of $X$ is contained in the domain outside of $\widehat Y$ support, $Y$ may be mixing when $X$ is non-mixing.

Special care must be taken in the case when we consider the convolution with $g$ which is not rapidly decaying, e.g. $g\in L^2(\R)$ but $g\notin L^1(\R)$. The Fourier transform of such $g$ has jumps. The well-known result from Fourier theory \citep{champeney} states that if $\widehat g$ has a jump at $\omega_0$, then $\int_{-T}^T\dd t\ g(t)\e^{\I\omega_0 t}$ converges as $T\to\infty$ to the value $(\widehat g(\omega_0^-)+\widehat g(\omega_0^+))/2$, i.e. to  the exact middle of the discontinuity of $\widehat g$. If the process $X$ has a spectral atom at the exact frequency $\omega_0$, then for the process $Y$ this spectral point will be modulated by $|\widehat g(\omega_0^-)+\widehat g(\omega_0^+)|^2/4$ as long as the filter is applied symmetrically as a limit of convolutions with functions $g$ which are supported on interval $[-T,T]$.

The harmonic representation of the convolution vastly increases the number of models for which it is easy to study ergodicity using Fourier methods, as using convolution is one of the most often chosen methods to model time-invariant linear responses of the system (see also next Sec. \ref{s:LangEq}). One of the most common examples that appear in practice is $g$ being a one or two sided exponent decaying with rate $\lambda$. These choices correspond to the  spectral responses  $1/(\lambda^2+\omega^2)$ or $4\lambda^2/(\lambda^2+\omega^2)$, respectively.

One other practical consequence is that one can filter out non-ergodicity from the data. Using estimators of power spectral density (e.g. periodogram \citep{BD}) the locations of spectral points can be estimated and the corresponding non-ergodicity removed by using any filter with zeros its spectral gain function at their frequencies. The simplest choice of such filter is the smoothing
\bgeq
\tilde X(t) = \int\limits_{t-\pi/\omega_0}^{t+\pi/\omega_0}\dd s\ X(s),
\eeq
which integrates the spectral component $\omega_0$ over its period, therefore removing it. It corresponds to the filter $g_{\omega_0}$ and spectral gain $\widehat g_{\omega_0}$
\bgeq
g_{\omega_0}(t) = \bg{cases}
1, & |t|\le \pi/\omega_0,\\
0, & |t|> \pi/\omega_0,
\end{cases}\quad \widehat g_{\omega_0}(\omega) =\sqrt{\f{2\omega_0^2}{\pi^3}}\sinc\lt(\f{\omega \pi}{\omega_0}\rt).
\eeq
For a multiple number of spectral points one may use the filter $g_{\omega_1}*g_{\omega_2}\ldots g_{\omega_N}$ or any other with a suitable spectral gain. The gain $|\widehat g|^2$ behaves like $(\omega-\omega_0)^2$ near $\omega_0$, therefore it removes the spectral point $\omega_0$ in a numerically stable manner. If one is more interested in sure removal of the non-ergodicity near the location $\omega_0$ than not distorting the spectrum, one can use a spectral gain which is more flat around $\omega_0$, e.g. using a triangular function filter guarantees the asymptotic behaviour $ (\omega-\omega_0)^4$. An other useful choice is spectral gain
\bgeq
\widehat g(\omega)=\bg{cases}
1-\e^{-(\omega-\omega_0)^2/c}, & |\omega|\le L;\\
0, & |\omega|>L,
\end{cases}
\eeq
which allows for the calibration of the level of distortion around $\omega_0$ (parameter $c$) and frequency cut-off (parameter $L$). The corresponding filter can be expressed using error, Gaussian and trigonometric functions, so it can be easily computed for the purpose of statistical usage.

The above approach can be understood as, instead of using original observables $f(X)$, to use the modified observable $\widetilde f(X)=f(Y)$ for which the time- and ensemble-averages coincide, even when this is not generally the case. Analysing the transformed process $Y$ instead of $X$ may be more difficult, as the properties of $X$ are distorted by filtering, albeit in a controlled manner. However, for a small number of spectral points it is manageable, moreover it can be used as an effective method of localising spectral points: if the observables of the filtered process behave like it is ergodic, it is a statistical verification of a good choice of these locations. Further analysis can be performed on the filtered process $Y$, which is ergodic, or by staying with the original $X$ and using methods from Section \ref{s:statProp}.

Identical reasoning applies to stochastic differential equations. The simplest case of the Langevin equation
\bgeq
\dd X=-\dd t\lambda X +\dd B
\eeq
is solved by the Ornstein-Uhlenbeck process, as discussed in Section \ref{s:elSol}. It has a spectral density $s_X(\omega) = (\lambda^2+\omega^2)^{-1}$ so it is clearly mixing and that would be the case also for equation with more general mixing force than $\dd B$, e.g. $\dd B_H$.
This result is much more general. Any stationary solution of the linear differential system
\bgeq
a_k\df{^k}{t^k}X(t)+a_{k-1}\df{^{k-1}}{t^{k-1}}X(t)+\ldots+ a_0 X(t)=F(t),
\eeq
has a casual solution given by a proper Green's function and spectral representation
\bgeq
X(t) = \int_\R \dd S_F(\omega)\  \f{1}{\sum_{j=1}^k a_j(\I\omega)^j}\e^{\I\omega t},
\eeq
It is clear that $X$ the inherits ergodic properties of $F$. In contrast to the case when $X$ was an effect of a linear filter on $Y$, in this case $X$ cannot even be  ergodic if  $F$ is not, as the rational function $1/|\sum_j a_j(\I\omega)^j|^2$ does not have zeros.

Very similar reasoning applies to systems with  richer memory structure, the most prominent being our main subject of interest, the generalized Langevin equation, which we will here analyse in its full form
\bgeq\label{eq:GLEfull2}
m\df{^2}{t^2}X(t)=-\kappa X-\int_{-\infty}^t\!\!\!\dd \tau\ \df{}{t}X(\tau)K(t-\tau) + F(t).
\eeq
Its stationary solution has the form
\bgeq\label{eq:GLEfullHarm}
X(t) = \int_\R \dd S_F(\omega)\ \f{1}{-m\omega^2 +\kappa+\I\omega\widehat K(\omega)},
\eeq
where $\dd S_F$ is the spectral process of the force $F$.  The GLE with this form has a visible similarity to the Newton equation and can model subdiffusion in the case when $F$ is fractional Gaussian noise \citep{vinales,kou}, and the kernel is $K(t)=t^{-\alpha}, 0<\alpha<2$. Such a kernel is outside $L^2$, but the harmonic representation above is still valid if we interpret $\widehat K$ in the generalized sense, as commented in Section \ref{s:genGauss}. Using this approach we can use tables of generalized Fourier transform to obtain $\widehat K(\omega) = |\omega|^{\alpha-1}\Gamma(1-\alpha)\e^{\I\,\text{sgn}(\omega)\pi (\alpha-1)/2}$ \citep{champeney}.

For $\kappa>0$ the fraction in Eq. \Ref{eq:GLEfullHarm} behaves like a  constant near $0$ and like $\omega^{-2}$ in the limit $\omega\to\pm\infty $. Therefore the spectral measure $\sigma_X$ is generally finite. When there is no external potential, $\kappa=0$, the particle is not confined, so physically speaking there should be no stationary solution. It is also visible from Eq. \Ref{eq:GLEfullHarm} in the fact that if there was a stationary solution, its spectral measure $\sigma_X$ would exhibit singularity $\omega^{-2}$ at $0$, which is a contradiction of $\sigma_X$ being a finite or truncated measure. In this case however the velocity would be a stationary process and $X$ could be represented as
\bgeq
X(t)=\int_\R \dd S_F \f{\e^{\I\omega t}-1}{-m\omega^2 +\I\omega\widehat K(\omega))}.
\eeq
In both cases the position $X$ or the velocity $V=\df{}{t} X$ inherit the ergodic properties of $F$.

The Fourier space approach gives also additional insight into the physical origin of the non-ergodicity. In Chapter \ref{ch:der} we presented a derivation of the GLE based on discrete and continuous field baths. For a pure field bath the spectral measure of $F$ was determined by the Fourier transform of the coupling function
\bgeq
\sigma_F(\dd\omega)=\dd\omega\ |\widehat\rho|^2,
\eeq
which lead to an immediate conclusion.
\bg{prp} The stationary solution of the Hamiltonian system from Section \ref{s:hamMod}, in which only the continuous heat bath is present, is mixing.
\end{prp}
Conversely, the force related to the discrete heat bath has a pure atomic spectrum supported on self-frequencies of the bath harmonic oscillators
\bgeq
\sigma_F(\dd\omega) = \sum_km_k\f{\gamma_k^2}{2\omega_k^2}\big(\delta(\dd\omega-\omega_k)+\delta(\dd\omega+\omega_k)\big).
\eeq

This fact immediately prohibits the ergodicity of the solution. Actually, if the fluctuation-dissipation theorem holds, even the stationary solution does not exist, as the integral $\int_{-\infty}^t\!\!\!\dd\tau\ X(\tau)K(t-\tau)$ does not converge for stationary $X$ and non-decaying $K$. The non-stationary  solutions of the equation with the finite-time integral $\int_0^t$ are still well-defined.

From the point of view of spectral theory, for non-decaying $K$ its Fourier transform can be defined only as a distribution and the existence of the stationary solution would require dividing by a distribution, which cannot be consistently defined. This insight is nothing surprising from the point of view of abstract ergodic theory, which prohibits ergodicity for systems governed by quadratic Hamiltonians such as of discrete heat bath \citep{70y}. The decay of the memory functions is such a ubiquitous physical assumption that it is understandable when the force $F$ is non-mixing, but the fluctuation-dissipation does not hold exactly and $K$ has some truncation or cut-off. In such situation there is a stationary solution, but there still cannot be an ergodic one. As an example, a natural form of such truncation is an exponential one, this truncated kernel is
\bgeq
K(t)=\sum_k m_k\f{\gamma_k^2}{\omega_k^2}\cos(\omega_k t)\e^{-\lambda_k t}.
\eeq
In this case the purely discrete spectral measure of the stationary solution of the GLE \Ref{eq:GLEfull2} is given by formula
\bgeq
\sigma_X(\dd\omega) = \sum_k\f{m^k \gamma_k^2\omega_k^{-2}}{\lt|-m\omega^2 +\kappa+\sum_j\f{\I\lambda_j\omega_k-\omega_k^2}{(\lambda_j+\I\omega_k)^2+\omega_j}\rt|^2}\big(\delta(\dd\omega-\omega_k)+\delta(\dd\omega+\omega_k)\big).
\eeq
The amplitudes of the spectral points are given by a rather complicated fraction, note however that the integrals of motion for the solution $X$ are determined solely by the set $\{\omega_k\}_k$, i.e. the dispersion relation from the discrete heat bath.

The presence of non-ergodicity does not depend on the number of oscillators in the heat bath, in our Hamiltonian model it could be infinite. During the derivation we have only kept the total energy finite which was reflected in the Hilbert space structure of the phase space. If the amplitudes of the single oscillators can be considered negligible, one could perform a rescaling procedure of $q_i,\omega_i,\gamma_i$, after which ergodicity would be regained \citep{ariel}. This is, however, effectively equivalent to using continuous heat bath if we keep the energy finite. For the rescaling which do not control the energy, the result could be different, but the physicality of such solutions can be doubted. This procedure surely demonstrates that various GLEs can be approximated using solutions of discrete heat bath models, but not that such GLEs have the corresponding Hamiltonian models. Therefore it appears that the ergodicity is determined by the type of the heat bath in the physical model: the discrete or the field one. The presence of any non-field degree of freedom prohibits the ergodicity of the solution.

\chapter{Superstatistical Langevin equations}\label{ch:superstat}

In this chapter we show how the Gaussian GLE can be used to model non-Gaussian distributions, which are observed in soft, biological, and active matter systems \citep{wangNG,matti17}. These non-Gaussian pdfs are present simultaneously with normal or anomalous diffusion. Examples include the Laplace distribution observed in the motion of messenger RNA
molecules in the cytoplasm of bacteria and yeast \citep{lampo,ralf_lampo}, and stretched Gaussian pdfs which were unveiled in the motion of lipids in
protein-crowded lipid bilayer systems \citep{jeon_prx}.

In our model both of these phenomena have a common origin, namely a random parametrisation of the stochastic force \citep{superstat}.  In statistics such an object is called a compound or mixture distribution \citep{mixMod}; in the analysis of diffusion processes this type of model is called superstatistical \citep{beck2} (which stands for ``superposition of statistics'')
or ``doubly stochastic'' \citep{doubleSts}, which is a term for stochastic models
generalised by replacing some parameter, for instance, the diffusion constant $D$, by a random process.

We perform a detailed analytical analysis demonstrating how various types of parameter distributions for the memory kernel result in the exponential, power law,
or power-log law tails of the memory functions. It proves that even the Langevin systems which are locally short memory, globally can exhibit properties characteristic for fractional dynamics.

The studied system is also shown to exhibit a further unusual property: the velocity has a Gaussian one point probability density but non-Gaussian joint
distributions. It is well-known that such processes exist, but are considered very atypical. In the considered model this property stems directly from Hamiltonian derivation of the GLE, i.e. fixed mean energy $k_B\mathcal T$.

This behaviour is reflected in the relaxation from Gaussian to
non-Gaussian distributions observed for the position variable. The limiting pdf can exhibit power-law tails, but for all finite $t$ a truncation is present, which causes the process to have all moments finite. During the analysis we show that
our theoretical results are in excellent agreement with Monte Carlo simulations.

\section{Compound Ornstein-Uhlenbeck process}
\label{s:cOU}

We start from considering the classical Langevin equation, which  can be considered as an approximation of the GLE
in which the covariance function of the stochastic force $r_F$ decays very rapidly in the relevant time scale. The superstatistical solution of the Langevin equation exhibits many properties typical to the superstatistical GLE in general. Further on we normalise the mass of the particle and the bath temperature, $m=k_B\mathcal T=1$, so the system is governed solely by the coefficient $\lambda$, proportional to viscosity. The former two parameters would only rescale the solution.
\bg{dfn}[Superastatistical Langevin equation]
We define the supersatistical generalisation of the Langevin equation as random coefficient stochastic It{\={o}} equation governed by the random parameter $\Lambda>0$ independent from the Brownian motion $B$
\bgeq 
\dd V(t)=-\dd t\Lambda V(t)+\sqrt{\Lambda} \dd B(t).
\eeq
The rescaling $\sqrt{\Lambda}\dd B(t)$ is chosen such that the fluctuation-dissipation relation is fulfilled.
\end{dfn}
The random variable $\Lambda$ can be interpreted as a local viscosity value, which varies from trajectory to trajectory, each confined (at least on relevant time scales) in a separate area with different local properties of the environment. This parameter determines the stationary solution
\begin{equation}
V_\Lambda(t)=\sqrt{\Lambda}\int_{-\infty}^t\!\!\!\dd B(\tau)\ \e^{-\Lambda(t-\tau)}.
\end{equation}
We call this solution a compound Ornstein-Uhlenbeck process. It
can also be represented in Fourier space in a natural way
\begin{equation}
V_\Lambda(t)=\int_\R \dd S(\omega)\ \f{1}{\I\omega+\Lambda}
\e^{\I\omega t}\dd\omega, \quad \E|\dd S(\omega)|^2=\dd\omega.
\end{equation}
Because the sampled Ornstein-Uhlenbeck is an 	AR(1) process for every $\lambda>0$ (see Proposition \ref{prp:sampledOU}), it is clear that the compound process will be random coefficient AR(1) time series which fulfils
\begin{align}
&V_\Lambda((k+1)\Delta t)=\e^{-\Delta t \Lambda} V_\Lambda(k\Delta t)+Z_k,
\nonumber\\
&Z_k=\sqrt{\Lambda}\int_{k\Delta t}^{(k+1)\Delta t}\!\!\!\dd B(\tau)\ \e^{-\Lambda((k+1)\Delta t-\tau)} \deq \f{1}{\sqrt{2}}\sqrt{1-\e^{-2\Delta t \Lambda}}\xi_k.
\end{align}
The noise $Z_k$ has the same distribution as a Gaussian discrete white noise $\xi_k$ multiplied by a random constant. So the series $Z_k$ is, conditionally on $\Lambda$, independent from past values $V_\Lambda(j\Delta t)$ with $j< k$. When there are only few distinct populations and $\Lambda$ has only few possible values, they can even be recognised on the phase plot of $y=V_\Lambda((k+1)\Delta t)$ versus $x=V_\Lambda(k\Delta t)$, see Fig. \ref{f:plotAR}. There, two distinct populations with different autoregressive coefficients can be distinguished. Both have Gaussian distribution, but each one has a distinct elliptical shape. The total distribution, as a mixture of two ellipsoids, is not Gaussian, nor even elliptical. The projection of the joint
distribution on the $x$ or $y$ axis are the pdf of $V_\Lambda(t)$ and are Gaussian, thus, one needs at least a two-dimensional phase plot to reveal the non-Gaussianity of $V_\Lambda$. For a larger number of populations the phase plot would be much less clear, but the huge advantage of this method is that it works even for trajectories of very short length.

\begin{figure}
\centering
\includegraphics[width=14cm]{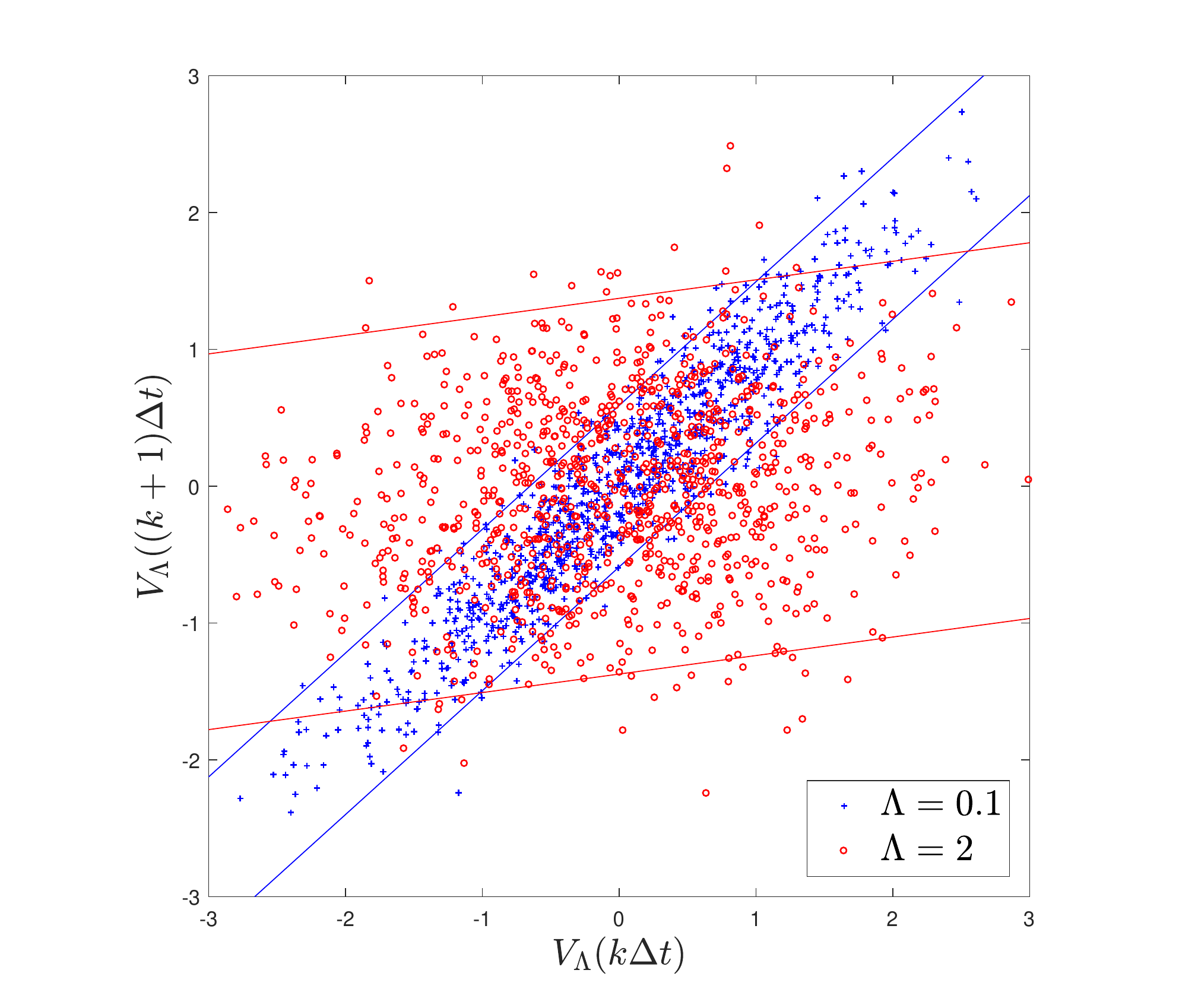}
\caption{Phase plot of the compound Ornstein-Uhlenbeck process with $\pr(\Lambda=
	1/10)=\pr(\Lambda=2)=1/2$. We took $\Delta t=1$. Solid lines correspond to 95\%
	conditional quantiles of the noise $Z_k$ in both populations.}
\label{f:plotAR}
\end{figure}

The situation becomes more complex and interesting when $\Lambda$ assumes a
continuous distribution. If $\Lambda$ has the pdf
$p_\Lambda$, the covariance function of $V_\Lambda$ is
\begin{equation}
r_{V_\Lambda}(t)=\f{1}{2}\int_0^\infty\!\!\dd\lambda\ p_\Lambda(\lambda) \e^{-\lambda t},
\end{equation}
so it is the Laplace transform of $p_\Lambda$: in probabilistic language this
quantity would be called a moment generating function of the variable $-\Lambda$.
For instance, if $\Lambda$ is a stable subordinator with index $0<\alpha<1$
\citep{taqqu} the covariance function is the stretched exponential
\begin{equation}
r_{V_\Lambda}(t)=\f{1}{2}\e^{-\sigma_\alpha t^\alpha},
\end{equation}
which is a common relaxation model \citep{kakalios,weronSExp},
sometimes referred to as Kohlrausch-Williams-Watts relaxation \citep{
kohlrausch,wwSExp}.

If $\Lambda$ can be decomposed into a sum of two independent random variables,
$\Lambda=\Lambda_1+\Lambda_2$, the corresponding covariance function is a product,
\begin{equation}
\label{eq:cOUmultipl}
r_{V_\Lambda}(t)=2r_{V_{\Lambda_1}}(t)r_{V_{\Lambda_2}}(t).
\end{equation}
Therefore, in this model various kinds of truncations of the kernel correspond
to a decomposition of $\Lambda$, for instance, if $\Lambda=\lambda+\Lambda'$
with deterministic $\lambda>0$, the covariance function $r_{V_\Lambda}$ will be
truncated by $\exp(-\lambda t)$.

In order to study the tails of the resulting covariance, let us introduce the asymptotic notation, which we will use further on.
\bg{dfn}[Asymptotic notation]
Let $f,g$ be two functions and $x_\infty$ a fixed number or $\pm\infty$. We consider a neighbourhood $x\to x_\infty$ and introduce the following symbols:
\bg{itemize}
\item[i)] We write $f\sim g$ when
\bgeq
\f{f(x)}{g(x)}\xrightarrow{x\to x_\infty} 1.
\eeq
For $g$ which has zeros near $x_\infty$ we say that the limit above should be valid for all sequences $x_k\to x_\infty$ for which $|g(x_k)|>\epsilon$ given any fixed $\epsilon > 0$.
\item[ii)]  We write $f\lesssim g$ when $f\sim h\le g$ for some function $h$.
\item[iii)] We write $f = \mathcal O(g)$ when $f\lesssim C g$ for some constant $C>0$.
\item[iv)] We write $f\asymp g$ when  $C_1g\lesssim f \lesssim C_2 g$ for some constants $0<C_1\le C_2$.
\end{itemize}
\end{dfn}
Using these notions some general observations about the behaviour of $r_{V_\Lambda}$ can be made.
\bg{prp}
When $\Lambda$ has a distribution supported on an interval $[\lambda_1,\lambda_2]$, and its pdf has no singularity, then
\bgeq
r_{V_\Lambda}\asymp t^{-1}\e^{-\lambda_1 t}.
\eeq
When $\Lambda$ is distributed uniformly, we observe the stronger asymptotic ``$\sim$'' with scaling constant $1/(\lambda_2-\lambda_1)$.
\end{prp}
\bg{proof}
In the considered situation the pdf is necessarily bounded,
that is, $m\le p_\Lambda(\lambda)\le M$. So
\begin{equation}
\label{eq:LbdUnif}
\f{m}{2}\int_{\lambda_1}^{\lambda_2}\dd\lambda\ \e^{-\lambda t}\le r_{V_\Lambda}(t)
\le\f{M}{2}\int_{\lambda_1}^{\lambda_2}\dd\lambda\ \e^{-\lambda t}.
\end{equation}
The integrals on the left and right have asymptotics of the form
\begin{equation}
\int_{\lambda_1}^{\lambda_2}\dd\lambda\ \e^{-\lambda t} =\f{1}{t}\lt(\e^{-\lambda_1
t}-\e^{-\lambda_2 t}\rt)\sim\f{1}{t}\e^{-\lambda_1 t}.
\end{equation}
We can join the above results, obtaining
\begin{equation}
\label{eq:cOUBoundedAs}
\f{m}{2t}\e^{-\lambda_1 t}\lesssim r_{V_\Lambda}(t)\lesssim \f{M}{2t}
\e^{-\lambda_1 t}.
\end{equation}
The ``$\asymp$'' asymptotic is thus established. When $\Lambda$ is distributed uniformly, $m=M=(\lambda_2-\lambda_1)^{-1}$. Now note that $ g\lesssim f$ and $f\lesssim g$ implies $g\sim f$.
\end{proof}
The  uniform distribution of
$\Lambda$ is important from a practical standpoint, because it is a maximal
entropy distribution supported on the interval $[\lambda_1,\lambda_2]$, so it can be interpreted as the choice taken using the weakest possible assumptions.

Heavier tails of $r_{V_\Lambda}$ may be observed  when the distribution of
$\Lambda$ is concentrated around $0^+$.The most significant case of such a
distribution is a power law of the form $p_\Lambda(\lambda)\sim \lambda^{\alpha-1}$,
with $\lambda\to 0^+$ and $\alpha>0$. For any distribution of this type Tauberian theorems guarantee that the covariance has a power law tail \citep{feller}
\begin{equation}
r_{V_\Lambda}(t)\sim\f{\Gamma(\alpha)}{2} t^{-\alpha},\quad t\to\infty.
\label{eq:cOUPowerLaw}
\end{equation}
For $\alpha<1$ the process $V_\Lambda$ exhibits a long memory. This observation
can be refined using the more general variant of the Tauberian theorem  which states that if the pdf of $\Lambda$
contains a slowly-varying factor $L$, then the tail of the covariance contains
the factor $L(t^{-1})$. One example of such a slowly-varying factor is $|\ln(
\lambda)|^\beta,\beta>0$, so heavy tails of the covariance of the power law form
$t^{-\alpha}\ln(t)^\beta$ can also be present for the compound Ornstein-Uhlenbeck
process if the distribution of $\Lambda$ exhibits a logarithmic behaviour at
$0^+$. This observation proves that this equation can also describe ultra-slow
diffusion and can be considered as an alternative to more complex models
based on distributed order fractional derivatives \citep{eab}.

In the description of Figure \ref{f:plotAR} we noted that the distribution of the process $V_\Lambda$ is not Gaussian. However, the marginal distributions of $V_\Lambda$ are Gaussian at any time $t$, because $\E[V_\lambda(t)^2]=1/2$ for any $\lambda$. Thus, only the joint multidimensional distributions are not Gaussian. This fact is easy to observe studying the two point characteristic
function. Let us fix $V_\Lambda(\tau)$, $V_\Lambda(\tau+t)$ and define
\begin{equation}
\varphi_\Lambda(\bd\theta,t)\defeq\E\lt[\e^{\I(\theta_1 V_\Lambda(\tau)+\theta_2
V_\Lambda(\tau+ t))}\rt],\quad \bd\theta=[\theta_1,\theta_2].
\end{equation}
For any deterministic $\lambda$ this function is determined by the covariance
matrix $\Sigma_t$ of the pair $V_\lambda(\tau),V_\lambda(\tau+t)$,
\begin{equation}
\varphi_\lambda(\bd\theta,t)=\e^{-\f{1}{2}\bd\theta^{\mathrm T}\Sigma_t\bd\theta},
\quad \Sigma_t=\f{1}{2}\bgmx 1, &\e^{-\lambda t} \\ \e^{-\lambda t}, & 1\emx,
\end{equation}
so in the superstatistical case it is an average over the conditional average
\begin{equation}
\label{eq:phi}
\varphi_\Lambda(\bd\theta,t)=\e^{-\f{1}{4}
\theta_1^2}\e^{-\f{1}{4}\theta_2^2}\E\lt[\e^{-\f{1}{2}\theta_1\theta_2\e^{
-\Lambda t}}\rt].
\end{equation}
The marginal factors $\exp(-\theta_1^2/4),\exp(-\theta_2^2/4)$ are
indeed Gaussian, but the cross factor describing the interdependence is not. The
function $\varphi_\Lambda$ would describe a Gaussian distribution if and only if the factor $\E\lt[\exp(\theta_1\theta_2\e^{-\Lambda t}/2)\rt]$ had the form $\exp(a \theta_1\theta_2)$. But we see that it is in fact a moment generating function
of the variable $\exp(-\Lambda t)$ at point $\theta_1\theta_2/2$, which is an
exponential if and only if $\Lambda$ equals one fixed value with probability
unity. The compound Ornstein-Uhlenbeck process is never Gaussian for
non-deterministic $\Lambda$.

This property is also evident if we calculate the conditional msd of $X_\Lambda$, that is (see Eq. \Ref{eq:msdclassLang})
\begin{equation}
\label{eq:condmsd}
\delta^2_{X_\Lambda}(t|\Lambda)\defeq \E[X_\Lambda(t)^2|\Lambda]=\f{1}{2\Lambda}t+\f{1}{2\Lambda^2}\lt(\e^{
-\Lambda t}-1\rt).
\end{equation}
Similarly we can consider the conditional covariance function $r_{V_\Lambda}(t|\Lambda)\defeq \E[V_{\Lambda}(\tau+t)V_\Lambda(\tau)|\Lambda)$. The unconditional msd can be calculated as $\delta^2_{X_\Lambda}(t)=\E[\delta^2_{X_\Lambda}(t|\Lambda)]$, similarly $r_{V_\Lambda}(t)=\E[r_{V_\Lambda}(t|\Lambda)]$. Using the conditional msd we can prove the following.
\bg{prp}\label{prp:GLEasympPDF}
For a position process $X_\Lambda$ which is an integral of the compound Ornstein-Uhlenbeck process:

\bg{itemize}
\item[i)] For small $t$ the process is asymptotically a Gaussian ballistic motion, that is
\bgeq
\f{X_{\Lambda}(t)}{t}\xrightarrow{\ d\ } \mathcal N(0,1/4), \quad t\to 0^+.
\eeq
\item[ii)] For any $t$ $p_{X_\Lambda(t)}(x)= \mathcal O(x^{-\infty})$, i.e. the pdf decays faster than any power of $x$.
\item[iii)] For large $t$ and $p_\Lambda(\lambda)\sim \lambda^{\alpha-1}, \lambda\to 0^+$ the limiting pdf of $X_\Lambda(t)/\sqrt{t}$ has power law tails $\pi^{-1/2}\Gamma(\alpha+1/2)x^{-2\alpha-1}$.
\end{itemize}
\end{prp}
\bg{proof}
For i) first note that $\delta^2_{X_\Lambda}(t|\Lambda) = t^2/4+\mathcal O(t^2)$ which stems from the Taylor expansion of $\exp(-\Lambda t)$. Now, consider the characteristic function of $X_\Lambda(t)/t$
\bgeq
\phi_{X_\Lambda(t)/t}(\theta) = \E\lt[\exp\lt(-\f{\delta^2_{X_\Lambda}(t|\Lambda)}{2 t^2}\theta^2\rt)|\Lambda\rt].
\eeq
Using the monotone convergence theorem we can switch taking the limit $t\to 0^+$ and averaging, obtaining
\bgeq
\phi_{X_\Lambda(t)/t}(\theta)\xrightarrow{t\to0^+} \exp\lt(-\f{1}{8}\theta^2\rt).
\eeq
For ii) We will show that all moments of $X_\Lambda(t)$ are finite, which implies $p_{X_\Lambda(t}(x)=\mathcal O(x^{-\infty})$ for all $t$. The function $\lambda\mapsto \delta^2_{X_\lambda}(t)$ is bounded, because it is continuous and
\bgeq
\lim_{\lambda\to 0^+}\delta^2_{X_\lambda}(t)=t^2/4,\quad  \lim_{\lambda\to
\infty}\delta^2_{X_\lambda}(t)=0.
\eeq
Therefore the msd of $X_\Lambda$ must be finite for any distribution of $\Lambda$. Moments of higher even order can be expressed as
\begin{equation} \label{eq:bound}
\E[X_\Lambda(t)^{2n}]=\E[\E[X_\Lambda(t)^{2n}|\Lambda]]=\prod_{k=2}^n(2k-1) \E\lt[(\delta^2_{X_\Lambda}(t|\Lambda)\big)^n\rt],
\end{equation}
where we used the formula for even moments of a Gaussian variable. The last result is an average over a bounded function, so the result is finite for any $n$.

For iii) we may write the resulting pdf as
\begin{align}
p_{X_\Lambda(t)/\sqrt{t}}(x)&=\f{1}{\sqrt{\pi}}\E\lt[\f{\sqrt{\Lambda}}{\sqrt{1+\f{1}{\Lambda t}\lt(\e^{-\Lambda t}-1\rt)}}\exp\lt(-\f{x^2\Lambda}{1+\f{1}{\Lambda t}\lt(\e^{-\Lambda t}-1\rt)}\rt)\rt]\nonumber\\
&\xrightarrow{t\to\infty} \f{1}{\sqrt{\pi}}\E\lt[\sqrt{\Lambda}\exp(-x^2\Lambda)\rt],
\end{align}
where commuting the limit and average was possible due to the fact that the function $1/\delta^2_{X_\lambda}(t)$ bounded for $t>t_0>0$, so the whole density is bounded as a function of $(\lambda,t)$ for large $t$. The last obtained result is the Laplace transform of the variable $\sqrt{\Lambda}$ at point $x^2$, so variable  $\Lambda$ with power law pdf
$p_\Lambda(\lambda)\sim \lambda^{\alpha-1},\lambda\to 0^+$ will yield
power law tails $\sim \pi^{-1/2}\Gamma(\alpha+1/2)x^{-2\alpha-1}$ of the limiting distribution.

\end{proof}

The behaviour of the msd itself also can be studied in general. If $\E[\Lambda^{-1}]<\infty$, the term $t/(2\Lambda)$ is dominating for large $t$,
\bgeq
\delta^2_{X_\Lambda}(t)=\E\lt[\f{1}{2\Lambda}t+\f{1}{2\Lambda^2}\lt(\e^{
-\Lambda t}-1\rt)\rt]\sim \f{\E[\Lambda^{-1}]}{2}t,\quad t\to\infty
\eeq
and the process is normal diffusion with effective diffusion coefficient $\E[\Lambda^{-1}]/2$. This situation
occurs when the distribution $\Lambda$ is not highly concentrated around
$0^+$. When $\Lambda$ has a power-law singularity as in \Ref{eq:cOUPowerLaw},
that is $\lambda^{\alpha-1}$ at $0^+$, $0<\alpha<1$, this condition is not
fulfilled: $\E[\Lambda^{-1}]=\infty$. But in this situation the assumptions
required for the Tauberian theorem hold and we can apply it twice: first
for relation \Ref{eq:cOUPowerLaw}, to show that 
\bgeq 
r^\#_{V_\Lambda}(s)\sim
2^{-1}\Gamma(\alpha)\Gamma(1-\alpha) s^{\alpha-1},\quad s\to0^+
\eeq
and the second time for the Laplace transform of msd $\delta^{2\#}_{X_\Lambda}(s)=2s^2r^\#_{V_\Lambda}(s)$ to prove that
\begin{equation}\label{eq:msdPL}
\delta_{X_\Lambda}^2(t)\sim\f{2\Gamma(\alpha)}{(1-\alpha)(2-\alpha)} t^{2-\alpha},
\quad t\to\infty.
\end{equation} 
In this regime the system is superdiffusive. The transition from superdiffusion
($0<\alpha<1$) to normal diffusion ($1\le\alpha$) is unusual among diffusion
models. Fractional Brownian motion and fractional Langevin equation undergo transitions from super- to subdiffusion at a
critical point of the control parameter. This is so as in these models
the change of the diffusion type is caused by the change of the memory type from
persistent to antipersistent. But the Ornstein-Uhlenbeck process models only
persistent dependence, so the mixture of such motions also inherits this
property. For $1\le \alpha$ (and any other case when $\E[\Lambda^{-1}]<\infty$)
this dependence is weak enough for the process to be normally diffusive, for
smaller values of $\alpha$ it induces superdiffusion.

In real systems it is commonly observed that  the position process exhibits a double exponential pdf \citep{wangNG,bhattacharya}, also called Laplace distribution. Therefore, it is interesting to check what compound Ornstein-Uhlenbeck process can model such observations. This exact distribution of $X_\Lambda(t)$ is observed when the diffusion coefficient $D$ of normal, conditionally Gaussian diffusion is random and has the exponential distribution $\mathcal E(\beta)$ with pdf
\begin{equation}
p_D(d)=\beta\e^{-\beta d}, \quad d>0.
\end{equation}
For the corresponding compound Ornstein-Uhlenbeck process the corresponding distribution of
$\Lambda$ is given by $\Lambda=(4D)^{-1}$. For this model the covariance function
of the velocity process is
\begin{align}
\label{eq:covLaplpdf}
\nonumber
r_{V_{(4D)^{-1}}}(t)&=\f{1}{2}\E\lt[\e^{-\f{t}{4D}}\rt]\\
\nonumber
&=\f{\beta}{2}\int_0^\infty\!\!\dd d\ \e^{-\f{t}{4d}}\e^{-\beta d}\\
\nonumber
&\overset{\f{t}{4d}\to \tilde d}{=}\f{\beta t}{8}\int_0^\infty\!\!\dd \tilde d\ \e^{-\tilde d-\beta t(4\tilde d)^{-1}}
\f{1}{\tilde d^2}\\
&=\f{\sqrt{\beta t}}{2}K_1(\sqrt{\beta t}),
\end{align}
where we used one of the integral representations of the modified Bessel function
of the second kind $K_1$ (see \cite{DLMF}, formula 10.32.10). This function has
the asymptotic $K_1(z)\sim \sqrt{\pi/2}\exp(-z)z^{-1/2}$, $z\to\infty$ (\cite{DLMF},
formula 10.40.2), so the covariance function behaves like
\begin{equation}
\label{eq:laplA}
r_{V_{(4D)^{-1}}}(t)\sim \sqrt{\f{\pi}{8}}(\beta t)^{1/4}\e^{-\sqrt{\beta t}},
\quad t\to\infty.
\end{equation}
This behaviour is shown in Figure \ref{f:laplCov}, where we present the covariance
function corresponding to the Laplace distributed $X_\Lambda(t)$ with random
diffusion coefficient $D\deq \mathcal E(2)$. We do not present the Bessel function
\Ref{eq:covLaplpdf}, as it appears to be indistinguishable from the result of
the Monte Carlo simulation. Figure \ref{f:laplCov} also shows how to distinguish
this behaviour from an exponential decay on a semi-logarithmic scale: the
covariance function and its asymptotic are concave, which is mostly visible for
short times $t$.

\begin{figure}
\centering\includegraphics[width=16cm]{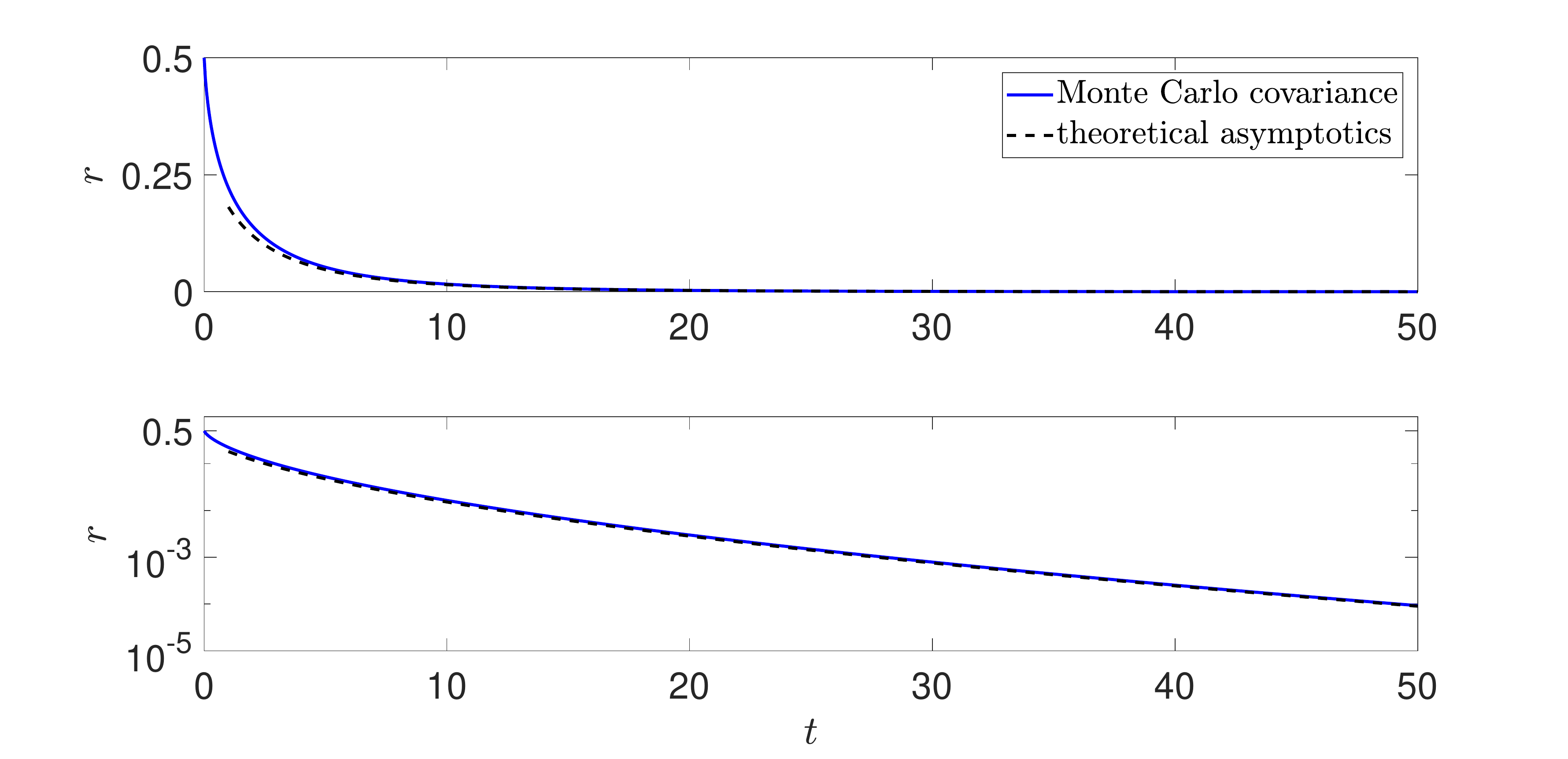}
\caption{Covariance function from Monte Carlo simulations for a system with
Laplace PDF of the position process, together with the theoretical asymptotics.
The sample size is $10^6$, $(2\Lambda)^{-1}\deq \mathcal E(1)$. The covariance
is shown on normal and semi-logarithmic scale. The convergence to the predicted
asymptotic behaviour is excellent. The full solution \Ref{eq:covLaplpdf} is not
shown, it fully overlaps with the simulations results.}
\label{f:laplCov}
\end{figure}

Analysing the shape of the covariance function can serve as a method to
distinguish between a superstatistic introduced by a local effective
temperature and the distribution of mass from superstatistics caused by the
randomness of the viscosity $\Lambda$. In the former case the resulting decay
is exponential (as in the non-superstatistical Langevin equation) or even zero
for a free Brownian particle, in the latter case it is given by relation
\Ref{eq:covLaplpdf}.

\section{Model with gamma distributed viscosity $\Lambda$}\label{s:gammaOU}
It is worth to consider a simple model with one particular choice for the distribution of $\Lambda$, calculating explicitly all related quantities. Our choice of distribution is the gamma distribution
$\mathcal G(\alpha,\beta)$ with the pdf
\begin{equation}
\label{eq:gammaDistr}
p_\Lambda(\lambda)=\f{\beta^\alpha}{\Gamma(\alpha)}\lambda^{\alpha-1}\e^{-\beta
\lambda}, \quad \alpha,\beta>0.
\end{equation}
This corresponds to a power law at $0^+$ which is truncated by an exponential decay. As the conditional covariance function is an exponential too, many integrals which in general would be hard to calculate, in this present case turn out to be surprisingly simple.

The gamma distribution is also a convenient choice because many of its special cases are well established in physics. The Erlang distribution is the special case of expression \Ref{eq:gammaDistr} when $\alpha$ is a natural number. An Erlang variable with $\alpha=k$ and $\beta$ can be represented as the sum of $k$ independent exponential variables $\mathcal E(\beta)$, in particular, for $k=1$ it is the exponential distribution itself. The chi-square distribution $\chi^2(k)$ is also a special case of expression \Ref{eq:gammaDistr} where $\alpha=k/2,\beta = 1/2$. The Maxwell Boltzmann distribution corresponds to the square root of $\chi^2(3)$, and the Rayleigh distribution to the square root of $\chi^2(2)$.

We already know from relation \Ref{eq:cOUPowerLaw} that $r_{V_\Lambda}$ has a
power tail $\sim 2^{-1}\beta^\alpha t^{-\alpha}$, more specifically, direct
integration yields
\begin{equation}
\label{eq:rcOU}
r_{V_\Lambda}(t)=\f{1}{2}\f{1}{\lt(1+t/\beta\rt)^\alpha}.
\end{equation}
This is solely a function of the ratio $t/\beta$ which shows that the parameter
$\beta$ changes the time scale of the process. Indeed, for any $\lambda$ the
process $V_\lambda(bt)$ is equivalent to $V_{b\lambda}(t)$, because the
Gaussian process is determined by its covariance function, which in both
cases is the same. Therefore, also the compound process $V_\Lambda(bt)$ is
equivalent to $V_{b\Lambda}(t)$ and $b\Lambda$ has the distribution $\mathcal
G(\alpha,\beta/b)$.

The function \Ref{eq:rcOU} would be observed if we calculated the ensemble
average of $V_\Lambda(\tau)V_\Lambda(\tau+t)$ for some $\tau$. If instead the
covariance function would be estimated as a time average over individual
trajectories, the Birkhoff's theorem determines that the result would be a
random variable, equal to the conditional covariance
\begin{equation}
\label{eq:taCov}
\overline{r}_{V_\Lambda}(t)\defeq \lim_{T\to\infty}\f{1}{T}\int_0^{T}\!\dd\tau\ V_\Lambda(
\tau)V_\Lambda(\tau+t)=\f{1}{2}\e^{-\Lambda t}.
\end{equation}
It is straightforward to calculate the pdf of this distribution,
\begin{equation}
\label{eq:expLbdpdf}
p_{\overline r_{V_\Lambda}}(x,t)=\f{2}{\Gamma(\alpha)}\lt(\beta/t\rt)^\alpha
|\ln(2x)|^{\alpha-1}(2x)^{\beta/t-1},\quad 0<x<1/2.
\end{equation}
The mean value of this quantity is given by Eq. \Ref{eq:rcOU}. This function is zero
in the point $x=1/2$ if $\alpha>1$ but has a logarithmic singularity at $x=(1/2)^-$
if $\alpha<1$, that is in the long-memory case. It is zero in $x=0$ for $t<\beta$
as in expression \Ref{eq:expLbdpdf} any power law dominates any power of the
logarithm. For $t>\beta$ there is a singularity at $x=0^+$ which approaches the
asymptotics $x^{-1}|\ln x|^{\alpha-1}$ as $t\to\infty$. This behaviour can be
observed in Figure \ref{f:pdfCov}, illustrating how the probability mass moves
from $(1/2)^-$ to $0^+$ as time increases.

\begin{figure}
\centering
\includegraphics[width=13cm]{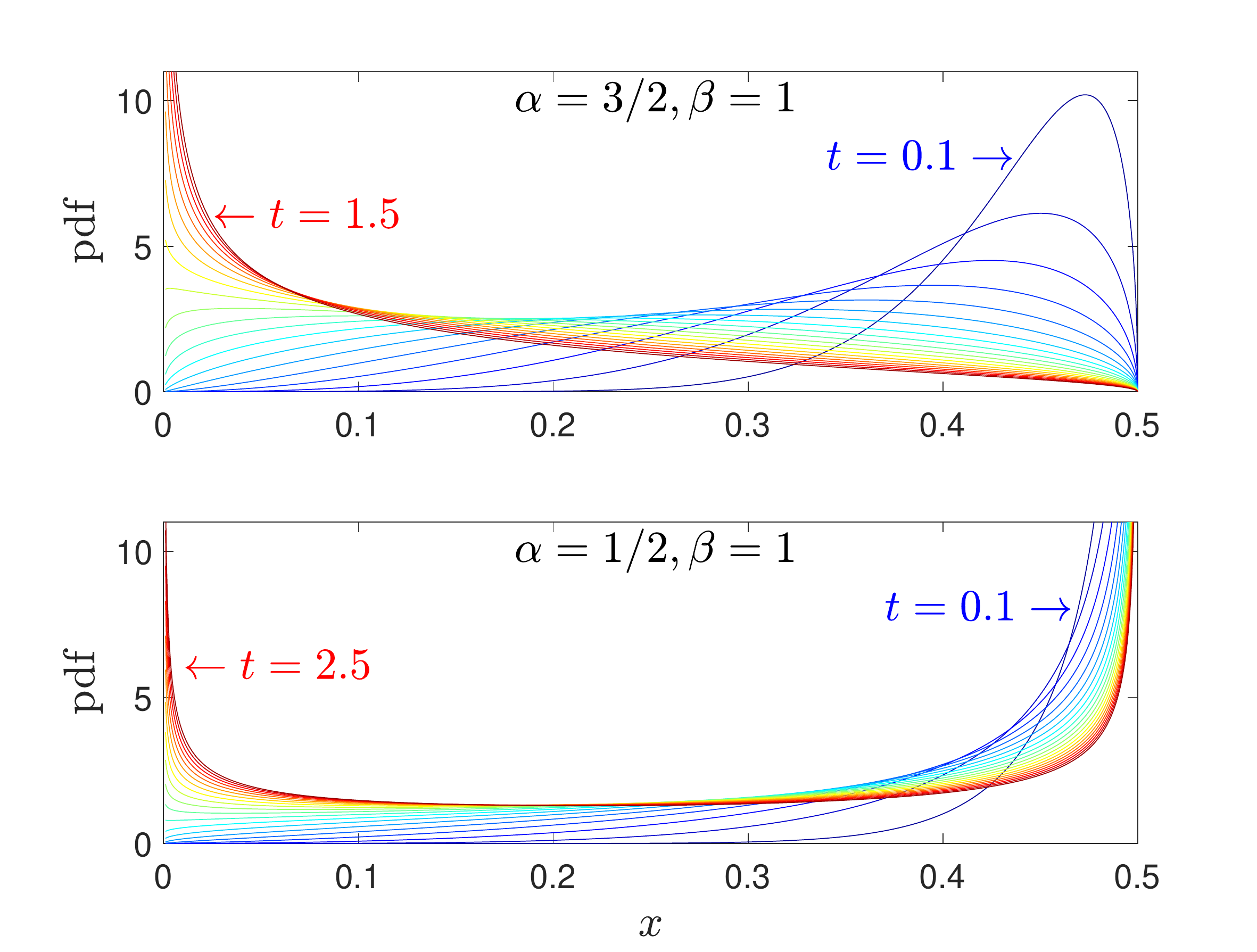}
\caption{Twenty pdfs of the covariance function $2^{-1}\exp(-\Lambda t), \Lambda
\deq\mathcal G(\alpha,\beta)$ for time $t$ changing linearly in the short memory
(top) and long memory (bottom) regimes.}
\label{f:pdfCov}
\end{figure}

As we already know, the compound Ornstein-Uhlenbeck process is non-Gaussian. Let
us follow up on this property in more detail. To study the characteristic function
we need to calculate the average in Eq. \Ref{eq:phi}, which is actually the
moment generating function for the random variable $\exp(-\Lambda t)$. Some
approximations can be made. First, let us assume that $\Lambda t$ is small in the
sense that the probability that this variable is larger than some small $\epsilon>0$
is negligible. In this regime we can approximate $\exp(-\Lambda t)\approx1-\Lambda
t$ and find
\begin{align}
\label{eq:apprt0}
\nonumber
\varphi_\Lambda(\bd\theta,t)&\approx \e^{-\f{1}{4}\theta_1^2}\e^{-\f{1}{4}\theta_2^2}
\E\lt[\e^{-\f{1}{2}\theta_1\theta_2(1-\Lambda t)}\rt]\\
&=\e^{-\f{1}{4}(\theta_1+\theta_2)^2}\E\lt[\e^{\f{1}{2}\theta_1\theta_2\Lambda
t}\rt]\nonumber\\
&=\e^{-\f{1}{4}(\theta_1+\theta_2)^2}\f{1}{(1-t\theta_1\theta_2/(2\beta))^{\alpha}},
\quad t\to 0^+.
\end{align}
The first factor describes a distribution of $V_\Lambda(\tau)=V_\Lambda(\tau+t)$.
So in our approximation we assume that the values in the process between short
time delays are nearly identical and the multiplicative correction $(1-t\theta_1
\theta_2/(2\beta))^{-\alpha}$ is non-Gaussian.

The second type of approximation can be made for long  times $t$ when $\exp(-\Lambda
t)\approx 0$. In this case
\begin{align}
\nonumber
\varphi_\Lambda(\bd\theta,t)&\approx\e^{-\f{1}{4}\theta_1^2}\e^{-\f{1}{4}\theta_2^2}
\E\lt[1-\f{1}{2}\theta_1\theta_2\e^{-\Lambda t}\rt]\\
&=\e^{-\f{1}{4}\theta_1^2}\e^{-\f{1}{4}\theta_2^2}\lt(1-\f{1}{2}\theta_1\theta_2
\E\lt[\e^{-\Lambda t}\rt]\rt)\nonumber\\
&=\e^{-\f{1}{4}\theta_1^2}\e^{-\f{1}{4}\theta_2^2}\lt(1-\f{\theta_1\theta_2}{2
(1-t/\beta)^\alpha}\rt),\quad t\to\infty.
\end{align}
Now we treat the values $V_\Lambda(\tau)$ and $V_\Lambda(\tau+t)$ as
nearly independent, the small correction is once again non-Gaussian. Apart from
the approximations, the exact formula for $\varphi_\Lambda$ can be provided using
the series
\begin{equation}
\label{eq:phiSeries}
\E\lt[\e^{-\f{1}{2}\theta_1\theta_2\e^{-\Lambda t}}\rt]=\sum_{k=0}^\infty
\f{(-1)^k}{2^kk!}(\theta_1\theta_2)^k\E\lt[\e^{-k\Lambda t}\rt]=\sum_{k=0}
^\infty \f{(-1)^k}{2^kk!}(\theta_1\theta_2)^k\f{1}{(1+kt/\beta)^\alpha},
\end{equation}
which is absolutely convergent.

Note that for the specific choice $\theta_1=\theta$, $\theta_2=-\theta$ the
function $\varphi_\Lambda$ is a characteristic function of
the increment $\Delta V_\Lambda(\tau,t)\defeq V_\Lambda(\tau)-V_\Lambda(\tau+t)$,
which therefore equals
\begin{equation}
\label{eq:apprtinfty}
\phi_{\Delta V_\Lambda(\tau,t)}(\theta) = \e^{-\f{\theta^2}{2}}\sum_{k=0}^
\infty \f{\theta^{2k}}{2^k k!}\f{1}{(1+kt/\beta)^\alpha}.
\end{equation}
Clearly, any increment of $V_\Lambda$ is non-Gaussian. This is demonstrated
in Figure \ref{f:fCharSS}, where we show $\sqrt{-\ln(\phi_{\Delta V_\Lambda(
\tau,1)}(\theta))}$ on the y-axis. In this choice of scale Gaussian distributions
are represented by straight lines. The concave shape of the empirical estimator
calculated using Monte Carlo simulation shows that the process $V_\Lambda$ is
indeed non-Gaussian. In the same plot we present the two types of approximations
of $\phi_{\Delta V_\Lambda(\tau,1)}$: for $t\to 0^+$ we have Eq. \Ref{eq:apprt0}, which reflects well the tails $\theta\to\pm\infty$, and
for $t\to\infty$ we see that with several terms of the series \Ref{eq:apprtinfty}
a good fit for $\theta\approx 0$ is obtained.
\begin{figure}
\centering
\includegraphics[width=16cm]{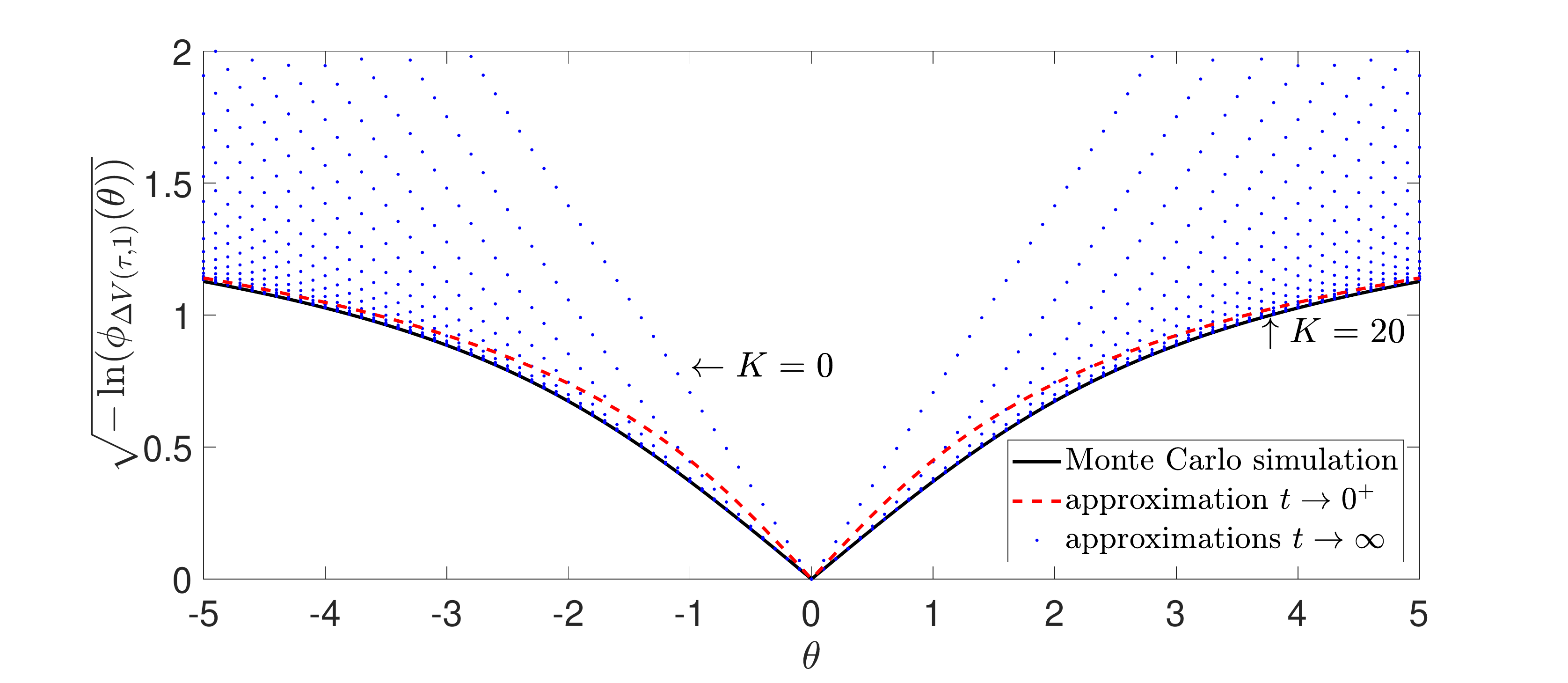}
\caption{Empirical characteristic function (the black line) calculated from
Monte Carlo simulated for $\Delta V_\Lambda(\tau,1),\Lambda\deq\mathcal G(1/2,1)$;
sample size was $10^6$. The red dashed line represents approximation
\Ref{eq:apprt0}, the blue dotted lines are the approximations based on Eq. \Ref{eq:apprtinfty} for $K=0,1,\ldots$, where $20$ terms in the Taylor
series were taken along.}
\label{f:fCharSS}
\end{figure}

It may appear counter-intuitive that the values $V_\Lambda(t)$, which are all
exactly Gaussian, are sums of non-Gaussian variables. If the increments were
independent that would be impossible, here their non-ergodic dependence
structure allows for this unusual property to emerge. However, the they are still
conditionally Gaussian with variance
\begin{equation}
\E\lt[\Delta V_\Lambda(\tau,t)^2|\Lambda\rt]= \lt(1-\e^{-\Lambda t}\rt).
\end{equation}
The non-Gaussianity is prominent for short times $t$. As $t$ increases, the
distribution of $\Delta V_\Lambda(\tau,t)$ converges to a Gaussian with
unit variance.

The non-Gaussian memory structure of the velocity $V_\Lambda$ affects also
the distribution of the position $X_\Lambda$, which, using results from Proposition \ref{prp:GLEasympPDF}, for large $t$ becomes
\begin{align}
\label{eq:Xpdf}
p_{X_\Lambda(t)/\sqrt{t}}(x)&\xrightarrow{t\to\infty}\f{1}{\sqrt{\pi}}\E\lt[\sqrt{\Lambda}\e^{-x^2\Lambda}
\rt]=\f{\beta^\alpha}{\Gamma(\alpha)\sqrt{\pi}}\int_0^\infty\!\!\dd\lambda\ \lambda^{\alpha-1/2}
\e^{-(x^2+\beta)\lambda}\nonumber\\
&=\f{\Gamma(\alpha+1/2)}{\sqrt{\pi}\Gamma(\alpha)}\f{\beta^{\alpha}}{(x^2+\beta)^{\alpha+1/2}}.
\end{align}
The above formula is a pdf of the Student's t-distribution
type, although unusual in the sense that most often it arises in statistics where
it is parametrised only by positive integer values of $\alpha$. It may therefore seem that for $\alpha\le 1/2$ the process $X_\Lambda$ may not
have a finite second moment, however Proposition \ref{prp:GLEasympPDF} excludes this possibility by stating that the pdf always has tails $\mathcal O(x^{-\infty})$. It means that for all finite $t$ a truncation is present; as time increases it is moved more away into $x=\pm\infty$. This property of $X_\Lambda(t)$ is illustrated in Figure
\ref{f:pdfX}, where we show the pdfs of the rescaled position position $X_\Lambda(t)/\sqrt{t}$ simulated with $\alpha=1/2,\beta=1$ and calculated using the kernel density estimator. In agreement with Eq. \Ref{eq:Xpdf}, the limiting distribution is of Cauchy type.

\begin{figure}
\centering
\includegraphics[width=14cm]{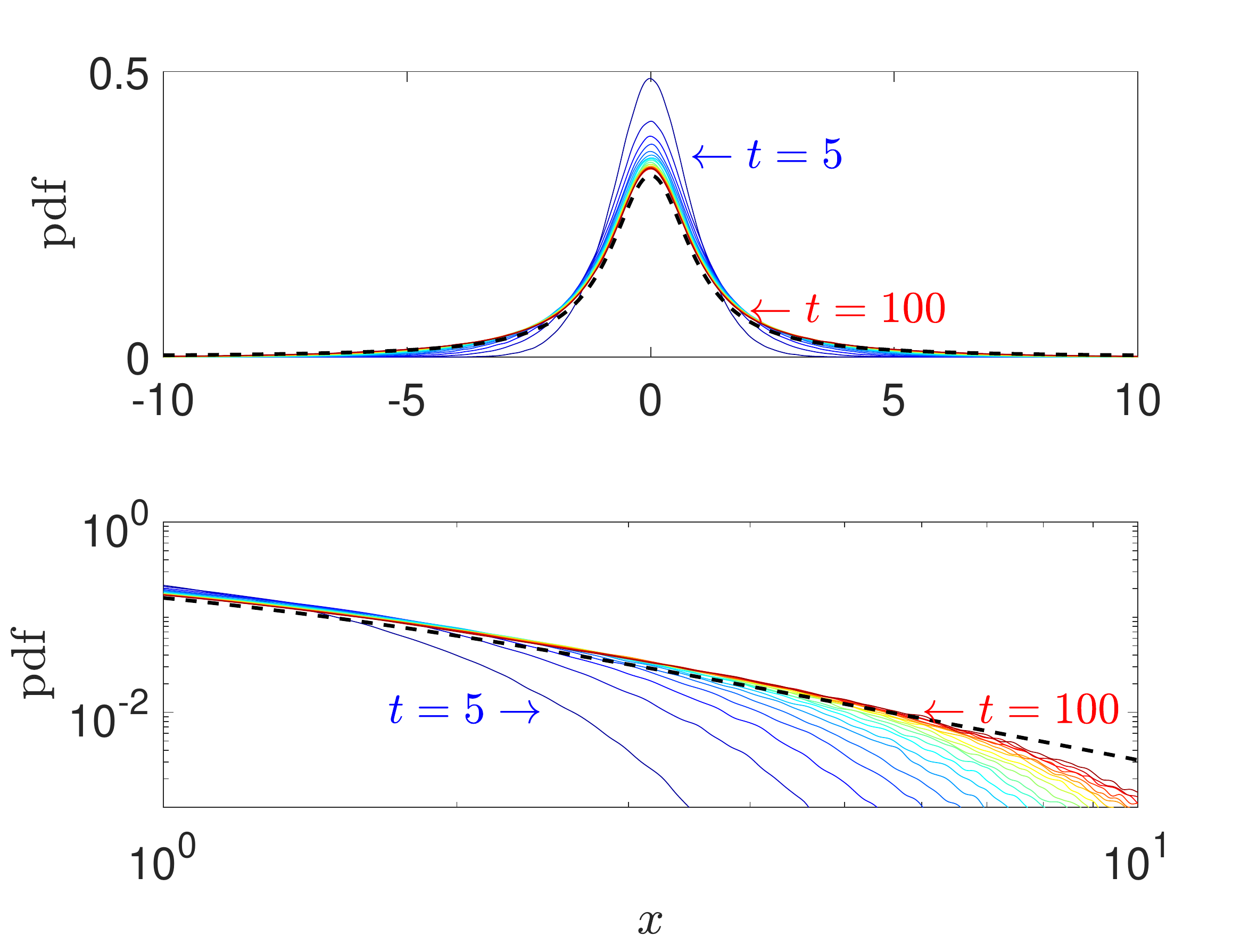}
\caption{Kernel density of variables $X_\Lambda(t)/\sqrt{t}$ estimated
for $t=5,10,\ldots,100$ (solid lines) and $\Lambda\deq\mathcal G(1/2,1)$
versus the Cauchy pdf (dashed line). The sample size was $10^6$. Both convergence to Cauchy distribution and the $\mathcal O(x^{-\infty})$ truncation of the tails can be observed.}
\label{f:pdfX}
\end{figure}

It may be expected that the relaxation to a power law pdf should affect the moments of $X_\Lambda$. Integrating the covariance function \Ref{eq:rcOU} twice we obtain the an exact formula for the msd
\begin{equation}
\delta^2_{X_\Lambda}(t)=\E[\delta^2_{X_\Lambda}(t|\Lambda)]=\f{\beta^2}{2}\f{(1+
t/\beta)^{2-\alpha}+(\alpha-2)t/\beta-1}{(1-\alpha)(2-\alpha)}.
\end{equation}
It describes superdiffusion for $0<\alpha<1$ and normal diffusion for $1\le
\alpha$ in agreement with the more general result \Ref{eq:msdPL}.

Similarly, a somewhat longer calculation yields
\begin{align}\label{eq:4momentX}
& \E\lt[\big(\delta^2_{X_\Lambda}(t|\Lambda)\big)^2\rt]=\f{\beta^4}{4}\f{1}{
(\alpha-4)(\alpha-3)(\alpha-2)(\alpha-1)}\nonumber\\
&\cdot\Big((\alpha-4) (\alpha-3) (t/\beta)^2-2 (\alpha-4) t/\beta+1\nonumber\\
& + 2 (\alpha-4) (t/\beta) (1+t/\beta)^{3-\alpha}-2
(1+t/\beta)^{4-\alpha}+(1+2 t/\beta)^{4-\alpha}\Big).
\end{align}
This formula can be used to determine the asymptotic behaviour of the popular measure used to study ergodicity breaking \citep{ergPar,weakErgBreak}.
\bg{dfn}[Ergodicity breaking parameter]
For a given stochastic process $X$ the ergodicity breaking parameter EB is defined by formula
\bgeq\label{eq:EB}
\text{EB}(t)\defeq \f{\var\lt[\overline\delta{}^2_X(t)\rt]}{\E\lt[\overline
\delta{}^2_X(t)\rt]^2} =\f{\E\lt[ \overline\delta{}^2_X(t)^2\rt]}{\E\lt[\overline
\delta{}^2_X(t)\rt]^2}-1.
\eeq
\end{dfn}
For the compound Ornstein-Uhlenbeck process $\overline\delta{}^2_{X_\Lambda}(t)=\delta^2_{X_\Lambda}(t|\Lambda)$ and
\bgeq
\text{EB}(t)=\f{\E\lt[\big(\delta^2_{X_\Lambda}(t|\Lambda)\big)^2\rt]}{\E\lt[\delta
^2_{X_\Lambda}(t|\Lambda)\rt]^2}-1\neq 0,
\eeq
for any $t>0$, as the mean of a square equals the square of a mean only for deterministic variables. It also does not converge to zero even for large $t$. Precisely, using Eq. \Ref{eq:4momentX} we get
\begin{equation}
\text{EB}(t)\sim\bg{cases}
\f{(1-\alpha)(2-\alpha)}{(4-\alpha)(3-\alpha)}\big(2
\alpha-10+2^{4-\alpha}\big)\big(\f{t}{\beta}\big)^{\alpha}, & \alpha<1\\
\f{(\alpha-1)}{(4-\alpha)(3-\alpha)(2-\alpha)}\big(10-2
\alpha-2^{4-\alpha}\big)\big(\f{t}{\beta}\big)^{2-\alpha}, & 1<\alpha<2\\
\f{1}{\alpha-2}, & 2<\alpha
\end{cases}
\end{equation}
as $t\to\infty$. This quantity has a practical value, because it can be effectively calculated from the data and used to distinguish between regimes of the parameter $\alpha$.

Additionally, in this model it is easy to check that $3
(\text{EB}(t)+1)$  is the kurtosis of $X_\Lambda(t)$, that is $\E[X_\Lambda(t)^4]
/(\E[X_\Lambda(t)^2])^2$. This is one of the measures of the thickness of the tails of a distribution; for any one-dimensional Gaussian distribution it equals 3. The divergence of the kurtosis for $\alpha<2$ reflects that the pdf converges to a thick power law. For $\alpha>2$ the tails of the limiting pdf are decaying faster than $|x|^{-5}$, so the fourth moment is bounded.

\section{Superstatistical GLE}

Motivated by the example considered above, we want to introduce a more general model of diffusion in a non-homogeneous viscoelastic medium.

\bg{dfn}[Superstatistical GLE] By the superstatistical solution of the generalized Lagevin equation we understand the stationary process which fulfils the GLE governed by random kernel $K_C$ and conditionally (on $C$) Gaussian stationary process $F_C$,
\begin{align}
\label{eq:ssGLE}
\nonumber
\df{}{t} V_C &= -\int_{-\infty}^t\!\!\!\dd\tau\ V_C(\tau)K_C(t-\tau)+F_C(t),\\
r_{F_C}(t|C)&=K_C(t).
\end{align}
The first equation is the GLE controlled by a random parameter $C$, a one dimensional variable or a random vector. The second equation is a local fluctuation-dissipation relation, which links $K_C$ and the conditional covariance function.
\end{dfn}
We tacitly assume that the introduction of the random parameter
$C$ does not change the spatially local structure of the GLE.
In the Hamiltonian model from which the GLE can be derived (see Section \ref{s:hamMod})  the studied variables $V,X$ interact with the heat bath only in their neighbourhood. The bath degrees of freedom themselves
do not interact with each other directly, which prohibits spatial long-range
correlations. Long-time correlations can still be present, but they result from
the interactions between $X$ and the bath degrees of freedom, which ``store'' the
memory structure for a long time, but do so only locally. That means for each
fixed, deterministic value $C=c$ the fluctuation-dissipation theorem should still hold, which is a second equation in the definition above.

In other words it is a model of particles which are confined in different, non-interacting areas, each with different properties of the local environment, denoted by $C$.

The stationary (mild) solution of equation \Ref{eq:ssGLE} is given by 
\bgeq
V_C(t)=\int_{-\infty}^t\!\!\!\dd\tau\ F_C(\tau)G_C(t-\tau),
\eeq
where the random Green's function $G_C$ is a randomized solution of the equation
\begin{equation}
\label{eq:defGc}
\df{}{t} G_c(t)=-\int_0^t\dd\tau\ G_c(\tau)K_c(t-\tau)+\delta(t), \quad G_c(0)=1,
\end{equation}
which should hold for every deterministic $c$.

From the previous considerations in Chapter \ref{ch:der} and \ref{ch:erg}, if we assume that for every choice of $c$ the covariance
function of $F_c$ decays to zero, that is $r_{F_c}(t)\to 0$ as $t\to\infty$ and $F$ is mixing, then the stationary solution $V_c$ is  mixing as well. However, the superstatistical solution $V_C$ (for non-trivial $C$) cannot be ergodic as averaging over one trajectory one cannot gain insight into the distribution of $C$. The process $V_C$ is still stationary because
\begin{align}
\E\lt[\e^{\I\sum_k\lambda_k V_C(t_k+\tau)}\rt] &= \E\lt[\E\lt[\e^{\I\sum_k\lambda_k V_C(t_k+\tau)}|C\rt]\rt]=\E\lt[\E\lt[\e^{\I\sum_k\lambda_k V_C(t_k)}|C\rt]\rt]\nonumber\\
&=\E\lt[\e^{\I\sum_k\lambda_k V_C(t_k)}\rt].
\end{align}
Therefore we can use the Birkhoff's theorem to determine all time-averages. The $\sigma$-algebra of invariant sets is generated by the values of $r_{V_C}(t|C)=G_C(t)$. If the parametrisation by $C$ is ``reasonable'', that is there is a one to one correspondence between $c$ and $t\mapsto G_c(t)$, this $\sigma$-algebra is simply $\sigma(C)$, and
\begin{equation}
\overline f(V_C)=\lim_{T\to\infty}\f{1}{T}\int_0^{T}\dd\tau\ 
f(\mathcal S_\tau V_C)=\E[f(V_C)|C].
\end{equation}
This generalisation of Eq. \Ref{eq:taCov} is intuitively reasonable: given a trajectory evolving with $C=c$ all time averaged statistics converge to the values corresponding to the solution of the GLE with $K_c$ and $F_c$, which
are exactly the conditional expected values $\E[f(V_C)|C=c]=\E[f(V_c)]$. For the msd this implies that 
\begin{equation}
\overline\delta{}_{X_C}^2(t) = \E[X_C(t)^2|C]=\delta^2_{X_C}(t|C),
\end{equation}
One consequence of
this fact is that the ergodicity breaking parameter does not
equal zero,
\begin{equation}
\text{EB}(t)=\f{\E\lt[\big(\delta^2_{X_C}(t|\Lambda)\big)^2\rt]}{\E\lt[\delta
^2_{X}(t|C)\rt]^2}-1\neq 0,
\end{equation}
which is the same behaviour as for the compound Ornstein-Uhlenbeck process. The difference is that for a very specific choice of $C$ parameter EB can converge to 0 as $t\to\infty$; it is possible for environments in which the inhomogeneity affects only the short-time properties of $X_C$.

Some general remarks about the properties of $V_C,X_C$ can be made.
\bg{prp}
Let the velocity $V_C$ be a solution of the superstatistical GLE and $X_C$ be the corresponding position process
\bg{itemize}
\item[i)] The variable $V_C(t)$  is Gaussian for any $t$.
\item[ii)] The motion is subballistic, $\delta^2_{X_C}(t)\le t^2$.
\item[iii)] All moments of $X_C(t)$ are finite.
\end{itemize}
\end{prp}
\bg{proof}
Point i) follows from the fact that $G_c(0)=1$ for every $c$ and in Proposition \ref{prp:covGreen} we proved that $r_{V_C}(t|C)=G_C(t)$. Therefore, the variable $V_C(t)$ is a mixture of Gaussian variables $\mathcal N(0,1)$. i.e. variable $\mathcal N(0,1)$. Moreover, using the results from the same proposition and the elementary inequality $r_{V_C}(t|C)\le r_{V_C}(0|C)$,
\bgeq
\delta^2_{X_C}(t)\le \E\lt[2\int_0^t\dd\tau_1\int_0^{\tau_1}\!\dd\tau_2|r_{V_C}(t|C)|\rt]\le \E\lt[ G_C(0) t^2\rt]=t^2,
\eeq
which proves ii). The last point is a consequence of the fact that
\bgeq
\E[X_C(t)^{2n}] = \prod_{k=2}^n(2k-1)\E\lt[\lt(\delta^2_{X_C}(t|C)\rt)^n\rt]\le t^{2n}.
\eeq
\end{proof}
These properties generalize the observations made for the compound Ornstein-Uhlenbeck process. It is interesting that the superstatistical Langevin equation preserves the finiteness of moments. It can be considered a consequence of the Hamiltonian derivation of the GLE, which bounds the average energy of the system to $k_B\mathcal T$, constant for all local environments. Because the superstatistical GLE can model  power law tails of the observed distribution, it naturally reconciles this notion with a finite second moment, by naturally introducing a truncation moving to $\pm\infty$ as $t\to\infty$, in the same manner as for the compound Ornstein-Uhlenbeck process.

Note also that process $V_C$ must exhibit a non-Gaussian memory structure, otherwise its non-ergodicity would contradict the Maruyama's theorem. Solutions of the superstatistical GLE are interesting physical examples of objects which have Gaussian marginals, but non-Gaussian memory structure. Such processes can be constructed easily using the copula theory, but this construction can be
considered artificial and without physical meaning. The unusual non-Gaussianity of $V_C$ here arises naturally from the physical model. The process $V_C$ could be very misleading during the analysis of measured data, using only basic statistical
methods it will seem Gaussian. Some techniques which can be used to analyse this behaviour were presented in Section~\ref{s:gammaOU}.

\section{Examples}
As we study superstatistical GLE with practical applications in mind, it is worthwhile analysing in detail some specific cases, which could be a model of real data. In Section \ref{s:elSol} we analysed 3 important cases: Dirac delta kernel, exponential kernel and power law kernel. The superstatistical generalisation of the first is the compound Ornstein-Uhlenbeck process, which we already described.

The second case was also briefly studied in Section \ref{s:physAppl} below Eq. \Ref{eq:expGLeF}, because of its relation to ARMA processes. Let us briefly repeat the most important information about this model. We assume that the covariance function has the conditional form
\begin{equation}
r_{F_{A,B}}(t|A,B)= B^2\e^{-2At},\quad A,B>0,
\end{equation}
therefore the stochastic force $F_{A,B}$ in this model is actually the compound
Ornstein-Uhlenbeck process, additionally
rescaled by the random coefficient $B^2$.  The conditional covariance function and msd were calculated using Laplace methods, and are
\begin{align}\label{eq:rABFull}
r_{V_{A,B}}(t|A,B)&=\f{1}{2}\lt(1-\f{A}{\sqrt{A^2-B^2}}\rt)\e^{-(A+\sqrt{A^2
-B^2})t}\nonumber\\
&+\f{1}{2}\lt(1+\f{A}{\sqrt{A^2-B^2}}\rt)\e^{-(A-\sqrt{A^2-B^2})t}
\end{align}
and
\begin{align}
&\delta^2_{X_{A,B}}(t|A,B) = 4\f{A}{B^2} t -\f{8A^2}{B^4}+2B^2\nonumber\\
&+\f{1}{\sqrt{A^2-B^2}}\e^{-At}\lt(\f{\sqrt{A^2-B^2}-A}{(A+\sqrt{A^2-B^2})^2}
\e^{\sqrt{A^2-B^2}t}+\f{\sqrt{A^2-B^2}+A}{(A-\sqrt{A^2-B^2})^2}\e^{-\sqrt{A^2
-B^2}t}\rt).
\end{align}
As we can see the asymptotic behaviour of the covariance function and the msd at $t=0^+$ and $t=\infty$
is very similar to that of the compound Ornstein-Uhlenbeck process. For
$\E[A/B^2]<\infty$ this GLE models normal diffusion with a random diffusion coefficient. When this condition is not fulfilled it may model superdiffusion. The behaviour of this system greatly depends on whether $A<B,A=B$ or $A>B$, so we will study all these cases separately.

\textbf{Critical regime $A=B$.} Taking the limit $A\to B$ in Eq. \Ref{eq:rABFull} we can determine
the form of the conditional covariance within the critical regime,
\begin{equation}
r_{V_{A}}(t|A)=(1+At)\e^{-A t}.
\end{equation}
The behaviour of the resulting solution $V_A$ is very similar to that of the
compound Ornstein-Uhlenbeck process. The differences are only slight. For
example, if $A=A_1+A_2$ for some independent $A_1$ and $A_2$, then
\begin{equation}
r_{V_{A_1+A_2}}(t)=r_{V_{A_1}}(t)r_{V_{A_2}}(t)-t^2\E\lt[A_1\e^{-A_1t}\rt]\E\lt
[A_2\e^{-A_2t}\rt].
\end{equation}
Therefore, for instance if $A>a_0$ we can write $A=A'+a_0, A'>0$ and
the covariance function becomes truncated by $a_0t\exp(-a_0 t)$.

The formula for $r_{V_{A}}$ consist of two terms. Function $A t\exp(-A t)$
has a thicker tail, but the asymptotic behaviour of $r_{V_A}$ is determined by
the distribution of small values $A\approx 0$, so it is not clear which term
is most important in that regard. If we assume $p_A(a)\sim a^{\alpha-1}, a\to 0^+$, then
\begin{equation}
r_{V_{A}}(t)\sim \Gamma(\alpha)t^{-\alpha}+t\Gamma(\alpha+1)t^{-\alpha-1}=
(\alpha+1)\Gamma(\alpha)t^{-\alpha},
\end{equation}
so actually both terms have comparable influence over the resulting tails
of the covariance.

\textbf{Exponential decay regime $A>B$.}
In this case the covariance function is a sum of two decaying exponentials.   The first one has a negative amplitude,
the second one a positive amplitude. In addition the second exponential always
has a heavier tail, as its exponent includes the difference of positive terms,
$A-\sqrt{A^2-B^2}$, whereas the other exponent includes a sum. Thus we expect
that the exponent with $A+\sqrt{A^2-B^2}$ cannot lead to a slower asymptotic than the one containing $A-\sqrt{A^2-B^2}$.

Given this reasoning let us change the variables in the form
\begin{equation}
A' = A-\sqrt{A^2-B^2}=\f{B^2}{A+\sqrt{A^2-B^2}},\quad A=\f{B^2+A'^2}{2A'}.
\end{equation}
The new parameter $A'$ attains the value $A'=B$ for $A=B$ and decays monotonically to 0 as $A\to\infty$. Note that for small values of $A'$, $A'\approx B^2/(2A)$,
so the tail behaviour of $A$ determines the distribution of $A'$ at $0^+$; in
particular, a power law shape of the former is equivalent to a power law shape
of the latter: for fixed $B=b$, $p_{A'}(a)\sim a^{\alpha-1}, a\to 0^+$ if and only if $p_A(a)\sim 2^{-\alpha}b^{\alpha}a^{-1-\alpha},a\to\infty$. Using the parameters $A'$ and $B$ the covariance function can be
expressed as
\begin{equation}\label{eq:covGLEexpDecay}
r_{V_{A', B}}(t|A',B)=\f{B^2}{B^2-A'^2}\e^{-A' t}-\f{A'^2}{B^2-A'^2}
\e^{-\f{B^2}{A'}t}.
\end{equation}
As $A'<B$ the variables $A'$ and $B$ cannot be independent unless $B$ is
deterministic. Dealing with dependent $A'$ and $B$ is complicated, so in the following result we introduce some simplifying assumptions.
\bg{prp} Let $A'$ have power law distribution $p_{A'}\sim a^{\alpha-1},a\to 0^+$. For the averaged covariance function \Ref{eq:covGLEexpDecay} the following asymptotic limits hold
\begin{itemize}
\item[i)] If the variable $B$ is deterministic, that is $B=b$, then
\bgeq
r_{V_{A',b}}(t)\sim \Gamma(\alpha)t^{-\alpha},\quad t\to\infty.
\eeq
\item[ii)] Let us assume that $B'\defeq A'/B$ is independent from $A'$ and $\E[(1-B'^2)^{-1}]<\infty$. Then, for any bounded distribution of $A'$, $m\le p_{A'}(a)\le M$ supported on some interval $[a_1,a_2]$,
\bgeq
\label{eq:expDecBoundedAs}
\f{m}{t}\e^{-a_1 t}\lesssim r_{V_{A',B'}}(t)\lesssim
\E\lt[\f{1}{1-B'^2}\rt]\f{M}{t}\e^{-a_1 t}.
\eeq
\item[iii)] Let $B'$ be as in ii) and assume that $A'$ has the power law distribution $p_{A'}(a)\sim a^{\alpha-1},a\to 0^{+}$. Then
\bgeq
\label{eq:expDecPowerLaw}
r_{V_{A', B'}}(t)\sim \Gamma(\alpha)\E\lt[\f{1-B'^{2\alpha+2}}{1-B'^2}\rt]
t^{-\alpha}.
\eeq
\end{itemize}
\end{prp}
\bg{proof}
Let us start with i). Eq. \Ref{eq:covGLEexpDecay} has two terms, averaging the right one we get
\bgeq
\int_0^b\dd a\ p_{A'}(a)\f{a^2}{b^2-a^2}\e^{-\f{b^2}{a}t}=\int_\f{1}{b}^\infty\!\!\dd\tilde a\ p_{A'}\lt(\f{1}{\tilde a}\rt)\tilde a^{-2}\f{1}{\tilde a^2 b^2-1}\e^{-b^2\tilde a t},
\eeq
which decays no slower than $\exp(-bt)$. The left term is the important one,

\bgeq
r_{V_{A',b}}(t)\sim\E\lt[\f{b^2}{b^2-A'^2}\e^{-A't}\rt]=\int_0^b\dd a\ p_{A'}(a)\f{b^2}{
b^2-a^2}\e^{-at}\sim \Gamma(\alpha)t^{-\alpha}.
\eeq
For ii) note that the conditional covariance can now be transformed into
\bgeq\label{eq:covA'B'}
r_{V_{A', B'}}(t|A',B')=\f{1}{1-B'^2}\e^{-A' t}-\f{B'^2}{1-B'^2}\e^{-\f{A'}{B'^2}t}.
\eeq
We use the simple inequality
\bgeq
\f{1}{1-B'^2}\e^{-A' t}-\f{B'^2}{1-B'^2}\e^{-A't}\le r_{V_{A', B'}}(t|A',B')\le \f{1}{1-B'^2}\e^{-A' t},
\eeq
which follows from the fact that $0\le B'\le 1$. Averaging both sides over $p_{A'}$ and using the inequality $m\le p_{A'}(a)\le M$ proves the result. For iii) let us fix $B'$ and average over $A'$. We obtain
\begin{equation}
r_{V_{A', B'}}(t|B')\sim\f{1}{1-B'^2}\Gamma(\alpha)t^{-\alpha}-\f{B'^2}{1-B'^2}
\Gamma(\alpha)\lt(\f{t}{B'^2}\rt)^{-\alpha} =\f{1-B'^{2+2\alpha}}{1-B'^2}\Gamma(\alpha)t^{-\alpha} ,
\end{equation}
so averaging over $B'$ yields the desired formula.
\end{proof}

These results show that in the considered cases the distribution of $B$ does not change the asymptotics of the covariance. Our assumption that $B'=A'/B$ independent from $A'$ is limiting, although it has a simple interpretation: for power law distributions it can be considered as a form of independence between $A'$ and $B$ for small values of $A'$. To see that consider the joint pdf 
\bgeq
p_{A',B}(a,b) = p_{A',B'}(a, a/b) = p_{A'}(a)p_{B'}(a/b).
\eeq
Now if pdf $B'$ has a power law at $0^+$, as we assumed above, the joint pdf factorises into a function of $a$ multiplied by a function of $b$ for $a\to 0^{+}$. 

Note also that the obtained asymptotic behaviour is similar to the behaviour of the compound
Ornstein-Uhlenbeck process, only with different scaling. For the same reason
$r_{V_{A', B'}}(t)$ is truncated under the same conditions as before, if $A'$
is a sum of independent $A'_1$ and $A'_2$
\begin{equation}
\label{eq:expDecmlpl}
r_{V_{A', B'}}(t)\sim \const\cdot r_{V_{A'_1, B'}}(t)r_{V_{A'_2, B'}}(t),\quad t\to\infty,
\end{equation}
see Eq. \Ref{eq:covA'B'}. The scaling constant depends on the distribution of $A'$ and $B'$. In particular $A>a_0$ results in an exponential truncation by $\exp(-a_0t)$ of the associated covariance.

\textbf{Oscillatory decay regime $A<B$.}
When the square root $\sqrt{A^2-B^2}$ is imaginary we can express the covariance
function as
\bgeq\label{eq:rOscReg}
r_{V_{A,B}}(t|A,B)=\lt(\cos\big(\sqrt{B^2-A^2}t\big)+\f{A}{\sqrt{B^2-A^2}}
\sin\big(\sqrt{B^2-A^2}t\big)\rt)\e^{-At}.
\end{equation}
This represents a trigonometric oscillation truncated by the factor $\exp(-At)$.
When calculating the unconditional covariance, this function acts as an integral kernel on the distribution of $A$ and $B$. The exponential factor acts similarly to the Laplace transform, but oscillations introduce Fourier-like behaviour of this transformation. It can be observed in the solutions of the corresponding GLE, which we show below.

\bg{prp}\label{prp:oscAs}
Let $r_{V_{A,B}}$ be a covariance function \Ref{eq:rOscReg}. Then the following asymptotic properties hold:
\bg{itemize}
\item[i)] For $A$ with bounded pdf $p_A(a)\le M$ supported on the interval $[a_1,
a_2]$, $a_2<B$ and independent of $B$, there exists the asymptotic bound
\begin{equation}\label{eq:oscAsInterval}
|r_{V_{A,B}}(t)|\lesssim \E\lt[\f{1}{\sqrt{1-\f{a_2^2}{B^2}}}\rt]\f{M}{t}
\e^{-a_1 t},\quad t\to\infty.
\end{equation}
\item[ii)] If additionally $A$ exhibits a power law behaviour at $a_1^+$, that
is, $A=a_1+A', p_{A'}(a)\sim a^{\alpha+1}$ for $a\to 0^+$, the asymptotic bound
can be refined to
\bgeq\label{eq:oscAsTrunc}
|r_{V_{A,B}}(t)|\lesssim\E\lt[\f{1}{\sqrt{1-\f{a_1^2}{B^2}}}\rt]\Gamma(\alpha)
t^{-\alpha}\e^{-a_1t}.
\eeq
\item[iii)]For $p_{A}(a)\sim a^{\alpha+1}$ at $a\to 0^+$ and deterministic $B=b$
the asymptotic limit of the covariance function is
\begin{equation}\label{eq:oscpl}
r_{V_{A,b}}(t)\sim \Gamma(\alpha)\cos(bt)t^{-\alpha},\,\,\, t\to\infty.
\end{equation}
We remind the readers that the asymptotics  ``$\sim$'' is understood as a limit of ratios for all sequences of $t_k\to\infty$ which do not target zeros of
$\cos(bt)$, that is $|bt_k-l\pi+\pi/2|>\epsilon$ for all $k,l\in\N$ and some
$\epsilon>0$.
\end{itemize}
\end{prp}

\bg{proof}
We start from the simple inequality
\begin{align}
|r_{V_{A,B}}(t|A,B)|&=\lt|\sqrt{1-\lt(\f{A}{B}\rt)^2}\cos\big(\sqrt{B^2-A^2}t
\big)+\f{A}{B}\sin\big(\sqrt{B^2-A^2}t\big)\rt|\nonumber\\
&\cdot\f{1}{\sqrt{1-\lt(\f{A}{B}\rt)^2}}\e^{-At}\nonumber\\
&=\lt|\cos\lt(\sqrt{B^2-A^2}t-\arcsin\lt(\f{A}{B}\rt)\rt)\rt|\f{1}{\sqrt{1-
\lt(\f{A}{B}\rt)^2}}\e^{-At}\nonumber\\
&\le \f{1}{\sqrt{1-\lt(\f{A}{B}\rt)^2}}\e^{-At}.
\end{align}
This allows us to prove i), namely:
\begin{equation}
|r_{V_{A,B}}(t|B)|\le M\int_{a_1}^{a_2}\!\dd a\f{1}{\sqrt{1-\f{a^2}{B^2}}}\e^{-at}
\le\f{M}{\sqrt{1-\f{a_2^2}{B^2}}}\int_{a_1}^{\infty}\!\!\dd a\ \e^{-at}
=\f{M}{\sqrt{1-\f{a_2^2}{B^2}}}\f{1}{t}\e^{-a_1 t}.
\end{equation}
Averaging over $B$ yields the result. For $B$ with a distribution concentrated at
$a_2^+$ it may happen that
\begin{equation}
\E\lt[\f{1}{\sqrt{1-\f{a_2^2}{B^2}}}\rt]=\infty
\end{equation} 
and in this case point i) is a trivial statement. However, it is sufficient that
$B>a_2+\epsilon,\epsilon>0$ for this average to be finite and $\le 2/\epsilon$.

Proof of point ii) is similar,
\begin{equation}
|r_{V_{A,B}}(t|B)|\le \int_0^\infty\!\!\dd a \f{p_{A'}(a)}{\sqrt{1-\f{(a+a_1)^2}{B^2}}}
\e^{-(a+a_1)t}\sim\f{\Gamma(\alpha)}{\sqrt{1-\f{a_1^2}{B^2}}}t^{-\alpha}
\e^{-a_1t}.
\end{equation}

Proving iii) requires a more delicate reasoning. We write $r_{V_{A,b}}(t)$
as an integral and change variables $at\to a$, so that
\begin{align}
r_{V_{A,b}}(t)&=\int_0^\infty\!\!\dd a\ p_A(a)\lt(\cos\lt(\sqrt{b^2-a^2} t\rt)+\f{a}{
\sqrt{b^2-a^2}}\sin\lt(\sqrt{b^2-a^2}t\rt)\rt)\e^{-at}\nonumber\\
&=\f{1}{t}\int_0^\infty\!\!\dd a\ p_A\lt(\f{a}{t}\rt)\lt(\cos\lt(\sqrt{b^2t^2-a^2}\rt)
+\f{a}{\sqrt{b^2t^2-a^2}}\sin\lt(\sqrt{b^2t^2-a^2}\rt)\rt)\e^{-a}.
\end{align}
After change of variables the Fourier oscillations depend on the variable
$\sqrt{b^2t^2-a^2}$. In the limit $t\to\infty$ they converge to oscillations
with frequency $b$,
\begin{equation}
\big|\sqrt{b^2t^2-a^2}-bt\big|=\f{a^2}{\sqrt{b^2t^2-a^2}+bt}\xrightarrow{t\to
\infty} 0.
\end{equation}
It is crucial that this frequency does not depend on $a$. The cosine function
also converges to a cosine with frequency $b$,
\begin{align}
\f{\cos(\sqrt{b^2t^2-a^2})}{\cos(bt)}&=\f{\cos(\sqrt{b^2t^2-a^2}-b t+bt)}{\cos(b
t)}\nonumber\\
&=\f{\cos(\sqrt{b^2t^2-a^2}-b t)\cos(bt)+\sin(\sqrt{b^2t^2-a^2}-b t)\sin(bt)}{
\cos(bt)}\nonumber\\
&\xrightarrow{t\to\infty}\cos(0)\cdot 1+\sin(0)\cdot \tan(bt)=1,\quad|\tan(bt)|
<\f{1}{\epsilon}.
\end{align}
Substituting this result into the integral for $r_{V_{A,b}}$ we obtain the
asymptotic
\begin{align}
\f{r_{V_{A,b}}(t)}{\cos(bt) t^{-\alpha}}&=\int_0^\infty\dd a\ p_A\lt(\f{a}{t}\rt)
t^{\alpha-1}\f{\cos(\sqrt{b^2t^2-a^2})+\f{a}{\sqrt{b^2t^2-a^2}}\sin(\sqrt{b^2
t^2-a^2})}{\cos(bt)}\e^{-a}\nonumber\\
&\xrightarrow{t\to\infty} \int_0^\infty\!\!\dd a\ a^{\alpha-1}\e^{-a}=\Gamma(\alpha).
\end{align}
\end{proof}

The behaviour shown in iii) can be seen in Fig. \ref{f:oscCov} which demonstrates that the convergence to the limit is fast. During the Monte Carlo simulation the
parameter $B$ was fixed as $B=\pi$ and $A$ was taken from the gamma distribution
$\mathcal G(1/2,1)$. For this distribution there exists a $98.8\%$ chance that $A<\pi=B$ and the system is in the oscillatory regime, so it is indeed dominating the result, as shown in Figure \ref{f:oscCov}.

\begin{figure}
\centering
\includegraphics[width=16cm]{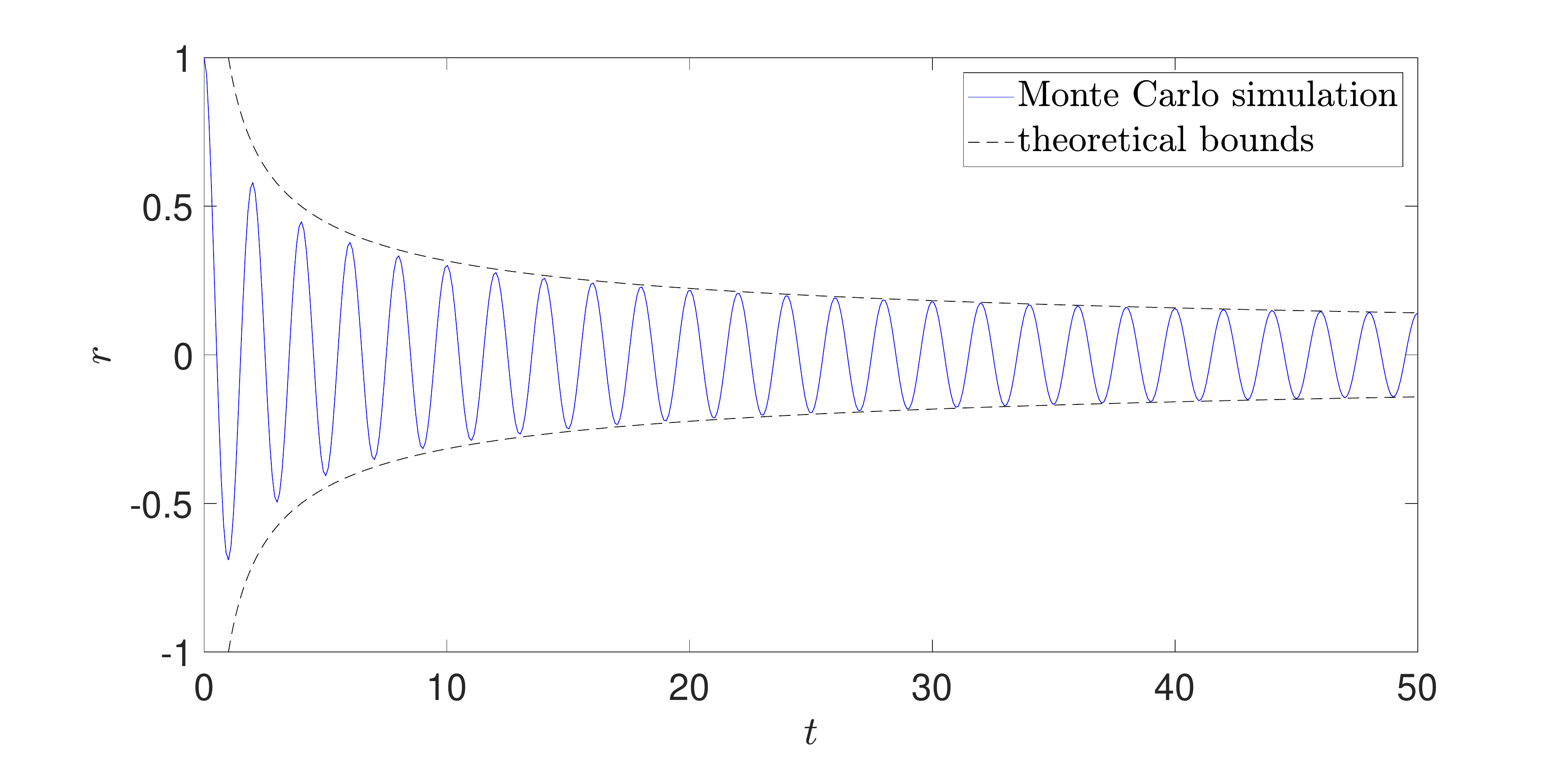}
\caption{Covariance function of the GLE (solid line) calculated using Monte Carlo
simulations, the sample size was $10^6$. Parameters are $B=\pi$, $A=\mathcal G
(1/2,1)$, the theoretical bounds are $\pm t^{-1/2}$ (dashed lines) as given in Eq. \Ref{eq:oscpl}.}
\label{f:oscCov}
\end{figure}

It is worth adding that the results from the above proposition can be slightly generalised by introducing any slowly varying factor in the distribution. One can also wonder that if for any deterministic $b$ the asymptotic is $\Gamma(\alpha)\cos(bt)t^{-\alpha}$, then for random $B$ it could be $\Gamma(\alpha)\E[\cos(Bt)]t^{-\alpha}$. For a variable $B$ with  a finite number of possible values, $\pr(B=b_k)=p_k$, this is clearly the case, but in general such a behaviour requires strong assumptions about the distribution of $B$.

\textbf{Power law kernel GLE}
The last case considered in Section \ref{s:elSol} was the GLE in which the force was fractional Gaussian noise. It has a power law conditional covariance function, namely
\begin{equation}
r_{F_{H,Z}}(t|H,Z)=\f{Z}{\Gamma(2H-1)}t^{2H-2},\quad 0<H<1.
\end{equation}
We already calculated that
\begin{equation}
r_{V_{H,Z}}(t|H,Z)=E_{2H}(-Zt^{2H})\sim \f{1}{Z\Gamma(1-2H)}t^{-2H},
\quad t\to\infty,
\end{equation}
moreover
\begin{equation}
\delta^{2}_{X_{H,Z}}(t|H,Z)=t^2E_{2H,3}(-Zt^{2H})\sim\f{1}{Z\Gamma(3-2H)}
t^{2-2H},\quad t\to\infty.
\end{equation}
From this formulas we see that the distribution of $Z$ should not have an
influence on the covariance and msd asymptotics. Further on we will assume that $Z$
is independent from $H$ and $\E[Z^{-1}]<\infty$.
A more interesting situation occurs when $H$ is distributed according to
a power law. Noting that $t^{-2H}=\e^{-2H\ln(t)}$ one may suspect that
the resulting covariance would exhibit power-log tails. This intuition is true, which we show in a  simple lemma.

\bg{lem}\label{lem:powerLaw}
Let $L$ be a slowly varying function at $0^+$ and $H$ a random variable of the form $H=h_1+H'$ with $h_1>0$. Then
\begin{equation} 
p_{H'}(h)\sim h^{\alpha-1} L(h),\quad h\to0^+
\end{equation}
implies that the mean value of a power law satisfies
\begin{equation}
\E\lt[t^{-2H}\rt]\sim \f{\Gamma(\alpha)}{2^\alpha}t^{-2h_1}L(\ln(t)^{-1})
\ln(t)^{-\alpha},\quad t\to\infty.
\end{equation}
\end{lem}
\bg{proof}
We write
the integral for $\E[t^{-2(h_1+H')}]$, reformulate it as a Laplace transform
using $t^{-2H'}=\e^{-2H'\ln(t)}$, change variables and calculate the limit
\begin{align}
&\f{\E\lt[t^{-2(h_1+H')}\rt]}{t^{-2h_1}L(\ln(t)^{-1})\ln(t)^{-\alpha}}=\f{1}{
t^{-2h_1}L(\ln(t)^{-1})\ln(t)^{-\alpha}}\int_0^{1-h_1}\!\! \dd h\ p_{H'}(h)t^{-2(h_1+h)}\nonumber\\
\nonumber\\
&=\int_0^{1-h_1}\!\!\dd h\f{p_{H'}(h)}{L(\ln(t)^{-1})\ln(t)^{-\alpha}}\e^{-2h\ln(t)}=
\int_0^{\ln(t)(1-h_1)}\!\!\dd h\f{p_{H'}\lt(\f{h}{\ln t}\rt)}{L(\ln(t)^{-1})\ln(t)^{1-
\alpha}}\e^{-2h}
\nonumber\\
&\xrightarrow{t\to\infty}\int_0^{\infty}\!\!\dd h\ h^{\alpha-1}\e^{-2h}=\f{\Gamma(
\alpha)}{2^\alpha}.
\end{align}

\end{proof}
This result will be useful in proving point ii) of the following proposition, which determines the asymptotics of the averages over Mittag-Leffler functions.

\bg{prp}\label{prp:MLas}
Let $H$ and $Z$ be independent, $\E[Z^{-1}]<\infty$ and $\beta\ge 1$. Then the
following asymptotic properties of $\E[E_{2H,\beta}(-Zt^{2H})]$ hold
\bg{itemize}
\item[i)] If $H$ is supported on $[h_1,h_2]$ with $0<h_1<h_2\le 1$ and its pdf is bounded, $m\le p_H(h)\le M$, then
\begin{equation}\label{eq:plInterval}
\f{m\E[Z^{-1}]}{2\Gamma(\beta-2h_1)}t^{-2h_1}\ln(t)^{-1}\lesssim \E[E_{2H,\beta}
(-Zt^{2H})] \lesssim \f{M\E[Z^{-1}]}{2\Gamma(\beta-2h_1)}t^{-2h_1}\ln(t)^{-1}.
\end{equation}
As usual, when the distribution is uniform $m=M=1/(h_2-h_1)$ and the asymptotics is stronger ``$\sim$''.

\item[ii)] If additionally $H$ exhibits a power law behaviour at $h_1^+$, that
is, $H=h_1+H'$, $p_{H'}(h)\sim h^{\alpha+1}$ with $h\to 0^+$, a much stronger
asymptotic property holds,
\begin{equation}
\E[E_{2H,\beta}(-Zt^{2H})]\sim\f{\E[Z^{-1}]\Gamma(\alpha)}{2^\alpha\Gamma(\beta
-2h_1)}t^{-2h_1}\ln(t)^{-\alpha},\quad t\to \infty.
\end{equation}
\end{itemize}
\end{prp}
\bg{proof}
Because
\begin{equation}
\label{eq:MLasympt2}
E_{2H,\beta}(-Zt^{2H})\sim \f{1}{Z\Gamma(\beta-2H)}t^{-2H}
\end{equation}
the left hand side is a function which has constant sign. For some small
$\epsilon>0$ and large enough $t$ we observe the inequality
\begin{equation}
\f{m(1-\epsilon)}{Z}\int_{h_1}^{h_2}\!\dd h\f{t^{-2h}}{\Gamma(\beta-2h)}
\le \E[E_{2H,\beta}(-Zt^{2H})|Z]\le\f{M(1+\epsilon)}{Z}\int_{h_1}^{h_2}\!\dd h
\f{t^{-2h}}{\Gamma(\beta-2h)}.
\end{equation}
Now we check the asymptotics of the integral above,
\begin{align}
&\f{\ln(t)}{t^{-2h_1}}\int_{h_1}^{h_2}\!\dd h\f{t^{-2H}}{\Gamma(\beta-2h)}
=\ln(t)\int_0^{h_2-h_1}\!\!\!\dd h\f{\e^{-2h\ln(t)}}{\Gamma(\beta-2h_1-2h)}
\nonumber\\
&=\int_0^{\ln(t)(h_2-h_1)}\!\!\!\dd h\f{\e^{-2h}}{\Gamma(\beta-2h_1-2h/\ln(t))}
\xrightarrow{t\to\infty}\f{1}{2\Gamma(\beta-2h_1)}.
\end{align}
Taking the limit $\epsilon\to 0$ and averaging over $Z$ proves point i).

For ii) let us first study the behaviour of the power law asymptotic itself,
\begin{align}
&\E\lt[\f{t^{-2(h_1+H')}}{Z\Gamma(\beta-2(h_1+H'))}\f{1}{t^{-2h_1}\ln(t)^{
-\alpha}}|Z\rt]
\nonumber\\
&=\f{1}{Zt^{-2h_1}\ln(t)^{-\alpha}}\int_0^{1-h_1}\!\!\!\dd h\ p_{H'}(h)\f{t^{-2(h_1+h)}}{
\Gamma(\beta-2(h_1+h))}
\nonumber\\
&=\f{1}{Z}\int_0^{\ln(t)(1-h_1)}\!\!\!\dd h\ p_H\lt(\f{h}{\ln t}\rt)\ln(t)^{\alpha-1}\f{
\e^{-2h}}{\Gamma(\beta-2(h_1+h/\ln(t)))}
\nonumber\\
&\xrightarrow{t\to\infty}\f{1}{Z}\int_0^{\infty}\!\!\dd h\ h^{\alpha-1}\f{\e^{-2h}}{
\Gamma(\beta-2h_1)}=\f{\Gamma(\alpha)}{Z2^\alpha\Gamma(\beta-2h_1)}.
\end{align}
Now, because of asymptotic \Ref{eq:MLasympt2} for every $\epsilon>0$ there exist
a $T_H$ such that for $t>T_H$
\begin{equation}\label{eq:MLineq}
\f{1-\epsilon}{Z\Gamma(\beta-2H)}t^{-2H}<E_{2H,\beta}(-Zt^{2H})<\f{1+\epsilon}{
Z\Gamma(\beta-2H)}t^{-2H}.
\end{equation}
This inequality holds for any $H$ in a closed interval $[h_1,h_2]$ and fixed
$Z$. As the Mittag-Leffler function and the power function are continuous with
respect to $H$ in this range, we can find $T$ sufficiently large such that this
inequality will hold for $t>T$ and all $H\in [h_1,h_2]$ simultaneously. Otherwise
we could take $T_k\to\infty$ and corresponding $H_k\in[h_1,h_2]$ for which
it does not hold and obtain a contradiction with continuity of $H\mapsto
E_{2H,\beta}(-Zt^{2H})$ or asymptotic \Ref{eq:MLasympt2} at an accumulation
point of the sequence $H_k$.

We may divide \Ref{eq:MLineq} by 
\begin{equation}
l(t)\defeq\f{\Gamma(\alpha)}{Z2^\alpha \Gamma(\beta-2h_1)}t^{-2h_1}\ln(t)^{-\alpha}
\end{equation}
and consider some large $t>T$ in order to obtain 
\begin{equation}
\label{eq:powerLawCovA}
1-\epsilon\le\liminf_{t\to\infty}\f{\E[E_{2H,\beta}(-Zt^{2H})|Z]}{l(t)},\quad
\limsup_{t\to\infty}\f{\E[E_{2H,\beta}(-Zt^{2H})|Z]}{l(t)}\le 1+\epsilon.
\end{equation}
Taking the limit $\epsilon\to0$ and averaging over $Z$ yields the desired asymptotic
\begin{equation}
\E[E_{2H,\beta}(-Zt^{2H})]\sim \E[l(t)]=\f{\E[Z^{-1}]\Gamma(\alpha)}{2^\alpha
\Gamma(\beta-2h_1)}t^{-2h_1}\ln(t)^{-\alpha}.
\end{equation}
\end{proof}
Using the above result, the tails of the covariance for a power-law distributed $H', H=h_0+H'$ exhibit
power-log factor
\begin{equation}
\label{eq:powerLawAs}
r_{V_{H,Z}}(t)\sim\f{\E[Z^{-1}]\Gamma(\alpha)}{2^\alpha\Gamma(1-2h_0)}
t^{-2h_0}\ln(t)^{-\alpha}.
\end{equation}
As we can take $h_0$ arbitrarily close to $0^+$ in this model we can
obtain tails which are very close to a pure power-log shape. However we cannot reach the limit $h_0=0$, which would cause the covariance function to diverge. This corresponds to the fact that for small $h$ the solutions of the GLE become increasingly irregular and they do not converge to a well-defined process for $h\to 0^+$.

For the asymptotic of $\delta^{2}_{X_{H,Z}}(t)$ the identical argument
as for the covariance can be used, so in this model the msd of the form
\begin{equation}
\label{eq:msdAs}
\delta^{2}_{X_{H,Z}}(t)\sim\f{\E[Z^{-1}]\Gamma(\alpha)}{2^\alpha\Gamma(3-2h_0)}
t^{2-2h_0}\ln(t)^{-\alpha}
\end{equation}
is present for $0< h_0<1$ and $0<\alpha\le 1$. A numerical evaluation of
this behaviour is shown in Figure \ref{f:powerLawMSD} where we consider the
subdiffusive case $H=3/10+H'$, with $p_{H'}(h)=\alpha5^{-\alpha} h^{\alpha-1}$
and $\alpha=3/4$, $0<H'<1/5$. The factor $t^2$ in expression \Ref{eq:msdAs}
does not depend on the particular form of the dynamics, so we divided all shown
functions by this factor to highlight the influence of $H$. As we can see the
convergence to the asymptotic behaviour is much slower than in the previous
examples, which stems from the fact that the Mittag-Leffler function converges
slowly to the power law
\begin{equation}
E_{2H,3}(-Zt^{2H})=\f{t^{-2H}}{Z\Gamma(3-2 H)}+\mathcal O(t^{-4H}).
\end{equation}
The inclusion of the power-log law is significant, but may be difficult
to determine on the log-log scale. It is demonstrated in the two lower panels
in Fig. \ref{f:powerLawMSD}. The asymptotic msd is concave in log-log
scale, but the effect is not very prominent, and for the msd estimated from
Monte Carlo simulations can be detected only on very long time scales. The
difference from  a power law is more visible if the lines are shown without
the factor $t^{-2h_0}$ (bottom panel), but $h_0$ may not be easy to estimate from real data. This comparison shows that different possible forms of decay
can be easily mistaken, so one should exert caution when analysing data suspected
to stem from such systems.

\begin{figure}
\centering
\includegraphics[width=16cm]{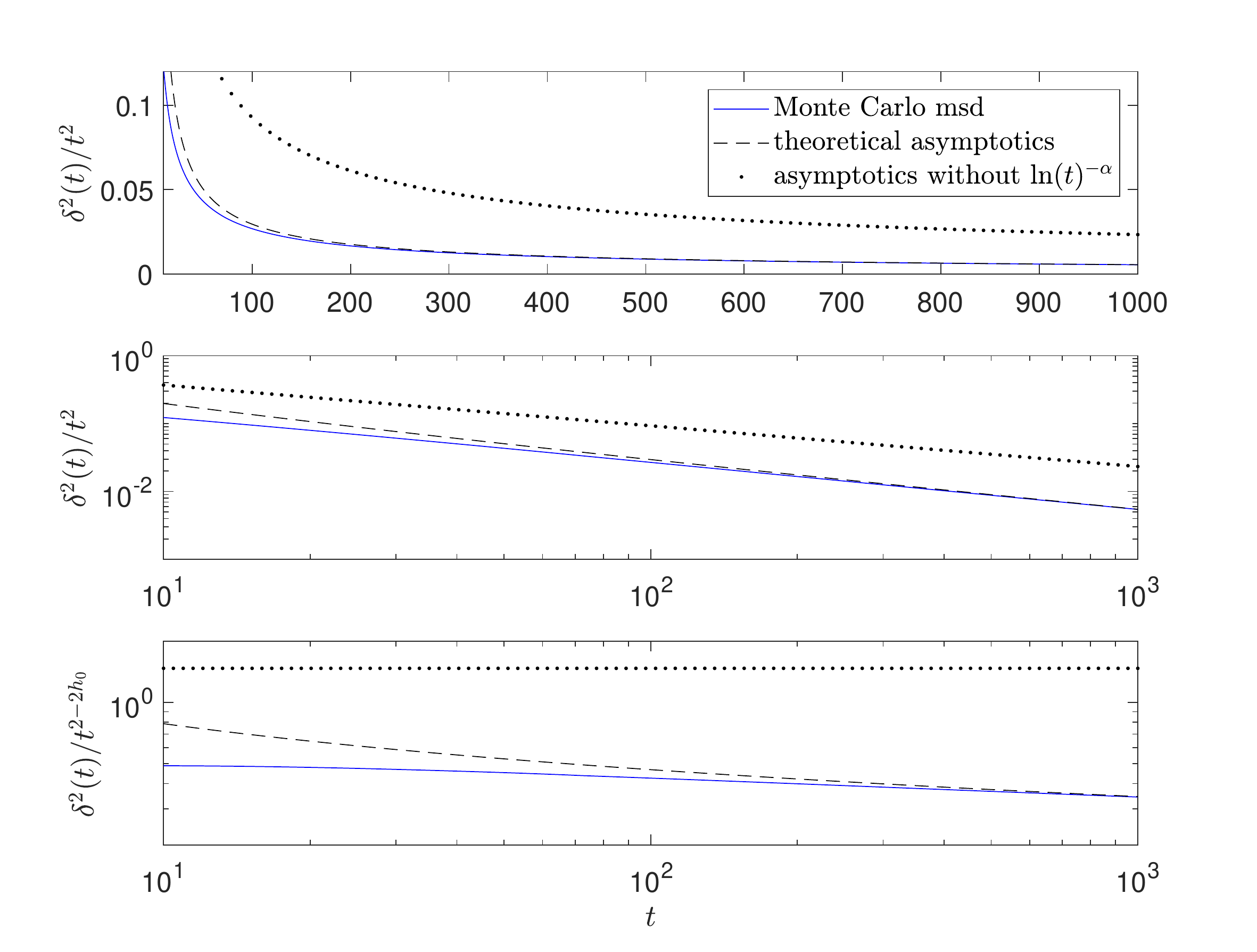}
\caption{Mean square displacement from Monte Carlo simulations (solid blue lines) for power law GLE
with fixed $Z=1$ and $0.3<H<0.5$ which has the power law form $p_{H}(0.3+h)\sim
\mathrm{const}\cdot h^{\alpha-1}$ with $\alpha=3/4$. The sample size was $10^3$.
The result is show on different scales, together with the asymptotic
\Ref{eq:powerLawAs} (dashed lines) and the same asymptotics without the factor
$\ln(t)^{-3/4}$ (dotted lines).}
\label{f:powerLawMSD}
\end{figure}

Let us compare all previously considered variants of the superstatistical GLE. We show the corresponding tails of the covariance in Table \ref{t:covBeh}. These models can describe normal or anomalous non-Gaussian diffusion. Even the ensembles of short memory processes can describe fractional, power law dynamics. In this regard the superstatistical GLE is similar to models in which the force is a superposition of other, more elementary processes \citep{eab,sandev}. On the other hand it differs considerably from this approach, using a local, not global fluctuation-dissipation relation. Modelling diffusion by a mixture of Gaussian variables instead of superposition of Gaussian variables leads to already discussed non-Gaussian behaviour, which has the same origin as the asymptotics of the covariance tails. Therefore, given experimental data, it is possible to compare the observed pdfs and tails of the covariance  function (as given in Table \ref{t:covBeh}) or the asymptotics of the msd, in this way effectively distinguishing between different variants of the model. A stronger statistical verification is possible when a number of long trajectories is available, which allows for the calculation of conditional statistics, such as $r_{V_C}(t|C)$, using time averages.

\begin{table}[ht]
\centerline{\begin{tabular}{|c|c|c|c|}
\cline{2-4}
\multicolumn{1}{c}{}  & \multicolumn{3}{|c|}{Shape parameter distribution} \\
\hline
$r_F$ & bounded on interval & power law at $0^+$ & power law + $c$, $c<1$\\
\hline
$\Lambda\delta(t)$   & $\asymp \e^{-\lambda t}\cdot t^{-1}$  \Ref{eq:cOUBoundedAs} &$\sim t^{-\alpha}$ \Ref{eq:cOUPowerLaw}&$\sim t^{-\alpha}\cdot\e^{-\lambda t}$  \Ref{eq:cOUmultipl}\\
$\e^{-\f{1+A'^2}{A'} t}$ &$\asymp \e^{-\lambda t}\cdot t^{-1}$ \Ref{eq:expDecBoundedAs}    &$\sim t^{-\alpha}$ \Ref{eq:expDecPowerLaw} &$\sim t^{-\alpha}\cdot\e^{-\lambda t}$ \Ref{eq:expDecmlpl} \\
$\e^{-2A t}$ &  $\lesssim \e^{-\lambda t}\cdot t^{-1}$ \Ref{eq:oscAsInterval} &$\sim\cos(t)\cdot t^{-\alpha}$ \Ref{eq:oscpl}& $\lesssim t^{-\alpha}\cdot\e^{-\lambda t}$ \Ref{eq:oscAsTrunc}\\
$t^{2H-2}$ &$\asymp t^{-\alpha}\cdot \ln(t)^{-1}$  \Ref{eq:plInterval}& ill-defined&$\sim t^{-\alpha}\cdot\ln(t)^{-\beta}$ \Ref{eq:powerLawAs}\\
\hline
\end{tabular}}
\caption{Different asymptotics of the covariance function $r_V$ for different
GLEs: memoryless, exponential kernel in decay regime (under convenient
parametrisation), exponential kernel in the oscillating exponential decay regime,
and power law. Different distributions of the shape parameters $\Lambda,A',A,H$
are considered. For brevity we omitted all scaling factors. We show the numbers of equations in the text, in which the full results are shown.}
\label{t:covBeh}
\end{table}

\chapter*{Summary}
\addcontentsline{toc}{chapter}{Summary}

In this work we reached into the rich world of diffusion phenomena in order to better understand the role and meaning of the models based on the classical and generalized Langevin equation. Our goal was to present these models from many different perspectives, each with its own advantages and limitations. Langevin equations can be interpreted as a reduced form of Hamilton's equations, which links macroscopic and microscopic level of physical description. Looking at them as  discrete-time linear filters stresses the properties of raw data and provides a link between physical models, statistical methods, and experimental observations. But, measurements may not be necessarily be interpreted as time dependent sequences. When analysing data in Fourier space, as a spectrum, the dynamics disappeared, and Langevin equations were viewed as multiplicative operators. We discussed the somewhat surprising relation of this fact to their ergodic properties. We also explored relations between time- and ensemble- averages from a different angle: using the superstatistical approach we rendered Gaussian and ergodic Langevin equations non-Gaussian, non-ergodic, and thus suitable as models of such types of systems.

Our methods were primarily based on the theory of Gaussian and second order processes, which was an intentional choice: we were interested in how models can be distinguished and verified. For this reason we concentrated our study on properties which are commonly estimated from the data: the covariance function, mean square displacement and power spectral density; similarly we advocated the use of the characteristic function as a tool for analysing non-ergodic and non-Gaussian properties. Through this approach we managed to illustrate even our more abstract results, such as the generalisation of the Maruyama's theorem, using Monte Carlo simulations and plots of measurable quantities.

As the final point, we want to stress that there are many ways in which the presented results could be improved and extended. From a mathematical perspective, the theory of the stationary solutions of the generalized Langevin equation is an open topic; in this work we were concerned only with the solutions in the mild sense and the cases where its existence was easily assured. The relations between Hamiltonian systems, their stationary solutions, and diffusion, are also intriguing; the model considered in this work was local and did not allow for many types of long-range correlations.

From a practical standpoint, the empirical validation is the most relevant goal. The proposed techniques and models were designed with experimental data in mind; therefore using them in applications is the most natural next step. It it certain that it would help to further refine the presented ideas and to find new unforeseen connections and interpretations.

\phantomsection\newpage\addcontentsline{toc}{chapter}{Bibliography}
\bibliography{mybib}

\begin{thebibliography}{127}
\providecommand{\natexlab}[1]{#1}
\providecommand{\url}[1]{\texttt{#1}}
\expandafter\ifx\csname urlstyle\endcsname\relax
  \providecommand{\doi}[1]{doi: #1}\else
  \providecommand{\doi}{doi: \begingroup \urlstyle{rm}\Url}\fi

\bibitem[Ariel and Vanden-Eijnden(2009)]{ariel}
G.~Ariel and E.~Vanden-Eijnden.
\newblock A strong limit theorem in the {Kac-Zwanzig} model.
\newblock \emph{Nonlinearity}, 22\penalty0 (1):\penalty0 145--162, 2009.
\newblock \doi{10.1088/0951-7715/22/1/008}.

\bibitem[Arsenault et~al.(2009)Arsenault, Sun, Baua, and Goldman]{molMot}
M.~Arsenault, Y.~Sun, H.~Baua, and Y.~Goldman.
\newblock Using electrical and optical tweezers to facilitate studies of
  molecular motors.
\newblock \emph{Phys. Chem. Chem. Phys.}, 11:\penalty0 4834--4839, 2009.
\newblock \doi{10.1039/b821861g}.

\bibitem[Banks and Fradin(2005)]{banks}
D.~S. Banks and C.~Fradin.
\newblock Anomalous diffusion of proteins due to molecular crowding.
\newblock \emph{Biophys. J.}, 89\penalty0 (5):\penalty0 2960 -- 2971, 2005.
\newblock \doi{10.1529/biophysj.104.051078}.

\bibitem[Barbuti et~al.(2011)Barbuti, Caravagna, Maggiolo-Schettini, and
  Milazzo]{barbuti}
R.~Barbuti, G.~Caravagna, A.~Maggiolo-Schettini, and P.~Milazzo.
\newblock \emph{Delay Stochastic Simulation of Biological Systems: A Purely
  Delayed Approach}, pages 61--84.
\newblock Springer-Verlag, 2011.
\newblock \doi{10.1007/978-3-642-19748-2_4}.

\bibitem[Beck and Cohen(2003)]{beck2}
C.~Beck and E.~G. Cohen.
\newblock Superstatistics.
\newblock \emph{Physica A}, 322:\penalty0 267 -- 275, 2003.
\newblock \doi{10.1016/S0378-4371(03)00019-0}.

\bibitem[Bhattacharya et~al.(2013)Bhattacharya, Sharma, Saurabh, De, Sain,
  Nandi, and Chowdhury]{bhattacharya}
S.~Bhattacharya, D.~Sharma, S.~Saurabh, S.~De, A.~Sain, A.~Nandi, and
  A.~Chowdhury.
\newblock Plasticization of poly(vinylpyrrolidone) thin films under ambient
  humidity: Insight from single-molecule tracer diffusion dynamics.
\newblock \emph{J. Phys. Chem. B}, 117\penalty0 (25):\penalty0 7771--7782,
  2013.
\newblock \doi{10.1021/jp401704e}.

\bibitem[Box and Jenkins(1994)]{BJ}
J.~Box and G.~Jenkins.
\newblock \emph{Time Series Analysis: Forecasting and Control}.
\newblock Prentice-Hall, 1994.

\bibitem[Br{\'e}maud(2014)]{procFourier}
P.~Br{\'e}maud.
\newblock \emph{Fourier Analysis and Stochastic Processes}.
\newblock Springer-Verlag, 2014.

\bibitem[Brockwell et~al.(2007)Brockwell, Davis, and Yang]{BrockwellCAR}
P.~Brockwell, R.~Davis, and Y.~Yang.
\newblock Continuous-time {Gaussian} autoregression.
\newblock \emph{Statist. Sinica}, 17:\penalty0 63--80, 2007.

\bibitem[Brockwell and Davis(2006)]{BD}
P.~J. Brockwell and R.~A. Davis.
\newblock \emph{Time Series: Theory and Methods}.
\newblock Springer-Verlag, 2006.

\bibitem[Budaev et~al.(2006)Budaev, Takamura, Ohno, and Masuzaki]{budaev}
V.~P. Budaev, S.~Takamura, N.~Ohno, and S.~Masuzaki.
\newblock Superdiffusion and multifractal statistics of edge plasma turbulence
  in fusion devices.
\newblock \emph{Nucl. Fusion}, 46\penalty0 (4):\penalty0 S181, 2006.
\newblock \doi{10.1088/0029-5515/46/4/S10}.

\bibitem[Burnecki and Weron(2014)]{ARFIMAalg}
K.~Burnecki and A.~Weron.
\newblock Algorithms for testing of fractional dynamics: a practical guide to
  {ARFIMA} modelling.
\newblock \emph{J. Stat. Mech. Theory Exp}, 2014\penalty0 (10):\penalty0
  P10036, 2014.
\newblock \doi{10.1088/1742-5468/2014/10/P10036}.

\bibitem[Burnecki et~al.(2012)Burnecki, Sikora, and Weron]{ARFIMunif}
K.~Burnecki, G.~Sikora, and A.~Weron.
\newblock Fractional process as a unified model for subdiffusive dynamics in
  experimental data.
\newblock \emph{Phys. Rev. E}, 86:\penalty0 041912, 2012.
\newblock \doi{10.1103/PhysRevE.86.041912}.

\bibitem[Caspi et~al.(2000)Caspi, Granek, and Elbaum]{caspi}
A.~Caspi, R.~Granek, and M.~Elbaum.
\newblock Enhanced diffusion in active intracellular transport.
\newblock \emph{Phys. Rev. Lett.}, 85:\penalty0 5655--5658, 2000.
\newblock \doi{10.1103/PhysRevLett.85.5655}.

\bibitem[Champeney(1987)]{champeney}
D.~C. Champeney.
\newblock \emph{A handbook of Fourier theorems}.
\newblock Cambridge University Press, 1987.

\bibitem[Chandler(1987)]{chandler}
D.~Chandler.
\newblock \emph{Introduction to Modern Statistical Mechanics}.
\newblock Oxford University Press, 1987.

\bibitem[Cherstvy et~al.(2013)Cherstvy, Chechkin, and Metzler]{ergPar}
A.~G. Cherstvy, A.~V. Chechkin, and R.~Metzler.
\newblock Anomalous diffusion and ergodicity breaking in heterogeneous
  diffusion processes.
\newblock \emph{New J. Phys.}, 15\penalty0 (8):\penalty0 083039, 2013.
\newblock \doi{10.1088/1367-2630/15/8/083039}.

\bibitem[Cornfel et~al.(1982)Cornfel, Fomin, and Sinai]{ergTh}
I.~Cornfel, S.~Fomin, and Y.~Sinai.
\newblock \emph{Ergodic Theory}.
\newblock Springer-Verlag, 1982.

\bibitem[Courtault et~al.(2000)Courtault, Kabanov, Bru, Cr{\'e}pel, Lebon, and
  Le~Marchand]{courtault}
J.-M. Courtault, Y.~Kabanov, B.~Bru, P.~Cr{\'e}pel, I.~Lebon, and
  A.~Le~Marchand.
\newblock {L}ouis {B}achelier on the centenary of {T}h{\'e}orie de la
  sp{\'e}culation.
\newblock \emph{Math. Fin.}, 10\penalty0 (3):\penalty0 339--353, 2000.
\newblock \doi{10.1111/1467-9965.00098}.

\bibitem[Da~Prato and Zabczyk(2014)]{SDEinfDim}
G.~Da~Prato and J.~Zabczyk.
\newblock \emph{Stochastic Equations in Infinite Dimensions}.
\newblock Cambridge University Press, 2014.

\bibitem[{\relax DLMF}()]{DLMF}
{\relax DLMF}.
\newblock {\it NIST Digital Library of Mathematical Functions}.
\newblock Release 1.0.15 of 2017-06-01.
\newblock URL \url{http://dlmf.nist.gov/}.
\newblock F.~W.~J. Olver, A.~B. {Olde Daalhuis}, D.~W. Lozier, B.~I. Schneider,
  R.~F. Boisvert, C.~W. Clark, B.~R. Miller and B.~V. Saunders, eds.

\bibitem[Doob(1942)]{DoobOryg}
J.~L. Doob.
\newblock The {Brownian} movement and stochastic equations.
\newblock \emph{Ann. Math.}, 43\penalty0 (2):\penalty0 351--369, 1942.
\newblock \doi{10.2307/1968873}.

\bibitem[Doob(1990)]{doob}
J.~L. Doob.
\newblock \emph{Stochastic Processes}.
\newblock Wiley-Interscience, 1990.

\bibitem[Drobczy\'{n}ski and \'{S}l\k{e}zak(2015)]{drobczynskiSlezak}
S.~Drobczy\'{n}ski and J.~\'{S}l\k{e}zak.
\newblock Time-series methods in analysis of the optical tweezers recordings.
\newblock \emph{Appl. Opt.}, 54\penalty0 (23):\penalty0 7106--7114, 2015.
\newblock \doi{10.1364/AO.54.007106}.

\bibitem[Dym and McKean(1976)]{dym}
H.~Dym and H.~P. McKean.
\newblock \emph{Gaussian Processes, Function Theory and the Inverse Spectral
  Problem}.
\newblock Academic Press, 1976.

\bibitem[Eab and Lim(2011)]{eab}
C.~Eab and S.~Lim.
\newblock Fractional {Langevin} equation of distributed order.
\newblock \emph{Phys. Rev. E}, 83\penalty0 (031136), 2011.
\newblock \doi{10.1103/PhysRevE.83.031136}.

\bibitem[Einstein(1905)]{einstein}
A.~Einstein.
\newblock {{\"U}ber die von der molekularkinetischen Theorie der W{\"a}rme
  geforderte Bewegung von in ruhenden Fl{\"u}ssigkeiten suspendierten
  Teilchen}.
\newblock \emph{Ann. der Physik}, 322\penalty0 (8):\penalty0 549--560, 1905.
\newblock \doi{10.1002/andp.19053220806}.

\bibitem[Embrechts and Mejima(2002)]{embrechts}
P.~Embrechts and M.~Mejima.
\newblock \emph{Selfsimilar Processes}.
\newblock Princeton University Press, 2002.

\bibitem[Enders(2009)]{enders}
W.~Enders.
\newblock \emph{Applied Econometric Time Series}.
\newblock Wiley, 2009.

\bibitem[Feller(1968)]{feller}
W.~Feller.
\newblock \emph{An Introduction to Probability Theory and Its Applications}.
\newblock Wiley, 3rd edition, 1968.

\bibitem[Fick(1995)]{fick}
A.~Fick.
\newblock On liquid diffusion.
\newblock \emph{J. Membrane Sci.}, 100\penalty0 (1):\penalty0 33 -- 38, 1995.
\newblock \doi{10.1016/0376-7388(94)00230-V}.
\newblock The early history of membrane science selected papers celebrating
  vol. 100.

\bibitem[Ford et~al.(1965)Ford, Kac, and Mazur]{kacGLE}
G.~Ford, M.~Kac, and P.~Mazur.
\newblock Statistical mechanics of assemblies of coupled oscillators.
\newblock \emph{J. Math. Phys.}, 6\penalty0 (4):\penalty0 504--515, 1965.
\newblock \doi{10.1063/1.1704304}.

\bibitem[Fox(1978)]{fox}
R.~F. Fox.
\newblock Gaussian stochastic processes in physics.
\newblock \emph{Phys. Rep.}, 48\penalty0 (3):\penalty0 179--283, 1978.
\newblock \doi{10.1016/0370-1573(78)90145-X}.

\bibitem[Franke(1985)]{frankeEntr}
J.~Franke.
\newblock {ARMA} processes have maximal entropy among time series with
  predescribed autocovariance and impulse response.
\newblock \emph{Adv. in Appl. Probab.}, 17:\penalty0 810--840, 1985.
\newblock \doi{10.2307/1427089}.

\bibitem[Golding and Cox(2006)]{goldingCox}
I.~Golding and E.~C. Cox.
\newblock Physical nature of bacterial cytoplasm.
\newblock \emph{Phys. Rev. Lett.}, 96:\penalty0 098102, 2006.
\newblock \doi{10.1103/PhysRevLett.96.098102}.

\bibitem[Gorenflo et~al.(2014)Gorenflo, Kilbas, Mainardi, and
  Rogosin]{gorenflo}
R.~Gorenflo, A.~Kilbas, F.~Mainardi, and S.~Rogosin.
\newblock \emph{Mittag-Leffler Functions, Related Topics and Applications}.
\newblock Springer-Verlag, 2014.

\bibitem[Goychuk et~al.(2014)Goychuk, Kharchenko, and
  Metzler]{goychukmotorpccp}
I.~Goychuk, V.~O. Kharchenko, and R.~Metzler.
\newblock Molecular motors pulling cargos in the viscoelastic cytosol: how
  power strokes beat subdiffusion.
\newblock \emph{Phys. Chem. Chem. Phys.}, 16:\penalty0 16524--16535, 2014.
\newblock \doi{10.1039/C4CP01234H}.

\bibitem[Grenander(1950)]{grenander}
U.~Grenander.
\newblock Stochastic processes and statistical inference.
\newblock \emph{Ark. Mat.}, 1:\penalty0 195--277, 1950.
\newblock \doi{10.1007/BF02590638}.

\bibitem[Haubold et~al.(2011)Haubold, Mathai, and Saxena]{haubold}
H.~Haubold, A.~Mathai, and R.~Saxena.
\newblock Mittag-leffler functions and their applications.
\newblock \emph{J. Appl. Math.}, 2011, 2011.
\newblock \doi{10.1155/2011/298628}.

\bibitem[Hilfer(1999)]{hilfer}
R.~Hilfer, editor.
\newblock \emph{Applications of Fractional Calculus in Physics}.
\newblock World Scientific, 1999.

\bibitem[H{\"o}fling and Franosch(2013)]{anTrans}
F.~H{\"o}fling and T.~Franosch.
\newblock Anomalous transport in the crowded world of biological cells.
\newblock \emph{Rep. Prog. Phys.}, 76\penalty0 (4):\penalty0 046602, 2013.
\newblock \doi{10.1088/0034-4885/76/4/046602}.

\bibitem[Hurst(1951)]{hurst}
H.~E. Hurst.
\newblock Long-term storage capacity of reservoirs.
\newblock \emph{Trans. Amer. Soc. Civ. Eng.}, 116:\penalty0 770--808, 1951.

\bibitem[It{\=o}(1944)]{ito44}
K.~It{\=o}.
\newblock Stochastic integral.
\newblock \emph{Proc. Imp. Acad.}, 20\penalty0 (8):\penalty0 519--524, 1944.
\newblock \doi{10.3792/pia/1195572786}.

\bibitem[It{\=o}(1946)]{ito46}
K.~It{\=o}.
\newblock On a stochastic integral equation.
\newblock \emph{Proc. Japan Acad.}, 22\penalty0 (2):\penalty0 32--35, 1946.
\newblock \doi{10.3792/pja/1195572371}.

\bibitem[Janson(1997)]{gaussHS}
S.~Janson.
\newblock \emph{Gaussian Hilbert Spaces}.
\newblock Cambridge University Press, 1997.

\bibitem[Javanainen et~al.(2017)Javanainen, Martinez-Seara, Metzler, and
  Vattulainen]{matti17}
M.~Javanainen, H.~Martinez-Seara, R.~Metzler, and I.~Vattulainen.
\newblock Diffusion of integral membrane proteins in protein-rich membranes.
\newblock \emph{J. Phys. Chem. Lett}, 8\penalty0 (17):\penalty0 4308--4313,
  2017.
\newblock \doi{10.1021/acs.jpclett.7b01758}.

\bibitem[Jeon et~al.(2011)Jeon, Tejedor, Burov, Barkai, Selhuber-Unkel,
  Berg-S\o{}rensen, Oddershede, and Metzler]{weakErgJeon}
J.-H. Jeon, V.~Tejedor, S.~Burov, E.~Barkai, C.~Selhuber-Unkel,
  K.~Berg-S\o{}rensen, L.~Oddershede, and R.~Metzler.
\newblock In vivo anomalous diffusion and weak ergodicity breaking of lipid
  granules.
\newblock \emph{Phys. Rev. Lett.}, 106:\penalty0 048103, 2011.
\newblock \doi{10.1103/PhysRevLett.106.048103}.

\bibitem[Jeon et~al.(2013)Jeon, Leijnse, Oddershede, and Metzler]{lene1}
J.-H. Jeon, N.~Leijnse, L.~B. Oddershede, and R.~Metzler.
\newblock Anomalous diffusion and power-law relaxation of the time averaged
  mean squared displacement in worm-like micellar solutions.
\newblock \emph{New J. Phys.}, 15\penalty0 (4):\penalty0 045011, 2013.
\newblock \doi{10.1088/1367-2630/15/4/045011}.

\bibitem[Jeon et~al.(2016)Jeon, Javanainen, Martinez-Seara, Metzler, and
  Vattulainen]{jeon_prx}
J.-H. Jeon, M.~Javanainen, H.~Martinez-Seara, R.~Metzler, and I.~Vattulainen.
\newblock Protein crowding in lipid bilayers gives rise to non-{Gaussian}
  anomalous lateral diffusion of phospholipids and proteins.
\newblock \emph{Phys. Rev. X}, 6:\penalty0 021006, 2016.
\newblock \doi{10.1103/PhysRevX.6.021006}.

\bibitem[Johnson(1928)]{johnson}
J.~B. Johnson.
\newblock Thermal agitation of electricity in conductors.
\newblock \emph{Phys. Rev.}, 32:\penalty0 97--109, 1928.
\newblock \doi{10.1103/PhysRev.32.97}.

\bibitem[Kakalios et~al.(1987)Kakalios, Street, and Jackson]{kakalios}
J.~Kakalios, R.~A. Street, and W.~B. Jackson.
\newblock Stretched-exponential relaxation arising from dispersive diffusion of
  hydrogen in amorphous silicon.
\newblock \emph{Phys. Rev. Lett.}, 59:\penalty0 1037--1040, 1987.
\newblock \doi{10.1103/PhysRevLett.59.1037}.

\bibitem[Karczewska(2007)]{karczewska}
A.~Karczewska.
\newblock \emph{Convolution type stochastic Volterra equations}.
\newblock Juliusz Schauder Center for Nonlinear Studies, 2007.

\bibitem[Kimme(1965)]{kimme}
E.~Kimme.
\newblock Distributions of phases of sinusoidal components of ergodic
  trigonometric series.
\newblock \emph{SIAM Review}, 7\penalty0 (1):\penalty0 88--99, 1965.

\bibitem[Kneller et~al.(2011)Kneller, Baczynski, and
  Pasenkiewicz-Gierula]{kneller}
G.~R. Kneller, K.~Baczynski, and M.~Pasenkiewicz-Gierula.
\newblock Consistent picture of lateral subdiffusion in lipid bilayers:
  molecular dynamics simulation and exact results.
\newblock \emph{J Chem Phys.}, 135\penalty0 (14):\penalty0 141105, 2011.
\newblock \doi{doi: 10.1063/1.3651800}.

\bibitem[Kohlrausch(1854)]{kohlrausch}
R.~Kohlrausch.
\newblock {Theorie des elektrischen R{\" u}ckstandes in der Leidener Flasche}.
\newblock \emph{Ann. der Physik}, 167\penalty0 (2):\penalty0 179--214, 1854.
\newblock \doi{10.1002/andp.18541670203}.

\bibitem[Kou(2008)]{kou}
S.~C. Kou.
\newblock Stochastic modelling in nanoscale physics: Subdiffusion within
  proteins.
\newblock \emph{Ann. Appl. Stat.}, 2\penalty0 (2):\penalty0 501--535, 2008.
\newblock \doi{10.1214/07-AOAS149}.

\bibitem[Kubo(1957)]{kuboOld}
R.~Kubo.
\newblock Statistical-mechanical theory of irreversible processes. {I. General}
  theory and simple applications to magnetic and conduction problems.
\newblock \emph{J. Phys. Soc. Jpn.}, 12:\penalty0 570--586, 1957.
\newblock \doi{10.1143/JPSJ.12.570}.

\bibitem[Kubo(1966)]{kubo}
R.~Kubo.
\newblock The fluctuation-dissipation theorem.
\newblock \emph{Rep. Prog. Phys.}, 29\penalty0 (1):\penalty0 255--284, 1966.
\newblock \doi{10.1088/0034-4885/29/1/306}.

\bibitem[Lampo et~al.(2017)Lampo, Stylianido, Backlund, Wiggins, and
  Spakowitz]{lampo}
T.~J. Lampo, S.~Stylianido, M.~P. Backlund, P.~A. Wiggins, and A.~J. Spakowitz.
\newblock Cytoplasmic {RNA}-protein particles exhibit non-{Gaussian}
  subdiffusive behavior.
\newblock \emph{Biophys. J.}, 112:\penalty0 532--542, 2017.
\newblock \doi{10.1016/j.bpj.2016.11.3208}.

\bibitem[Lemons and Gythiel(1997)]{langevinEng}
D.~Lemons and A.~Gythiel.
\newblock {Paul {Langevin's} 1908 paper ``On the Theory of {Brownian} Motion''
  [``Sur la th{\'e}orie du mouvement brownien,'' C. R. Acad. Sci. (Paris) 146,
  530--533 (1908)]}.
\newblock \emph{Am. J. Phys.}, 65\penalty0 (11):\penalty0 1079--1081, 1997.
\newblock \doi{10.1119/1.18725}.

\bibitem[Leonard(1916)]{lucretius}
W.~E. Leonard.
\newblock \emph{Of the nature of things; a metrical translation}.
\newblock Dutton, 1916.

\bibitem[L{\' e}vy(1948)]{levy}
P.~L{\' e}vy.
\newblock \emph{Processus stochastiques et mouvement brownien}.
\newblock Jacques Gabay, 1948.

\bibitem[Li and Raizen(2013)]{BmotShortT}
T.~Li and M.~G. Raizen.
\newblock Brownian motion at short time scales.
\newblock \emph{Ann. der Physik}, 525\penalty0 (4):\penalty0 281--295, 2013.
\newblock \doi{10.1002/andp.201200232}.

\bibitem[Lindsay(1995)]{mixMod}
B.~G. Lindsay.
\newblock Mixture models: Theory, geometry and applications.
\newblock \emph{NSF-CBMS Reg. Conf. Ser, in Prob. Stat.}, 5:\penalty0 i--163,
  1995.

\bibitem[Lutz(2001)]{lutz}
E.~Lutz.
\newblock Fractional {Langevin} equation.
\newblock \emph{Phys. Rev. E}, 64:\penalty0 051106, 2001.
\newblock \doi{10.1103/PhysRevE.64.051106}.

\bibitem[Lyons(1995)]{70y}
R.~Lyons.
\newblock Seventy years of {Rajchmann} measures.
\newblock \emph{J. Fourier Anal. Appl. Special Issue}, pages 363--377, 1995.

\bibitem[Mandelbrot(2002)]{mandelbrot}
B.~Mandelbrot.
\newblock \emph{Gaussian Self-Affinity and Fractals}.
\newblock Springer-Verlag, 2002.

\bibitem[Maniglia and Rhandi(2004)]{gaussHilb}
S.~Maniglia and A.~Rhandi.
\newblock \emph{Gaussian Measures on Separable Hilbert Spaces and
  Applications}.
\newblock Quaderni di Matematica, 2004.

\bibitem[Maruyama(1970)]{maruyama}
G.~Maruyama.
\newblock Infinitely divisible processes.
\newblock \emph{Theory Probab. Appl.}, 15:\penalty0 1--22, 1970.

\bibitem[Metzler(2017)]{ralf_lampo}
R.~Metzler.
\newblock Gaussianity fair: The riddle of anomalous yet non-{Gaussian}
  diffusion.
\newblock \emph{Biophys. J.}, 112:\penalty0 413--415, 2017.
\newblock \doi{10.1016/j.bpj.2016.12.019}.

\bibitem[Metzler and Klafter(2000)]{rwGuide}
R.~Metzler and J.~Klafter.
\newblock The random walk's guide to anomalous diffusion: a fractional dynamics
  approach.
\newblock \emph{Phys. Rep.}, 339\penalty0 (1):\penalty0 1 -- 77, 2000.
\newblock ISSN 0370-1573.
\newblock \doi{10.1016/S0370-1573(00)00070-3}.

\bibitem[Metzler et~al.(2014)Metzler, Jeon, Cherstvy, and Barkai]{weakErgBreak}
R.~Metzler, J.-H. Jeon, A.~G. Cherstvy, and E.~Barkai.
\newblock Anomalous diffusion models and their properties: non-stationarity,
  non-ergodicity, and ageing at the centenary of single particle tracking.
\newblock \emph{Phys. Chem. Chem. Phys.}, 16\penalty0 (44):\penalty0
  24128--24164, 2014.
\newblock \doi{10.1039/C4CP03465A}.

\bibitem[Meyer et~al.(2006)Meyer, Marshall, Bush, and Furst]{OTcoloid}
A.~Meyer, A.~Marshall, B.~Bush, and E.~Furst.
\newblock Laser tweezer microrheology of a colloidal suspension.
\newblock \emph{J. Rheol}, 50\penalty0 (1):\penalty0 77--92, 2006.
\newblock \doi{10.1122/1.2139098}.

\bibitem[Mills(1990)]{mills}
C.~Mills.
\newblock \emph{Time Series Techniques for Economists}.
\newblock Cambridge University Press, 1990.

\bibitem[Mori(1965)]{mori}
H.~Mori.
\newblock Transport, collective motion and {Brownian} motion.
\newblock \emph{Progr. Theoret. Phys.}, 33\penalty0 (3):\penalty0 423--455,
  1965.
\newblock \doi{10.1143/PTP.33.423}.

\bibitem[Muralidhar et~al.(1990)Muralidhar, Ramkrishna, Nakanishi, and
  Jacobs]{muralidhar}
R.~Muralidhar, D.~Ramkrishna, H.~Nakanishi, and D.~Jacobs.
\newblock Anomalous diffusion: A dynamic perspective.
\newblock \emph{Physica A}, 167\penalty0 (2):\penalty0 539--559, 1990.
\newblock \doi{10.1016/0378-4371(90)90132-C}.

\bibitem[Neuman and Block(2004)]{OTintr}
K.~Neuman and S.~Block.
\newblock Optical trapping.
\newblock \emph{Rev. Sci. Instrum}, 75\penalty0 (9):\penalty0 2787--2809, 2004.
\newblock \doi{10.1063/1.1785844}.

\bibitem[Nyquist(1928)]{nyquist}
H.~Nyquist.
\newblock Thermal agitation of electric charge in conductors.
\newblock \emph{Phys. Rev.}, 32:\penalty0 110--113, 1928.
\newblock \doi{10.1103/PhysRev.32.110}.

\bibitem[Onsager(1931{\natexlab{a}})]{onsager1}
L.~Onsager.
\newblock Reciprocal relations in irreversible processes. i.
\newblock \emph{Phys. Rev.}, 37:\penalty0 405--426, 1931{\natexlab{a}}.
\newblock \doi{10.1103/PhysRev.37.405}.

\bibitem[Onsager(1931{\natexlab{b}})]{onsager2}
L.~Onsager.
\newblock Reciprocal relations in irreversible processes. ii.
\newblock \emph{Phys. Rev.}, 38:\penalty0 2265--2279, 1931{\natexlab{b}}.
\newblock \doi{10.1103/PhysRev.38.2265}.

\bibitem[Panja(2010)]{panja}
D.~Panja.
\newblock Generalized {Langevin} equation formulation for anomalous polymer
  dynamics.
\newblock \emph{J. Stat. Mech.}, 2010\penalty0 (02):\penalty0 L02001, 2010.
\newblock \doi{10.1088/1742-5468/2010/02/L02001}.

\bibitem[Pavliotis(2014)]{pavliotis}
G.~Pavliotis.
\newblock \emph{Stochastic Processes and Applications}.
\newblock Springer-Verlag, 2014.

\bibitem[Perri et~al.(2015)Perri, Zimbardo1, Effenberger, and
  Fichtner]{perriSun}
S.~Perri, G.~Zimbardo1, F.~Effenberger, and H.~Fichtner.
\newblock Parameter estimation of superdiffusive motion of energetic particles
  upstream of heliospheric shocks.
\newblock \emph{A{\&} A}, 578, 2015.
\newblock \doi{10.1051/0004-6361/201425295}.

\bibitem[Perrin(1908)]{perrin1}
J.~Perrin.
\newblock L'agitation mol{\'e}culaire et le mouvement brownien.
\newblock \emph{Compt. Rend. (Paris)}, 146:\penalty0 967, 1908.

\bibitem[Perrin(1913)]{perrin2}
J.~Perrin.
\newblock \emph{Les atomes}.
\newblock Librairie F{\' e}lix Alcan, 1913.

\bibitem[Pfaff(2008)]{pfaff}
B.~Pfaff.
\newblock {VAR, SVAR and SVEC} models: Implementation within r package vars.
\newblock \emph{J. Stat. Softw.}, 27\penalty0 (4), 2008.
\newblock \doi{10.18637/jss.v027.i04}.

\bibitem[Phillips(1957)]{phillips}
A.~W. Phillips.
\newblock The estimation of parameters in systems of stochastic differential
  equations.
\newblock \emph{Biometrika}, 46\penalty0 (1/2):\penalty0 67--76, 1957.
\newblock \doi{10.2307/2332809}.

\bibitem[Pipiras and Taqqu(2000)]{intFBM}
V.~Pipiras and M.~Taqqu.
\newblock Integration questions related to fractional {Brownian} motion.
\newblock \emph{Prob. Th. Rel. Fields}, 118:\penalty0 1121--291, 2000.
\newblock \doi{10.1007/s440-000-8016-7}.

\bibitem[Porr{\`a} et~al.(1996)Porr{\`a}, Wang, and Masoliver]{porra96}
J.~M. Porr{\`a}, K.~G. Wang, and J.~Masoliver.
\newblock Generalized langevin equations: Anomalous diffusion and probability
  distributions.
\newblock \emph{Phys. Rev. E}, 53:\penalty0 5872--5881, 1996.
\newblock \doi{10.1103/PhysRevE.53.5872}.

\bibitem[Postnikov(1980)]{taub}
A.~Postnikov.
\newblock \emph{Tauberian Theory and Its Applications}.
\newblock American Mathematical Society, 1980.

\bibitem[Reverey et~al.(2015)Reverey, Bao, Jeon, Leippe, Metzler, and
  Selhuber-Unkel]{reverey}
J.~F. Reverey, H.~Bao, J.-H. Jeon, M.~Leippe, R.~Metzler, and
  C.~Selhuber-Unkel.
\newblock Superdiffusion dominates intracellular particle motion in the
  supercrowded cytoplasm of pathogenic acanthamoeba castellanii.
\newblock \emph{Sci. Rep.}, 5\penalty0 (11690), 2015.
\newblock \doi{10.1038/srep11690}.

\bibitem[Rey-Bellet(2006)]{rey-bellet}
L.~Rey-Bellet.
\newblock Open classical systems.
\newblock In \emph{Quantum Dissipative Systems II}, pages 41--78.
  Springer-Verlag, 2006.

\bibitem[Richardson(1926)]{richardson}
L.~F. Richardson.
\newblock Atmospheric diffusion shown on a distance-neighbour graph.
\newblock \emph{Proc. Roy. Soc.}, 110\penalty0 (756):\penalty0 709--737, 1926.
\newblock \doi{10.1098/rspa.1926.0043}.

\bibitem[Risken(1989)]{risken}
H.~Risken.
\newblock \emph{The Fokker-Planck Equation: Methods of Solution and
  Applications}.
\newblock Springer-Verlag, 2nd edition, 1989.

\bibitem[Rouse~Jr.(1953)]{rouse}
P.~E. Rouse~Jr.
\newblock A theory of the linear viscoelastic properties of dilute solutions of
  coiling polymers.
\newblock \emph{J. Chem. Phys.}, 21\penalty0 (7):\penalty0 1272--1280, 1953.
\newblock \doi{10.1063/1.1699180}.

\bibitem[Rozanov(1967)]{rozanov}
Y.~Rozanov.
\newblock \emph{Stationary Random Processes}.
\newblock Holden-Day, 1967.

\bibitem[Salem(1963)]{salem}
R.~Salem.
\newblock \emph{Algebraic numbers and Fourier analysis}.
\newblock Heath, 1963.

\bibitem[Samko et~al.(1993)Samko, Kilbas, and Marichev]{samko}
S.~G. Samko, A.~A. Kilbas, and O.~I. Marichev.
\newblock \emph{Fractional Integrals and Derivatives: Theory and Applications}.
\newblock CRC Press, 1993.

\bibitem[Samorodnitsky and Taqqu(1994)]{taqqu}
G.~Samorodnitsky and M.~S. Taqqu.
\newblock \emph{Stable Non-Gaussian Random Processes}.
\newblock Chapman \& Hall, 1994.

\bibitem[Sandev and Tomovski(2014)]{sandev}
T.~Sandev and {\v{Z}}.~Tomovski.
\newblock Langevin equation for a free particle driven by power law type of
  noises.
\newblock \emph{Phys. Lett. A}, 378:\penalty0 1--9, 2014.
\newblock \doi{10.1016/j.physleta.2013.10.038}.

\bibitem[Sat{\= o}(1969)]{satoGauss}
H.~Sat{\= o}.
\newblock Gaussian measure on a {Banach} space and abstract {Wiener} measure.
\newblock \emph{Nagoya Math. J.}, 36:\penalty0 65--81, 1969.

\bibitem[Seisenberger et~al.(2001)Seisenberger, Ried, Endre{\ss}, B{\"u}ning,
  Hallek, and Br{\"a}uchle]{seisenberger}
G.~Seisenberger, M.~U. Ried, T.~Endre{\ss}, H.~B{\"u}ning, M.~Hallek, and
  C.~Br{\"a}uchle.
\newblock Real-time single-molecule imaging of the infection pathway of an
  adeno-associated virus.
\newblock \emph{Science}, 294\penalty0 (5548):\penalty0 1929--1932, 2001.
\newblock \doi{10.1126/science.1064103}.

\bibitem[Silverman(1986)]{denEst}
B.~W. Silverman.
\newblock \emph{Density Estimation for Statistics and Data Analysis}.
\newblock Chapman and Hall, 1986.

\bibitem[\'{S}l\k{e}zak(2017)]{gaussErg}
J.~\'{S}l\k{e}zak.
\newblock Asymptotic behaviour of time averages for non-ergodic gaussian
  processes.
\newblock \emph{Ann. Phys.}, 383:\penalty0 285 -- 311, 2017.
\newblock \doi{10.1016/j.aop.2017.05.015}.

\bibitem[{\' S}l\k{e}zak and Weron(2015)]{slezakWeron}
J.~{\' S}l\k{e}zak and A.~Weron.
\newblock From physical linear systems to discrete-time series. {A} guide for
  analysis of the sampled experimental data.
\newblock \emph{Phys. Rev. E}, 91:\penalty0 053302, 2015.
\newblock \doi{10.1103/PhysRevE.91.053302}.

\bibitem[\'{S}l\k{e}zak et~al.(2014)\'{S}l\k{e}zak, Drobczy\'{n}ski, Weron, and
  Masajada]{slezakDrobczynski}
J.~\'{S}l\k{e}zak, S.~Drobczy\'{n}ski, K.~Weron, and J.~Masajada.
\newblock Moving average process underlying the holographic-optical-tweezers
  experiments.
\newblock \emph{Appl. Opt.}, 53\penalty0 (10):\penalty0 B254--B258, 2014.
\newblock \doi{10.1364/AO.53.00B254}.

\bibitem[\'{S}l\k{e}zak et~al.(2017)\'{S}l\k{e}zak, Metzler, and
  Magdziarz]{superstat}
J.~\'{S}l\k{e}zak, R.~Metzler, and R.~Magdziarz.
\newblock Superstatistical generalised langevin equation: non-gaussian
  viscoelastic anomalous diffusion.
\newblock 2017.
\newblock Accepted for publication in New J. Phys.

\bibitem[Smith et~al.(1996)Smith, Cui, and Bustamante]{poly2}
S.~Smith, Y.~Cui, and C.~Bustamante.
\newblock Overstretching {B-DNA}: the elastic response of individual
  double-stranded and single-stranded dna molecules.
\newblock \emph{Science}, 271\penalty0 (5250):\penalty0 795--799, 1996.
\newblock \doi{10.1126/science.271.5250.795}.

\bibitem[Smoluchowski(1906)]{smol}
M.~Smoluchowski.
\newblock {Zur kinetischen Theorie der Brownschen Molekularbewegung und der
  Suspensionen}.
\newblock \emph{Ann. der Physik}, 326\penalty0 (14):\penalty0 756--780, 1906.
\newblock \doi{10.1002/andp.19063261405}.

\bibitem[Sokolov and Klafter(2005)]{sokolovAnDiff}
I.~M. Sokolov and J.~Klafter.
\newblock From diffusion to anomalous diffusion: A century after {Einstein’s
  Brownian} motion.
\newblock \emph{Chaos}, 15\penalty0 (2):\penalty0 026103, 2005.
\newblock \doi{10.1063/1.1860472}.

\bibitem[Szymanski and Weiss(2009)]{szymanski}
J.~Szymanski and M.~Weiss.
\newblock Elucidating the origin of anomalous diffusion in crowded fluids.
\newblock \emph{Phys. Rev. Lett.}, 103:\penalty0 038102, 2009.
\newblock \doi{10.1103/PhysRevLett.103.038102}.

\bibitem[Tabei et~al.(2013)Tabei, Burov, Kim, Kuznetsov, Huynh, Jureller,
  Philipson, Dinner, and Scherer]{tabei}
S.~M.~A. Tabei, S.~Burov, H.~Y. Kim, A.~Kuznetsov, T.~Huynh, J.~Jureller,
  L.~Philipson, A.~Dinner, and N.~Scherer.
\newblock Intracellular transport of insulin granules is a subordinated random
  walk.
\newblock \emph{Proc. Natl. Acad. Sci. USA}, 110\penalty0 (13):\penalty0
  4911--4916, 2013.
\newblock \doi{10.1073/pnas.1221962110}.

\bibitem[Taylor(1953)]{taylor}
S.~J. Taylor.
\newblock The {Hausdorff} $\alpha$-dimentional measure of {Brownian} paths in
  $n$-space.
\newblock \emph{Proc. Camb. Philos. Soc.}, 49:\penalty0 31--39, 1953.
\newblock \doi{10.1017/S0305004100028000}.

\bibitem[Tj{\o}stheim(1986)]{doubleSts}
D.~Tj{\o}stheim.
\newblock Some doubly stochastic time series models.
\newblock \emph{J. Time Ser. Anal.}, 7\penalty0 (1):\penalty0 51--72, 1986.
\newblock \doi{10.1111/j.1467-9892.1986.tb00485.x}.

\bibitem[Uhlenbeck and Ornstein(1930)]{OUoryg}
G.~E. Uhlenbeck and L.~S. Ornstein.
\newblock On the theory of the {Brownian} motion.
\newblock \emph{Phys. Rev.}, 36:\penalty0 823--841, 1930.
\newblock \doi{10.1103/PhysRev.36.823}.

\bibitem[Vi\~nales and Desp\'osito(2006)]{vinales}
A.~D. Vi\~nales and M.~A. Desp\'osito.
\newblock Anomalous diffusion: Exact solution of the generalized {Langevin}
  equation for harmonically bounded particle.
\newblock \emph{Phys. Rev. E}, 73:\penalty0 016111, 2006.
\newblock \doi{10.1103/PhysRevE.73.016111}.

\bibitem[Walters(1982)]{ergWalters}
P.~Walters.
\newblock \emph{Introduction to Ergodic Theory}.
\newblock Springer-Verlag, 1982.

\bibitem[Wang et~al.(2012)Wang, Kuo, Bae, and Granick]{wangNG}
B.~Wang, J.~Kuo, S.~C. Bae, and S.~Granick.
\newblock When {Brownian} diffusion is not {Gaussian}.
\newblock \emph{Nat. Mat.}, 11\penalty0 (481), 2012.
\newblock \doi{10.1038/nmat3308}.

\bibitem[Wang(1992)]{KGWang92}
K.~G. Wang.
\newblock Long-time-correlation effects and biased anomalous diffusion.
\newblock \emph{Phys. Rev. A}, 45:\penalty0 833--837, 1992.
\newblock \doi{10.1103/PhysRevA.45.833}.

\bibitem[Wang and Tokuyama(1999)]{KGWang99}
K.~G. Wang and M.~Tokuyama.
\newblock Nonequilibrium statistical description of anomalous diffusion.
\newblock \emph{Physica A}, 265\penalty0 (3):\penalty0 341--351, 1999.
\newblock \doi{10.1016/S0378-4371(98)00644-X}.

\bibitem[Wang et~al.(1997)Wang, Yin, Landick, Gelles, and Block]{wang}
M.~Wang, H.~Yin, R.~Landick, J.~Gelles, and S.~Block.
\newblock Stretching {DNA} with optical tweezers.
\newblock \emph{Biophys. J.}, 72\penalty0 (3):\penalty0 1335--1346, 1997.
\newblock \doi{10.1016/S0006-3495(97)78780-0}.

\bibitem[Weiss et~al.(2003)Weiss, Hashimoto, and Nilsson]{weissAnDiff}
M.~Weiss, H.~Hashimoto, and T.~Nilsson.
\newblock Anomalous protein diffusion in living cells as seen by fluorescence
  correlation spectroscopy.
\newblock \emph{Biophys. J.}, 84:\penalty0 4043--4052, 2003.
\newblock \doi{10.1016/S0006-3495(03)75130-3}.

\bibitem[Weron and Kotulski(1996)]{weronSExp}
K.~Weron and M.~Kotulski.
\newblock {On the Cole-Cole relaxation function and related Mittag-Leffler
  distribution}.
\newblock \emph{Physica A}, 232\penalty0 (1):\penalty0 180--188, 1996.
\newblock \doi{10.1016/0378-4371(96)0020}.

\bibitem[Wiener(1923)]{wiener23}
N.~Wiener.
\newblock Differential-space.
\newblock \emph{J. Math. \& Phys.}, 2\penalty0 (1--4):\penalty0 131--174, 1923.
\newblock \doi{10.1002/sapm192321131}.

\bibitem[Williams and Watts(1970)]{wwSExp}
G.~Williams and D.~C. Watts.
\newblock Non-symmetrical dielectric relaxation behaviour arising from a simple
  empirical decay function.
\newblock \emph{Trans. Faraday Soc.}, 66:\penalty0 80--85, 1970.
\newblock \doi{10.1039/TF9706600080}.

\bibitem[Yaglom(1987)]{yaglom}
A.~Yaglom.
\newblock \emph{Correlation Theory of Stationary and Related Random Functions}.
\newblock Springer-Verlag, 1987.

\bibitem[Zwanzig(1973)]{zwanzigGLE}
R.~Zwanzig.
\newblock Nonlinear generalized {Langevin} equations.
\newblock \emph{J. Stat. Phys}, 9\penalty0 (3):\penalty0 215--220, 1973.
\newblock \doi{10.1007/BF01008729}.

\end{thebibliography}

\end{document}